# Quantum algorithms:

## A survey of applications and end-to-end complexities

Alexander M. Dalzell[*,1], Sam McArdle[*,1], Mario Berta[1,2,3] Przemysław Bienias[1]
Chi-Fang Chen[1,4] András Gilyén[5] Connor T. Hann[1] Michael J. Kastoryano[1,6]
Emil T. Khabiboulline[1,7,8,9] Aleksander Kubica[1,10] Grant Salton[1,11] Samson Wang[1,3]
and Fernando G. S. L. Brandão[1,4]

[1] *AWS Center for Quantum Computing, Pasadena, CA, USA*
[2] *Institute for Quantum Information, RWTH Aachen University, Aachen, Germany*
[3] *Imperial College London, London, UK*
[4] *Institute for Quantum Information and Matter, Caltech, Pasadena, CA, USA*
[5] *HUN-REN Alfréd Rényi Institute of Mathematics, Budapest, Hungary*
[6] *IT University of Copenhagen, Copenhagen, Denmark*
[7] *Department of Physics, Harvard University, Cambridge, MA, USA*
[8] *Joint Center for Quantum Information and Computer Science, NIST and University of Maryland
College Park, Maryland, USA*
[9] *Joint Quantum Institute, NIST and University of Maryland College Park, Maryland, USA*
[10] *Department of Applied Physics, Yale University, New Haven, CT, USA*
[11] *Amazon Quantum Solutions Lab, Seattle, WA, USA*

## Abstract

The anticipated applications of quantum computers span across science and industry, ranging from quantum chemistry and many-body physics to optimization, finance, and machine learning. Proposed quantum solutions in these areas typically combine multiple quantum algorithmic primitives into an overall quantum algorithm, which must then incorporate the methods of quantum error correction and fault tolerance to be implemented correctly on quantum hardware. As such, it can be difficult to assess how much a particular application benefits from quantum computing, as the various approaches are often sensitive to intricate technical details about the underlying primitives and their complexities. Here we present a survey of several potential application areas of quantum algorithms and their underlying algorithmic primitives, carefully considering technical caveats and subtleties. We outline the challenges and opportunities in each area in an "end-to-end" fashion by clearly defining the problem being solved alongside the input-output model, instantiating all "oracles," and spelling out all hidden costs. We also compare quantum solutions against state-of-the-art classical methods and complexity-theoretic limitations to evaluate possible quantum speedups.

The survey is written in a modular, wiki-like fashion to facilitate navigation of the content. Each primitive and application area is discussed in a standalone section, with its own bibliography of references and embedded hyperlinks that direct to other relevant sections. This structure mirrors that of complex quantum algorithms that involve several layers of abstraction, and it enables rapid evaluation of how end-to-end complexities are impacted when subroutines are altered.

---

*These authors contributed equally. Corresponding emails: dalzell@amazon.com, sammcard@amazon.com

# Contents

# Introduction

The first half of the twentieth century witnessed the foundation of three pillars of modern science: quantum mechanics, information theory, and computer science. In the latter half of the century, scientists began to connect these fields, first by exploring the implications of information itself being quantum—leading to the birth of quantum information theory. The nonclassical features of quantum information, such as no-cloning and entanglement, were identified as resources for novel applications, for instance, information theoretically secure communication. Alongside these developments were the observations of Benioff [1], Feynman [2], and Manin [3] that models of computation and simulation could be formulated within quantum mechanics—and that in some cases these models appeared exponentially challenging to simulate using a classical computer.

In 1985, Deutsch [4] further developed these early models of quantum computation and presented what was essentially the first quantum algorithm—a simple procedure that, with just one black-box query, could accomplish a task that classically requires two queries. Over the next decade, larger black-box separations were discovered, such as the Deutsch–Jozsa [5], Bernstein–Vazirani [6], and Simon's [7] algorithms, and finally, in 1994, the first truly *end-to-end* quantum algorithm was developed: Shor's algorithm [8] for factoring integers and computing discrete logarithms, bringing extensive ramifications for cryptography. This breakthrough demonstrated that quantum computers could not only speed up the solution of contrived black-box problems but, at least in theory, could provide faster solutions to important real-world problems. The discovery of Shor's algorithm transformed the field of quantum algorithms from a relatively niche topic into a major research area.

During the three decades since Shor's seminal discovery, the field of quantum algorithms matured significantly. For example, our knowledge of upper and lower bounds on the quantum query complexity of black-box problems—often deduced through sophisticated, nonconstructive mathematical arguments—has been greatly expanded. Moreover, many additional quantum algorithms and subroutines—for example, primitives for quantum simulation and linear algebra—have been discovered, optimized, and subsequently generalized multiple times. Meanwhile, advances in hardware and the theory of fault-tolerant quantum computation have reached the point where it is conceivable that (some of) these algorithms might soon become implementable at scales large enough to surpass what can be done classically.

Nevertheless, the magnitude of available quantum speedups for real-world applications is often hard to assess and can be obscured by technical caveats, assumptions, and limitations in the underlying quantum algorithmic primitives. Despite being one of the oldest, Shor's algorithm for factoring arguably remains the cleanest example of a substantial quantum speedup with minimal caveats that targets a problem of significant real-world relevance. This survey aims to elucidate the true resource requirements of end-to-end quantum computing applications, and thereby aid in identifying the most likely applications for fault-tolerant quantum computers.

Through this distinct perspective, the survey is intended to complement the wealth of existing quantum algorithms resources, including a number of review articles, lecture notes, textbooks, and the quantum algorithm zoo [9].

We highlight both the opportunities and challenges of currently known quantum algorithms. To truly understand the potential advantage of a quantum algorithm, it is necessary to consider its resource requirements in an end-to-end fashion. By this, we mean the cost of solving the full problem of interest to the user, not only the cost of running a given quantum circuit that is a subroutine of the full solution. One must consider all quantum and classical overheads: keeping track of classical precomputation and postprocessing, explicitly instantiating quantum oracles and data access structures, and ideally computing the constant prefactors of all quantum subroutines (including those overheads associated with fault-tolerant protocols and quantum error correction). We note, however, that this task is a major undertaking for complex quantum algorithms, and so has only been achieved for a minority of quantum algorithms in the literature. In addition to studying end-to-end quantum complexities, it is also necessary to compare any quantum results to the state-of-the-art classical solutions of the same problem, as well as known complexity-theoretic limitations.

We summarize the end-to-end complexities of several leading quantum application proposals (by which we mean quantum algorithms applied to a well-defined, real-world problem). The complexities of these applications are deduced from the complexities of their underlying primitives, which we review in detail. The modular structure of the survey aids the high-level understanding of the costs and tradeoffs coming from the various choices one makes when designing and compiling a quantum algorithm, as well as identifying the bottlenecks for a given application. On the technical front, this survey does not attempt to advance the state of the art; rather, it aims to collect, synthesize, and contextualize key results in the literature. We consider algorithms in the quantum circuit model, which is arguably the best-studied model for quantum computation and renders the presented complexities hardware agnostic (although the overhead associated with executing these circuits in a fault-tolerant fashion can, of course, depend on details of the hardware). In order to obtain concrete bounds, we require oracles to be explicitly instantiated. We generally assume that quantum error correction of some form will be necessary, due to unavoidable imperfections inherent to all known quantum hardware modalities. As such, we typically consider the non-Clifford cost of quantum algorithms as the dominant cost, in keeping with leading quantum fault-tolerance schemes. Due to the general lack of application-scale experimental data, we focus on elucidating provable speedups, and we only mention noisy, intermediate-scale quantum (NISQ) algorithms in passing, where appropriate, since they are typically heuristic.

Throughout this survey, we attempt to be thorough, but not exhaustive in presentation; we only aim to give a representative collection of references, rather than providing a complete list. Generally, we try to explain how asymptotic complexity statements arise from their underlying primitives, but technical results are typically presented without explicit derivation or proofs, for which we refer the reader to the cited references. Additionally, we often quote resource estimates from the literature without covering all of the application-specific optimizations to the underlying primitives that are required to arrive at the reported constant prefactors. We survey a number of quantum applications, primitives, and fault-tolerance schemes; however, the omission of other approaches does not indicate that they are unimportant. Also, the primary scope of this work excludes substantial topics, such as quantum sensing or communications, measurement-based quantum computing, adiabatic quantum computing and quantum annealing, analog quantum simulators, quantum-inspired ("dequantization") methods, and tensor network algorithms—comments on these topics are provided in instances where they are relevant to our primary discussion.

An overarching takeaway of this survey is that the current literature generally lacks fully end-to-end analyses for concrete quantum applications. Consequently, in several parts of this survey, a fully satisfactory end-to-end accounting is not achieved. In part, this is due to certain technical aspects of the relevant quantum algorithms being underexplored, and in some cases also due to a lack of specific details on how the output of the quantum algorithm will integrate into concrete computational workflows for future quantum computing users. Quantum algorithms research often works upward from algorithmic primitives to identify computational tasks with maximal quantum speedups, but these may not align with the tasks most relevant to the user. On the other hand, potential users themselves may not yet know exactly how they would use a new capability to advance their high-level goals. Yet, we find ourselves at a point in the history of quantum computing at which it behooves us to fill in these details and adopt this end-to-end lens. As more end-to-end applications are found, and with small fault-tolerant quantum computers now on the horizon, we expect the story to continue to evolve—this survey provides a snapshot of the state of play in roughly mid-2024. While improved quantum algorithms and approaches to quantum error correction and fault tolerance are likely to be discovered, classical computers continue to grow in scale and speed, and classical algorithms are also constantly refined and developed, thereby moving the goalposts for end-to-end quantum speedups. We hope the reader will find this survey a valuable guide for navigating this complex and dynamic landscape.

**How to use this survey**

This survey does not need to be read from cover to cover. Instead, it has a modular, wiki-like structure, which enables readers to directly explore the applications and primitives relevant for their use case. To the extent possible, each numbered section or subsection has been written in a self-contained fashion and can be read independently from the rest of the document. Rather than scrolling through the survey to locate a certain section, readers are encouraged to utilize the hyperlinks embedded throughout the document as well as those in the header of every page, which direct back to the tables of contents. To facilitate usage of the survey in this fashion, we include an independent bibliography for each numbered section and subsection of the document. A consolidated bibliography in alphabetical order appears at the end of the survey, along with back references to the pages in which each reference is cited.

Readers looking for a quick introduction to (or refresher on) the common notation, conventions, and background concepts that underlie the technical exposition in the main text are advised to begin with the Appendix, where we provide information on quantum mechanics, bra-ket notation, quantum circuits (and quantum computing more generally), big-$\mathcal{O}$ notation, and complexity theory. Readers may also find utility in the index, which organizes mentions of the important topics discussed in the book, including computational tasks and problems, quantum algorithmic tools and primitives used to solve those tasks, and finally competing classical methods for those tasks (in addition to other miscellaneous topics).

In April 2025, this version of the survey was published as a book with the same title by Cambridge University Press [10]. The book is available as an open access e-book and in hard copy form.

**Acknowledgments**

We thank Joao Basso, J. Kyle Brubaker, Christopher Chamberland, Andrew Childs, Isabel Franco Garrido, Helmut G. Katzgraber, Eric M. Kessler, Robin Kothari, Péter Kutas, Yi-Kai Liu, Pavel Lougovski, Carl Miller, Oskar Painter, Nicola Pancotti, Simone Severini, Sophia Simon, James D. Whitfield, and Xiaodi Wu for helpful comments and conversations on various aspects of this survey.

After the first version of this survey was released online in October 2023, we conducted a self-managed, nonanonymized peer-review process. We are grateful to colleagues who agreed to help us by reviewing a subset of the survey. At the beginning of each section of the survey, we acknowledge those who reviewed that section as part of this process, and we also list them alphabetically here: Dong An, Eric Anschuetz, Ryan Babbush, Matthew Campagna, Earl Campbell, Marco Cerezo, Andrew Cross, Zohreh Davoudi, Vedran Dunjko, Glen Evenbly, Di Fang, Joshua Goings, Johnnie Gray, Sander Gribling, Thomas Häner, Matthew Hastings, Samuel Jaques, Robin Kothari, Richard Kueng, Lin Lin, Daniel Malz, Ashley Milsted, Ashley Montanaro, John Preskill, Patrick Rall, Patrick Rebentrost, Rolando Somma, Nikitas Stamatopoulos, Damian Steiger, Yuan Su, Ewin Tang, Ronald de Wolf, Nobuyuki Yoshioka, and Xiao Yuan.

Finally, we acknowledge the staff from across the AWS Center for Quantum Computing that enabled this project, with special thanks to Mike Sadowitz and Wendy Yu for their legal support. We are also grateful to the Institute for Quantum Information and Matter at Caltech, which is an NSF Physics Frontier Center. We thank Harry Atwater, Fiona Harrison, Tom Rosenbaum, and David Tirrell at Caltech, and Nafea Bshara, Peter DeSantis, James Hamilton, Andy Jassy, Simone Severini, and Bill Vass at AWS, for their involvement and support of the research activities at the AWS Center for Quantum Computing.

# Areas of application

To provide benefit, quantum computers must solve computational problems where the solutions are simultaneously valuable to the user and also difficult to obtain classically. Simply developing a quantum algorithm with a theoretical quantum speedup is not sufficient to meet these criteria; we must directly compare the performance of classical and quantum algorithms for concrete problems of interest.

In this part, we survey a number of specific computational problems where quantum algorithms have been proposed, organized by application area. We present an overview of these algorithms through an end-to-end lens, noting clearly the actual end-to-end problem that is being solved and the dominant resource cost/complexity (derived from the algorithmic primitives that are being used), and emphasizing noteworthy caveats. We list known resource estimates for implementing these algorithms on fault-tolerant quantum computers (we also comment in passing on NISQ implementations), and we compare to classical complexities for the same problem, both in a practical and asymptotic sense. The list of applications presented is not exhaustive, but represents a broad spectrum of the most well-studied applications proposed in the literature.

**This part contains:**

# 1 Condensed matter physics

Condensed matter physics constructs and studies the behavior of simplified models designed to capture the universal physics of material systems. Phenomena of interest include magnetism, phase transitions, superconductivity, frustrated systems, topological phases, and the interplay of thermalization and many-body localization in closed systems. While many seminal models can be studied analytically in certain limits (e.g., the 1D and 2D classical Ising model), a number of seemingly innocuous models have proven exceedingly difficult to solve. This has led to some models, such as the Fermi–Hubbard model, becoming a proving ground for classical numerical methods. While there has been significant progress in recent decades in understanding the physics of these models through numerical simulation, it is still a challenging problem for many models and parameter regimes. As observed by Feynman [1], quantum computers have a natural advantage over their classical counterparts for simulating the simple Hamiltonians studied in condensed matter physics. While Feynman's proposal was more focused on analog simulation, digital quantum simulation of condensed matter systems has evolved into a major research direction. In this section, we focus on models whose end-to-end complexities have been well studied in the literature: the Fermi–Hubbard model, spin models, and the Sachdev–Ye–Kitaev (SYK) model.

*The authors are grateful to Ashley Montanaro and Nobuyuki Yoshioka for reviewing this section of the survey.*

**This application area contains:**

## 1.1 Fermi–Hubbard model

**Overview**

The Fermi–Hubbard model was originally introduced as a simplified model of electrons in materials [1], closely related to the tight-binding model. It displays a wide range of behaviors including metallic, insulating, and antiferromagnetic phases. The model has more recently found applicability in studying high-temperature superconductivity. The 2D Fermi–Hubbard model has a complex phase diagram that appears to reproduce universal (rather than chemical-specific) features of the phase diagram of cuprate high-temperature superconductors.

General analytic solutions are not known beyond 1D chains or specific parameter regimes—see [2] for a discussion—which has motivated the use of numerical methods to understand the physics of the Fermi–Hubbard model. More recently, there has been increased interest in understanding the nonequilibrium properties of the model such as its behavior following a quench.

Based on the current estimates, quantum simulation of Fermi–Hubbard models requires considerably fewer resources than simulations of molecules or solving optimization problems. This makes the Fermi–Hubbard model a promising candidate for early demonstrations of quantum advantage.

A detailed case study on the Fermi–Hubbard model is presented in [3], including descriptions of the parameters to probe open scientific questions and estimates of the utility of these computational capabilities.

**Actual end-to-end problem(s) solved**

The Fermi–Hubbard Hamiltonian on $M/2$ lattice sites is given by

$$H = -t \sum_{\sigma \in \{\uparrow,\downarrow\}} \sum_{\langle i,j \rangle} (c_{i\sigma}^\dagger c_{j\sigma} + c_{j\sigma}^\dagger c_{i\sigma}) + U \sum_i n_{i\uparrow} n_{i\downarrow}, \tag{1}$$

where $c_{i\sigma}$ are fermionic operators and $n_{i\sigma} \equiv c_{i\sigma}^\dagger c_{i\sigma}$ is the number operator, with $t$ denoting the strength of the kinetic term, $U$ the onsite interaction strength, and $\langle i,j \rangle$ a sum over nearest-neighbor lattice sites, given a lattice geometry. It is also possible to consider longer-range hopping terms, the inclusion of site-dependent chemical potentials, or additional "orbitals" per site.

Quantum simulation provides insights into both equilibrium and nonequilibrium physics. With regards to equilibrium physics, the primary computational task is to resolve and probe the properties of the phase diagram of the Fermi–Hubbard model, as a function of lattice geometry, parameter values $(t, U)$, doping (the expected number of fermions divided by the number of sites), and temperature. This is achieved by preparing the thermal state $\rho \propto \mathrm{e}^{-\beta H}$ (with $\beta = 1/k_B T$, where $k_B$ is the Boltzmann constant and $T$ the temperature) or at zero temperature the ground state $|E_0\rangle$ for the Fermi–Hubbard Hamiltonian instantiated by the given parameters, and measuring the expectation values of a set of physical observables to error $\epsilon$. A thorough discussion of this end-to-end problem (at zero temperature) is provided in [4], where it is shown how to perform the following steps:

- Prepare mean-field states in a given phase, for example, a BCS superconducting ground state.

- Adiabatically evolve from the mean-field Hamiltonian to the final Fermi–Hubbard Hamiltonian. The absence of a phase transition confirms the predicted phase.

- Measure observables, including density correlation functions $(n_{i\uparrow} + n_{i\downarrow})(n_{j\uparrow} + n_{j\downarrow})$, pair correlation functions $c_{i\sigma}^\dagger c_{j\sigma'}^\dagger c_{k\sigma'} c_{l\sigma}$, and dynamical correlation functions $\langle E_0 | e^{iHt} A e^{-iHt} B | E_0 \rangle$ (for operators $A, B$ and ground state $|E_0\rangle$).

The difficulty of this problem depends on the parameter regime under consideration. The ground state in the weak coupling regime of $U < 4t$ is well understood, but questions remain in the intermediate ($4t \leq U \leq 6t$) and strong ($U > 6t$) regimes [5]. Challenges include precisely determining the phase boundaries and understanding the nature of the superconducting phase [6]. Progress has been made on this latter question in recent years, for example, by showing the absence of a superconducting phase at the physically relevant parameters of $U \sim 8t$ and 1/8th doping (see [5] for a more detailed discussion). Calculations are made challenging by small energy differences between competing phases, as well as the need to extrapolate from finite simulations to the thermodynamic limit.

The simulation of nonequilibrium quantum dynamics is of interest for modeling materials driven by an external field (e.g., an ultrafast laser pulse or an applied voltage), or following a quench in the Hamiltonian. Classically simulating nonequilibrium quantum dynamics has so far proven challenging and is a less well-studied problem than probing the equilibrium physics of the model. Example applications include modeling ultrafast spintronics (whereby lasers are used to manipulate spin degrees of freedom to control and store information) [7], understanding photo-induced phase transitions [8], and clarifying the nature of thermalization in isolated quantum systems following a quench [9].

**Dominant resource cost/complexity**

**Mapping the problem to qubits:** Simulation of the Fermi–Hubbard model is most naturally performed in the second-quantized representation, as the regime of interest is usually close to half-filling (for comparison, we refer to Section 2 on simulating molecules). The Jordan–Wigner mapping between fermions and qubits is typically used. Locality-preserving mappings have also been developed, which map fermionic operators to qubit operators acting on a constant number of qubits [10, 11]. For an $L \times L$ lattice, we require $M = 2L^2$ qubits to simulate the spinful Fermi–Hubbard model using the Jordan–Wigner mapping.

**Accessing the Hamiltonian:** Quantum algorithms for simulating the Fermi–Hubbard model require access to the Hamiltonian. This is typically provided by block-encoding or Hamiltonian simulation.[1] The structure in the Fermi–Hubbard Hamiltonian reduces the costs of these subroutines. For example, performing a block-encoding using the linear combinations of unitaries (LCU) technique requires access to a PREPARE unitary and a SELECT unitary (we refer to Section 10.2 for definitions). The PREPARE unitary requires preparing a quantum state from classical data. Because the Fermi–Hubbard Hamiltonian has a small number of unique coefficients, the cost of this unitary can be reduced. Combining the results of [12, 13, 14], one can implement an $(M(2t + U/8), \mathcal{O}(\log(M)), \epsilon)$-block-encoding (see Eq. (43) for definition) of the

---

[1]Hamiltonian simulation is used to explicitly simulate dynamics but can also be used implicitly to provide access to the Hamiltonian for use in static calculations, for example, in quantum phase estimation.

Fermi–Hubbard Hamiltonian using

$$\mathcal{O}(M + \log(M/\epsilon))$$

non-Clifford gates.

As another example, the costs of product formula approaches for Hamiltonian simulation can exploit the fact that many terms in the Fermi–Hubbard Hamiltonian commute, due to their locality. We will explicitly discuss these costs below.

**State preparation:**

- Classical trial states: Approximate eigenstates obtained from a classical calculation can be prepared as quantum trial states; examples include Slater determinant states [4], linear combinations of $D$ Slater determinants (with complexity $\widetilde{\mathcal{O}}(D)$ [15]–$\mathcal{O}(MD)$ [16]), and matrix product states with bond dimension $\chi$ (with complexity $\mathcal{O}(M\chi^2)$ [15]).

- Quantum trial states: Parameterized quantum circuits, in conjunction with variational quantum algorithms, have been proposed for preparing approximate energy eigenstates (see §NISQ implementations, below). Like classical trial states, the states prepared by these circuits can be used as inputs to other quantum algorithms that further refine the initial state, such as eigenstate filtering, or quantum phase estimation. Initial resource estimates for the Fermi–Hubbard model can be found in [17].

- Eigenstate preparation: There exist quantum algorithms that can prepare energy eigenstates using QSVT-based eigenstate filtering [18], where the cost scales as $1/\gamma$ with $\gamma$ the overlap of the initial state with the desired eigenstate. Alternatively, adiabatic state preparation can be used, with a cost that depends on the gap between energy levels along the adiabatic path. Adiabatic state preparation was proposed as a method of classifying the phase diagram of the Fermi–Hubbard model [4]. A discrete version of the adiabatic approach based on qubitization and quantum phase estimation (QPE) was numerically investigated in the context of preparing ground states of the Fermi–Hubbard model [19], and showed promising results for the small system sizes considered (see also [16]).

- Thermal states: A number of algorithms have been developed for preparation of thermal states, also known as Gibbs states. The most promising variants of these "Gibbs sampling" algorithms depend on the mixing time of a Markov chain (similar to classical Monte Carlo approaches for preparing Gibbs states), which is currently undetermined for the Fermi–Hubbard model.

**Time evolution:**

- As discussed above, Trotter approaches for Hamiltonian simulation can exploit beneficial features of the Fermi–Hubbard Hamiltonian, such as locality, fixed particle number, and commutativity of the terms [20, 21, 22]. For a Fermi–Hubbard model with $\eta$ fermions on $M$ lattice sites, $p$th-order Trotter methods can simulate time evolution for time $\tau$ up to error $\epsilon$ using

$$\mathcal{O}\left(\frac{5^p M \eta^{1/p} \tau^{1+1/p}}{\epsilon^{1/p}}\right)$$

gates. Explicit gate counts for Trotterization can be obtained from [23, 21, 14, 24], which have focused on constant prefactors for low-order product formulas, rather than the asymptotic scaling.

Post-Trotter methods, such as [25], using quantum signal processing as a building block, can achieve similar scaling in $M$ and $\tau$. A suboptimal approach (i.e., not using the method of [25]), briefly discussed in [26], has a gate complexity of approximately

$$44M^2(2t + 3U/8)\tau$$

$T$ gates to simulate time evolution for time $\tau$ using quantum signal processing, neglecting logarithmic dependence on the error of the simulation.[2]

**Measuring observables:**

- Energies: QPE can be used to measure the energy eigenvalues of the Fermi–Hubbard Hamiltonian. Given access to (i) an initial state $|\psi\rangle$ that has sufficient overlap $\gamma = |\langle\psi|E_j\rangle|$ with the target eigenstate $|E_j\rangle$ and (ii) a unitary $U = f(H)$ that encodes the eigenspectrum of the Hamiltonian with a known, classically invertible relationship $f$, we can use QPE to project into the desired eigenstate and provide an estimate of the eigenphase $\phi_i$ of $U$—which can then be converted into an estimate of the eigenenergy of $H$ using $\phi_i = f(E_i)$. QPE makes

$$\mathcal{O}\big(\gamma^{-2}\|f'(H)\|^{-1}\epsilon^{-1}\log(\theta^{-1})\big)$$

  calls to the unitary $U$ encoding the spectrum of the Hamiltonian, where $\theta$ is the failure probability, and $\epsilon$ is the desired precision in the eigenenergy of $H$.[3]

  A common choice for the unitary encoding the Hamiltonian is $U \approx e^{iHt}$ approximated via quantum algorithms for Hamiltonian simulation, where the approximation error must be balanced against the error from QPE. Using $U \approx e^{iHt}$ implemented via a second-order product formula results in a $T$ gate count of $\mathcal{O}(M^{3/2}/\Delta E^{3/2})$ to resolve the energy of the Fermi–Hubbard model to precision $\Delta E$, neglecting the cost of initial state preparation and the dependence on the overlap and failure probability [23, 14]. Another common choice is to perform QPE on a quantum walk operator $W(H)$ which acts like $e^{i\arccos(H/\alpha)}$, where $\alpha$ is the normalization of the block-encoding of $H$. The operator $W(H)$ can be implemented exactly via qubitization [28, 27]. This results in a $T$ gate scaling of $\mathcal{O}(M^2/\Delta E)$, also neglecting the cost of initial state preparation and the dependence on the overlap and failure probability [12].

- Other observables: There have been few studies considering the costs of measuring observables other than the ground state energy using fault-tolerant quantum algorithms. In general, it is important to minimize the number of calls to the unitary $U_\psi$ that prepares the desired state, as this is typically considered the dominant cost. Reference [4] discussed methods for measuring density correlation functions $(n_{i\uparrow} + n_{i\downarrow})(n_{j\uparrow} + n_{j\downarrow})$, pair correlation functions $c_{i\sigma}^\dagger c_{j\sigma'}^\dagger c_{k\sigma'} c_{l\sigma}$, and dynamical correlation functions $\langle E_0|e^{iHt}Ae^{-iHt}B|E_0\rangle$ (for operators $A, B$ and ground state $|E_0\rangle$), including approaches for nondestructively measuring some of these observables. Some of these approaches can now be reframed as performing amplitude estimation [29] on $U_O$, a unitary block-encoding of the observable $O$ with subnormalization factor $\alpha_O$ [30]. The measurement of similar observables using these modern algorithmic tools was studied in [31].

---

[2]Note that in [26], $M$ is defined as the number of lattice sites, and so corresponds to $M/2$ here.

[3]It is possible to improve the complexity to $\mathcal{O}(\gamma^{-1}\|f'(H)\|^{-1}\epsilon^{-1}\log(\theta^{-1}))$ using amplitude amplification if a sufficiently precise estimate of the eigenvalue is known, or to $\mathcal{O}((\gamma^{-2}\Delta^{-1} + \epsilon^{-1})\|f'(H)\|^{-1}\log(\theta^{-1}))$ by exploiting knowledge of the gap $\Delta$ between the energy eigenstates to perform rejection sampling [27].

A recent approach [32, 33] based on the quantum gradient estimation algorithm of [34] simultaneously computes the value of $K$ (noncommuting) observables $O_j$. The algorithm makes $\widetilde{\mathcal{O}}(K^{1/2}/\epsilon)$ calls to $U_\psi$ and $U_\psi^\dagger$ (or $R_\psi = I - 2|\psi\rangle\langle\psi|$) and either $\widetilde{\mathcal{O}}(K^{3/2}/\epsilon)$ calls to gates of the form $\mathrm{e}^{\mathrm{i}x O_j}$ [32] or $\widetilde{\mathcal{O}}(K/\epsilon)$ calls to a block-encoding of the observables [33]. The algorithm also requires $\mathcal{O}(K\log(1/\epsilon))$ additional qubits. This approach has been considered in the context of measuring fermionic reduced density matrices and dynamic correlation functions [32].

**Existing resource estimates**

There have been a number of logical resource estimates for algorithms targeting both static and dynamic properties of the Fermi–Hubbard model. In Table 1, we present approximate resource estimates for simulations of the 2D $10 \times 10$ spinful Fermi–Hubbard model. The table presents the number of logical qubits and gates required to run the algorithm; these can be converted into physical resource estimates via methods for fault-tolerant quantum computation.

References [12, 13] applied qubitization-based QPE to calculate the ground state energy to constant additive error. For a lattice with $M$ spin sites, using the compilation of [12], the number of $T$ gates scales as approximately (neglecting the dependence on the overlap and failure probability) [12, Eq. (61)]

$$\#T \propto \frac{(4t+U)M^2}{\Delta E},$$

and the number of logical qubits scales as approximately [12, Eq. (62)]

$$\#\mathrm{Qubits} \sim M + \log\left(\frac{(2t+0.5U)M^4}{\Delta E}\right).$$

References [23, 14] applied second-order Trotter-based QPE to calculate the ground state energy. In both references, rigorous but potentially loose upper bounds on the Trotter error are computed. For a lattice with $M$ spin orbitals, using the compilation of [14], the number of $T$ gates scales as approximately (neglecting the dependence on the overlap and failure probability) [14, Eqs. (C3), (D6), (D10), (E17), (F10)]

$$\#T \propto t\sqrt{t+U}\left(\frac{M}{\Delta E}\right)^{3/2},$$

and the number of logical qubits scales as approximately [14, Table II]

$$\#\mathrm{Qubits} \sim (1+\kappa)M,$$

where $\kappa$ is a free parameter that controls the number of ancilla qubits used for a compilation technique known as Hamming weight phasing (which reduces the cost of applying identical arbitrary angle rotation gates in parallel) [35, 23], set to $\kappa = 0.25$ in [14] and in our Table 1.

The methods described above for encoding the Hamiltonian spectra (qubitization and Trotter) can also be used to simulate the dynamics of the Fermi–Hubbard model. Trotter methods can be applied directly, while qubitization can be combined with quantum signal processing (QSP) to perform Hamiltonian simulation. In [26], a comparison was made between fourth-order Trotterization and qubitization+QSP for simulating time evolution of a $10 \times 10$ Fermi–Hubbard model. Trotter was determined to be the more efficient method, although this conclusion hinges on a Trotter decomposition with large steps (justified via numerical simulations). We note that the Trotter decompositions and analyses in [14, 26] are different, which hampers an immediate comparison. It may also be fruitful to compare with Hamiltonian simulation algorithms designed explicitly for simulating local Hamiltonians [25] (see discussion in [12]).

| Problem and method | # $T$ gates | # Logical qubits | Parameters |
|---|---|---|---|
| Ground state energy via qubitized QPE [12, 13] | $\sim 10^8$ | $\sim 236$ | $U/t = 4$ and $\Delta E = 0.01t$ |
| Ground state energy via Trotterized QPE [14, 23] | $\sim 5 \times 10^6$ | $\sim 250$ | $U/t = 8$ and $\Delta E = 0.005E_{\mathrm{tot}}$ |
| Dynamics via fourth-order Trotter [26] | $4.6 \times 10^5$ | $200$ | $T = 10/t$, $U = t$, and $\epsilon \leq 1\%$ |

Table 1: Logical resource estimates for quantum phase estimation (QPE) and dynamics simulation applied to a 2D $10 \times 10$ Fermi–Hubbard model. The QPE circuits target an energy error of $\Delta E$. In the second row, $E_{\mathrm{tot}}$ denotes the ground state energy. The dynamics simulation runs for time $T$, and targets an error of less than 1% in a spatially averaged intensive observable, with Trotter errors bounded numerically via extrapolated small-scale simulations. The presented gate counts are for a single run of the circuit. For QPE, the number of required runs depends on the overlap between the initial state and the ground state (inverse polynomial dependence), as well as the desired failure probability of QPE. For dynamics simulations, the number of circuit repetitions depends on the precision to which one wants to estimate a given observable. The parameters for each problem vary between different rows of the table, and so cannot be directly compared (although the different methods for the same problem, e.g., ground state energy estimation, could be compared by adjusting the analyses in the original papers to the desired matching parameter values).

### Caveats

In general, preparing the ground state of the Fermi–Hubbard model is known to be a hard problem, even for a quantum computer. This task has been proven QMA-hard for the Fermi–Hubbard model with a site-dependent magnetic field [36] and for the Fermi–Hubbard model with a site-dependent kinetic term strength (i.e., $t \to t_{ij}$ in Eq. (1)) [37]. While the complexity class of the canonical Fermi–Hubbard model is not yet known, when preparing the ground state via QPE or eigenstate filtering methods, it is necessary to prepare an initial state with an overlap that decays no worse than polynomially with system size; otherwise, the overall complexity will be superpolynomial. While numerical simulations on small system sizes have shown encouraging results [16, 19], it is still an open question as to whether this property holds for sufficiently large system sizes to enable extrapolation to the thermodynamic limit.

It is also important to note that this extrapolation of measured properties, computed at a range of finite system sizes, to the thermodynamic limit, has been observed to contribute a significant proportion of the uncertainty and errors in classical methods [38], and will also afflict quantum simulations.

Finally, it will be necessary to repeat simulations a large number of times. In order to measure a single observable to precision $\epsilon$ we require $\mathcal{O}(1/\epsilon^2)$ incoherent repetitions of the simulation, or $\mathcal{O}(1/\epsilon)$ using methods based on amplitude estimation. To map out and compute properties of the phase diagram or extract the phase following a quench, we may need to measure a large number of observables. In some cases, it may be necessary to re-prepare the initial state for each observable.

### Comparable classical complexity and challenging instance sizes

The Fermi–Hubbard model has been a fertile environment for the development and testing of classical numerical methods for both static and dynamical properties. To the best of our knowledge, the largest exact diagonalization calculations performed to date are on systems with 17 fermions in 22 sites (44 spin sites), requiring 7.1 terabytes of memory [39]. State-of-the-art approximate methods for computing the phase diagram include quantum Monte Carlo methods (variational QMC, determinantal QMC, diagrammatic MC, auxiliary-field QMC, diffusion MC), density matrix renormalization group (DMRG), coupled cluster methods, and impurity methods (dynamical mean-field theory, density matrix embedding theory), among others. These methods typically have an approximation parameter (e.g., the excitation degree in coupled cluster or the bond dimension in DMRG) which influences the scaling of the algorithm and the accuracy of the simulation. Modern numerical studies of the Fermi–Hubbard model typically cross-validate using a number of simulation methods [38, 40]. For example, [38] benchmarked a range of methods and performed sufficiently large and accurate simulations for extrapolation to the thermodynamic limit. That work concluded that "the ground-state properties of a substantial part of the Hubbard model phase space are now under numerical control," but that some uncertainties still remain for $4t \leq U \leq 8t$ and dopings near half-filling. For a recent review of numerical simulations of the Fermi–Hubbard model, we refer the reader to [5]. We also refer to [41], which benchmarked a number of variational classical methods on a range of condensed matter systems, including the Fermi–Hubbard model.

The simulation of *dynamics* of the Fermi–Hubbard model appears to be more challenging for classical methods. For example, [42, 26] concluded that simulating the dynamics of a $10 \times 10$ lattice would be infeasible for tensor network techniques. Other classical approaches for simulating time evolution of the Fermi–Hubbard model include nonequilibrium extensions of dynamical mean-field theory [43] or Floquet methods [8].

### Speedup

The speedup of quantum algorithms for computing static properties of the Fermi–Hubbard model, such as its ground state energy, is difficult to determine. In general, we know that closely related models are QMA-hard (see §Caveats, above) and so should be exponentially difficult for both classical and quantum computers. Assuming an initial state that has overlap with the target eigenstate that decays no faster than polynomially, then QPE can be used to efficiently measure the eigenenergy and project into the desired eigenstate. It does so with cost $\text{poly}(M/\Delta E)$, where the precise scaling depends on the quantum algorithm used. Exact classical methods such as exact diagonalization have a cost that scales exponentially with $M$ or $1/\Delta E$. Approximate classical methods scale with an approximation parameter (e.g., bond dimension, number of excitations) which will depend on both $M$ and $\Delta E$. For example, [44, Fig. 4] shows the convergence of a tensor network calculation for the 2D Fermi–Hubbard model as a function of bond dimension and system size. For the small systems studied (up to $16 \times 4$ sites) the plots are consistent with the bond dimension scaling polynomially in $1/\Delta E$, with a weak dependence on the system size. If this holds for larger system sizes and across a range of system parameters, this would suggest that quantum algorithms provide only a polynomial speedup for computing the ground state energy.

Simulating the dynamics of the Fermi–Hubbard Hamiltonian requires polynomial resources using quantum algorithms, scaling almost linearly both in $M$ and in the evolution time $\tau$. By using carefully engineered interactions (e.g., deviating significantly from a square lattice) it can be shown that simulating the dynamics of the Fermi–Hubbard model on a planar graph is a BQP-complete problem and so is expected to be hard for classical computers in the worst case [45]. Supporting this observation, all known classical methods appear to scale exponentially in system size and simulation accuracy. For example, [26] used tensor network (matrix product state) approaches for simulating the dynamics of the Fermi–Hubbard model following a quench. When truncating the bond dimension to facilitate efficient classical simulation, they found that errors in the observables grew exponentially with time. While this supports the conclusion of an exponential quantum speedup, we note that classical approaches will likely continue to improve and be applied to increasingly large system sizes.

### NISQ implementations

There have been a number of proposals (and experimental demonstrations) for simulating the Fermi–Hubbard model on NISQ hardware. Ground state calculations can be performed using the variational quantum eigensolver (VQE) [46, 47, 48, 49, 17], and experimental demonstrations have been carried out on lattices of size $1 \times 8$ and $2 \times 4$ using 16 superconducting qubits, yielding qualitative agreement with theoretical expectation [50].

Dynamics can be simulated using Hamiltonian simulation (typically Trotter methods) [21] and have been demonstrated for an $8 \times 1$ lattice on 16 superconducting qubits [51].

The simple Hamiltonian of the Fermi–Hubbard model makes it well suited to realization in analog quantum simulators, including ultracold atoms in optical lattices, trapped ions, and neutral atom arrays. It has been argued that some local observables can be robust to errors in the simulation [52, 26], enabling analog simulations to already surpass classical methods for simulating dynamics. Nevertheless, it can be challenging to cool the analog fermionic system to its ground state. We refer the reader to [42, 53] for additional discussion on analog simulation.

Reference [21] considered an approach with flavors of both digital and analog simulation, moving to a cost model based on evolution time, rather than number of gates. This reduces the cost of simulating Trotterized dynamics of the Fermi–Hubbard model, compared to purely gate-based approaches.

### Outlook

The Fermi–Hubbard model provides a long-standing and physically relevant computational challenge. The low gate counts and modest number of logical qubits required to compute ground state energies could make quantum algorithms competitive with leading classical approaches in challenging regimes. We note that further research is required to ascertain the costs for initial state preparation for these calculations. For the less well-studied task of simulating the dynamics of the Fermi–Hubbard model, quantum algorithms currently provide an exponential speedup over known classical algorithms. Nevertheless, as the Fermi–Hubbard Hamiltonian is sufficiently simple to be realized in many controlled physical systems, future fault-tolerant quantum computers will also have to compete against analog quantum simulators.

## 1.2 Spin models

**Overview**

Classical and quantum spin systems are prototypical models for a wide range of physical phenomena including magnetism, neuron activity, simplified models of materials and molecules, and networks. Studying the properties of spin Hamiltonians can also provide useful insights in quantum information science.

A number of scientific and industrial problems can be mapped onto finding the ground or thermal states of classical or quantum spin models, for example, solving combinatorial optimization problems, training energy-based models in machine learning, and simulating low-energy models of quantum chemistry [1].

Simulating the *dynamics* of quantum spin models is primarily of interest for quantum information science, as well as condensed matter physics or chemistry. For instance, such simulations are relevant for interpreting nuclear magnetic resonance (NMR) [2, 3] or related spectroscopy experiments [4, 5].

Because of the natural mapping between spin-1/2 systems and qubits, as well as the fact that interactions in spin models are typically local, the resources required to simulate simple spin models using quantum algorithms can be much lower than for problems in areas like quantum chemistry or cryptography.

While our discussion will focus on quantum algorithms designed to be run on fault-tolerant quantum computers, the simple Hamiltonians of spin models are naturally realized in many physical systems. This has led to the use of analog simulators [6, 7], such as arrays of trapped ions or neutral atoms, for simulating the static and dynamic properties of interesting spin models. We will comment briefly on this below.

**Actual end-to-end problem(s) solved**

The most commonly studied spin models are those with pairwise interactions, referred to as 2-local Hamiltonians. We note that the interactions are not necessarily geometrically local, although this will be present in many models of physical systems. Given a graph $\mathcal{G}$ with $N$ vertices $\{i\}$ and $L$ edges $\{[i,j]\}$, we associate a classical or quantum spin with each vertex, and an interaction between spins with each edge. We can also add one-body interactions acting on individual spins. The Hamiltonian can then be written as

$$H = \sum_i \sum_{\alpha \in \{x,y,z\}} B_i^\alpha \sigma_\alpha^i + \sum_{[i,j]} \sum_{\alpha,\beta \in \{x,y,z\}} J_{ij}^{\alpha\beta} \sigma_\alpha^i \sigma_\beta^j \,, \tag{2}$$

where $\{\sigma_x^i, \sigma_y^i, \sigma_z^i\}$ denote the Pauli operators $X_i, Y_i, Z_i$ acting on site $i$, and $\{B_i^\alpha\}, \{J_{ij}^{\alpha\beta}\}$ are coefficients. For classical spin Hamiltonians, the sums are restricted to $Z$ operators. The Hamiltonian in Eq. (2) encompasses a wide range of spin models, including the following:

- The classical Ising model

$$H = \sum_i B_i Z_i + \sum_{ij} J_{ij} Z_i Z_j \,, \tag{3}$$

  which also describes the Hamiltonians arising from quadratic unconstrained binary optimization (QUBO) problems.

- The (quantum) transverse-field Ising model (TFIM)

$$H = B \sum_i X_i + J \sum_{ij} Z_i Z_j. \tag{4}$$

- The Heisenberg model with a site-dependent magnetic field, defined in 1D with nearest-neighbor interactions by

$$H = \sum_j \left( B_j Z_j + J^x X_j X_{j+1} + J^y Y_j Y_{j+1} + J^z Z_j Z_{j+1} \right). \tag{5}$$

Across the different models, we can vary the dimension, locality of interactions (e.g., nearest-neighbor vs. fully connected vs. power-law), and values of the site-dependent coefficients in comparison to the interaction terms. The models can be extended beyond 2-local by considering couplings of 3 or more spins—see, for example, $p$-spin models, which are $p$-local [8]. The above definitions can be extended from spin-1/2 systems to higher-dimensional spin operators by generalizing the Pauli operators with their higher-dimensional counterparts.

For classical spin models, we seek to prepare the ground or thermal states of the model, as these may encode, for example, the solution to a combinatorial optimization problem, or a probability distribution that can be used for generative modeling. For quantum spin models, we similarly seek to compute ground or thermal states. However, because these are not classical states that can be easily extracted, we typically wish to sample observables with respect to these states. Examples include the energy, the magnetization of the system, and correlations between sites. In dynamics simulations of quantum systems, we seek to determine how observables of interest vary as a function of evolution time. Examples include the magnetization (used to infer the Hamiltonian in nuclear magnetic resonance [9] or related [10] experiments), or the growth of correlations between sites to probe thermalization. Digital Hamiltonian simulation can additionally extract certain quantities that may be hard to directly measure in experiments (e.g., time reversal) [11]. Since studies of quench dynamics often require preparation of simple states, such as product states or the ground states of classically solvable Hamiltonians, and the measurement of local observables, propagation under the Hamiltonian typically dominates the simulation cost. For lattice systems with $N$ spins in $D$ spatial dimensions, it is conventional to consider evolution times that scale as $\Omega(N^{1/D})$, as the system must evolve for at least this long in order for information to propagate across the system due to the Lieb–Robinson bound [12].

Reference [13] details a number of applications of simulating quantum spin models, with relevance to systems studied at Los Alamos National Laboratory, including the parameters of end-to-end simulations. Reference [14] presents an end-to-end assessment of using quantum computers to simulate quantum spin Hamiltonians for applications in NMR spectroscopy.

**Dominant resource cost/complexity**

For a system of $N$ spin-$\frac{1}{2}$ particles, we require $N$ qubits to represent the state of the system. For $N$ spin-$S$ particles, the problem can be mapped to qubits in different ways, for example, using $N\lceil \log_2(2S+1) \rceil$ qubits [15] or using $2NS$ qubits [5].

Quantum algorithms for preparing the ground or Gibbs states of classical spin systems are discussed in detail in Section 4 (combinatorial optimization), and Section 9.2 (energy-based machine learning models), respectively. We will restrict our discussion to the resources required for performing time evolution of quantum spin models. The reason for this is that quantum algorithms for preparing ground or thermal states require similar primitives for Hamiltonian access to algorithms for time evolution (e.g., block-encodings or Hamiltonian simulation itself) and use these in conjunction with either (i) eigenstate filtering approaches [16, 17] based on quantum singular value transformation, (ii) adiabatic state preparation, (iii) quantum phase estimation from a trial state, or (iv) quantum algorithms for thermal state preparation, that is, Gibbs sampling. More detailed discussions of these algorithms and their caveats can be found in the corresponding sections, as well as in the discussion of quantum algorithms for simulating molecules and materials (Section 2.1) or the Fermi–Hubbard model (Section 1.1), where preparing (approximate) eigenstates is the primary topic of interest. All of these algorithms depend on either an overlap between the trial state and the target state, the minimum gap along an adiabatic path, or the mixing time of a Markov chain—all of which are difficult to bound in the general case.

When simulating the time evolution of spin systems via Hamiltonian simulation, the most efficient algorithms exploit the locality of interactions in the Hamiltonian, and the resulting commutation structure. For 2-local spin-1/2 systems on a $D$-dimensional lattice with nearest-neighbor geometric locality, algorithms with almost optimal gate complexity are known for performing time evolution. Reference [18] showed that the gate complexity of the $(2k)$th-order product formula scales as

$$\mathcal{O}\Big((Nt)^{1+1/2k}/\epsilon^{1/2k}\Big)$$

to simulate time evolution for time $t$ to accuracy $\epsilon$, using a Hamiltonian given in the Pauli access model. Note that this expression suppresses the $5^{2k}$ constant prefactor present in $(2k)$th-order Trotter (see Eq. (49)). Similarly, [19] gave an algorithm with complexity $\mathcal{O}(Nt \cdot \mathrm{polylog}(Nt/\epsilon))$ for Hamiltonians given in the sparse access model. In contrast, note that approaches that are asymptotically optimal in the black-box setting, such as quantum signal processing, have a gate complexity of $\mathcal{O}(N^2t + N\log(1/\epsilon))$ using a block-encoding based on linear combinations of unitaries (LCU)—the normalization of the block-encoding is $\alpha = \mathcal{O}(N)$ and the gate complexity to implement the block-encoding is also $\mathcal{O}(N)$.

Spin Hamiltonians with power-law interactions were studied in [20, 21], that is, where the interaction strength between spins $i$ and $j$ depends inversely on a power of the distance between the spins, denoted by $\|i - j\|_2$. (This implies that the model is translationally invariant.) For a $D$-dimensional lattice with 2-local interactions with interaction strengths scaling as $1/\|i - j\|_2^\alpha$, $(2k)$th-order Trotter gives a gate complexity scaling as (as above, suppressing the $5^{2k}$ constant prefactor present in $(2k)$th-order Trotter) [21]

$$\widetilde{\mathcal{O}}\bigg( \begin{array}{ll} N^{3-\frac{\alpha}{D}(1+1/2k)+1/k}t^{1+1/2k}\epsilon^{-1/2k} & \text{if } 0 \le \alpha < D, \\ N^{2+1/2k}t^{1+1/2k}\epsilon^{-1/2k} & \text{if } \alpha \ge D \end{array} \bigg).$$

Further improvements are possible in cases where the Hamiltonian coefficients are efficiently computable by an oracle, or if certain truncations can be performed [22]. Focusing on the $D = 1$ case, if one were to directly apply quantum signal processing based on a block-encoding via the linear combination of unitaries approach, the scaling of the gate complexity would be

$$\widetilde{\mathcal{O}}\big(N^2t + N\log(1/\epsilon)\big).$$

These asymptotic complexities are complemented by the constant prefactor analyses discussed in the following section.

For estimating expectation values of observables to precision $\epsilon$, one can either consider directly sampling and then re-preparing the state of interest (scaling as $\mathcal{O}(1/\epsilon^2)$), or coherent approaches based on amplitude estimation (scaling as $\mathcal{O}(1/\epsilon)$, but requiring a longer coherent circuit depth). Measurements of simple observables, such as the magnetization, can be obtained through the computational basis measurements on single qubits. For more complicated observables, one can consider the approaches in [23, 24, 25], discussed in more detail in Section 2 (quantum chemistry).

**Existing resource estimates**

A number of logical resource estimates for simulating the dynamics of spin systems and for finding their ground states via quantum phase estimation have been reported in the literature. In such calculations, it is necessary to optimize the constant prefactor contributions from implementing the algorithmic primitives used. A detailed comparative study on simulating the dynamics of a 1D nearest-neighbor Heisenberg model (Eq. (5)) was reported in [26], comparing the logical qubit and $T$ gate counts of product formulas, Taylor series, and quantum signal processing. The two most efficient approaches are shown in the first two rows of Table 2.

| Problem | Method | # Spins | # $T$ gates | # Logical qubits | Parameters |
|---|---|---|---|---|---|
| 1D Heisenberg dyn. | QSP | 50 | $2.4 \times 10^9$ | 67 | $B_j \in [-1, 1], J^x = J^y = J^z = 1,$ $t = N, \epsilon = 10^{-3}$ [26] |
| 1D Heisenberg dyn. | Trotter (sixth order) | 50 | $1.8 \times 10^8$ | 50 | $B_j \in [-1, 1], J^x = J^y = J^z = 1,$ $t = N, \epsilon = 10^{-3}$ [26] |
| 2D NN TFIM[a] dyn. | Trotter (fourth order) | 100 | $1.7 \times 10^5$ | 100 | $t = 10/J, B = J, \epsilon = 10^{-2}$ [27, 28] |
| 2D $1/r^2$ TFIM dyn. | Trotter (fourth order) | 100 | $1.5 \times 10^7$ | 100 | $t = 10/J, B = J, \epsilon = 10^{-2}$ [27] |
| 2D Heisenberg ground state with nearest- and next-nearest-neighbor interactions | Qubitized QPE | 100 | $10^8$ | 112 | $\epsilon = 10^{-2}, J_1 = 1, J_2 = 0.5, B_j = 0$ [29] |

Table 2: Logical resource estimates for quantum phase estimation (QPE) and dynamics simulation (dyn.) applied to different spin models. The presented gate counts are for a single run of the circuit. The results presented in rows 1 and 2 can be compared to each other, and both target an error of $\epsilon = 10^{-3}$ in the operator norm distance between the ideal and implemented time evolution unitary. While [26] presents both analytic and empirical Trotter error bounds, the gate count presented in the table is that resulting from the empirical bound, though we remark that more recent analytic bounds are close to matching the empirical bounds [21]. The results presented in rows 3 and 4 can be compared to each other, and determine the number of Trotter steps used empirically by targeting an error of $\epsilon = 10^{-2}$ in a particular spatially averaged local observable, and then extrapolating this behavior to larger system sizes.

---

[a] 2D nearest-neighbor transverse-field Ising model.

On a fault-tolerant quantum computer, arbitrary angle rotation gates must be synthesized using a sequence of $T$ and Clifford gates [30]. The number of $T$ gates to synthesize a group of parallel rotation gates can be reduced if they share the same angle [31, 32, 33], a method known as Hamming weight phasing. This technique can be exploited in fault-tolerant compilations of algorithms simulating physical spin systems, which often exhibit features such as translational invariance.

In addition to the entries given in Table 2, fault-tolerant approaches to simulating NMR [3, 14] and muon spectroscopy [5] experiments, which are effectively spin model simulations, have been considered.

## Caveats

When formulated as a decision problem, determining the ground state energy for 2-local classical and 2-local quantum spin models is NP-complete [34, 35] and QMA-complete [36], respectively. As such, we do not expect quantum algorithms to provide efficient solutions to these problems in the general case. Nevertheless, given the success of classical heuristics for these problems, one may hope to observe a similar benefit from quantum heuristic algorithms, such as Monte Carlo–style Gibbs sampling algorithms.

In contrast, simulating the dynamics of spin models is a BQP-complete problem [37]; it is likely one of the most simple beyond-classical calculations that could be performed on a future fault-tolerant quantum computer. While such a computation would be of great scientific interest, providing new insights in quantum information and many-body physics, it is currently unclear whether dynamics simulations of large systems will have a direct impact on industrially relevant applications.

## Comparable classical complexity and challenging instance sizes

Exact classical simulations of quantum spin models are exponentially costly in system size. Exact simulations that consider a time evolution long enough for information to propagate across the system (as per the Lieb–Robinson bound) are limited to around 50 spins using the largest classical supercomputers [38, 26].

Approximate classical algorithms for studying quantum spin systems include tensor network approaches and quantum Monte Carlo (QMC) methods. These methods provide empirically accurate results for computing the ground states of physically motivated spin systems, in particular those with local interactions, in low dimensions. For example, the ground states of local, gapped 1D Hamiltonians have area law entanglement [39]; thus, they can be efficiently represented by matrix product states, a type of tensor network. The situation in higher dimensions is less clear, as an area law for gapped 2D local Hamiltonians has not been proven, and such a relation does not imply efficient representation via classical tensor network approaches such as projected entangled pair states (PEPS) [40]. Indeed, the 2D local Hamiltonian problem with the constraint that the ground state obeys area laws is QMA-complete [41]. Other approximate classical methods, including QMC approaches, can also be effective for preparing low energy states of spin systems. For example, [42] provides benchmarks for both tensor network and QMC-based variational methods applied to spin systems in a range of lattice geometries.

In contrast, these methods are less accurate when performing simulations of quantum spin dynamics [43, 44]. In many of these systems, the entanglement entropy grows linearly with time [45], resulting in a cost that grows exponentially with time for tensor network approaches targeting fixed accuracy (see, e.g., [46, 47] for counterexamples). For example, it was claimed in [27] that simulations of the dynamics of the 2D TFIM for $N = 100$ spins would be far beyond the current capabilities of tensor network methods [27].

Many physical systems are subject to strong interactions with their environment, which limits their coherence times. In these cases, the behavior of the system can often be reproduced by simulating a smaller number of spins (e.g., $N \leq 30$) and accounting for the interactions with the environment through physically motivated heuristics [48]. Such simulations (accessible via open-source software libraries) are used to analyze NMR [9] and muon spectroscopy experiments [10].

**Speedup**

The speedup for computing the ground states of quantum spin Hamiltonians over classical approximate methods (such as tensor networks or QMC) is currently an open research question; it depends on the complexity of being able to prepare good approximations of the ground state using quantum algorithms for cases where classical trial states are unable to efficiently and accurately represent the ground state [49].

The simulation of quantum spin dynamics appears to be exponentially costly using all known classical methods. As such, quantum algorithms for Hamiltonian simulation would provide an exponential speedup for this task. This would likely provide insights in quantum information and many-body physics. As an example, such systems could study the competition and interplay between thermalization and many-body localization in quantum systems.

**NISQ implementations**

Quantum spin models are commonly used as benchmark systems for NISQ algorithms—for example, finding ground states [50], simulating dynamics [51], and probing thermalization [52]. For instance, recent variational algorithms have been applied to spin systems with up to 24 qubits, showing rapid convergence to the ground state as the number of layers of the variational circuit was increased [53, 54].

The Hamiltonians of spin models are also naturally realized in a wide range of physical systems, including trapped ions or neutral atoms [6, 7]. For example, recent experiments in neutral atom systems have studied the dynamics of on the order of 200 spins, which went beyond the capabilities of classical simulation via matrix product state approaches [55, 56]. Analog simulators are already an important tool providing new scientific insights, and they set a high bar for the future performance of fault-tolerant approaches to simulating spin systems. Nevertheless, analog simulators are restricted in the complexity of the models that they can simulate (e.g., it may be more challenging to study site-dependent impurity models or composite spin–fermion systems), and they are susceptible to errors from miscalibration and interactions with the environment.

**Outlook**

Simulating the behavior of spin systems is arguably one of the most natural tasks for quantum computers, while being exponentially costly using all known classical methods. Such simulations can provide important insights into questions in quantum information science and many-body physics. Spin system simulations are also relevant to condensed matter physics and chemistry, since spin systems can act as models for more complex systems in those fields.

Logical resource estimates for quantum algorithms simulating spin systems are among the lowest known for beyond-classical tasks. Nevertheless, analog quantum simulators are already able to natively simulate the dynamics of hundreds of spins. In order to surpass these capabilities, digital approaches may need to consider more complex observables or target better accuracies that are only achievable using devices capable of quantum error correction.

In addition, for many systems of scientific interest in related fields, such as chemistry or condensed matter physics, decoherence-inducing interactions with the environment often limit the required simulation sizes. Identifying applications where accurate dynamics simulation of large spin models is required would increase the impact and applicability of quantum algorithms in this area.

## 1.3 SYK model

### Overview

The Sachdev–Ye–Kitaev (SYK) model [1, 2] is a simplified model of a quantum black hole that is strongly coupled and "maximally chaotic," but still analytically tractable. This remarkable and, to date, unique combination of properties has led to great activity surrounding SYK. It has applications in high-energy physics through its connections to black holes and quantum gravity, and it has applications in condensed matter physics as a model of quantum chaos and scrambling, which sheds light on phases of matter in strongly coupled metals [3, 4]. While many interesting properties of the SYK model can be computed analytically in certain limits, not all properties qualify, and questions remain about the behavior of the model outside of these limits—these questions can potentially be addressed numerically by a quantum computer.

### Actual end-to-end problem(s) solved

The SYK model has many variants; a common version to consider is the four-body ($q = 4$) $N$-site Majorana fermion Hamiltonian with Gaussian coefficients

$$H_{\text{SYK}} = \frac{1}{4 \cdot 4!} \sum_{i,j,k,\ell=1}^{N} g_{ijk\ell} \, \chi_i \chi_j \chi_k \chi_\ell \,, \tag{6}$$

where $\chi_i$ denote Majorana fermion mode operators obeying the anticommutation relation $\chi_i \chi_j + \chi_j \chi_i = 2\delta_{ij} I$ (where $I$ denotes the identity operator, and $\delta_{ij}$ the Kronecker delta symbol), and $g_{ijk\ell}$ are coefficients drawn independently at random from a Gaussian distribution with zero mean and variance $\sigma^2 = 3! g^2 / N^3$ (with $g$ the tunable coupling strength). The normalization of Eq. (6) matches the convention of [5] and ensures that the ground state energy of $H_{\text{SYK}}$ is extensive (i.e., growing linearly in $N$).

In the limit of a large number of local degrees of freedom $N \to \infty$ and at strong coupling $\beta g \gg 1$ (where $\beta$ is the inverse of the temperature), analytic predictions can be computed for the asymptotic behavior of some properties. While these arguments are not mathematically rigorous, in practice they provide a consistent picture for the SYK model and provide insights into quantum gravity and quantum chaos. However, questions remain about the wealth of properties out of reach by taking limits or the nonasymptotic regime of parameters. For example, it has been challenging to rigorously calculate the density of states at arbitrary energies and the ground state energy in the large-$N$ limit [6, 5, 7]. These problems can potentially be probed numerically on a quantum computer.

Generally speaking, this often reduces to performing the following task on the quantum computer: given as input an instance of $H_{\text{SYK}}$ (generated by choosing the couplings $g_{ijk\ell}$ at random) and an observable $O$, estimate the expectation value $\text{tr}(\rho O)$, where $\rho$ could be, for instance, (i) the ground state of $H_{\text{SYK}}$, (ii) the thermal state $\rho \propto e^{-\beta H_{\text{SYK}}}$, or (iii) a time-evolved state $\rho = e^{iH_{\text{SYK}}t} |0\rangle\langle 0| e^{-iH_{\text{SYK}}t}$ from an easy-to-prepare initial state $|0\rangle$, among other possibilities. The observable $O$ could be a local operator or even $H_{\text{SYK}}$ itself. Another case is for $O$ to be composed of $t$-dependent time-evolution unitaries $e^{iH_{\text{SYK}}t}$.

For example, computing the ground state energy corresponds to taking $\rho$ to be the ground state of $H_{\text{SYK}}$ and $O$ to be $H_{\text{SYK}}$, and computing a 4-point out-of-time-ordered correlation function corresponds to taking $\rho$ to be the thermal state at inverse temperature $\beta$ and $O$ to be $A e^{iH_{\text{SYK}}t} B e^{-iH_{\text{SYK}}t} A e^{iH_{\text{SYK}}t} B e^{-iH_{\text{SYK}}t}$, where $A$ and $B$ are few-body operators [8]. In another example, [9, 10] give a detailed proposal to "simulate quantum gravity in the lab" via computing expectation values of observables and states formed via simulation of the SYK model.

Depending on the ultimate end-to-end goal, one may need to repeat this calculation for many different $O$ or for many instances of $H_{\text{SYK}}$, for example, to compute an ensemble average.

**Dominant resource cost/complexity**

**Mapping the problem to qubits:** To simulate the SYK model on a quantum computer, the Majorana operators are represented by strings of Pauli operators according to the Jordan–Wigner representation (e.g., [11]). As a result, the Hamiltonian $H_{\text{SYK}}$ on $N$ Majoranas becomes a linear combination of multi-qubit Pauli operators over $N/2$ qubits. Methods for Hamiltonian simulation given a classical description of the Hamiltonian as a linear combination of Pauli strings (the Pauli access model) typically introduces into the complexity a dependency on the number of terms, $N^4$, and on the 1-norm of Pauli coefficients, denoted by $\lambda$, which for typical SYK instances is seen to be $\lambda = \mathcal{O}(gN^{5/2})$ (see [5, Eq. (16)]).

**State preparation:** To solve the problem of estimating $\text{tr}(\rho O)$, one must be able to prepare the $(N/2)$-qubit state $\rho$. In some cases, $\rho$ could simply be a product state, which is trivial to prepare. If $\rho$ is the thermal state at inverse temperature $\beta$, then algorithms for Gibbs sampling would be used to prepare the state. Due to the chaotic properties of SYK and the fact that the system is expected to thermalize quickly in nature, one expects that Monte Carlo–style Gibbs samplers (e.g., [12, 13, 14, 15, 16]) have a favorable poly($N$) gate complexity, but the exact performance is unknown. If $\rho$ is the ground state of $H_{\text{SYK}}$, there are several methods for preparing $\rho$, including projection onto $\rho$ by measuring (and postselecting) an ansatz state $\phi$ in the energy eigenbasis using quantum phase estimation (QPE), or by adiabatic state preparation. The cost of either of these methods depends on details such as which ansatz state is used (in particular, its overlap with $\rho$), the adiabatic path, and the spectrum of $H_{\text{SYK}}$—in both cases, in the absence of evidence to the contrary, the scaling can be exponential in $N$. In [7], a poly($N$)-time quantum algorithm for preparing states $\rho$ achieving a constant-factor approximation to the ground state energy of $H_{\text{SYK}}$ was given, which could be used as $\rho$ to probe low-energy properties of the system.

**Time evolution:** The calculation also requires simulating time evolution by $H_{\text{SYK}}$. This can be because $O$ is a time-evolved operator, because the state $\rho$ corresponds to a time-evolved state, or simply as a subroutine for QPE or Gibbs sampling, mentioned above. Reference [11] proposed a scheme for simulating time evolution using a first-order product formula approach to Hamiltonian simulation. That is, it implements the unitary $e^{iH_{\text{SYK}}t}$ to precision $\epsilon$, with gate complexity $\mathcal{O}(N^{10}g^2t^2/\epsilon)$. However, this steep scaling with $N$ suggests that accessing large system sizes will be difficult with this method. Reference [5] later gave a method with better $N$ dependence, achieving gate complexity $\mathcal{O}(N^{7/2}gt + N^{5/2}gt\,\text{polylog}(N/\epsilon))$, leveraging qubitization with quantum signal processing. This gate complexity grows more slowly than the number of terms in $H_{\text{SYK}}$ (i.e., $\mathcal{O}(N^4)$), a feat that is only possible because the simulation method generates the SYK coupling coefficients pseudorandomly: to construct the block-encoding of $H_{\text{SYK}}$, they perform the PREPARE step in the linear combination of unitaries using a shallow quantum circuit composed of polylog($N$) random two-qubit gates, producing a state for which the $N^4$ amplitudes are distributed approximately as independent Gaussians. Further reduction in the gate count would be bottlenecked by the 1-norm $\lambda$ of the coefficients of $H_{\text{SYK}}$; however, note that [17] suggests gravitational features may remain even if the Hamiltonian is substantially sparsified, which could reduce the number of terms and the value of $\lambda$.

**Measuring observables:** Finally, given the ability to prepare a purification of $\rho$ and supposing $O$ is unitary (if it is not, it could be decomposed into a sum of unitaries and each constituent computed separately), estimating the expectation value $\text{tr}(\rho O)$ to precision $\epsilon$ can be done by overlap estimation, costing $\mathcal{O}(1/\epsilon)$ calls to the routine that prepares $\rho$ and to the routine that applies $O$. If the purification of $\rho$ cannot be prepared, the cost is $\mathcal{O}(1/\epsilon^2)$.

**Existing resource estimates**

Reference [5] compiled the dominant contributions in their approach to Hamiltonian simulation into Clifford + $T$ gates, and they found that at $N = 100$, implementing $e^{iHt}$ requires fewer than $10^7 gt$ $T$ gates, and at $N = 200$, it requires fewer than $10^8 gt$ $T$ gates. Both of these figures are for a single circuit; that is, they do not include the cost of averaging over many SYK instances. The $T$-count can be turned into an estimate of the running time and number of physical qubits; see the discussion in Section 17 on fault-tolerant quantum computation.

**Caveats**

Existing resource estimates only focus on simulating the dynamics of SYK models, but the proposed classically challenging problems involve static properties such as density of states and properties of thermal states. Probing these static properties in an end-to-end fashion would likely require preparing thermal states, ground states, or other kinds of low-energy states, in addition to being able to implement $e^{iHt}$. The cost of preparing these states is unknown and difficult to assess analytically. Another caveat is that the gate counts quoted above do not take into account the $\mathcal{O}(1/\epsilon)$ scaling of reading out an observable to precision $\epsilon$, or any repetitions for different instances of $H_{\text{SYK}}$ required for making inferences about the physics of SYK.

**Comparable classical complexity and challenging instance sizes**

As mentioned above, one of the reasons that the SYK model is appealing is that many properties can be computed analytically in certain limits. Other properties that would be of interest to numerically compute on a quantum computer require poorly scaling classical methods. Exact diagonalization of systems consisting of more than roughly 50 Majoranas would be challenging due to the exponential growth of the Hilbert space, which has dimension $2^{N/2}$. For example, [6] and [18] gave a variety of numerical results based on exact diagonalization up to $N = 34$ and $N = 36$, respectively.

**Speedup**

Hamiltonian simulation has $\text{poly}(N)$ runtime, an exponential speedup over exact diagonalization, which is the go-to method for classical simulation of SYK-related problems. However, Hamiltonian simulation does not alone solve the same end-to-end problem as exact diagonalization; the persistence of the exponential speedup requires identifying specific interesting properties where the relevant initial states can also be prepared in $\text{poly}(N)$ time, which is currently less clear.

**NISQ implementations**

Experimental realizations of the SYK model have been proposed on several different experimental platforms [19, 20, 21]. However, even if these demonstrations can be realized, we do not expect this approach to scale in the absence of quantum error correction.

**Outlook**

Simulating time evolution of the SYK model on a quantum computer has relatively mild gate cost, due to the model's straightforward mapping to a qubit Hamiltonian. At the same time, it is difficult to simulate the SYK model on a classical computer, owing to its chaotic and strongly coupled nature. However, further work is needed to understand the entire end-to-end pipeline. It has not yet been identified which properties would be most valuable to compute on a quantum computer and how costly they will be. Computing these properties will likely involve far more than a single run of time evolution on a single instance of the SYK model, so the overall cost is likely to be much larger than what initial gate counts in the literature suggest.

# 2 Quantum chemistry

Computational chemistry seeks to use computational methods to predict the physical properties and behaviors of atoms, molecules, and materials, and includes methods such as first-principles simulation, classical molecular dynamics, and cheminformatics. We will restrict our focus to first-principles simulations that treat chemical systems quantum mechanically (noting that these methods may be used within a larger workflow incorporating other techniques). Despite the apparent exponential cost of exact classical methods for this task, scientists have made incredible progress over the last century via increasingly sophisticated approximate methods. As a result, quantum chemistry is now a core part of several applications, including the analyses of chemistry experiments, the pharmaceutical drug discovery pipeline, and the optimization of materials for catalysts and batteries. Given the inherently quantum mechanical nature of these problems, it follows that several quantum algorithms have been proposed for computational chemistry [1]. In this section, we focus on simulating the electrons and vibrations of nuclei in molecules and materials. For further reviews of quantum computing for chemistry, we refer readers to [2, 3, 4, 5].

*The authors are grateful to Ryan Babbush, Joshua Goings, and Ashley Montanaro for reviewing this section of the survey.*

**This application area contains:**

## 2.1 Simulating electrons in molecules and materials

**Overview**

We seek the energy eigenstates, thermal states (i.e., statistical ensembles of eigenstates at a given temperature), or dynamics corresponding to the Hamiltonian used to describe the electrons in molecules or material systems. The electrons interact with each other, in addition to fields produced by the nuclei and any external applied fields. In many systems it is appropriate to use the Born–Oppenheimer approximation, which treats the nuclei classically and fixes their spatial positions, separating the nuclear and electronic degrees of freedom.

Material systems can be described by a periodically repeating (i.e., translationally invariant) finite-size computational cell of interacting atoms. By simulating increasingly large computational cells and mitigating finite-size effects, we can extrapolate simulation results to the thermodynamic limit. This enables the measurement of bulk properties, such as magnetization, tensile strength, and thermal or electrical conductivity.

**Actual end-to-end problem(s) solved**

The Hamiltonian of a system consisting of $K$ nuclei and $\eta$ electrons interacting via the Coulomb interaction is (in atomic units)

$$H = -\sum_{i=1}^{\eta} \frac{(\nabla_i)^2}{2} - \sum_{I=1}^{K} \frac{(\nabla_I)^2}{2M_I} - \sum_{i,I} \frac{Z_I}{|r_i - R_I|} + \sum_{i \neq j} \frac{1}{2|r_i - r_j|} + \sum_{I \neq J} \frac{Z_I Z_J}{2|R_I - R_J|},$$

where $\nabla$ is the gradient operator, $r_i$ gives the position of the $i$-th electron, and $R_I$ and $Z_I$ give the position and charge of the $I$-th nucleus. This Hamiltonian can be discretized using a basis set $\{\phi_i(r)\}_{i=1}^{N}$ of electron spin orbital and $\{\chi_i(r)\}_{i=1}^{M}$ of nuclear orbital functions, or grid points, and can either be used with the time-dependent Schrödinger equation to simulate dynamics, or with the time-independent Schrödinger equation to obtain energy eigenstates. When simulating dynamics, it is necessary to use a basis set that is sufficiently flexible (or adaptive) to accurately describe the states at all times, as many chemical basis sets are highly optimized for ground state calculations and so are less suitable for dynamics calculations. It is often appropriate to make the Born–Oppenheimer approximation, fixing the positions of the nuclei, which are treated as classical point charges. The resulting electronic Hamiltonian at a fixed nuclear configuration is given by

$$H(\{R_I\}) = -\sum_{i} \frac{(\nabla_i)^2}{2} - \sum_{i,I} \frac{Z_I}{|r_i - R_I|} + \frac{1}{2} \sum_{i \neq j} \frac{1}{|r_i - r_j|} + V(\{R_I\}), \tag{7}$$

where $V(\{R_I\})$ is the constant offset from the nuclear repulsion energy. In this case, the Hamiltonian is discretized using a basis set $\{\phi_i(r)\}_{i=1}^{N}$ of electron spin orbital functions or grid points. For many molecules at room temperature, the ground state of the electronic structure Hamiltonian is a good approximation for the thermal state $\rho \propto e^{-\beta H}$ (with $\beta = 1/k_B T$, where $k_B$ is the Boltzmann constant and $T$ the temperature), as the electronic energy levels are well separated with respect to $k_B T$. This can be contrasted with the vibrational structure of molecules, where vibrational energies are on the order of $k_B T$, and so excited states are also populated at room temperature.

The electronic eigenstates (or thermal states) often provide a good starting description of a wide range of system properties, which then can be corrected by, for example, vibrational, rotational, relativistic (e.g., spin-orbit coupling) or entropic contributions. Preparing the desired electronic state for a given nuclear configuration is typically the first step in learning properties of the system. We then measure the expectation values of observables with respect to these states. Moreover, the electronic response to weak or slowly varying perturbations can often be described by a sequence of static calculations, for example, linear response theory for radiation absorption, or Born–Oppenheimer molecular dynamics [1] where one iteratively solves the electronic time-independent Schrödinger equation to obtain the forces on the nuclei, whose positions can then be updated using Newton's laws. Static calculations can be used to probe:

- Energy values (potentially across a range of nuclear configurations)—for electronic excitation energies at a fixed nuclear geometry, for determining molecular geometries by computing the electronic ground state energy at different geometries, and for finding reaction pathways and rates by computing energy differences between a sequence of geometries involved in a reaction. To obtain accurate predictions, the electronic energy values must be corrected by other contributions to the free energy.

- Determining transition probabilities between states—for reactions and optical properties.

- Differential changes in electronic energy in response to an applied field, for example, electronic or magnetic dipole moments, polarizability.

- Calculating forces on the nuclei, for use in molecular dynamics calculations—used in a range of applications, including protein folding and calculating drug molecule binding affinities.

- Orbital occupancies and correlation functions, as well as real- and imaginary-time Green's functions.

Properties of interest for materials include the following:

- Energy densities for given system parameters, to determine phase diagrams.

- Bulk properties, such as magnetization, thermal or electrical conductivity, and tensile strength.

- Particle densities and correlation functions between sites, as well as real- and imaginary-time Green's functions.

In order to understand how these observables vary as the system parameters (nuclear positions, atomic doping, temperature, applied field, etc.) are changed, the desired state may need to be prepared and measured a number of times.

Simulations of dynamics may be used to explicitly probe many of the equilibrium phenomena implicitly being probed above—for example, chemical reactions occurring at thermodynamic (quasi-)equilibrium—as well as additional nonperturbative or nonequilibrium phenomena that are difficult to implicitly describe as a sequence of static calculations, such as scattering from collisions, absorption of UV and X-ray radiation, charge-transfer dynamics, and optimal control. As a result, many of the same observables described above are still of interest and can be monitored as a function of time, including the following:

- Changes in kinetic or potential energy.

- Changes in particle densities or orbital occupancies.

- Changes in charge or spin densities.

**Dominant resource cost/complexity**

**Mapping the problem to qubits:** We discretize the electron positions using a basis of $N$ spin orbital functions. For many basis sets, the discretization error decays as $1/N$ [2, 3] and is limited by the resolution of singularities in the Coulomb interaction from charge coalescence. A variety of functional forms have been considered for the electron orbitals (see Table 3 for a list of orbitals commonly considered in quantum computing). The optimal choice will be system dependent and must consider the following non-exhaustive list of factors:

- The resolution of the orbital, improved by matching the character of local vs. delocalized physics in the system to that of the orbital.

- The cost of computing the Hamiltonian, either in classical precomputation or (if required) coherently on a quantum device (see §Accessing the Hamiltonian, below).

- The properties of the resulting Hamiltonian (number of terms, norm, locality of terms, etc.) which determine the cost of accessing the Hamiltonian in algorithms.

| Representation | First-quantized | Second-quantized |
|:---:|:---:|:---:|
| Gaussians | [4][a] [5] | [6] |
| Plane waves | [7] | [8] |
| Bloch/Wannier functions | Not yet studied | [9, 10] |
| Grids | [11] | [8, App. A] |
| Pseudo-spectral / Discrete variable representations | [12, 13, 14] | [8, 15] |

Table 3: Representative references (chosen based on their discussion of their choice of representation) showing the use of different basis functions in quantum algorithms for the electronic structure problem.

[a]This reference is not technically a first-quantized representation, as antisymmetry is stored in the operators rather than the wavefunction, but it stores states in an analogously compressed way to first-quantized representations.

The commonly used "Galerkin discretization scheme" discretizes the Hamiltonian via integrals over the basis functions, with one- and two-electron integrals

$$h_{ij} = \int \mathrm{d}r \, \phi_i^*(r) \left( -\frac{(\nabla)^2}{2} - \sum_I \frac{Z_I}{|r - R_I|} \right) \phi_j(r)$$
$$h_{ijkl} = \int \mathrm{d}r_1 \mathrm{d}r_2 \, \frac{\phi_i^*(r_1)\phi_j^*(r_2)\phi_k(r_2)\phi_l(r_1)}{|r_1 - r_2|} \, . \tag{8}$$

Hamiltonians defined using grids, pseudo-spectral representations, or discrete-variable representations are not obtained from integrals over basis functions, as specified in Eq. (8), and the values of $h_{ij}, h_{ijkl}$ are instead defined using finite difference formulas, and/or by their values at discrete grid points. An attractive feature of the Galerkin discretization scheme is that the discretization error is strictly positive. We refer readers to [14, 8, 16] for a more complete discussion.

We can represent electronic states on a quantum computer using either first or second quantized representations.

- For $\eta$ electrons in $N$ spin orbitals, first quantization uses $\eta$ registers, which each contain $\log_2(N)$ qubits; each register enumerates which orbital its corresponding electron is in, and the wavefunction must then be antisymmetrized to respect fermionic constraints [17]. The Hamiltonian of Eq. (7) in first quantization can be written as

$$H = \sum_{\alpha=1}^{\eta} \sum_{i,j=1}^{N} h_{ij}|i\rangle\langle j|_\alpha + \frac{1}{2}\sum_{\alpha\neq\beta}\sum_{i,j,k,l=1}^{N} h_{ijkl}|i\rangle\langle l|_\alpha \otimes |j\rangle\langle k|_\beta \,,$$

  where $\alpha, \beta$ index which of the electron registers the operators act on.

- In second quantization, antisymmetry is stored in the operators, which obey fermionic anticommutation relations: $\{a_i, a_j^\dagger\} = \delta_{ij}I$ and $\{a_i, a_j\} = \{a_i^\dagger, a_j^\dagger\} = 0$ (where $\delta_{ij}$ denotes the Kronecker delta symbol, $I$ denotes the identity operator, and $\{u,v\} = uv + vu$). The Hamiltonian of Eq. (7) in second quantization can be written as

$$H = \sum_{i,j=1}^{N} h_{ij}a_i^\dagger a_j + \frac{1}{2}\sum_{i,j,k,l=1}^{N} h_{ijkl}a_i^\dagger a_j^\dagger a_k a_l.$$

  A number of fermion-to-qubit mappings have been studied; see [18] for discussion. Under the commonly used Jordan–Wigner mapping we require $N$ qubits, where each qubit stores the occupancy of the corresponding spin orbital. These mappings induce a mapping of the Hamiltonian (and other observables) to qubit operators.

**Accessing the Hamiltonian:** Quantum algorithms for both static and dynamic simulations require access to the Hamiltonian. This is typically provided by block-encoding or Hamiltonian simulation.[4] A common block-encoding strategy for chemistry Hamiltonians is the linear combinations of unitaries (LCU) block-encoding, whereby the Hamiltonian is expressed as a linear combination of unitary operators $\sum_{i=0}^{L-1} c_i U_i$ (e.g., $U_i$ could be products of Pauli matrices), and the block-encoding is then realized using the oracles PREPARE and SELECT[5] that act on the main register and a $\lceil\log_2(L)\rceil$ ancilla system as

$$\text{PREPARE}|0^{\lceil\log_2(L)\rceil}\rangle = \frac{1}{\sqrt{\lambda}}\sum_{j=0}^{L-1}\sqrt{|c_j|}|j\rangle$$

$$\text{SELECT} = \sum_{j=0}^{2^{\lceil\log_2(L)\rceil}-1}|j\rangle\langle j| \otimes \text{sign}(c_j)U_j \,,$$

where $\lambda = \sum_{i=0}^{L-1}|c_i|$. Then the sequence PREPARE$^\dagger$·SELECT·PREPARE is a $(\lambda, \lceil\log_2(L)\rceil, 0)$-block-encoding of the Hamiltonian written as an LCU. The oracle SELECT can be implemented using the unary iteration method [19] or the approach of [20]. The oracle PREPARE can be implemented by coherently loading coefficients stored in memory [19, 21, 22] or by computing coefficients on-the-fly using quantum arithmetic [12, 4, 23, 7]. In many cases, loading the coefficients for PREPARE is the bottleneck, with cost—in terms of the number of non-Clifford

---

[4]Hamiltonian simulation is used to explicitly simulate dynamics, but can also be used implicitly to provide access to the Hamiltonian for use in static calculations, for example, in quantum phase estimation.

[5]To be precise for $j \notin \{0, 1, \ldots, L-1\}$ we define $\text{sign}(c_j)U_j := I$.

gates—scaling linearly with $\widetilde{L}$, the number of unique coefficients in the Hamiltonian (this scaling can be reduced to $\mathcal{O}(\widetilde{L}^{1/2})$ using $\mathcal{O}(\widetilde{L}^{1/2})$ ancilla qubits [21]). As a result, a number of algorithms have reduced the quantity of data to load by using compressed representations of the Coulomb Hamiltonian, achieved through tensor factorizations [22, 24, 25, 10] (see [26] for a recent unifying perspective on these approaches).

**State preparation:** Simulating the behavior of electrons in molecules and materials reduces to the task of preparing a desired state and measuring observables. The state to be prepared is typically an energy eigenstate, a thermal state, or a time-evolved state.

- Energy eigenstates: In the following discussion, we refer to the overlap $\gamma = |\langle\psi|E_j\rangle|$ between a desired eigenstate $|E_j\rangle$ and a given initial state $|\psi\rangle$, and the minimum gap $\Delta$ between the desired energy eigenvalue and other energy eigenvalues. Below, we list several methods for preparing energy eigenstates, or approximations to them.

  – Classical trial states: Approximate eigenstates obtained from a classical calculation can be prepared as quantum trial states, including Slater determinant states [27], linear combinations of $D$ Slater determinants (with complexity $\widetilde{\mathcal{O}}(D)$ [28]–$\mathcal{O}(ND)$ [29]), and matrix product states (MPSs) with bond dimension $\chi$ (with complexity $\mathcal{O}(N\chi^2)$ [28, 30]). In [30] it was observed that MPS with modest bond dimension could have large overlaps with chemical systems of interest. Several of these methods have been adapted to simulations performed in first quantization [31, 32].

  – Quantum trial states: Parameterized quantum circuits, in conjunction with variational quantum algorithms, have been proposed for preparing approximate energy eigenstates (see §NISQ implementations, below). Like classical trial states, the states prepared by these circuits can be used as inputs to other quantum algorithms that further refine the initial state, such as eigenstate filtering, or quantum phase estimation. Initial resource estimates for models of materials systems can be found in [33].

  – Eigenstate filtering: Methods such as those in [34, 35] filter out undesired eigenstates using spectral window functions applied via quantum singular value transformation (QSVT) to a block-encoding of the Hamiltonian. The complexity to prepare the ground state (to infidelity $\epsilon$, with failure probability less than $\theta$) using this approach scales as

  $$\widetilde{\mathcal{O}}\left(\frac{\alpha}{\gamma\Delta}\log\left(\theta^{-1}\epsilon^{-1}\right)\right)$$

  calls to an $(\alpha, m, 0)$-block-encoding of the Hamiltonian, where $\alpha \geq \|H\|$ is a normalization factor of the block-encoding. For comparison to related methods, we refer the reader to [35, 36]. These algorithms can also be adapted for the case where access to the Hamiltonian is provided by Hamiltonian simulation [37].

  – Adiabatic state preparation (ASP): ASP can be used to prepare a target eigenstate (typically the ground state) by evolving from the corresponding easy-to-prepare eigenstate of an initial Hamiltonian $H(0)$ to the full Hamiltonian $H(1)$. Time evolution can be implemented using algorithms for Hamiltonian simulation. The total evolution time is typically chosen according to the heuristic

  $$T \gg \max_{0\leq s\leq 1}\frac{\|\frac{\mathrm{d}H}{\mathrm{d}s}\|}{\Delta(s)^2},$$

where $s$ describes the adiabatic path $H(s)$ and $\Delta(s)$ is the spectral gap of $H(s)$. It is difficult to analytically bound this (highly system-dependent) complexity for molecular systems (see, e.g., [38]) motivating numerical studies on small molecules [39, 40, 41, 42].

– Quantum phase estimation (QPE): The above techniques all provide methods of preparing approximate eigenstates, in some cases using promises on the gap $\Delta$, or by exploiting pre-existing knowledge of the energy eigenvalue. Given an approximate eigenstate, and a unitary $U = f(H)$ that encodes the eigenspectrum of the Hamiltonian (with a known, classically invertible relationship $f$), we can use QPE to project into the desired eigenstate and provide an estimate of the eigenphase $\phi_i$ of $U$, which can then be converted into an estimate of the eigenenergy of $H$ using $\phi_i = f(E_i)$. QPE makes

$$\mathcal{O}\big(\gamma^{-2}\|f'(H)\|^{-1}\epsilon^{-1}\log\big(\theta^{-1}\big)\big)$$

calls to the unitary $U(H)$ encoding the spectrum of the Hamiltonian, where $\gamma = |\langle\psi|E_j\rangle|$ is the overlap between the state $|\psi\rangle$ input to QPE and the desired energy eigenstate $|E_j\rangle$, $\theta$ is the failure probability, and $\epsilon$ is the desired precision in the eigenenergy of $H$. It is possible to improve the complexity to

$$\mathcal{O}\big(\gamma^{-1}\|f'(H)\|^{-1}\epsilon^{-1}\log\big(\theta^{-1}\big)\big)$$

using amplitude amplification if a sufficiently precise estimate of the eigenvalue is known, or to

$$\mathcal{O}\big((\gamma^{-2}\Delta^{-1} + \epsilon^{-1})\|f'(H)\|^{-1}\log\big(\theta^{-1}\big)\big)$$

by exploiting knowledge of the gap $\Delta$ between the energy eigenstates to perform rejection sampling [17]. The unitary encoding the Hamiltonian is often chosen to be $U(H) \approx e^{iHt}$ approximated via quantum algorithms for Hamiltonian simulation. In this case, the approximation error in $U$ must be balanced against the error from QPE. Alternatively, it is common to encode the Hamiltonian using a quantum walk operator $W(H)$ which acts like $e^{i\arccos(H/\alpha)}$ and can be implemented exactly via qubitization [43, 17]. The costs to implement $U(H)$ are inherited from the method used, based on the properties (e.g., commutativity, locality, number of terms, norm, cost of coherently calculating coefficients) of the Hamiltonian in the chosen spin orbital basis and representation. As indicated by the complexities presented above, QPE incurs an overhead from imperfect overlap with the target eigenstate, as well as from needing to suppress the failure probability of the method. We refer to Section 13 on QPE for a more detailed discussion of the latter issue, which can either be mitigated by repeating the calculation and using methods for probability amplification [44, 45], or by using window functions [46, 47, 48] to guarantee a desired confidence interval. The latter strategy appears to require fewer resources in practice, especially when considering the imperfect overlap with the target state [30].

• Thermal states: Several quantum algorithms have been proposed for preparing thermal states, also known as Gibbs states. The most efficient algorithms proceed by simulating the dissipative open system dynamics, and typically make repeated calls to a block-encoding of the Hamiltonian. The complexity of these methods for concrete electronic structure problems of interest has not yet been determined and depends on the spectral gap of the Lindbladian considered, at the desired temperature. Thermal states could be used as an approximation to the ground state, by choosing the temperature to be sufficiently low compared to the gap between the ground and first excited state [49]. Quantum algorithms for simulating open systems dynamics can also be adapted to directly prepare approximations to the ground state [50].

- Time-evolved states: A time-evolved state can be prepared using Hamiltonian simulation algorithms, which approximate the propagator to error $\epsilon$ (which provides an upper bound on the error in the resulting state). The cost of Hamiltonian simulation depends both on the algorithm used and the details of the Hamiltonian being simulated. Plane wave, grid, and pseudo-spectral (DVR) basis sets are well suited to simulations of dynamics, as they treat all points in space on an equal footing. For both the plane wave basis and pseudo-spectral DVR in first quantization, the scaling is [51, 7]

$$\widetilde{\mathcal{O}}\Big(\eta^{8/3}N^{1/3}t + \eta \log(1/\epsilon)\Big)$$

using Hamiltonian simulation in the interaction picture [52], or

$$\widetilde{\mathcal{O}}\Big((\eta^{8/3}N^{1/3} + \eta^{4/3}N^{2/3})t + \eta \log(1/\epsilon)\Big)$$

using qubitization with quantum signal processing [51, 7]. In the pseudo-spectral DVR, the cost scales as

$$\widetilde{\mathcal{O}}\left((\eta^{7/3}N^{1/3} + \eta^{4/3}N^{2/3})\frac{t^{1+o(1)}N^{o(1)}}{\epsilon^{o(1)}}\right)$$

using high-order product formulas [31].

**Measuring observables:** Many proposed algorithms consider the ground or excited state energy of the chemical system as the observable of interest. This can be calculated directly using QPE, as discussed above. QPE (and related methods) can also be adapted for other application, for example, calculating absorption spectra of molecules [53].

In other applications, it may be necessary to measure observables other than the energy. In a fault-tolerant computation, it can be preferable to measure these observables through phase-estimation-like approaches, rather than direct measurement averaging, as the former is asymptotically more efficient and can be made robust to logical errors through repetition and probability amplification. Measurement schemes have been developed which achieve this using overlap estimation [54] (see Section 14.2 on amplitude estimation) or the approach of [55, 56] based on the quantum gradient estimation algorithm of [57]. Both approaches require access to a state preparation unitary $U_\psi$, and its inverse.[6] The algorithm based on overlap estimation can be formulated as performing amplitude estimation on $U_O$, a unitary block-encoding of the observable $O$ with subnormalization factor $\alpha_O$. The complexity to compute the expectation value to precision $\epsilon$ is $\mathcal{O}(\alpha_O/\epsilon)$ calls to $U_O$ and $U_\psi$ (or the reflection $R_\psi = I - 2|\psi\rangle\langle\psi|$) and their inverses. This approach has been considered in the context of measuring correlation functions, density of states, and linear response properties (all in [58]), as well as energy gradients with respect to various parameters, which can be used to compute forces or dipole moments, and for which a range of estimation strategies are possible [59, 60].

---

[6]Note that it can be substantially cheaper to directly execute the reflection $R_\psi = I - 2|\psi\rangle\langle\psi|$ used in both methods, rather than through the use of $U_\psi$, as the complexity of $R_\psi$ does not depend on the overlap $\gamma$ that appears in state preparation—see [35] for additional discussion.

The gradient-based algorithm simultaneously computes the value of $M$ potentially noncommuting observables $O_j$ by making $\widetilde{\mathcal{O}}(M^{1/2}/\epsilon)$ calls to $U_\psi, U_\psi^\dagger$ (or $R_\psi$) and either $\widetilde{\mathcal{O}}(M^{3/2}/\epsilon)$ calls to gates of the form $e^{i x O_j}$ [55] or $\widetilde{\mathcal{O}}(M/\epsilon)$ calls to a block-encoding of the observables [56]. The algorithm also requires $\mathcal{O}(M \log(1/\epsilon))$ additional qubits. This approach has been considered in the context of measuring nuclear forces [59], fermionic reduced density matrices [55], and dynamic correlation functions [55].

**Existing resource estimates**

There are a large number of resource estimates for performing phase estimation to learn the ground state energies of molecular or material systems, which we list in Table 4 and Table 5. These resource estimates use compilation methods described in Section 17 on fault-tolerant quantum computing. We also note the existence of a software package that provides features for calculating the non-Clifford costs of QPE for the electronic structure problem [61].

| Molecule(s) & references | Algorithm | Logical qubits | $T$/Toffoli gates per sample | Number of samples |
|---|---|---|---|---|
| FeMo-co (nitrogen fixation) [38, 22, 24, 25, 62, 61] | 2nd Q, THC qubitization, Gaussians | 2196 [25] | $3.2 \times 10^{10}$ [25] | 7 |
| | 2nd Q, randomized compilation, Gaussian | ~193 [62] | ~$3 \times 10^{12}$ [62] | ~600 |
| Cytochrome P450 (biological drug metabolizing enzyme) [63] | 2nd Q, THC qubitization, Gaussians | 1434 | $7.8 \times 10^9$ | 7 |
| Lithium-ion battery molecules [64, 7] | 2nd Q, DF qubitization, Gaussians | $10^4$–$10^5$ [64] | $10^{12}$–$10^{14}$ [64] | 7 |
| | 1st Q, qubitization, plane waves | 2000–3000 [7] | $10^{11}$–$10^{12}$ [7] | 7 |
| Chromium dimer [65] | 2nd Q, sparse qubitization, Gaussians | ~1300 | ~$10^{10}$ | 7 |
| Ruthenium catalyst ($CO_2$ fixation) [24] | 2nd Q, DF qubitization, Gaussians | ~4000 | ~$3 \times 10^{10}$ | 7 |
| Ibrutinib (drug molecule) [66] | 2nd Q, sparse qubitization, Gaussians | 2207 | $1.1 \times 10^{10}$ | 7 |
| Molybdenum catalysts (nitrogen fixation) [67] | 2nd Q, DF qubitization, Gaussians | 1000–8000 | $10^{11}$–$10^{12}$ | 3–9[a] |
| Amyloid beta binding site fragment (Alzheimer's disease) [68] | 2nd Q, DF qubitization, Gaussians | 5000 | $10^{14}$ | 7 |
| Fullerene-encapsulated cyclic ozone (rocket fuel) [69] | 2nd Q, DF qubitization, Gaussians | $10^3$ | $10^{12}$–$10^{13}$ | 7 |
| | 2nd Q, sparse qubitization, plane wave dual | $10^4$ | $10^{15}$ | |

Table 4: Logical resource estimates for quantum phase estimation (QPE) applied to a range of molecular systems to compute a single energy eigenvalue. We list state-of-the-art resource estimates and refer to the references therein for prior estimates. The presented gate counts are for a single run of the phase estimation circuit. The number of samples is determined by the desired maximum failure probability, taken here as 0.1. In most cases, the number of samples is calculated using [44, Lemma 1] which assumes that the median value is taken from a number of incoherent repetitions of QPE. The success probability of a single run of QPE is $8/\pi^2$ [70]. QPE must be run a number of times if the overlap $\gamma \leq 1$, in general contributing a multiplicative cost of $\mathcal{O}(1/\gamma^2)$ (though this may be reduced, as described in the main text). "Algorithm" denotes the phase estimation unitary considered (e.g., Trotterization, Qubitization) as well as details about the quantization scheme (1st or 2nd), basis used, or factorization method used to compile the unitary (sparse [22], SF = single factorized [22], DF = double factorized [24], THC = tensor hypercontraction [25]). The resource estimates presented can be for different numbers of electrons and orbitals, accuracies, and can have differing classical simulation complexities. As such, the results may not be directly comparable, even within a single row of the table.

---

[a]This resource estimate assumed an overlap of $\gamma < 1$, and a lower failure probability of 0.01.

| Material(s) & references | Algorithm | Logical qubits | $T$/Toffoli gates per sample |
|---|---|---|---|
| Electron gas (prototypical model) [19, 71, 72, 7] | 1st Q, qubitization, plane waves | 1500–5000 [7] | $10^9$–$10^{14}$ [7] |
| | 2nd Q, qubitization [72]/Trotter [72], plane wave (dual) | 100–1000 [19, 72] | $10^8$–$10^{11}$ [19, 72] |
| Lithium-ion battery materials [73, 74, 10, 75] | 2nd Q, sparse/SF/DF/THC qubitization, Bloch orbitals | $10^5$–$10^6$ [10] | $10^{12}$–$10^{14}$ [10] |
| | 1st Q, qubitization, plane waves (w. pseudopotential) | ∼1000 [75] | ∼$10^{14}$ [75] |
| Transition metal catalysts nickel/palladium oxide [9, 5] | 2nd Q, sparse qubitization, Bloch/Wannier orbitals | $10^4$–$10^5$ [9] | $10^{10}$–$10^{13}$ [9] |
| | 1st Q, sparse qubitization, plane wave (dual) | 17,505 [5] | $10^{15}$ [5] |
| Magnesium/niobium alloys (corrosion resistant) [76] | 2nd Q, qubitization, plane wave (dual) | 9000–500,000 | $10^{14}$–$10^{19}$ |
| Nitrogen vacancy center in diamond (quantum sensing) [77] | 2nd Q, DF qubitization, plane wave (projector augmented-wave method) | 5000–150,000 | $10^{12}$–$10^{14}$ |

Table 5: Logical resource estimates for quantum phase estimation (QPE) applied to a range of material systems. We list state-of-the-art resource estimates and refer to the references therein for prior estimates. The presented gate counts are for a single run of the phase estimation circuit. The number of samples for all listed systems is 7, determined by the desired maximum failure probability, taken here as 0.1 (calculated using [44, Lemma 1] which assumes that the median value is taken from a number of incoherent repetitions of QPE. The success probability of a single run of QPE is $8/\pi^2$ [70]). QPE must be run a number of times if the overlap $\gamma \leq 1$, in general contributing a multiplicative cost of $\mathcal{O}(1/\gamma^2)$ (though this may be reduced, as described in the main text). "Algorithm" denotes the phase estimation unitary considered (e.g., Trotterization, qubitization) as well as details about the quantization scheme (1st or 2nd), basis used, or factorization method used to compile the unitary (sparse [22], SF = single factorized [22], DF = double factorized [24], THC = tensor hypercontraction [25]). The resource estimates presented can be for different numbers of electrons and orbitals, and can have differing classical simulation complexities. As such, the results may not be directly comparable, even within a single row of the table.

There have been comparatively few studies of the logical resources required for the simulation of chemical dynamics. Recent work has computed the resources required to calculate the energy loss of charged particles moving through a medium ("stopping power"), as pertaining to nuclear fusion experiments [78]. End-to-end resource estimates were determined, including the costs of initial state preparation, measurement of observables, and repetitions across a range of parameters. The resource estimates for the end-to-end task ranged from roughly 2000 logical qubits and order-$10^{13}$ Toffoli gates to roughly 30,000 logical qubits and order-$10^{17}$ Toffoli gates.

## Caveats

Existing resource estimates typically consider only a single run of phase estimation and assume that we have access to the desired energy eigenstate. As outlined above, both phase estimation and eigenstate filtering scale as $\Omega(\gamma^{-1}\Delta^{-1})$ when we have a lower bound on the gap. The "orthogonality catastrophe" suggests that the overlap of simple trial states with the desired eigenstate will decay exponentially as a function of system size. Although simulations will always be performed on finite-size systems, it is an open question as to when asymptotic behavior becomes problematic and whether initial states with sufficient overlaps can be prepared for systems of interest [29, 41, 30]. This issue may become more pressing for materials systems as we scale to the thermodynamic limit. In general, we know that the problem of finding the ground state of electronic structure Hamiltonians is QMA-hard [79], but it is not yet known if these complexity-theoretic statements provide intuition for physically realistic Hamiltonians.

As noted above, to accurately resolve the system, a large basis set must be used, as the discretization error decays as $1/N$ where $N$ is the number of spin orbitals considered. In practice, one typically repeats the calculation using increasingly accurate basis sets and then extrapolates to the continuum limit. Many quantum resource estimates consider active spaces with a small number of active orbitals, and so underestimate the resources required to achieve sufficiently accurate results to be informative. It is an active area of research to develop methods for increasing the resolution without increasing the basis set size, such as perturbative approaches, downfolding techniques, and embedding theories.

The end-to-end applications typically solved in the electronic structure problem can require between tens (e.g., structure determination, low temperature properties) and millions (e.g., biologically relevant molecular dynamics) of energy evaluations—each with different Hamiltonian parameters that may require preparing a new state to be measured. For example, a recent analysis of quantum algorithms applied to pharmaceutical chemistry [80] highlighted that to calculate the binding affinity between a drug molecule and its target (free energy differences) requires sampling a range of thermodynamic configurations, resulting in millions to billions of single-point energy evaluations. This introduces a large overhead when preparing a different state for each configuration and measuring its energy [59], although alternative approaches may provide more favorable scaling [81].

**Comparable classical complexity and challenging instance sizes**

The cost of exact diagonalization of the electronic structure Hamiltonian scales exponentially with the number of electrons and basis set size. As such, classical approaches to the electronic structure problem typically utilize a range of approximations that reduce their complexity to polynomial in the system size but introduce a (potentially uncontrolled) deviation from the exact ground state, leading to a bias in energy estimates and/or the expectation values of other observables. Approaches include Hartree–Fock (HF), density functional theory (DFT), perturbation theory, configuration interaction (CI) methods, coupled cluster (CC) methods, quantum Monte Carlo (QMC) techniques, and tensor network approaches. The cheapest approaches can be applied to thousands of orbitals but can be qualitatively inaccurate for strongly correlated systems. The most expensive approaches are more effective for strongly correlated systems, but their higher computational cost limits their applicability to roughly 100 spin orbitals. For example, [63] found that a density matrix renormalization group (DMRG) calculation performed on an 86 spin orbital active space of the cytochrome P450 enzyme molecule referenced in Table 4 required around 50 hours, using 32 threads, 48 gigabytes of RAM, and 235 gigabytes of disk memory. We also refer to [82] for a comparison of 20 first-principles many-body electronic structure methods applied to a test set of seven transition metal atoms and their ions and monoxides.

Due to their extended nature, material systems are most commonly targeted with DFT. DFT can be applied to systems with thousands of electrons and orbitals but can lead to uncontrolled energy bias in strongly correlated systems. QMC and tensor network methods have been successfully applied to prototypical models of material systems and are becoming increasingly practical for more realistic models. We refer to [83, 84, 85, 86] for benchmarks of classical electronic structure methods on hydrogen chains and Hubbard models scaling to the thermodynamic limit, which act as simplified models for real materials.

Many of the techniques discussed above for computing ground and excited states of chemical systems have been extended to explicitly simulate the time dynamics of the electronic Hamil-tonian. These include time-dependent HF, real-time time-dependent DFT, and time-dependent CI & CC methods. In general, the errors from the approximations made in these approaches are larger than for their static counterparts. We refer readers to [87] for a more detailed discussion of classical methods for real-time time-dependent electronic structure theory.

**Speedup**

It is nontrivial to determine the speedup of quantum algorithms for the electronic structure problem over their classical counterparts. If we consider the subtask of determining energy eigenstates, then for speedup greater than polynomial to be achieved, we require:

- The ability to prepare a trial state with nonexponentially vanishing overlap with the ground state as the system size increases.

- Polynomially scaling (with system size) classical algorithms having an exponential growth in their approximation parameter (e.g., bond dimension, number of excitations) as the system size increases.

Whether these two requirements can coexist in systems of interest is an active area of research [41, 30]. Even if exponential speedups are not available, it may be the case that quantum algorithms provide polynomial speedups over exact classical algorithms—and potentially over approximate classical algorithms [88].

From a complexity-theoretic viewpoint, we know that simulating the dynamics of a quantum system is a BQP-complete problem [89]. Combined with the observed difficulty of classically simulating the time evolution of electronic structure Hamiltonians, this may be taken as evidence for the possibility of an exponential speedup when simulating dynamics. In [31], quantum algorithms for simulating the fully correlated dynamics of electrons in a pseudo-spectral DVR or plane-wave basis [12, 51, 7] were compared against classical methods for mean-field dynamics. Large polynomial speedups were observed, ranging from superquadratic to seventh power in the salient parameters, depending on the relation between $N$ and $\eta$.

**NISQ implementations**

Solving the electronic structure problem is one of the most widely studied and promoted NISQ applications. The primary NISQ approach is the variational quantum eigensolver (VQE). There have been a number of experimental demonstrations on small molecules, for example, [90, 91], as well as proposals to simulate material systems [92, 93, 33]. Related methods, such as quantum computing–assisted QMC methods [94] have also been developed. Nevertheless, current device noise rates are too high to enable the running of circuits sufficiently deep that they can outperform classical electronic structure methods, and the number of circuit repetitions required to measure energy expectation values can be impractically large [95]. As such, there are several challenges that must be overcome if heuristic NISQ approaches are to scale to classically intractable system sizes and provide advantage over classical methods. There have also been proposals to simulate the electronic structure problem using analog quantum simulators [96], though to the best of our knowledge, these have not yet been experimentally demonstrated and are limited by the high-precision requirements of the electronic structure problem.

**Outlook**

Simulating the behavior of electrons in molecules and materials has repeatedly been identified as one of the most promising applications of quantum computing. Nevertheless, the discussion above highlights several challenges for current quantum approaches to become practical. Most notably, after incorporating the costs of initial state preparation and measuring observables, using larger active spaces to capture dynamic correlation, and including algorithmic repetitions to account for nonzero failure probabilities and sampling a range of parameters, a large number of logical qubits and total $T$/Toffoli gates may be required. The success of approximate classical methods for a wide range of chemical problems sets a high bar for quantum simulations to achieve advantage and encourages continued focus on resource estimations for end-to-end applications.

## 2.2 Simulating vibrations in molecules and materials

**Overview**

We seek the energy eigenstates, thermal states (i.e., statistical ensembles of eigenstates at a given temperature), or dynamics corresponding to the Hamiltonian that describes the vibrations of the nuclei in a molecule or material around their equilibrium positions. This Hamiltonian contains the kinetic energy of the nuclei and the effective potential that they move on, which is determined by the electronic potential energy surface (i.e., the electronic energy expressed as a function of the nuclear coordinates). It is also possible to consider nonadiabatic couplings between the vibrational degrees of freedom and electronic ("vibronic") or rotational ("ro-vibrational") degrees of freedom.

**Actual end-to-end problem(s) solved**

Classically solving the Schrödinger equation while treating electrons and nuclei on an equal footing has prohibitively high computational cost for all but the smallest systems. For systems where it is valid to separate the electronic and nuclear motions (the Born–Oppenheimer approximation), we can imagine the nuclei moving on the electronic potential energy surface (PES). For molecules composed of light atoms (where relativistic effects can be neglected), the vibrations of the nuclei around their equilibrium positions provide a first-order correction to the electronic energies and influence photo-emission/absorption properties. For a system with $K$ classical nuclei at equilibrium positions $\{R_I\}$, the vibrational Hamiltonian can be written as

$$H = -\sum_I \frac{\nabla_I^2}{2M_I} + V_e(\{R_I\}),$$

where $V_e(\{R_I\})$ denotes the nuclear potential determined by the electronic potential energy surface, obtained by first solving the electronic Hamiltonian for a range of nuclear positions. Computing vibrational eigenstates can be made classically tractable by modeling $V_e$ as a harmonic potential, which reduces the problem to solving a number of coupled quantum harmonic oscillators. The harmonic approximation can also be used when simulating vibronic transitions between vibrational energy levels on different PESs. However, due to the coordinate change between the PESs—which acts as a squeezing and displacement operation on the vibrational modes—exact simulation is #P-hard, and therefore inefficient for both classical and quantum algorithms. Nevertheless, vibronic spectra can be efficiently approximated using classical [1, 2] and quantum [3] algorithms in many regimes of interest.

To accurately describe nonrigid molecules or highly excited vibrational states, additional anharmonic terms are required in the potential. These can be obtained by performing higher-order fits of the potential, for example, by expanding the potential $V_e$ to degree $d$. Computing accurate solutions of this Hamiltonian is prohibitively costly for many systems of interest. Probing certain phenomena, such as vibronic spectra, internal conversion, intersystem crossings, and conical intersections, additionally requires the consideration of vibrations on multiple PESs and may require a description of the nonadiabatic couplings between the different PESs (which must often be explicitly determined [4]). We seek to prepare eigenstates or thermal states, or simulate the dynamics of the anharmonic vibrational Hamiltonian, and then measure the expectation values of observables with respect to these states. Properties of interest include the following:

- The vibrational energy at the minimum of the PES, which provides a first-order correction to the electronic energies—for calculating excitation energies, determining stable molecular structures, or finding reaction pathways and rates.

- Determining transition probabilities between states and transition dipole moments—for calculating infrared/Raman spectra between vibrational levels of the same electronic state or vibronic spectra between vibrational levels of different electronic states.

- Measuring the occupancy of vibrational modes as a function of time following excitation, to understand vibrational energy transfer and relaxation in chemical systems (i.e., internal conversion and intersystem crossings).

Thermal states $\rho \propto e^{-\beta H}$ (with $\beta = 1/k_B T$, where $k_B$ is the Boltzmann constant and $T$ the temperature) are often of greater interest in the vibrational case than in the electronic case; vibrational energies are on the order of $k_B T$ and so excited vibrational states are populated even at room temperature. This can be contrasted with the electronic structure problem, where the larger electronic energy gaps of many molecules mean that ground states are typically of primary interest at room temperature.

**Dominant resource cost/complexity**

A molecule with $K$ atoms has $M = 3K - 6$ ($M = 3K - 5$ for linear molecules) vibrational modes. Each vibrational mode excitation is treated as a distinguishable particle and so the wavefunction does not need to be explicitly symmetrized. The Hamiltonian is discretized using a basis set of vibrational modal functions $\{\chi_i\}_{i=1}^N$, for example, the truncated eigenfunctions of the quantum harmonic oscillator Hamiltonian, pseudo-spectral (Fourier) discrete variable representations, or grids.

The vibrational wavefunction can be stored in a first-quantized representation using $M \log(N)$ qubits, where the basis function of each vibrational mode is specified in binary (or an equivalent representation, such as the Gray code [5]). Alternatively, one can use second-quantized representation using $MN$ qubits [6].

Preparing the desired eigenstate or thermal state, or simulating the dynamics can be achieved using a range of quantum algorithms, including quantum phase estimation, quantum singular value transformation, adiabatic state preparation, variational quantum algorithms, Gibbs sampling, and Hamiltonian simulation. These methods are discussed in more detail for the electronic Hamiltonian, as the explicit costs of many of these methods have not yet been determined for simulating vibrations. Nonetheless, many of the same high-level considerations apply. The complexities of subroutines to prepare eigenstates and extract observables are determined by the following observations:

(i) All methods scale as $\Omega(1/\epsilon)$ to measure the desired observable to an error of $\pm\epsilon$. For the energy, we typically seek $\epsilon \sim (1-10)$ cm$^{-1} \approx (4.56 \times 10^{-6})$–$(4.56 \times 10^{-5})$ Hartree.[7] For comparison, the largest matrix elements in the vibrational Hamiltonian (the harmonic couplings) are typically on the order of 1000 cm$^{-1}$, and there are $\mathcal{O}(M)$ such terms [7]. As such, the ratio $\|H\|_1/\epsilon$ that features multiplicatively in the complexity of quantum phase estimation (at least, variants based on qubitization) can be on the order of $10^4$ (or larger) for modest system sizes with $M \approx 100$.

---

[7] Due to the close historical ties with spectroscopy, in vibrational chemistry it is common to see energies expressed as wavenumbers. Interconversion can be performed using the Planck relation.

(ii) To date, only product formula–based methods have been quantitatively studied in the context of providing coherent access to the vibrational Hamiltonian. If expanding the Hamiltonian as a linear combination of Pauli operators, the number of terms grows as $\mathcal{O}(M^d N^{2d})$ for a degree $d$ of anharmonic terms considered in the Hamiltonian (often at least 4th order). An alternative approach is to consider the Hamiltonian discretized by a real-space grid or Fourier pseudo-spectral basis, where the position and momentum operators can be easily applied [8, 9].

**Existing resource estimates**

To date, there have been no end-to-end resource estimates for the vibrational structure problem. In terms of initial steps in this direction, [5] considered the resources required to map vibrational operators to qubit operators, [7] compared the number and magnitude of terms in vibrational Hamiltonians to those in electronic structure Hamiltonians, and [10] estimated the number of terms and Trotter steps required to perform quantum phase estimation on polyyne molecules.

**Caveats**

Quantum algorithms and many (but not all [11]) classical algorithms for simulating the vibrations of nuclei require the availability of an electronic PES, from classical calculations. For a grid-based interpolation of the multidimensional PES with $h$ points per dimension, we require $\mathcal{O}(h^M)$ PES evaluations. Nevertheless, several interpolation techniques and adaptive methods have been developed to obtain high-accuracy PESs, at lower costs. Moreover, a few molecules with classically challenging vibrational spectra have been identified with classically easy-to-compute electronic PESs [7].

There has been less work on the number of vibrational basis states required to achieve a given accuracy than in the electronic case. While rigorous results exist for more simple bosonic Hamiltonians [12], the truncation level $N$ has not yet been established for anharmonic potentials.

**Comparable classical complexity and challenging instance sizes**

A hierarchy of approximate classical methods has been developed for computing vibrational eigenstates, which trade increased accuracy for increased cost. Vibrational states with a multireference nature—which are required to describe vibrational resonances that arise due to near degeneracies between different vibrational eigenstates, resulting from anharmonicities in the PES—require more accurate (and thus costly) methods. Moreover, nonrigid molecules require a higher-degree approximation of the PES, leading to an increased cost for classical methods—and potentially increasing the complexity of the resulting eigenstates. For such challenging systems, accurate classical results have been obtained for molecules with $G = 20$–30 atoms [13, 14, 15, 16].

Recently, a number of quantum-inspired classical algorithms (see [2, 17] and references therein) have been developed for classically approximating the results of Gaussian boson sampling experiments. These experiments have been proposed as analog simulators of harmonic vibrational phenomena, including vibronic spectra and vibrational dynamics (see §NISQ implementations, below). This reduces the regime where quantum advantage may be possible and reinforces the necessity of considering anharmonicities in the vibrational Hamiltonian [2].

For a review of classical algorithms for simulating coupled vibrational and electronic degrees of freedom, we refer to [11, 18]. Commonly used algorithms include multiconfigurational time-dependent Hartree (MCTDH) and the ab initio multiple spawning method (AIMS).

**Speedup**

The speedup for quantum algorithms in computing vibrational eigenstates depends on the overlap and error convergence of classical trial states to the true eigenstate and energy. This has yet to be determined for systems of interest. Nevertheless, in spectroscopic calculations that start from a classically easy-to-prepare state, the overlap between the initial state and a desired excited state becomes a quantity of interest—corresponding to the absorption intensity—rather than a limiting factor on the calculation. For example, in [19] it was proposed to use quantum phase estimation to project from the initial state into other eigenstates with probability given by the squared overlap between the states. However, while a single (exponentially costly) classical diagonalization of the vibrational Hamiltonian would provide complete access to the entire vibrational spectrum, a large number of repetitions of the quantum algorithm would be required to reconstruct the spectrum.

As discussed above, the development of recent quantum-inspired algorithms [2, 17] has reduced the prospect of achieving quantum advantage for vibrations in harmonic potentials, motivating the need to include anharmonicities in the vibrational Hamiltonian, or to consider more complex models that nonadiabatically couple vibrational and electronic degrees of freedom. We refer to [18] for a discussion of the classical complexity of simulating such models.

**NISQ implementations**

There have been proposals to apply variational algorithms to solve the vibrational structure problem [20, 6, 5, 7], but additional developments are required in order to implement sufficiently deep circuits to surpass classical methods, and the number of circuit repetitions required to measure energy observables is likely a bottleneck [7].

There have also been several proposals and experimental demonstrations for using analog quantum simulators to simulate molecular vibrations. Analog simulations have been performed for zero and finite temperature vibronic spectra [3, 21] as well as vibrational dynamics [22]. We note that these approaches use harmonic approximations for the vibrational potential, and can be approximated efficiently by classical algorithms in some regimes [2]. There have also been analog quantum simulations of systems with coupled electronic and vibrational degrees of freedom (typically via a linear vibronic coupling model) including simulations of conical intersections [23, 24, 25] and vibrational assisted energy transfer [26].

**Outlook**

Further work is required to identify target systems that are challenging to simulate classically, but that may be amenable to quantum algorithms. In addition, existing quantum algorithms need to be further optimized for the accuracy required in vibrational structure problems and the form of the vibrational Hamiltonian. This will enable resource estimates for end-to-end applications, such as estimating vibrational spectra or simulating vibrational dynamics.

# 3 Nuclear and particle physics

Simulating nuclear and particle physics is an inherently quantum problem. There have been proposals to use quantum computers to accelerate simulations of quantum field theories, nuclear physics, neutrino physics, and quantum gravity [1]. In this section, we focus on the simulation of quantum field theories and nuclear physics, as these have received the most attention in the literature to date and are the closest to having end-to-end resource estimates available. While not covered explicitly in this section, the building blocks of quantum algorithms for data analysis in high-energy physics [2] can be found in Section 20 on variational quantum algorithms and Section 9 on machine learning. For existing reviews of quantum computing for nuclear and particle physics, we direct the reader to [1, 3, 4, 5, 6, 7, 8].

*The authors are grateful to Zohreh Davoudi and John Preskill for reviewing this section of the survey.*

## This application area contains:

## 3.1   Quantum field theories

**Overview**

We seek the static and dynamic properties of quantum field theories, specifically gauge field theories and scalar field theories. Gauge field theories describe the interactions between matter and/or gauge degrees of freedom and can be classified by their symmetry groups, such as U(1) (describing quantum electrodynamics), SU(2) (the weak interaction), and SU(3) (quantum chromodynamics). Scalar field theories describe interactions between scalar fields, such as the Higgs field or $\phi^4$ theory.

Interacting quantum field theories are typically not analytically solvable, and techniques such as perturbation theory are only accurate in some parameter regimes. For example, low-energy quantum chromodynamics (QCD), relevant to quark confinement and hadron formation, cannot be treated perturbatively. As such, complex scattering processes at particle accelerators are currently treated with a combination of first-principles calculations and approximate phenomenological methods.

To tackle quantum field theories numerically from first principles, lattice field theory is employed. The Lagrangians arising from lattice field theory can be numerically solved on classical computers using Euclidean Monte Carlo methods, which have proven highly efficient and accurate for a number of static quantities, including hadron masses, static matrix elements of (primarily) time-local operators between hadronic states, and even certain properties of light nuclei [1, 2, 3]. However, these classical Monte Carlo methods become intractable due to a sign problem in two regimes: (i) at high fermion density (of considerable scientific interest for understanding the decomposition of neutron stars and large atomic nuclei) and (ii) in simulations of real-time dynamics (e.g., scattering problems). Hamiltonian formulations of these problems are challenging due to the size of the required Hilbert space. As such, there have been a number of proposals to use quantum computers for calculating the static and dynamic properties of matter described by scalar and gauge field theories. For further background, see [4, 5, 6] and references therein.

**Actual end-to-end problem(s) solved**

Classical computational methods for lattice field theories have produced a number of insights, including high-precision computations of fundamental quantities (such as the muon's magnetic moment and quark masses), tests of beyond-the-Standard-Model physics (such as charge conjugation and parity (CP) violation and beyond-Higgs theories), and nuclear cross sections with dark matter candidates or neutrinos. For a more complete and detailed list, we refer the reader to [2, Page 6].

We primarily focus on the case of lattice field theories in the Hamiltonian formulation, which explicitly separates temporal and spatial degrees of freedom [7] and discretizes the $d$-dimensional space using an $L^d$ lattice (which may be noncubic). Matter degrees of freedom (e.g., quarks, scalar fields) are placed on the vertices of the lattice. Gauge degrees of freedom (e.g., the value of the electromagnetic field) are placed on the links between lattice sites. Dynamical simulations proceed by initializing the system in a desired state [8], performing time evolution under the Hamiltonian, and measuring relevant observables. See [9] for an example of simulating scalar field theories. Static simulations aim to prepare a state of interest, such as the ground state of a collection of quarks representing a composite hadron, and then measure observables of interest, including binding energies, as well as structural and reaction properties.

Quantum simulations of lattice field theories may be incorporated as part of a larger (multiscale) computational workflow. For example, when studying scattering processes (such as those that occur at particle accelerators), it is not necessary to simulate the entire scattering process on a quantum (or classical) computer. Instead, the scattering cross section can be separated into short- and long-distance contributions, which can be computed using perturbative and nonperturbative (e.g., quantum or classical simulation of lattice gauge theories) methods, respectively [10, 11].

**Dominant resource cost/complexity**

In this section, we focus predominantly on the simulation of dynamics in lattice gauge theories (LGTs), as the majority of studies to date have considered this application. In the standard formulation, one allocates one qubit per fermion (or antifermion) type per site of an $N = L^d$ lattice. Each gauge degree of freedom (one in U(1), three in SU(2), eight in SU(3)) requires its own register associated with each edge between lattice sites. The quantum numbers associated to the gauge degrees of freedom are encoded in binary, up to a maximum cutoff value $\Lambda$, so the corresponding register requires $\log(\Lambda)$ qubits. It was shown in [12] that for time evolution performed with fixed lattice spacing, the cutoff can be set as $\Lambda = \Lambda_0 + \widetilde{\mathcal{O}}(T) \cdot \mathrm{polylog}(N/\epsilon)$, where $\Lambda_0$ is the maximum initial value of the gauge fields, $T$ is the time evolution duration, and $\epsilon$ is the resulting error in the final state. Hence, the overall number of qubits required to store the state of the system scales as

$$\mathcal{O}\left(L^d \log\left(\Lambda_0 + T\,\mathrm{polylog}\left(\frac{L^d}{\epsilon}\right)\right)\right).$$

Algorithms for implementing time evolution under LGT Hamiltonians are presented in [12, 13, 14, 15, 16, 17, 18]. It is necessary to (approximately) maintain gauge invariance during the simulation, which can be achieved either by the choice of formulation, by actively protecting symmetries [19, 20], or by detecting and eliminating the gauge-violating states [21, 22]. As an example of the first option, one can calculate the desired Hamiltonian matrix elements on the fly using Clebsch–Gordon coefficients [23], but this is expensive in terms of elementary quantum operations [14]. An alternative approach described in [16] encodes only the physical transitions in the SU(3) gauge theory. This method requires many controlled operations and a large classical precomputing overhead.

**Existing resource estimates**

The number of $T$ gates required to simulate instances of the lattice Schwinger model (U(1) LGT in $d = 1$ with both matter and gauge degrees of freedom) was studied in [13]. That work considered the resources required to perform Trotterized time evolution and estimate the electron-positron pair density. The most complex simulations analyzed (64 lattice sites, cutoff of $\Lambda = 8$) required $5 \times 10^{13}$ $T$ gates per shot, and 333 logical qubits. Such a circuit would need to be repeated $\mathcal{O}(1/\epsilon^2)$ times to estimate the pair density to accuracy $\epsilon$; this overhead could be improved to $\mathcal{O}(1/\epsilon)$ using quantum amplitude estimation at the expense of increased gate depth [13]. These estimates were later improved in the small system-size, long-time, or low-error regime using algorithms based on qubitization with quantum signal processing [24]. Note that a simulation of the 64-site lattice Schwinger model with $\Lambda = 8$ is well within the range of classical simulations [25, 26].

Reference [14] performed similar resource estimates for the simulation of dynamics in U(1), SU(2), and SU(3) LGTs for $d = 3$. These resource estimates were performed for synthesizing the time evolution operator, with choice of simulation parameters inspired by the following tasks: computing transport coefficients relevant to the study of quark-gluon plasmas, simulating heavy ion collisions, and computing the hadronic tensor of the proton—although we note that the costs of initial state preparation and observable measurements are not included in these resource estimates. Logical qubit counts ranged from $10^4$ to $10^8$, while $T$ gate counts ranged from $10^{17}$ to $10^{56}$. The large constant factors present in these resource estimates stem partly from the use of quantum arithmetic (e.g., constituting 99.998% of the gate count in the most expensive calculations [14]), and partly from the decomposition of the plaquette term, which is exponential in the number of colors. The large gate counts also arise from the value chosen for the error in the time evolution operators $\epsilon = 10^{-8}$, which may be overly conservative when viewed in conjunction with other sources of error.

These gate counts can be reduced significantly using the approach of [17], which investigated alternative ways of representing the simulation, allowing for an improved grouping of terms in the product formula. These ideas were illustrated for SU(2) in 1+1D, but can likely be generalized to more complex, higher-dimensional LGTs.

Nevertheless, we note that any implementation scaling as $\Omega(TL^3)$ (i.e., linearly in spacetime volume) already faces a factor of $10^8$ for $T = L = 100$, highlighting the potentially large resource counts of simulating quantum field theories. Recent work presented an algorithm for simulating the time evolution of LGTs that achieves this optimal complexity, up to polylogarithmic factors [18]. This work uses a number of subroutines for Hamiltonian simulation [27, 28, 29] and requires locality-preserving fermion- and boson-to-qubit mappings. Resource estimates were carried out for U(1), SU(2), and SU(3) LGTs. For simulating time evolution in an SU(3) LGT on a lattice with $T = L = 100$, the algorithm required approximately $10^{21}$ $T$ gates and $6 \times 10^7$ logical qubits. Despite the large improvements compared to [14], the significant discrepancy from the expected lower bound of $10^8$ discussed above suggests that there is further opportunity for optimization in the implementation of algorithms for simulating LGTs.

## Caveats

Additional investigation is required to better quantify the theoretical uncertainties arising from discretization, finite-volume, and Hilbert space truncation effects of quantum computing formulations, as well as the algorithmic errors present in quantum simulation algorithms applied to lattice and scalar field theories.

For example, discretization of the continuous field theory to the lattice setting introduces a number of nuances (which are also present in classical approaches but must be considered afresh in quantum calculations). As discussed in [30, 31], discretization of the fermion field breaks the Lorentz invariance of the fermion kinetic term, which introduces unphysical additional flavors of fermions (known as the fermion doubling problem). This issue can be mitigated in several established ways, each with their own merits and drawbacks for quantum simulation [32]. It is also necessary to carefully track other errors resulting from discretization and ensure that these vanish when scaling and extrapolating to the continuum limit [9, 33].

As noted in [5, Section 6b] and [16, 34], there are a number of possible representations/basis sets that can be used for the gauge degrees of freedom, and it is currently unclear which choice is optimal for quantum simulation.

**Comparable classical complexity and challenging instance sizes**

The end-to-end scattering processes typically considered at particle accelerators are too complex to be solved from first principles and are instead tackled using a range of approximate techniques [10]. These computations often include parameters obtained from first-principles LGT calculations on simpler systems, and they typically proceed through a Lagrangian formulation, rather than a Hamiltonian formulation. This leads to Monte Carlo sampling of a path integral in Euclidean spacetime, the application of which to dynamical problems or static problems with high fermion density is limited by the fermionic sign problem [35]. For example, the phase diagram of QCD and the existence of exotic phases at extreme densities, in and out of equilibrium, have eluded classical methods. Nevertheless, classical approaches have been very effective for static problems with low fermion density and for dynamical scattering problems at low energy and low inelasticity; for a review of current state-of-the-art computations and limitations, see [2, 36] and companion whitepapers referenced therein.

Recent work has begun to investigate using tensor network methods to simulate the Hamiltonian formulation of LGTs; see, for example, [25] ($d = 2, L = 16$, U(1) LGT with gauge field cutoff $\Lambda = 1$) and [26] ($d = 3, L = 8$, U(1) LGT with gauge field cutoff $\Lambda = 1$). Like quantum simulations, tensor network approaches are sign-problem free and thus may be of interest in regimes out of reach of conventional Monte Carlo–based approaches. However, tensor network approaches are currently limited to small system sizes, and often need to be verified by comparing to other methods, as they do not come with provable guarantees on the bond dimension required for capturing the entanglement structure of the states present in LGTs. For recent reviews on the use of tensor networks to simulate LGTs, we refer the reader to [37, 38].

**Speedup**

For simulations with a sign problem, classical Monte Carlo methods are exponentially costly in system size [39]. In addition, it was observed that the bond dimensions required for tensor network approaches increase rapidly with system size [26], suggesting the potential for exponential quantum speedups for dynamical problems. This suggestion is reinforced by the BQP-completeness of the simulation of certain field-theoretic processes [40]. Nevertheless, the constant prefactors for quantum simulations of LGTs are currently high, and we require the (currently underexplored) ability to efficiently prepare initial states of interest.

**NISQ implementation**

A number of works have investigated the simulation of scalar or gauge field theories on noisy digital, or analog, quantum simulators. A common strategy is to map the lattice field theory Hamiltonian to that of a bosonic system, such as cold atoms or trapped ions; see, for example, [5, 34, 41, 42] and references therein. While these techniques appear promising for simple Hamiltonians, such as the Schwinger model, it may be challenging to engineer the more complicated interactions required in nonabelian gauge field theories. There have also been works applying variational algorithms to lattice field theories, such as [43, 44, 45], as well as digital simulations of the time dynamics of the lattice Schwinger model [46, 47].

## Outlook

Investigations into how quantum computers can be used to complement classical methods for simulating lattice field theories are advancing rapidly. While quantum computers can, in principle, efficiently simulate the complex scattering experiments performed in particle accelerators, the resources required to do so would be impractical using currently known techniques. Future work must determine the best targets for quantum simulations and reduce asymptotic scaling as well as constant prefactors. In particular, the qubit encoding (currently scaling as $\mathcal{O}(L^d)$ qubits for a lattice in $d$ spatial dimensions with each dimension having $L$ sites) means that a large number of logical qubits will likely be required for computations of interest where, as illustrated by examples above, we may consider $L = 10$–$100$ to challenge classical approaches.

## 3.2 Nuclear physics

**Overview**

Nuclear physics describes the behavior of individual nuclei, as well as that of dense nucleonic matter, such as neutron stars. The structure of nuclei can be approximately described using the shell model (see [1] for an overview), a phenomenological model with parameters fitted to experimental observations. However, high-accuracy descriptions of nuclear structure, exotic nuclei, accurate scattering cross sections, and nonequilibrium phenomena require a first-principles treatment. Describing the properties of nuclei from first principles (e.g., lattice quantum chromodynamics simulations) is beyond the reach of analytic and current computational capabilities for all but the simplest nuclei [2, 3, 4]. Nevertheless, one can often integrate out the short-range physics to obtain effective field theories (EFTs) that describe the interactions of nucleons. The prototypical example is chiral effective field theory, which describes the interactions of nucleons and pions (pions, which have mass less than $6\times$ smaller than that of the proton, are mediators of the residual strong interaction between pairs of nucleons). The parameters of the EFT can be inferred from experiments or directly from lattice quantum chromodynamics (QCD) calculations, resulting in a many-body Hamiltonian that describes the formation, structure, and potential decay of nuclei.

**Actual end-to-end problem(s) solved**

An EFT provides a many-body Hamiltonian describing how nucleons interact. This quantum many-body problem can be tackled using a range of classical methods, which employ different mathematical approaches to approximately solve the nuclear many-body Schrödinger equation including quantum Monte Carlo (QMC) methods [5], the no-core shell model (NCSM) [6], the coupled cluster (CC) method [7], the self-consistent Green's function (SCGF) method [8], the in-medium similarity renormalization group (IMSRG) method [9], and nuclear lattice methods [10] (a related review article discussing inputs to these calculations is given by [11]).

A common problem is to prepare the ground state of a collection of nucleons, in order to compute nuclear binding energies and determine if a given nucleus is stable (e.g., determining the long lifetime of $^{14}$C [12, 7]). Simulations can also be used for computing scattering cross sections, in order to analyze experiments on nucleus-neutrino scattering [13], beta decay [14], and nuclear reactions. Reactions, such as nuclear fission and nuclear fusion, can also be studied using explicitly time-dependent approaches [15], although these have higher computational costs than static computations and are often based on semiclassical, mean-field, or other phenomenological models. Simulating both fusion and fission reactions has a number of use cases, such as an improved understanding of nuclear astrophysics, where reactions commonly occur at energies too high or too low to be replicated in experiments [16].

**Dominant resource cost/complexity**

The prototypical EFT for nuclear interactions is chiral effective field theory. At very low energies, the theory can be expressed as a convergent perturbative expansion (chiral perturbation theory). However, for the larger energies relevant to multi-nucleon systems, the theory becomes nonperturbative. Despite several notable successes, developing EFTs that converge across a wider range of scenarios remains an active area of research [11, 17].

At low energies (below the rest mass of the pion), it can be appropriate to apply pionless EFTs, which integrate out the pions, leading to implicit interactions between nucleons, including a 3-nucleon contact interaction required at leading-order by renormalization. At higher energies, it is necessary to explicitly account for the effect of pions. Pionfull EFTs are typically studied numerically, as it can be difficult to obtain analytic predictions [11]. In a formulation known as the one-pion-exchange Hamiltonian, pions are integrated out, leading to a long-range two-body interaction between nucleons, which decays exponentially with distance. In an alternative formulation, known as dynamical pion EFT, the (relativistic) pions and their interaction with nucleons are explicitly simulated.

A common formulation of EFT simulations is to project the problem onto a lattice in position or momentum space [18]. For quantum simulations formulated on a lattice, a typical second-quantized mapping uses 4 qubits per lattice site for nucleons (two isospin degrees of freedom and two spin degrees of freedom), although additional qubits may be required if using a fermion-to-qubit mapping that maintains the locality of the fermionic Hamiltonian (e.g., 6 qubits per site [19]). If simulated explicitly (e.g., dynamical pion EFT), the value of the spin-0 pion field at each lattice site can be stored using a number of qubits scaling logarithmically with the pion energy cutoff by storing its quantum number in binary.

An alternative approach is to project the EFT Hamiltonian onto a single-particle basis, commonly harmonic oscillator eigenfunctions [20]. In second quantization, a qubit is required per single-particle (iso)spin mode. However, this mapping can lead to long-range interactions between modes, and in the most general case, up to $\mathcal{O}(N^6)$ distinct terms for an $N$-mode system [21, 22].

Quantum algorithms that prepare energy eigenstates (or good approximations thereof) scale either as $1/\gamma$ (where $\gamma$ is the overlap of the initial state with the desired eigenstate) [23], or with the minimum gap size along an adiabatic or thermalizing path. If we are only interested in measuring the energy of the state, this can be obtained using the quantum phase estimation algorithm, which also projects the system into the corresponding energy eigenstate. The cost of this approach scales as $\mathcal{O}(1/\gamma^2)$ in terms of the original overlap $\gamma$, which can be improved to $\mathcal{O}(1/\gamma)$ using amplitude amplification at the expense of increased circuit depths. Once the desired state has been prepared, observables can be measured to precision $\epsilon$ with complexity $\mathcal{O}(1/\epsilon^2)$ (direct sampling) or $\mathcal{O}(1/\epsilon)$ (using amplitude estimation, also requiring coherent state preparation, e.g., via amplitude amplification).

The above algorithms for preparing states (and related algorithms for performing time evolution in dynamics simulations) require access to the Hamiltonian, typically implemented via block-encoding or Hamiltonian simulation.

**Existing resource estimates**

The gate costs of Hamiltonian simulation using product formulas for neutrino-nucleus scattering in pionless EFT were studied in [13]. The $T$-gate costs of Hamiltonian simulation using product formulas, as well as quantum phase estimation, were estimated in [19] for pionless EFT, one-pion-exchange EFT, and dynamical pion EFT (all formulated on a lattice). As an example, resources to simulate the dynamics of 40 nucleon system ranged from $4 \times 10^{12}$ $T$ gates and 6000 logical qubits for pionless EFT, to $2.0 \times 10^{24}$ $T$ gates and 6000 logical qubits for one-pion-exchange EFT, to $5.2 \times 10^{49}$ $T$ gates and 168,000 logical qubits for dynamical pion EFT (all to an error $\epsilon = 0.1$). The resources required for the latter EFT are considerably higher, due to the qubit overhead and worse error dependence that stems from explicitly simulating the pions.

Resource estimates for other parameter regimes can also be found in [19], where it is noted that higher-order EFTs are required for the accurate simulation of medium- and large-mass nuclei, which may further increase simulation costs.

## Caveats

Developing accurate nuclear EFT Hamiltonians describing heavier nuclei is an active research direction. Most studies are currently based on phenomenological models, which are limited in their predictive capabilities. In some systems, some higher-order EFTs provide sufficient accuracy, but only after fitting the EFT coefficients to experimental data. We refer the interested reader to [11] for a more detailed discussion.

## Comparable classical complexity and challenging instance sizes

Classical approaches use similar techniques to those developed for the electronic structure problem, such as perturbation theory, Monte Carlo methods, or coupled cluster approaches. References [5, 6, 7, 8, 9, 10, 24] provide excellent overviews of state-of-the-art approaches. Classical methods can provide outstanding agreement with experiments for the binding energies of small nuclei with 20–50 nucleons [24]. As a further example, recent high-accuracy simulations of the $^{100}$Sn nucleus have improved the agreement between theory and experiment for observed $\beta$-decay rates [14]. Time-dependent simulations of dynamics or nonequilibrium phenomena are more challenging and are an active area of research [15, 16]. We refer readers to [25] for a detailed analysis of the capabilities and requirements of exascale supercomputers in nuclear physics simulations.

## Speedup

The majority of classical approaches for the nuclear structure problem are designed to run in polynomial time with respect to the system size but introduce errors due to the use of approximations (e.g., truncating the expansion in coupled cluster methods) [24]. For quantum computers to achieve exponential speedups, one needs to identify systems where (i) classical methods require an exponential increase in resources to obtain accurate results and (ii) it is efficient to prepare an initial state for the quantum computation with at least inverse polynomially large (in terms of the system size) overlap with the desired state. Recent initial investigations have explored whether these requirements coexist in chemical systems [26]. We are not aware of similar work in nuclear physics, although it has been noted that the states obtained from classical methods could be used as initial states for quantum algorithms [27].

For simulating the dynamics of nuclear systems, classical methods typically proceed via mean-field methods, and exact simulations are limited to small system sizes due their exponential scaling. In contrast, quantum algorithms are able to simulate the dynamics of nuclear systems with a cost scaling polynomially with the system size and desired accuracy.

## NISQ implementation

Almost all of the work to date on demonstrating near-term quantum computing approaches for the nuclear structure problem has focused on variational algorithms, such as [28, 29, 30, 31], or small-scale simulations of dynamics [13, 32]. There is currently no evidence that near-term quantum devices will be able to implement sufficiently deep circuits to achieve advantage over their classical counterparts with these methods.

## Outlook

Nuclear physics presents a classically challenging quantum many-body problem that appears well suited to simulation on quantum computers. While there are similarities to the electronic structure problem in quantum chemistry, which has led to a transfer of several ideas, nuclear Hamiltonians are typically more complex, involving (depending on the formulation) long-range interactions, 3-body interactions, multiple species of nucleons and pions, and interactions with nontrivial spin and isospin dependence. The simulation of nuclear reaction dynamics appears a particularly interesting target, and future work should determine the resources required for end-to-end simulations, including state preparation and measurement of observables.

# 4 Combinatorial optimization

Combinatorial optimization problems are tasks where one seeks an optimal solution among a finite set of possible candidates. In industrial settings, combinatorial optimization arises via scheduling, routing, resource allocation, supply chain management, and other logistics problems, where it can be difficult to find optimal solutions that obey various desired constraints. The field of operations research—which came to prominence after its application to logistics problems faced by World War II–era militaries—applies methods of combinatorial optimization (as well as continuous optimization) to these problem areas for improved decision-making and efficiency in real-world problems.

Combinatorial optimization problems are also at the heart of classical theoretical computer science, where they are used to characterize and delineate complexity classes, such as P and NP. Typical combinatorial optimization problems have limited structure to exploit, and therefore quantum computing most often only provides polynomial speedups. In fact, it came as a surprise in the early days of quantum computing research that for a wide variety of such problems, quantum computers do offer up to quadratic speedups via Grover's search algorithm [1]. Subsequently, much effort was devoted to understanding how Grover search and its generalization, amplitude amplification, offer speedups for various combinatorial optimization problems.

In this section, we cover several distinct approaches to solving combinatorial optimization problems. First, we look at combinatorial optimization through its relation to search problems, where Grover's algorithm, or its generalizations, can be applied to give a quadratic or subquadratic speedup. Then, we cover several proposals—variational algorithms (viewed as an exact algorithm), the adiabatic algorithm, and the "short-path" algorithm [2, 3]—that have the potential to surpass the quadratic speedup of Grover's algorithm. We discuss the (limited) evidence that these approaches could generate significant advantages, as well as the associated caveats.

We do not specifically cover the large body of work on quantum approaches for *approximate* combinatorial optimization (typically variational quantum algorithms or quantum annealing). These algorithms usually translate the optimization problem to energy minimization of a spin system with a Hamiltonian that encodes the classical objective function. They apply some physically motivated heuristics to efficiently generate solutions that have low energy, and seek a better objective value than could be generated classically in a comparable amount of time. An advantage of these approaches is that they are often more compatible with noisy near-term hardware. While approximate optimization remains an interesting direction, these quantum algorithms are heuristic and there is a general scarcity of concrete evidence that they will deliver practical advantages.

We refer the reader to [4] for a comprehensive survey of quantum methods for combinatorial and continuous optimization.

*The authors are grateful to Ashley Montanaro for reviewing this section of the survey.*

**This application area contains:**

## 4.1   Search algorithms à la Grover

**Overview**

Grover's search algorithm [1] and its generalizations, such as amplitude amplification, are essential sources of quantum speedups. A straightforward application of Grover search in the spirit of optimization is quantum minimum finding [2, 3], which provides a quadratic speedup for finding the minimizer of a function on a given set of elements.

As search is a generic primitive, Grover's algorithm is widely applicable, and it can speed up many subroutines, especially in algorithms for combinatorial optimization. We list a few representative applications that demonstrate how Grover's algorithm may be applied to speed up combinatorial optimization.

**Actual end-to-end problem(s) solved**

The goal is to solve a search problem, that is, decide whether there is an element among a set of objects that satisfies some criterion, and if there is such an object, find one. Many combinatorial optimization problems are fundamentally search problems; a notable class of examples are graph problems, such as finding a maximal independent set, a $k$-coloring, a lowest weight Hamiltonian cycle[8] (called the traveling salesperson problem), or the shortest path between two vertices.

For conceptual clarity, here, we focus on the prototypical Boolean satisfiability problem, that is, SAT solving: given a Boolean formula in the so-called *conjunctive normal form*, decide whether it has a satisfying Boolean assignment (and if so, find one). A formula in this form consists of some constraints (called *clauses*) each containing the logical AND of some variables or their negation (called *literals*). We denote the number of Boolean variables by $n$ and the total number of literals of the formula by $\ell$ (typically $\ell \geq n$ since each variable should appear at least once).

**Dominant resource cost/complexity**

If there are at least $m$ marked elements among $N$ possible ones, then the search problem can be solved with high probability by using $\mathcal{O}(\sqrt{N/m})$ Grover iterations. Each Grover iteration requires generating a uniform superposition over the $N$ elements starting from the all $|0\rangle$ state and checking whether an element is marked (in superposition), which can be implemented with gate cost $\mathcal{O}(\ell + n)$. If the formula is satisfiable, then there is at least one solution, thus $\mathcal{O}(\sqrt{2^n})$ Grover iterations suffice, giving an overall complexity of $\mathcal{O}((\ell + n)\sqrt{2^n})$.

In some applications, it is useful to consider a generalization of Grover search, amplitude amplification, which enables working with an arbitrary prior distribution on the elements, unlike Grover's algorithm which effectively uses a uniform prior. The relevance of this extension can be seen through the example of 3-SAT, which is a restricted version of SAT where each clause has at most 3 literals. A clever application of amplitude amplification described by Ambainis [4] for solving 3-SAT more efficiently uses Schöning's algorithm [5] and thus generates a nontrivial prior distribution on the solutions.

The complexity of amplitude amplification is similar to that of Grover search in general. If $|\psi\rangle$ is the quantum state representing the prior distribution, so that measuring the state yields a marked element with probability at least $p$, then $\mathcal{O}(\sqrt{1/p})$ "Grover iterations" suffice to find a marked element with high probability. The algorithm requires preparing the initial state $|\psi\rangle$, and then each iteration consists of a reflection $2|\psi\rangle\langle\psi| - I$ around $|\psi\rangle$ and checking whether an element is marked (in superposition). The former reflection can be implemented with two uses of the circuit that prepares $|\psi\rangle$ from the all $|0\rangle$ state, and a reflection about the all $|0\rangle$ state.

---

[8]A Hamiltonian cycle in a graph is a cycle that visits each vertex once, not to be confused with a quantum Hamiltonian.

**Existing resource estimates**

There are several studies on the resource estimation of Grover-type (sub)quadratic speedups. Due to the wide range of these problems, we do not focus on explicit gate counts on any particular problem/implementation variant, but rather list some prominent articles and illustrate their findings at a high level [6, 7, 8, 9, 10, 11]. Unfortunately, these recent studies revealed that quadratic or smaller speedups alone are unlikely to be useful for the foreseeable future, unless the large overheads of current fault-tolerant quantum computing schemes can be greatly reduced. For example, [6] concluded that even if there is some reasonable advantage in quantum gate counts for solving the constraint satisfaction problems that they consider, the classical computation supporting the fault-tolerant quantum computation actually voids the speedup in practice. They state that "Even when considering only problem instances that can be solved within one day, we find that there are potentially large quantum speedups available. ... However, the number of physical qubits used is extremely large, ... . In particular, the quantum advantage disappears if one includes the cost of the classical processing power required to perform decoding of the surface code using current techniques." The most recent of the references listed above [11] estimates that achieving a quantum advantage via a quadratic speedup requires at least a month-long computation already if each iteration contains at least one floating-point operation. The situation looks more promising for cubic and quartic speedups, but unfortunately such improvements seem to require techniques beyond Grover search.

**Caveats**

Grover originally described his result as "A fast quantum mechanical algorithm for database search" [1]. If we work in the circuit model of quantum computation, then strictly speaking Grover search gives a slowdown for database search, as every Grover iteration needs to "touch" every element in the database. If we anyway need to touch all $N$ elements in the database, then the best we can do is to simply go over every element in linear time $\mathcal{O}(N)$. Grover's search circuit, on the other hand, would have gate complexity $\widetilde{\mathcal{O}}(N^{3/2})$, clearly worse than sequentially going through the entire dataset.

In the database scenario, we can only recover the quadratic speedup if we assume that we can use a quantum random access memory (QRAM), with constant (or logarithmic) cost for each database query. The analogous assumption regarding ordinary RAM is often made in classical computer science, simply because RAM calls are cheap in practice. However, since a RAM call should be able to touch every bit of the database, from a circuit complexity perspective a RAM call must have gate cost at least $N$. On the other hand, from a time complexity perspective, one can view a RAM call as a massively parallel piece of computation implementable in a binary tree structure with logarithmic depth.[9] While QRAM can also be implemented with a quantum circuit of $\mathcal{O}(\log(N))$ depth, a similar accounting might not be fair in the quantum case, depending on the eventual cost of hardware implementation—especially if error correction of the QRAM is necessary and the entire QRAM circuit is implemented in a fault-tolerant fashion.

---

[9]Viewing RAM as a low-depth circuit disregards issues regarding signal transmission. Considering that the speed of light is finite and we have only 3 dimensions to fit the memory cells into, a RAM call should asymptotically cost at least $\sqrt[3]{N}$ time. In fact, state-of-the-art clock speeds are already in a regime where the speed of light may be a bottleneck, so we might eventually need to reconsider how the time complexity of RAM is modeled.

Nevertheless, Grover's algorithm can provide a quadratic speedup without extra hardware assumptions when the elements of the list that we search over can be easily generated and checked "on the fly." For example, in the case of SAT, we search over the $2^n$ possible truth assignments, yet we can easily check whether an individual assignment is satisfactory by simply substituting the assignment into the formula and evaluating the resulting Boolean expression. This is the defining feature of problems in the complexity class NP, whose solutions are efficient to verify.

**Comparable classical complexity**

For the unstructured search problem, exhaustive search is essentially the best that can be done, with a running time $\sim \ell \cdot 2^n$. Of course, SAT seems to be far from unstructured, but under the Strong Exponential-Time Hypothesis [12, 13] the best classical algorithm for SAT has running time $2^{n-o(n)}$.

A similar argument holds for the generalized problem considered in the setting of amplitude amplification: if we have some prior distribution, we can classically find a marked element by sampling from this distribution roughly $1/p$ times. Since unstructured search is a special case of this problem, we cannot hope for a better classical algorithm in general.

**Speedup**

The speedup is quadratic in terms of the number of required iterations if we compare to corresponding naive classical algorithms. It can be shown that this speedup is optimal in the black-box query model [14]. Moreover, we do not expect that there would be a bigger than quadratic speedup in gate complexity [15] in the general (non-black-box) case.

**Outlook**

We have discussed how Grover search provides a quadratic speedup for SAT, and how amplitude amplification yields a quadratic speedup for Schöning's 3-SAT algorithm [5]. Since the best known 3-SAT solvers [16, 17] have complexity $\mathcal{O}(1.308^n)$—only slightly better than Schöning's $\mathcal{O}(1.334^n)$ complexity—this implies a close-to-quadratic quantum speedup. However, note that this relates to worst-case complexity, and on practical instances, the scaling can be much better.

We now comment on some of the other combinatorial optimization problems where Grover's algorithm can be used as a subroutine. One class of examples is graph-related problems. In the literature, these problems are most often studied in the query model, therefore, here we also only discuss their speedup in terms of query complexity. (Since these are (sub)quadratic speedups, we know that the fault-tolerant resource estimates will be unfavorable anyway, as discussed above.) For instance, the problem of finding the shortest paths from a single source $s$ in graph $G = (V, E)$ to all other vertices $v \in V$ can be solved classically using Dijkstra's algorithm in time $\mathcal{O}(|E| + |V| \log |V|)$ if the graph is provided with its adjacency list (and with query complexity $\mathcal{O}(|E|)$), whereas the quantum query complexity of this problem is $\tilde{\Theta}(\sqrt{|V||E|})$ [18]. Reference [18] determines the query complexity of several other graph problems such as deciding graph connectivity and strong connectivity as well as finding the minimum-weight spanning tree. For all of these problems, there is a similar (sub)quadratic quantum speedup.

One graph problem that is often mentioned in connection to quantum computation is the (in)famous traveling salesperson problem. However, for this problem, the best provable speedup is only subquadratic. The naive classical algorithm runs in time $\widetilde{\mathcal{O}}(n!)$, and Grover's algorithm offers a quadratic speedup over it. The best classical algorithm uses dynamic programming and runs in time $\widetilde{\mathcal{O}}(2^n)$. Ambainis et al. [19] showed how to obtain a speedup over this algorithm by combining classical precalculation with recursive applications of Grover's search resulting in time complexity $\widetilde{\mathcal{O}}(1.728^n)$ assuming that QRAM calls have unit costs. Considering the overheads coming from the implementation of QRAM and fault tolerance, the traveling salesperson problem seems to be one of the *least* likely candidates to achieve a practical quantum speedup when the nodes have large degree. For bounded degree graphs there is slightly more hope as quantum algorithms with close-to-quadratic speedups have been devised [20] that do not require QRAM.

Finally, let us mention quantum walk algorithms, which can also be viewed as a generalization of Grover search. However, quantum walks are more distant relatives of Grover search and can only be applied in more specific settings. They can be used for proving many nontrivial speedups in query complexity, however, the resulting algorithms are often not practical due to high space and/or gate complexity overheads, as is the case for the prototypical element distinctness problem. The query reduction is moderate $N \to N^{2/3}$ in the number of elements $N$, but the corresponding quantum algorithm [21] unfortunately uses a QRAM consisting of roughly $N^{2/3}$ registers; moreover, the QRAM must be able to store data in superposition.

There are nevertheless more practical quantum walk algorithms applicable, for example, to speed up backtracking algorithms [22, 23, 24, 25], which are among the most successful and widely used classical heuristics for solving SAT instances in practice. The quantum algorithm can achieve an essentially quadratic speedup compared to its classical backtracking variant. This approach is applicable to the traveling salesperson problem in the special case that the graph has degree at most 4 [20]. For resource estimates, see the earlier quoted reference [6]. A further extension of this algorithm is applicable to branch-and-bound algorithms [26, 27], and in some cases yields running times that are substantially better than what we know can be achieved by naively using Grover's algorithm. For example, it can find exact ground states for most instances of the Sherrington–Kirkpatrick model [28] in time $\mathcal{O}(2^{0.226n})$ [26], which means about a quadratic speedup compared to classical methods. Branch-and-bound-based speedups can also be applied to solve mixed-integer programs, which include certain formulations of the portfolio optimization problem [27].

There is a plethora of other applications of quantum search speedups, ranging from machine learning [29] to dynamic programming solutions of other NP-hard problems [19], which we do not discuss here for length constraints and due to discouraging resource estimates for (sub)quadratic quantum speedups.

## 4.2 Beyond quadratic speedups in exact combinatorial optimization

**Overview**

The discovery of Grover's algorithm [1] (later generalized to amplitude amplification) has long been the source of enthusiasm that quantum algorithms can be advantageous for combinatorial optimization, as it leads to quadratic asymptotic speedups for many concrete end-to-end search problems in this area. However, resource estimates indicate that early and intermediate-term fault-tolerant devices will fail to deliver practical advantages when the available speedup is only quadratic, due to intrinsic overheads of quantum computation compared to classical computation (see, e.g., [2, 3]). Thus, identifying whether beyond-quadratic speedups are available is of principal importance for identifying end-to-end practical advantages in combinatorial optimization. Despite the fact that Grover's algorithm is optimal in the black-box (unstructured) setting, superquadratic speedups could be possible when the combinatorial optimization problem has a certain structure that can be better exploited by a quantum algorithm than a classical algorithm.

Unfortunately, many proposals that could conceivably deliver superquadratic speedups lack rigorous theoretical performance guarantees. This includes the quantum adiabatic algorithm and variational quantum algorithms such as the quantum approximate optimization algorithm (QAOA) [4], which is typically formulated to give approximate solutions, but at higher cost could also be used to find exact solutions. Limited analytic and numerical work provides some evidence (e.g., [5, 6]) that QAOA could outperform a vanilla application of Grover's algorithm to the $k$-SAT problem, but provides no definitive conclusion on the matter. Alternatively, a line of work in [7, 8] studies a different algorithm (related in certain aspects to the quantum adiabatic algorithm) and provides rigorous running time guarantees that *slightly* surpass Grover's algorithm.

However, while these algorithms may have a speedup over Grover's algorithm, this does not entail a superquadratic speedup over the *best* classical algorithm, which can often exploit structure in other ways to do much better than exhaustive search. Overall, it remains an open question whether quantum algorithms can provide superquadratic speedups for useful problems in exact combinatorial optimization.

**Actual end-to-end problem(s) solved**

Combinatorial optimization problems ask to find which solution is optimal among a finite set of possible candidates. Here, we stick to binary optimization on $n$ bits, where the universe of possible candidates are bit strings $z = (z_1, z_2, \ldots, z_n) \in \{1, -1\}^n$. The input to the problem is a compact description of some cost function $C : \{1, -1\}^n \to \mathbb{R}$, and the desired output is the string $z^*$ for which $C$ is minimized. Let $E^* = C(z^*)$ denote the optimal value of the cost function. For simplicity we assume $z^*$ is unique and $E^*$ is known ahead of time.[10] This setting contrasts with that of *approximate* optimization, where the acceptable outputs include a much larger set of strings $z$ that are not necessarily optimal solutions, but are still good enough, for example, because they achieve a nontrivial approximation ratio $|C(z)|/|E^*|$ with the optimal cost value. Classical and quantum algorithms for approximate optimization are often heuristic, making it more difficult to systematically study the complexity of the algorithms and the possibility that quantum algorithms may provide a speedup.

---

[10]This assumption can often be relaxed at the expense of at most poly($n$) overhead, for example, by iterating over all possible values $E^*$ might take, which fall within a poly($n$)-size range when the cost function consists of only poly($n$) constant-size (integer-valued) terms.

Concrete examples can be formed by choosing the function $C(z)$ to be a low-degree polynomial in the bits of $z$. For example, if $C$ is a degree-2 polynomial in $z$, this is a quadratic unconstrained binary optimization (QUBO) problem, which is also equivalent to the classical Ising spin model from Eq. (3). If, furthermore, every term of $C$ has degree exactly 2 (no degree-1 or constant terms) and every coefficient is either 0 or 1, then the problem is equivalent to a MAX-CUT problem. Finally, if $C$ is a sum of degree-3 terms of the form

$$z_a z_b z_c + z_a z_b + z_a z_c + z_b z_c + z_a + z_b + z_c,$$

where

$$z_a, z_b, z_c \in \{z_1, -z_1, z_2, -z_2, \ldots, z_n, -z_n\},$$

then the problem is equivalent to a MAX-3-SAT instance in conjunctive normal form. To see this, note that if $z_a = z_b = z_c = 1$, the term evaluates to 7, and for any other setting, it evaluates to $-1$. Thus, the solution $z^*$ that optimizes $C$ represents the bit string that minimizes the number of "unsatisfied" clauses for which $z_a = z_b = z_c = 1$. This is easily generalized from MAX-3-SAT to MAX-$k$-SAT.

For a fixed instance $C$, the quantum algorithms must find $z^*$ with high probability over measurement outcomes. If it does so for every $C$ chosen from some class of problem, we say it succeeds in the worst case. Alternatively, we can consider ensembles of instances chosen from some class of problem; if for a large fraction of instances from the ensemble, the algorithm finds $z^*$ with high probability, then we say the algorithm succeeds in the average case.[11] A commonly considered average-case ensemble is the Sherrington–Kirkpatrick (SK) model [9], defined as

$$C(z) = \sum_{i=1}^{n} \sum_{j=i+1}^{n} J_{ij} z_i z_j \quad \text{where} \quad J_{ij} \sim \mathcal{N}(0,1), \tag{9}$$

where the coefficients $J_{ij}$ are drawn randomly from a standard Gaussian distribution $\mathcal{N}(0,1)$. The SK model is relevant in spin glass theory, and can be generalized to higher-degree interactions, where it is referred to as the $p$-spin model [10]. Another ensemble is the random MAX-$k$-SAT ensemble, where MAX-$k$-SAT instances are generated by choosing each clause uniformly at random with some fixed clause-to-variable ratio (see, e.g., [11]).

**Dominant resource cost/complexity**

A vanilla application of Grover's algorithm to binary optimization problems achieves $\mathcal{O}^*(2^{0.5n})$ running time, where notation $\mathcal{O}^*(2^{an})$ is shorthand for $\text{poly}(n)2^{an}$. We cover three approaches to solving binary optimization problems on a quantum computer that have some potential to improve upon this running time. Note that all of these algorithms require polynomial (in fact, linear $\mathcal{O}(n)$) space. However, their running time is expected to scale exponentially in $n$.

---

[11] A more typical definition of the average-case complexity of an algorithm is the expected runtime required for it to find the solution $z^*$, averaged over both choice of instance and internal algorithmic randomness (i.e., classical coin flips or quantum measurement outcomes). This definition is related to the convention we follow, but it is more coarse grained as it does not distinguish between the two types of randomness, the latter of which can be boosted by repetition.

- First, we consider variational quantum algorithms, using the QAOA [4] as a representative. These algorithms are typically studied as efficient (polynomial-time) quantum algorithms that produce approximate solutions, that is, strings $z \neq z^*$ for which $C(z)$ is small, but not optimal. However, they may also be viewed as exact algorithms, since, if repeated a sufficient number of times, they eventually produce the exactly optimal $z^*$. The QAOA fixes a depth parameter $p$ and variational parameters $\gamma = (\gamma_1, \ldots, \gamma_p)$ and $\beta = (\beta_1, \ldots, \beta_p)$ (sometimes these are set to some fixed instance-independent value, and sometimes they are variationally updated on subsequent repetitions of the algorithm). The QAOA starts in the $n$-qubit equal superposition state $|+\rangle^{\otimes n}$ and implements alternating rounds of rotations about the diagonal cost function $C$ and a "mixing" operator $X = \sum_i X_i$, where $X_i$ denotes the Pauli-$X$ gate about qubit $i$. The state produced by QAOA is thus given by

$$|\psi_{\gamma,\beta}\rangle = \mathrm{e}^{-\mathrm{i}\beta_p X} \mathrm{e}^{-\mathrm{i}\gamma_p C} \cdots \mathrm{e}^{-\mathrm{i}\beta_2 X} \mathrm{e}^{-\mathrm{i}\gamma_2 C} \mathrm{e}^{-\mathrm{i}\beta_1 X} \mathrm{e}^{-\mathrm{i}\gamma_1 C} |+\rangle^{\otimes n} \, .$$

  If one makes a computational basis measurement of $|\psi_{\gamma,\beta}\rangle$, one obtains $z^*$ with probability $|\langle z^*|\psi_{\gamma,\beta}\rangle|^2$. The expected number of repetitions required to obtain $z^*$ is the inverse of this probability, and this running time can be quadratically sped up by performing amplitude amplification on top of the QAOA protocol; thus, the QAOA unitary is applied $\mathcal{O}(|\langle z^*|\psi_{\gamma,\beta}\rangle|^{-1})$ times. Implementing the QAOA unitary typically requires only $p \cdot \mathrm{poly}(n)$ gates, as each of the rotations about $X$ and $C$ are efficient to implement. For hard combinatorial optimization problems such as typical MAX-$k$-SAT instances, the expectation is that the total running time required will be exponential. If the depth $p$ is chosen to be constant or even $\mathrm{poly}(n)$, the dominant cost will come from the $\mathcal{O}(|\langle z^*|\psi_{\gamma,\beta}\rangle|^{-1})$ repetitions required to amplify the $|z^*\rangle$ state. Alternatively, one can reduce the number of repetitions needed to $\mathcal{O}(1)$ at the expense of taking $p$ to be very large (at least exponentially large in $n$); indeed, for sufficiently large $p$, the QAOA can be viewed as a Trotterized simulation of the adiabatic algorithm [4].

  There is some analytic evidence that the QAOA may outperform Grover's algorithm at finding the exact solution for constant $p$ in certain cases. Reference [5] studied the QAOA applied to hard (i.e., near the satisfiability threshold) $k$-SAT instances with instance-independent choice of $\gamma$, $\beta$ for constant $p$, and developed an analytic formula for the expected success probability $|\langle z^*|\psi_{\gamma,\beta}\rangle|^2$ averaged over random instance in the limit $n \to \infty$. This formula was evaluated numerically and suggested, for example, that the average success probability behaves as $2^{-0.33n}$ for $p = 10$ on 8-SAT. One might be tempted to declare that this implies an overall average running time of $\mathcal{O}^*(2^{0.33n/2})$, substantially better than Grover, but such a conclusion is not analytically supported as the average of the inverse probability can be much larger than the inverse of the average probability. Nevertheless, it provides intriguing evidence in favor of such a conclusion. Further numerical evidence that QAOA may be effective as an exact algorithm was provided in [6], which numerically assessed the performance of QAOA on instances of the low autocorrelation binary sequences (LABS) problem up to $n = 40$, although compared to the best classical heuristic solver, the advantage appeared to be subquadratic.

- Second, we consider the quantum adiabatic algorithm [12, 13]. The standard approach, as applied to binary optimization problems, is to start in the state $|+\rangle^{\otimes n}$ and evolve by a Hamiltonian that interpolates along a path $H(s)$ parameterized by $s \in [0, 1]$, given by

$$H(s) = (1 - s)(-X) + sC \, . \tag{10}$$

It is important to note that the ground state of $H(0)$ is $|+\rangle^{\otimes n}$ and the ground state of $H(1)$ is $|z^*\rangle$. This evolution can be simulated on a fault-tolerant gate-based quantum computer using Hamiltonian simulation, and its running time is dominated by the inverse of the minimum spectral gap $\Delta_{\min}$ of $H(s)$. That is, the gate complexity to run the algorithm and produce $|z^*\rangle$ scales as at least $\Delta_{\min}^{-1}$ and possibly a larger power of $\Delta_{\min}^{-1}$. Much numerical work has been done on the performance of the adiabatic algorithm on small instances of combinatorial optimization problems, but it generally lacks analytical guarantees. The expectation is that $\Delta_{\min}$ will be exponentially small [14, 15, 16] in $n$ (or worse, see, e.g., [17, 18]), meaning the running time of the algorithm is exponentially large, but it remains possible that it surpasses the $\mathcal{O}^*(2^{0.5n})$ running time of Grover's algorithm in some cases, and could in principle deliver a superquadratic speedup.

- Third, we consider the *short-path* algorithm studied in [7, 19, 20] and a dual version of the algorithm studied in [8]. The goal of these algorithms was to be able to provide a rigorous guarantee that the algorithm can find $z^*$ in time $2^{(0.5-c)n}$ for some value of $c > 0$. Similar to the adiabatic algorithm, the short-path algorithm also considers a single-parameter family of Hamiltonians

$$H(s) = (1-s)f_X\left(-\frac{X}{n}\right) + sf_Z\left(\frac{C}{|E^*|}\right), \tag{11}$$

where $f_X, f_Z : \mathbb{R} \to \mathbb{R}$ are monotonic filter functions, and each term $X/n$ and $C/|E^*|$ are normalized to have minimum value $-1$. The idea of the short-path algorithm is to, rather than evolve smoothly from $s = 0$ to $s = 1$, perform a pair of discrete "jumps." The first jump goes from the ground state $|+\rangle^{\otimes n}$ at $s = 0$ to the ground state $|\psi_b\rangle$ of an intermediate point with $s = b$. The second jump goes from $|\psi_b\rangle$ to the ground state $|z^*\rangle$ at $s = 1$. The jumps are accomplished with quantum phase estimation (or more advanced versions utilizing the quantum singular value transformation) of the Hamiltonian $H_b$ combined with amplitude amplification. The running time of the algorithm is [8, Theorem 1]

$$\text{poly}(n) \cdot \frac{1}{\Delta} \cdot \left(\frac{1}{|\langle + |\psi_b\rangle|} + \frac{1}{|\langle \psi_b | z^*\rangle|}\right), \tag{12}$$

where $\Delta$ is the spectral gap of the Hamiltonian $H(b)$. The $\Delta^{-1}$ factor comes from the need to perform phase estimation at $\mathcal{O}(\Delta)$ resolution to successfully prepare $|\psi_b\rangle$, and the two additive inverse overlap terms represent the number of rounds of amplitude amplification for the first and second jumps, respectively. In [7], filter functions $f_X(x) = x^K$ for odd integers $K$ (e.g., $K = 3$) and $f_Z(x) = x$ were chosen, and $b$ was chosen close to 1, such that the first term of Eq. (11) could be viewed as a small perturbation of the second term. If $C$ is an instance of MAX-E$k$-LIN2, that is, if it is a polynomial for which all monomials are degree exactly $k$, then it was shown that certain conditions on the spectral density of $C$ near the optimal cost value imply sufficient analytic control of $\Delta$ and the other parameters in Eq. (12) such that the algorithm runs in time $\mathcal{O}^*(2^{(0.5-c)n})$ for $c > 0$. However, it remained unclear when these conditions were met. Inspired by [7], [8] proposed using the filter functions $f_X(x) = x$ and $f_Z(x) = \min(0, (x + 1 - \eta)/\eta)$ for a fixed choice of $\eta \in [0, 1]$, and chose a value of $s$ close to 0 (rather than close to 1). In this sense, the algorithm in [8] is dual to that of [7]. These modifications allowed additional statements to be proved. For example, it was unconditionally shown that the algorithm solves $k$-SAT (whether or not a formula has a fully satisfiable solution) in time upper bounded by $\mathcal{O}^*(2^{(0.5-c)n})$ for a (extremely small) constant $c > 0$, and that the same is true for typical instances of the SK model and its higher-body generalization ($p$-spin model), a polynomial speedup over Grover's algorithm and superquadratic advantage over classical exhaustive search.

**Existing resource estimates**

Reference [21] compiled resource estimates for various primitive tasks related to combinatorial optimization. For example, it estimated that for an $n = 512$ instance of the SK model, implementing a single QAOA step $e^{-i\beta_j X} e^{-i\gamma_j C}$ would require 577 logical qubits and $5.0 \times 10^5$ Toffoli gates. A similar estimate would hold for performing a single step of adiabatic evolution with a first-order product formula. The total logical estimate for finding $z^*$ would be the product of the depth of the circuit and any number of repetitions or rounds of amplitude amplification. An estimate of the physical resource cost could then be computed for a specific fault-tolerant architecture. Without knowing the number of repetitions, it is hard to give precise estimates, but a rough attempt was made in [3] for different speedup factors. There, under different possible assumptions on the amount of classical parallelism available, a breakeven point was estimated for different possible polynomial speedups (quadratic, cubic, and quartic). It was found that with a quartic speedup, the breakeven point could be reasonable (on the order of seconds to hours) even assuming the availability of classical parallelism.

**Caveats**

There are several caveats. The most salient one is that for most of the algorithms above, there is no provable beyond-Grover advantage. Meanwhile, in the case of [8], the size of the provable beyond-Grover advantage is miniscule. The prospect of these algorithms is thus left to extrapolations from numerical simulations carried out at very small instance sizes and speculation based on physical principles.

A second important caveat is that to deliver practical superquadratic speedups, the performance of the quantum algorithm needs to be compared to the best classical algorithm, which is often substantially better than the $\mathcal{O}^*(2^n)$ running time of exhaustive enumeration. For example, 3-SAT problems are classically solvable in $\mathcal{O}^*(2^{0.39n})$ time [22].

Along these lines, a third caveat is the existence of classical "quantum Monte Carlo" algorithms (see, e.g., [23, 24, 25, 26, 27]), which can, under certain conditions, classically simulate the quantum algorithms described above. This is because the Hamiltonians in Eqs. (10) and (11) are *stoquastic* Hamiltonians, defined by the property that their off-diagonal matrix elements are non-positive (when written in the computational basis). Stoquasticity implies that the ground state of the Hamiltonian can be written such that all amplitudes are non-negative real numbers [28], meaning that these Hamiltonians avoid the so-called "sign problem" enabling the potential application of quantum Monte Carlo techniques. To be clear, it remains possible that quantum algorithms for these combinatorial optimization problems involving stoquastic Hamiltonians can evade classical simulation—indeed, superpolynomial oracle separations have been shown between classical computation and adiabatic quantum computation restricted to stoquastic paths [29, 30]—but it is something to keep in mind when designing algorithms based on stoquastic Hamiltonians.

A final caveat is that the quantum algorithms described here are typically not amenable to parallelization, although in principle QAOA could be parallelized if one opts not to use am-plitude amplification (resulting in worse asymptotic complexity). This lies in stark contrast to many classical optimization algorithms for exact combinatorial optimization which are highly parallelizable, a feature that can be exploited to significantly reduce the running time of these classical algorithms on high-performance computers, making achieving practical quantum advantage more difficult [3].

**Comparable classical complexity and challenging instance sizes**

For many binary optimization problems, there exist classical algorithms that exploit the structure of the problem to perform significantly better than exhaustive search. For example, the best 3-SAT algorithm runs in time $\mathcal{O}^*(2^{0.39n})$ and in general $k$-SAT can be solved in time $2^{(1-\Omega(1/k))n}$ [22]. This running time suggests the solution will be impractical once $n$ is on the order of 100. The algorithm analyzed in [22] is designed for the worst case, and it is likely not the best practical algorithm for typical instances. For random instances, the hardness of $k$-SAT depends sensitively on the clause-to-variable ratio $\alpha$. Remarkably, heuristic algorithms can succeed at finding a satisfiable solution for typical instances with thousands or even tens of thousands of variables even very close to the satisfiability threshold $\alpha_c$ where most instances become unsatisfiable (e.g., [31]). However, these algorithms are expected to fail sufficiently close to the satisfiability threshold and in the worst case.

Similarly, the SK model admits a classical branch-and-bound algorithm guaranteed to run in time $2^{0.45n}$ (for a large fraction of instances) and likely better than that in practice [32]. However, once the interaction degree becomes larger than 2, the problem becomes significantly harder. The branch-and-bound algorithm is not known to generalize to the $p$-spin model, and for $p \geq 3$ there is no known classical algorithm that provably achieves $2^{(1-c)n}$ for any constant $c$ (although it has not garnered much attention, see [8]). Similarly, in contrast to $k$-SAT, the MAX-$k$-SAT problem (i.e., the version of the problem that asks for the optimal assignment even if it does not satisfy all the clauses) only has a $\mathcal{O}^*(2^{(1-c)n})$ time algorithm for $k = 2$, and, notably, this algorithm requires exponential space [33].

**Speedup**

As there are generally no rigorous running time guarantees for the quantum algorithms, the speedup cannot be estimated. However, it is worth emphasizing that for hard combinatorial optimization problems, the speedup could be superquadratic, but it is not expected to be superpolynomial.

The rigorous results of [8] establish a beyond-Grover running time, but the only case in which the speedup is beyond quadratic when compared with the best known classical algorithm is the $p$-spin model with $p \geq 3$ (here, the comparison benefits from little work on classical algorithms for the problem).

We also mention the result of [34], which studies a quantum algorithm for random instances of a QUBO-like combinatorial optimization problem with a "planted" optimal solution—the goal is to exactly or approximately find the planted solution, or alternatively to simply distinguish instances drawn from the ensemble with planted solutions from instances drawn from the ensemble without a planted solution. The algorithm generalizes the tensor PCA algorithm of [35] and gives a *quartic* speedup over its closest classical counterpart, although it is unclear if this speedup can extend to non-planted scenarios as well.

## NISQ implementation

The QAOA approach is amenable to NISQ implementation (assuming one opts not to apply amplitude amplification on top of it), since the quantum circuit one needs to implement is fairly shallow depth. In this case, the effect of uncorrected errors in the NISQ device may degrade the performance (and require more repetitions to extract the optimal bit string $z^*$). Similarly, on a NISQ quantum annealer [36, 13], one could run a noisy version of the quantum adiabatic algorithm and repeat until finding the optimal bit string $z^*$.

## Outlook

In contrast to algorithms for approximate optimization, which are often heuristic but run in polynomial time, algorithms for exact optimization are often more rigorous but run in exponential time. For quantum computers to be impactful for exact combinatorial optimization, we require great advancements in the estimated clock speeds of quantum hardware and the overheads of fault-tolerant quantum computing, or else the development of quantum algorithms that significantly improve upon existing (sub)quadratic Grover-type speedups—either quantitatively (bigger speedups) or qualitatively (e.g., requiring only shallow circuits). Although ideas have been proposed that could potentially deliver such improvements, they either come without provable guarantees, provide only minor superquadratic improvement, or only apply to artificial problems. Much more attention shall be devoted to studying these quantum algorithms and developing new ones if we are to leverage them into actual practical advantages, especially considering the extensive amount of work devoted to developing sophisticated classical algorithms for these problems.

# 5 Continuous optimization

Continuous optimization problems arise throughout science and industry. On their face, continuous optimization problems rarely seem quantum mechanical; nevertheless, quantum algorithms have been proposed for accelerating both convex and nonconvex continuous optimization. Most of the research on these algorithms thus far has been to develop and utilize the diverse set of primitive ingredients that give rise to potential quantum advantage in this space, without an eye toward the end-to-end practicality of the algorithms. Developing a better understanding of the practicality of these approaches should be a focus of future work.

We refer the reader to [1] for a comprehensive survey of quantum methods for continuous and combinatorial optimization.

*The authors are grateful to Sander Gribling for reviewing this section of the survey.*

**This application area contains:**

## 5.1   Zero-sum games: Computing Nash equilibria

**Overview**

In a two-player zero-sum game, each player independently chooses an action and then receives a "payoff" (such that the sum of the payoffs is always zero) that depends on which pair of actions was chosen. A Nash equilibrium is an optimal way of (probabilistically) choosing an action that maximizes a player's worst-case payoff. The problem of computing a Nash equilibrium is, in a certain sense, equivalent to solving a linear program (LP): computing a Nash equilibrium is a special case of LP, and conversely any LP can be reduced to computing a Nash equilibrium at the expense of introducing dependencies on a certain instance-specific "scale-invariant" precision parameter [1]. However, the quantum approach to solving LPs based on the multiplicative weights update method [1] is more efficient in the special case of computing Nash equilibria, and has fewer caveats—notably, it avoids the dependence on the difficult-to-predict scale-invariant precision parameter. It gives a potentially quadratic speedup over its classical counterpart.

**Actual end-to-end problem(s) solved**

A two-player zero-sum game is defined by an $n \times m$ matrix $A$ called the "payoff matrix," which specifies how much player 1 wins from player 2 when player 1 chooses action $i \in [n]$ and player 2 chooses action $j \in [m]$. A pure strategy is one in which the players deterministically choose one fixed action $i \in [n]$ (or $j \in [m]$) in each game. By contrast, a mixed strategy is one in which players randomly choose an action, according to some probability distribution. Assume the entries of $A$ are between $-1$ and $1$. A Nash equilibrium is an optimal (generally mixed) strategy that maximizes a player's worst-case payoff regardless of the other player's choice. That is, a distribution $y \in \Delta^m$, where $\Delta^m$ denotes the $m$-dimensional probability simplex, is an optimal strategy for player 2 if it is the argument that optimizes the equation

$$\lambda^* = \min_{y \in \Delta^m} \max_{i \in [n]} e_i^\mathsf{T} A y,$$

where $[n]$ denotes the set of actions available to player 1, and $e_i$ denotes a basis state associated with action $i$. The quantity $\lambda^*$ is the value of the game. This can be rewritten explicitly [1] as the following LP

$$\min_{y \in \mathbb{R}^m} \lambda$$

subject to        $Ay \leq \lambda \mathbf{1}, \qquad \sum_j y_j = 1, \qquad y_j \geq 0 \;\forall j,$

where $\mathbf{1}$ is the all-ones vector. The dual LP for the above then corresponds to computing the Nash equilibrium for player 1.

The end-to-end problem solved is to, given access to the entries of the matrix $A$ and an error parameter $\epsilon$, compute a probability vector $y$ such that

$$Ay \leq (\lambda^* + \epsilon)\mathbf{1} \,.$$

**Dominant resource cost/complexity**

The quantum algorithm builds on a classical algorithm based on the multiplicative weights update method from [2]. With probability at least $1 - \delta$, the classical algorithm finds a solution $y$ that approximates a Nash equilibrium to error $\epsilon$ after $\lceil 16\ln(nm/\delta)/\epsilon^2 \rceil$ iterations, where the cost per iteration is $n+m$ queries to the entries of the matrix $A$ and $\mathcal{O}(n+m)$ other arithmetic operations [1, Lemma 3]. An important subroutine of each iteration is a Gibbs sampling step for a diagonal matrix (a special case of the general quantum Gibbs sampling problem in which any Hermitian matrix is allowable). When the matrix $A$ is sparse, the number of queries per iteration can be reduced to $2s$, where $s$ is the maximum number of nonzero entries in a row or column of $A$, and the total time per iteration can be reduced to $\widetilde{\mathcal{O}}(s)$ [1, Lemma 4].

The quantum algorithm assumes coherent access to the matrix entries of $A$. Through amplitude amplification and the related subroutines of amplitude estimation and minimum finding, the quantum algorithm of [1] speeds up the Gibbs sampling task and reduces the maximum cost of an iteration to $\widetilde{\mathcal{O}}(\sqrt{n+m}/\epsilon)$ queries to the matrix elements of $A$ and an equal amount of time complexity, where $\widetilde{\mathcal{O}}$ notation suppresses logarithmic factors. In the case that the matrices are sparse, the maximum cost of an iteration is reduced to $\widetilde{\mathcal{O}}(\sqrt{s}/\epsilon^{1.5})$. The work of [3] introduces a technique called dynamic Gibbs sampling, which exploits the fact that the distribution to be sampled changes slowly from iteration to iteration and further reduces the iteration cost to $\widetilde{\mathcal{O}}(\sqrt{n+m}/\epsilon^{1/2} + 1/\epsilon)$ in the dense case. This gives a total query and time complexity roughly given by

$$\text{dense:} \quad \left( \frac{16\ln(nm)}{\epsilon^2} \text{ iters.} \right) \times \left( \widetilde{\mathcal{O}}\left( \frac{\sqrt{n+m}}{\sqrt{\epsilon}} + \frac{1}{\epsilon} \right) \text{ per iter.} \right) = \widetilde{\mathcal{O}}\left( \frac{\sqrt{n+m}}{\epsilon^{2.5}} + \frac{1}{\epsilon^3} \right)$$

$$\text{sparse:} \quad \left( \frac{16\ln(nm)}{\epsilon^2} \text{ iters.} \right) \times \left( \widetilde{\mathcal{O}}\left( \frac{\sqrt{s}}{\epsilon^{1.5}} \right) \text{ per iter.} \right) = \widetilde{\mathcal{O}}\left( \frac{\sqrt{s}}{\epsilon^{3.5}} \right).$$

This complexity assumes access to a quantum random access memory (QRAM). Without a QRAM, the cost per iteration increases by a factor $\widetilde{\mathcal{O}}(1/\epsilon^2)$.

See also [4], which independently from [1] gave a quantum algorithm that solves zero-sum games with slightly worse $\epsilon$ dependence, as well as [5], which gave quantum algorithms for generalizations of zero-sum games to other vector norms.

**Existing resource estimates**

There are no existing explicit resource estimates for this algorithm.

**Caveats**

- Due to poor dependence of the complexity on the error $\epsilon$, this algorithm is only likely to be useful in situations where it is not necessary to learn the optimal strategy to high precision. It is unclear when such situations arise in practice.

- As mentioned above, if no QRAM is available, the runtime suffers a $\widetilde{\mathcal{O}}(1/\epsilon^2)$ time slowdown.

- A fully end-to-end analysis should also consider the exact way that the queries to the matrix entries of $A$ are implemented. If they are given in a classical database, a large $\mathcal{O}(nm)$-size QRAM may also be required to implement the queries in polylog$(mn)$ time. Note that this would be separate from the $\widetilde{\mathcal{O}}(1/\epsilon^2)$-size QRAM the algorithm uses to reduce the time complexity. To avoid the QRAM requirement for implementing a query, it must be the case that the matrix entries are efficiently computable in some other way.

**Comparable classical complexity and challenging instance sizes**

The classical version of the quantum algorithm has time and query complexity given by [1, Section 2]

$$\text{dense:} \quad \left(\frac{16\ln(nm)}{\epsilon^2} \text{ iters.}\right) \times (\mathcal{O}(n+m) \text{ per iter.}) = \widetilde{\mathcal{O}}\left(\frac{n+m}{\epsilon^2}\right)$$

$$\text{sparse:} \quad \left(\frac{16\ln(nm)}{\epsilon^2} \text{ iters.}\right) \times \left(\widetilde{\mathcal{O}}(s) \text{ per iter.}\right) = \widetilde{\mathcal{O}}\left(\frac{s}{\epsilon^2}\right).$$

Alternatively, the problem could be solved using other approaches for solving the associated LP. Classical interior point methods for LPs can achieve $\mathcal{O}(n^\omega \log(1/\epsilon))$ runtime in the common case that $m = \mathcal{O}(n)$ [6], where $\omega < 2.37$ is the matrix multiplication exponent. This runtime exhibits better $\epsilon$ dependence at the expense of worse $n$ dependence. Note that quantum interior point methods have also been proposed for conic programs like LPs, but whether they could yield a speedup over classical interior point methods would depend on the scaling of certain instance-specific parameters.

**Speedup**

The quantum complexity has a quadratic improvement in complexity with respect to the parameter $n + m$, and a polynomial slowdown with respect to the parameter $\epsilon$.

**Outlook**

It is difficult to assess whether a practical advantage could be obtained in the setting of zero-sum games without further investigation of how queries to matrix elements are accomplished, an assessment of constant prefactors involved in the algorithm, and consideration of any additional overheads from fault-tolerant quantum computation. The theoretical speedup available is quadratic and may require a medium- or large-scale QRAM. This speedup may not be sufficiently large to overcome these overheads in practice.

It is perhaps instructive to compare the outlook of zero-sum games to conic programming more generally. On the one hand, unlike the algorithm for general SDPs and LPs, the algorithm for zero-sum games does not have a complexity dependence on instance-specific parameters denoting the size of the primal and dual solutions. This makes it easier to evaluate the runtime of the algorithm and more likely that it can be an effective algorithm. On the other hand, a core subroutine of the quantum algorithm is to perform *classical* Gibbs sampling quadratically faster than a classical computer can using techniques like amplitude amplification. However, it is not clear how the speedup could be made greater than quadratic, even in special cases. A similar subroutine is required in the multiplicative weights approach to solving SDPs, but in that case, the Gibbs state to be sampled is a truly quantum state (i.e., nondiagonal in the computational basis), rather than a classical state. Using more advanced methods for Gibbs sampling, it is possible that in some special cases there could be a superquadratic quantum speedup for SDPs that would not be available for the simpler case of LPs and zero-sum games.

## 5.2   Conic programming: Solving LPs, SOCPs, and SDPs

**Overview**

Conic programs are a specific subclass of convex optimization problems, where the objective function is linear and the convex constraints are restrictions to the intersection of affine spaces and certain cones within $\mathbb{R}^n$. Commonly considered cones are the positive orthant, the second-order cone ("ice-cream cone"), and the semidefinite cone, which give rise to linear programs (LPs), second-order cone programs (SOCPs), and semidefinite programs (SDPs), respectively. This framework remains quite general, and many real-world problems can be reduced to a conic program. However, the additional structure of the program allows for more efficient classical and quantum algorithms, compared to completely general convex problems.

Algorithms for LPs, SOCPs, and SDPs have long been a topic of study. Today, the best classical algorithms are based on interior point methods (IPMs) [1, 2, 3] and cutting-plane methods [4, 5], but other algorithms based on the multiplicative weights update (MWU) method [6, 7, 8] exist and can be superior in a regime where high precision is not required. Both of these approaches can be turned into quantum algorithms with potential to deliver asymptotic quantum speedup for general LPs, SOCPs, and SDPs. However, the runtime of the quantum algorithm typically depends on additional instance-specific parameters, which makes it difficult to produce a general apples-to-apples comparison with classical algorithms.

**Actual end-to-end problem(s) solved**

- Linear programs (LPs) are the simplest convex program. An LP instance is specified by an $m \times n$ matrix $A$, an $n$-dimensional vector $c$, and an $m$-dimensional vector $b$. The problem can then be written as

$$\min_{x \in \mathbb{R}^n} \langle c, x \rangle$$

$$\text{subject to } Ax = b$$

$$x_i \geq 0 \text{ for } i = 1, \dots, n \,,$$

  where notation $\langle u, v \rangle$ denotes the standard dot product of vectors $u$ and $v$. The function $\langle c, x \rangle$, which is linear in $x$, is called the objective function, and a point $x$ is called feasible if it satisfies the linear equality[12] constraints $Ax = b$ as well as the positivity constraints $x_i \geq 0$ for all $i$. We denote the feasible point that optimizes the objective function by $x^*$. Let $\epsilon$ be a precision parameter. The actual end-to-end problem solved is to take as input a classical description of the problem instance $(c, A, b, \epsilon)$ and output a classical description of a feasible point $x$ for which $\langle c, x \rangle \leq \langle c, x^* \rangle + \epsilon$. The set of points that obey the positivity constraints $x_i \geq 0$ forms the positive orthant of the vector space $\mathbb{R}^n$. This set meets the mathematical definition of a convex cone: for any points $u$ and $v$ in the set and any non-negative scalars $\alpha, \beta \geq 0$, the point $\alpha u + \beta v$ is also in the set.

- Second-order cone programs (SOCPs) are formed by replacing the positivity constraints in the definition of LPs with one or more second-order cone constraints, where the second-order cone of dimension $k$ is defined to include points $(x_0; x_1; \dots; x_{k-1}) \in \mathbb{R}^k$ for which $x_0^2 \geq x_1^2 + \dots + x_{k-1}^2$.

---

[12]Inequality constraints of the form $Ax \leq b$ can be converted to linear equality constraints and positivity constraints by introducing a vector of slack variables $s$ and imposing $Ax + s = b$ and $s_i \geq 0$ for all $i$. An analogous trick is possible for SOCP and SDP.

- Semidefinite programs (SDPs) are formed by replacing the $n$-dimensional vector $x$ in the definition of LPs with an $n \times n$ symmetric matrix $X$ and replacing the positive orthant constraint with the conic constraint that $X$ is a positive semidefinite matrix. Denote the set of $n \times n$ symmetric matrices by $\mathbb{S}^n$, and for any pair of matrices $U, V \in \mathbb{S}^n$, define the notation $\langle U, V \rangle = \operatorname{tr}(UV)$ (which generalizes the standard dot product). Then, an SDP instance is specified by matrices $C, A^{(1)}, A^{(2)}, \ldots, A^{(m)} \in \mathbb{S}^n$, as well as $b \in \mathbb{R}^m$, and can be written as

$$\min_{X \in \mathbb{S}^n} \langle C, X \rangle$$
$$\text{subject to } \langle A^{(j)}, X \rangle = b_j \text{ for } j = 1, \ldots, m$$
$$X \succeq 0,$$

  where $X \succeq 0$ denotes the constraint that $X$ is positive semidefinite.

In the LP or SDP case, we might also require as input parameters $R$ and $r$, where $R$ is a known upper bound on the size of the solution in the sense that $\sum_i |x_i| \leq R$ (LP) or $\operatorname{tr}(X) \leq R$ (SDP), and where $r$ is an analogous upper bound on the size of the solution to the *dual* program (not written explicitly here, see [9, 10, 11]).

**Dominant resource cost/complexity**

Two separate approaches to solving conic programs with quantum algorithms have been proposed in the literature. Both methods start with classical algorithms and replace some of the subroutines with quantum algorithms.

(i) Quantum interior point methods (QIPMs) for LPs [12], SOCPs [13, 14], and SDPs [12, 15, 16] have been proposed. In the standard approach, these methods start with classical interior point methods, for which the core step is solving a linear system, and simply replace the classical linear system solver with a quantum linear system solver (QLSS), combined with pure state quantum tomography. Given a linear system $Gu = v$, the QLSS produces a quantum state $|u\rangle$, and quantum tomography is subsequently used to gain a classical estimate of the amplitudes of $|u\rangle$ in the computational basis. The QLSS ingredient introduces complexity dependence on a parameter $\kappa = \|G\| \|G^{-1}\|$, the condition number of $G$, where $\|\cdot\|$ denotes the spectral norm. Additionally, the QLSS requires that the classical data defining $G$ be loaded in the form of a block-encoding, for which the standard construction introduces a dependence on the factor $\zeta = \|G\|_F \|G\|^{-1}$, where $\|\cdot\|_F$ denotes the Frobenius norm. Finally, the tomography ingredient introduces a complexity dependence on a parameter $\xi$, defined as the precision to which the vector $u$ must be classically learned, measured in $\ell_2$ norm. Assuming $m$ is on the order of the number of degrees of freedom (i.e., $\mathcal{O}(n)$ in the case of LP and SOCP, and $\mathcal{O}(n^2)$ in the case of SDP), the number of queries the QIPM makes to block-encodings of the input matrices is

$$\text{LP, SOCP [14]:} \qquad \widetilde{\mathcal{O}}\left( \frac{n^{1.5} \zeta \kappa}{\xi} \log(1/\epsilon) \right)$$

$$\text{SDP [12, 15]:} \qquad \widetilde{\mathcal{O}}\left( \frac{n^{2.5} \zeta \kappa}{\xi} \log(1/\epsilon) \right),$$

where the $\widetilde{\mathcal{O}}$ notation hides logarithmic factors. Note that depending on how $\xi$ is defined, extra factors of $\kappa$ may be required. Moreover, note that the complexity statements in [15] go further and analyze the worst-case dependence of $\xi$ on the overall error $\epsilon$, and additionally make the worst-case replacement $\zeta \leq \mathcal{O}(n)$—this explains the deviation in our presentation from the bounds in [15]. We do not include these worst-case assumptions on $\zeta$ and $\xi$ because it is possible they are not achieved in practice.[13] Generally speaking, the numerical values of $\kappa$, $\zeta$, and $\xi$ are not possible to determine in advance for a specific application; empirical investigations at small system sizes such as those of [17, 18] require an assumption that trends observed accurately extrapolate to other untested instances and to larger system sizes. The block-encoding queries can be executed in circuit depth polylog$(n + m, 1/\epsilon)$, which can also be absorbed into the $\widetilde{\mathcal{O}}$ notation (although it is important to note that the circuit *size* is generally $\mathcal{O}(n^2)$)—this is equivalent to an assumption of log-depth quantum random access memory (QRAM). If the input matrices are sparse or given in a form other than as a list of matrix entries, there may be other more efficient methods for block-encoding; in this case the parameter $\zeta$ might be replaced with another parameter $\alpha > 1$, whose value would depend on the block-encoding method. It should also be noted that in addition to the quantum complexity quoted above, the QIPM can require purely classical complexity on the order of $O(n^{2.5})$ for LP/SOCP and $O(n^{4.5})$ for SDP.

Alternatives to the standard approach above have been proposed. For "tall" LPs (where $m \gg n$), one can quantize interior point methods in a distinct way that avoids the QLSS and dependence on any condition numbers. Specifically, [19] gave an algorithm that runs in time $\widetilde{\mathcal{O}}(\sqrt{m} \log(1/\epsilon)) \cdot \text{poly}(n)$. The algorithm leverages primitives for spectral approximation (i.e., given a tall matrix $B$, finding a smaller matrix $\tilde{B}$ for which $\tilde{B}^\intercal \tilde{B} \approx B^\intercal B$), and approximate matrix-vector multiplication, as well as multivariate mean estimation [20] (which is related to quantum gradient estimation).

(ii) Quantum algorithms based on the multiplicative weights update (MWU) method have been proposed for SDP [9, 21, 10, 11] and LP [10, 22]. The quantum algorithm closely follows the classical algorithm based on MWU to iteratively update a candidate solution to the program. Each iteration is carried out using quantum subroutines, including Gibbs sampling, as well as Grover search and quantum minimum finding [23, 10] (a direct application of Grover search). Let $s$ denote the sparsity, that is, the maximum number of nonzero entries in any row or column of the matrices composing the problem input (thus, $s \leq \max(m, n)$). Then, the number of queries the algorithm makes to the matrix entries (assuming a sparse access input model) has been upper bounded by

$$\text{LP [24]:} \qquad \widetilde{\mathcal{O}}\left(\sqrt{s}\left(\frac{rR}{\epsilon}\right)^{3.5}\right)$$

$$\text{SDP [11]:} \qquad \widetilde{\mathcal{O}}\left(s\sqrt{m}\left(\frac{rR}{\epsilon}\right)^{4} + s\sqrt{n}\left(\frac{rR}{\epsilon}\right)^{5}\right),$$

where $r, R$ are the parameters related to the size of the primal and dual solutions, defined above. The sparse access queries can be implemented with quantum circuits of size polylog$(m, n)$ if the matrix entries are given by succinct formulas computable in polylog$(n, m)$ time. Otherwise, their implementation can be accomplished with circuits of polylog$(m, n)$ depth (but poly$(m, n)$ size) assuming availability of log-depth QRAM. In [11], the input model was generalized to a "quantum operator input model," based on block-encodings where $s$ is replaced by the block-encoding normalization factor $\alpha$ in the runtime expressions, but here again the full end-to-end complexity must account for the gate cost of implementing the block-encoding. Note that it is possible the $\epsilon$-dependence of the runtime for LP could be slightly improved by applying the dynamic Gibbs sampling method of [24] together with the reduction from LP to zero-sum games in [22].

---

[13]For example, numerical results in [17] for small instances of an SOCP formulation of the portfolio optimization problem suggested that the $\zeta$ parameter was upper bounded by a small constant, and similar numerical investigations in [18] suggest that $\xi$ can be independent of the target error $\epsilon$, at least for the instances that were simulated.

The runtime expressions for the QIPM approach and the MWU approach are not directly comparable, as the former depends on instance-specific parameters $\kappa$, $\zeta$, and $\xi$, while the latter depends on instance-specific parameters $r$ and $R$. However, note that the explicit $n$-dependence is better in the case of MWU than QIPM, while the $\epsilon$-dependence is worse.

**Existing resource estimates**

Neither of the approaches for conic programs have garnered study at the level of resource estimates for physical devices. Reference [18] performed a resource analysis for a QIPM at the logical level, but did not analyze additional overheads due to error correction. The goal of that analysis was to completely compile the QIPM for SOCP into Clifford gates and $T$ gates, and then to numerically estimate the parameters $\kappa$, $\zeta$, and $\xi$ for the particular use case of financial portfolio optimization, which can be reduced to SOCP. A salient feature of the QIPM is that $\mathcal{O}(n + m) \times \mathcal{O}(n + m)$ matrices of classical data must be repeatedly accessed by the QLSS via block-encoding, necessitating a large-scale QRAM with $\mathcal{O}(n^2)$ qubits. Accordingly, for SOCPs with $n = 500$ and $m = 400$ (which are still easily solved on classical computers) it was estimated that 8 million logical qubits would be needed. The total number of $T$ gates needed for the same instance size was on the order of $10^{29}$, which can be distributed over roughly $10^{24}$ layers. These estimates would likely be improved by incorporating subsequent improvements to the underlying primitives of tomography [25] and QLSS [26, 27].

We are not aware of an analogous logical resource analysis for the MWU approach to conic programming. Such an analysis would be valuable and should ideally choose a specific use case to be able to evaluate the size of all parameters involved. A use case that may fit this criteria is solving the SDP relaxation of binary quadratic optimization problems, where $r$ and $R$ can be bounded; quantum algorithms for this task have been studied in [28, 29].

**Caveats**

- The QIPM approach requires a large-scale QRAM of size $\mathcal{O}(n^2)$. This is a necessary ingredient to retain any hope of a speedup, and for relevant choice of $n$ the associated hardware requirements could be prohibitively large. Note that recent work of [30] gave a method for solving LPs that, like QIPMs, follows the central path to the optimal point, but does so in a distinct, non-iterative way, and has the potential for a small polynomial speedup without the need for a QRAM.

- The standard QIPM approach has a weak case for a large asymptotic speedup: even under optimal circumstances, the asymptotic speedup over classical interior point methods is less than quadratic. See Section 22 on the QIPM approach for more information.

- The MWU approach also requires a large-scale QRAM of size $\mathcal{O}(\min(ms, ns))$ to implement the queries to the arbitrary entries of the $s$-sparse input matrices. This could be avoided if the matrix elements are efficiently computable.

- Beyond the number of queries to the input data (and the cost to implement those queries), the MWU approach for LP and SDP requires some additional complexity deriving from a step in the algorithm where it prepares a state with $\mathcal{O}(R^2r^2/\epsilon^2)$ nonzero amplitudes stored in a classical database. The cost of state preparation from classical data is $\widetilde{\mathcal{O}}(R^2r^2/\epsilon^2)$ total gates, which can be parallelized to circuit depth polylog$(R^2r^2/\epsilon^2)$. If the cost of state preparation is taken to be equal to the circuit depth (similar to the assumption of access to a medium-scale QRAM), the additional complexity is on the same order as the number of queries. If the cost is taken to be equal to the circuit size, the additional complexity is a factor $\widetilde{\mathcal{O}}(R^2r^2/\epsilon^2)$ larger than the query complexity quoted above.

- The MWU approach has poor dependence on error $\epsilon$; for SDPs it is $\epsilon^{-5}$. Even at modest choices of $\epsilon$, this may lead the algorithm to be impractical pending significant improvements.

- A general caveat that applies to both approaches is that the appearance of instance-specific parameters makes it difficult to predict the performance of these algorithms for more specific applications.

**Comparable classical complexity and challenging instance sizes**

As in the quantum case, there are multiple distinct approaches in the classical case.

(i) Classical interior point methods (CIPMs): There exist fast IPM-based software implementations for solving conic programs, such as ECOS [31], MOSEK [32], Gurobi [33], SCIP [34], and CPLEX.[14] These solvers can solve instances with thousands of variables in a matter of seconds on a standard laptop (e.g., [31]). However, the runtime scaling is poor and scaling too far beyond this regime leads the solvers to be far less practical. Many variants of IPMs exist; the runtime of the best provably correct classical IPMs for the regime where the number of constraints is roughly equal to the number of degrees of freedom is

$$\text{LP [1]:} \qquad \widetilde{\mathcal{O}}(n^\omega \log(1/\epsilon))$$
$$\text{SOCP [2]:} \qquad \widetilde{\mathcal{O}}(n^{\omega+0.5} \log(1/\epsilon))$$
$$\text{SDP [3]:} \qquad \widetilde{\mathcal{O}}(n^{2\omega} \log(1/\epsilon)),$$

where $\omega < 2.37$ is the matrix multiplication exponent. It is plausible that, with some attention, the extra $n^{0.5}$ factor for SOCP could be eliminated with modern techniques. Additionally, the runtime can be somewhat reduced when the number of constraints is much less than the number of degrees of freedom; for example, the $n$-dependence of the complexity of the CIPM for SDP in [35] can be as low as $\widetilde{\mathcal{O}}(n^{2.5})$ when there are few constraints. On practical instances, employing techniques for fast matrix multiplication is often not beneficial, and Gaussian elimination–like methods are used, where $n^\omega$ is replaced with $n^3$. Note that, alternatively, by using iterative classical linear system solvers, such as the randomized Kaczmarz method [36], each $n^\omega$ factor could be replaced by a factor of $n$ at the cost of a linear dependence on $(\kappa\zeta)^2$, which could be superior if the matrices are well conditioned.

---

[14]In practice, these solvers are not solely based on IPMs and utilize many methods at once. Additionally, they employ heuristic preprocessing methods to transform and simplify inputs prior to applying an IPM. Note that they are useful also for nonconvex problems such as mixed-integer programs, where LP-solving can often be an important subroutine.

For tall LPs ($m \gg n$), CIPMs can achieve scaling nearly linear in the number of matrix entries: the algorithm of [37] runs in time $\mathcal{O}(mn + n^3)$.

We refer the reader to [38, 39, 40, 41] for additional information on CIPMs and their historical development.

(ii) Classical MWU methods: A classical complexity statement for LPs is inferred from the reduction in [22] from LPs to zero-sum games and the classical analysis that appears there. For the SDP case, references in the classical literature appear only to examine specific subclasses of SDPs (e.g., [6, 7]). A general statement of the classical complexity for SDPs appears alongside the quantum algorithm in [10, Section 2.4]:

$$\text{LP [22]:} \qquad \widetilde{\mathcal{O}}\left(s\left(\frac{rR}{\epsilon}\right)^{3.5}\right)$$

$$\text{SDP [10]:} \qquad \widetilde{\mathcal{O}}\left(snm\left(\frac{rR}{\epsilon}\right)^{4} + sn\left(\frac{rR}{\epsilon}\right)^{7}\right).$$

(iii) Cutting-plane methods: These classical methods are used for SDPs and can outperform IPMs when the number of constraints is small. The best algorithm, based on [4, 5], has runtime $\mathcal{O}(m(mn^2 + n^\omega + m^2)\log(1/\epsilon))$, which can be as low as $\mathcal{O}(n^\omega)$ when $m$ is small.

It is important to note that the algorithms with the best provable complexities may not be the ones that are most useful in practice.

## Speedup

For both the IPM approach and the MWU approach, there can be at most a polynomial quantum speedup: upper and lower bounds scaling polynomially with $n$ are known in both the classical and quantum cases [11]. The speedup of the QIPM method depends on the scaling of $\kappa$ with $n$, but the speedup cannot be more than quadratic. For the MWU method, if $m = \mathcal{O}(n)$, $s = \mathcal{O}(1)$, and $rR/\epsilon = \mathcal{O}(1)$, existing bounds on the classical and quantum complexities leave open the possibility of a quartic $\widetilde{\mathcal{O}}(n^2) \to \widetilde{\mathcal{O}}(\sqrt{n})$ speedup. However, it is unclear if the classical complexity quoted in [10, Section 2.4] is optimal (e.g., if the classical $\widetilde{\mathcal{O}}(mn)$ scaling could be improved to $\widetilde{\mathcal{O}}(m+n)$, the available quantum speedup would be at most quadratic). There is a possibility that the speedup could be larger in practice if the Gibbs sampling routine is faster on actual instances than its worst-case upper bounds suggest, perhaps by utilizing Monte Carlo–style approaches to Gibbs sampling.

## Outlook

It is very plausible that an asymptotic polynomial speedup can be obtained in problem size using the MWU method for solving LPs or SDPs, but the speedup appears only quadratic, and an assessment of practicality depends on the scaling of certain unspecified instance-specific parameters. Similarly, the standard QIPM method could bring a subquadratic speedup but only under certain assumptions about the condition number of certain matrices. The alternative QIPM method of [19] could deliver a nearly quadratic speedup without assumptions on the condition number in the case the LP constraint matrix is very tall. In any case, these quadratic and subquadratic speedups alone might be regarded as unlikely to yield practical speedups after error correction overheads and slower quantum clock speeds are considered. Future work should aim to find additional asymptotic speedups while focusing on specific practically relevant use cases that allow the unspecified parameters to be evaluated.

## 5.3   General convex optimization

**Overview**

A convex problem asks to minimize a convex function $f$ over a convex set $K$, where $K$ is a subset of $\mathbb{R}^n$. Here we examine the situation where the value of $f(x)$ and the membership of $x$ in the set $K$ can each be efficiently computed classically. However, we do not exploit/assume any additional structure that may be present in $f$ or $K$. This situation contrasts with that of solving conic programs, where $f$ is linear and $K$ is an intersection of convex cones and affine spaces, features that can be exploited to yield more efficient classical and quantum algorithms.

A so-called "zeroth-order" solution to this problem solves it simply by adaptively evaluating $f(x)$ and $x \in K$ for different values of $x$. For the zeroth-order approach, a quantum algorithm can obtain a quadratic speedup with respect to the number of times these functions are evaluated, reducing it from $\widetilde{\mathcal{O}}(n^2)$ to $\widetilde{\mathcal{O}}(n)$, where $\widetilde{\mathcal{O}}$ notation hides factors polylogarithmic in $n$ and other parameters. This could lead to a practical speedup only if the cost to evaluate $f(x)$ and $x \in K$ is large, and lack of structure rules out other, possibly faster, approaches to solving the problem.

**Actual end-to-end problem(s) solved**

Suppose we have classical algorithms $\mathcal{A}_f$ for computing $f(x)$ and $\mathcal{A}_K$ for computing $x \in K$ ("membership oracle"), which require $C_f$ and $C_K$ gates to perform with a reversible classical circuit, respectively. Suppose further we have an initial point $x_0 \in K$ and that we have two numbers $r$ and $R$ for which we know that $B(x_0, r) \subset K \subset B(x_0, R)$, where $B(y, t) = \{z \in \mathbb{R}^n : \|z - y\| \le t\}$ denotes the ball of radius $t$ centered at $y$. Using $\mathcal{A}_f$, $\mathcal{A}_K$, $x_0$, $r$, $R$, and $\epsilon$ as input, the output is a point $\tilde{x} \in K$ that is $\epsilon$-optimal, that is, it satisfies

$$f(\tilde{x}) \le \min_{x \in K} f(x) + \epsilon \,.$$

**Dominant resource cost/complexity**

The work of [1] and [2] independently establish that there is a quantum algorithm that solves this problem with gate complexity upper bounded by

$$\left[ (C_f + C_K)n + n^3 \right] \cdot \mathrm{polylog}(nR/r\epsilon) \,,$$

where the polylogarithmic factors were left unspecified. The rough idea behind the algorithm is to leverage the quantum gradient estimation algorithm to implement a *separation oracle*—a routine that determines membership $x \in K$ and when $x \notin K$ outputs a hyperplane separating $x$ from all points in $K$—using only $\mathcal{O}(1)$ queries to algorithm $\mathcal{A}_K$ and $\mathcal{A}_f$. It had been previously established that $\widetilde{\mathcal{O}}(n)$ queries to a separation oracle then suffice to perform optimization [3], where $\widetilde{\mathcal{O}}$ denotes that logarithmic factors have been suppressed.

**Existing resource estimates**

There have not been any explicit resource estimates for this algorithm. It may not make sense to perform such an estimate without a more concrete scenario in mind, as the estimate would highly depend on the complexity of performing the circuits for $\mathcal{A}_f$ and $\mathcal{A}_K$. The estimate would also require a more detailed accounting of the hidden polylogarithmic factors in the complexity statements above, and it would only be meaningful if the comparable classical complexity for solving the same problem using the best available algorithm were well understood.

**Caveats**

One caveat is that the quantum algorithm must coherently perform reversible implementations of the classical functions that compute $f(x)$ and $x \in K$. Compared to a nonreversible classical implementation, this may cost additional ancilla qubits and gates. Another caveat relates to the scenario where $f(x)$ and $x \in K$ are determined by classical data stored in a classical database. Such a situation may appear to be an appealing place to look for applications of this algorithm because when $f$ and $K$ are determined empirically rather than analytically, it becomes easier to argue that there is no structure that can be exploited. However, in such a situation, implementing $\mathcal{A}_f$ and $\mathcal{A}_K$ would require a large gate complexity, so $C_f$ and $C_K$ would scale with the size of the classical database. It would almost certainly be the case that a quantum random access memory (QRAM) admitting log-depth queries would be needed in order for the algorithm to remain competitive with classical implementations that have access to classical RAM, and the practical feasibility of building a large-scale log-depth QRAM has many additional caveats.

Another caveat is that there may not be many practical situations that are compatible with a quantum speedup by this algorithm. The source of the speedup in [1, 2] comes from a separation between the complexity of computing the gradient of $f$ classically vs. quantumly using calls to the function $f$. Classically, this requires at least linear-in-$n$ number of calls. Quantumly, it can be done in $\mathcal{O}(1)$ calls using the quantum algorithm for gradient estimation. In both the classical and the quantum case, the gradient can subsequently be used to construct a "separation" oracle for the set $K$, which is then used to solve the convex problem.

Thus, a speedup is only possible if there is no obvious way to classically compute the gradient of $f$ other than to evaluate $f$ at many points. This criterion is violated in many practical situations, which are often said to obey a "cheap gradient principle" [4, 5] that asserts that the gradient of $f$ can be computed in time comparable to the time required to evaluate $f$. For example, the fact that gradients are cheap is crucial for training modern machine learning models with a large number of parameters. When this is the case, the algorithms from [1, 2] do not offer a speedup. On the other hand, as observed in [2, Footnote 19] a nontrivial example of a problem where the cheap gradient principle may fail (enabling a possible advantage for these quantum algorithms) is the moment polytope problem, which has connections to quantum information [6].

When both the function $f$ and the gradient of $f$ can be evaluated at unit cost, this constitutes "first-order" optimization, which can be solved classically by gradient descent. However, gradient descent does not generally offer a quantum speedup, as general quantum lower bounds match classical upper bounds for first-order optimization, although a quantum speedup could exist in specific cases [7]. Indeed, for any $p$, there is no general quantum speedup for $p$th-order optimization, that is, the setting where an oracle provides access to the function and its first $p$ derivatives [8]. For a comprehensive exposition of classical methods for black-box optimization, see [9].

**Comparable classical complexity**

The best classical algorithm [10] in the same setting has complexity

$$\left[ (C'_f + C'_K)n^2 + n^3 \right] \cdot \text{polylog}(nR/r\epsilon) \,,$$

where $C'_f$ and $C'_K$ denote the classical complexity of evaluating $f$ and querying membership in $K$, respectively, without the restriction that the circuit be reversible.

**Speedup**

The speedup is greatest when quantities $C_f$ and $C_K$ are large compared to $n$ and roughly equal to $C_f'$ and $C_K'$. In this case, the quantum algorithm can provide an $\mathcal{O}(n)$ speedup, which is at best a polynomial speedup. The maximal power of the polynomial would be obtained if $C_f + C_K \approx C_f' + C_K'$ scales as $n^2$, corresponding to a subquadratic speedup from $\mathcal{O}(n^4)$ to $\mathcal{O}(n^3)$.

**Outlook**

The only analyses of this strategy are theoretical in nature, interested more so in the query complexity of solving this problem than any specific applications it might have. As such, the analysis is not sufficiently fine-grained to determine any impact from constant prefactors or logarithmic prefactors. While a quadratic speedup in query complexity is possible, the maximal speedup in gate complexity is smaller than quadratic. Moreover, there is a lack of concrete problems that fit into the paradigm of "structureless" quantum convex optimization. Together, these factors make it unlikely that a practical quantum advantage can be found in this instance.

## 5.4 Nonconvex optimization: Escaping saddle points and finding local minima

**Overview**

Finding the global minimum of nonconvex optimization problems is challenging because local algorithms get stuck in local minima. In analogy to physics where the objective function is the energy of the system, these local minima are stable configurations that locally optimize the energy but do not achieve the globally minimal energy. Often, there are many local minima and they are each separated by large energy barriers. Accordingly, instead of finding the global minimum, one may settle for finding a local minimum: local minima can often still be used effectively in situations such as training machine learning models. An effective approach to finding a local minimum is gradient descent, but gradient descent can run into the problem of getting stuck near saddle points, which are not local minima but nonetheless have a vanishing gradient. Efficiently finding local minima thus requires methods for escaping saddle points. Limited work in this area suggests a potential polynomial quantum speedup [1] in the dimension dependence for finding local minima, using subroutines for Hamiltonian simulation and quantum gradient estimation.

**Actual end-to-end problem(s) solved**

Suppose we have a classical algorithm $\mathcal{A}_f$ for (approximately) computing a function $f : \mathbb{R}^n \to \mathbb{R}$ which requires $C_f$ gates to perform with a reversible classical circuit. The amount of error tolerable is specified later. Following [1], suppose further that $f$ is $\ell$-smooth and $\rho$-Hessian Lipschitz, that is,

$$\|\nabla f(x_1) - \nabla f(x_2)\| \le \ell \|x_1 - x_2\| \qquad \forall x_1, x_2 \in \mathbb{R}^n$$
$$\|\nabla^2 f(x_1) - \nabla^2 f(x_2)\| \le \rho \|x_1 - x_2\| \qquad \forall x_1, x_2 \in \mathbb{R}^n \,,$$

where $\nabla f$ denotes the gradient of $f$ (a vector), $\nabla^2 f$ denotes the Hessian of $f$ (a matrix), and the norm notation $\|\cdot\|$ denotes the standard Euclidean norm for vector arguments and the spectral norm for matrix arguments.

The end-to-end problem solved is to take as input a specification of the function $f$, an initial point $x_0$, and an error parameter $\epsilon$, and to output an $\epsilon$-approximate second-order stationary point (i.e., approximate local minimum) $x$, defined as satisfying

$$\|\nabla f(x)\| \le \epsilon \qquad\qquad \lambda_{\min}(\nabla^2 f(x)) \ge -\sqrt{\rho\epsilon} \,,$$

where $\lambda_{\min}(\cdot)$ denotes the minimum eigenvalue of its argument. In other words, the gradient should be nearly zero, and the Hessian should be close to a positive-semidefinite matrix.

**Dominant resource cost/complexity**

The idea pursued in the quantum algorithm of [1] is to run normal gradient descent, which has gradient query cost independent of $n$, until reaching an approximate saddle point. Classical algorithms typically apply random perturbations to detect a direction of negative curvature and continue the gradient descent. Instead, the quantum algorithm constructs a Gaussian wavepacket localized at the saddle point, and evolves according to the Schrödinger equation

$$\mathrm{i}\frac{\partial}{\partial t}\Phi = \left(-\frac{1}{2}\Delta + f(x)\right)\Phi \,, \tag{13}$$

where $\Delta$ denotes the Laplacian operator. The intuition is that, in the directions of positive curvature, the particle stays localized (as in a harmonic potential), while in the directions of negative curvature, the particle quickly disperses. Thus, when the position of the particle is measured, it is likely to have escaped the saddle point in a direction of negative curvature, and gradient descent can be continued. The other technical ingredient is the quantum gradient estimation algorithm, which uses a constant number of (coherent) queries to the function $f$ to estimate $\nabla f$.

The main finding of [1] is that, to do the gradient descent and the Hamiltonian simulation, the algorithm need only query the circuit evaluating the function $f$ polylog($n$) times. Specifically, the algorithm performs the $C_f$-gate quantum circuit for coherently computing $f$ a number of times scaling as

$$\widetilde{\mathcal{O}}\bigg(\frac{\log(n)(f(x_0) - f^*)}{\epsilon^{1.75}}\bigg),$$

where $x_0$ is the initial point and $f^*$ is the global minimum of $f$. The evaluation of $f$ must be correct up to precision $\mathcal{O}(\epsilon^2/n^4)$. Note that the work of [1] initially showed a $\log^2(n)$ dependence, which was later improved to $\log(n)$ using the improved simulation method of [2, Corollary 8]. However, it is important to emphasize that the method has additional cost beyond the queries to the circuit for evaluating $f$, originating from the Hamiltonian simulation of the kinetic term $-\frac{1}{2}\Delta$ in Eq. (13). Specifically, the number of additional gates needed in the Hamiltonian simulation is seen from [2, Lemma 12 & Corollary 8] to scale as $\widetilde{\mathcal{O}}(n(f(x_0) - f^*)/\epsilon^{1.75})$, although we remark that it is possible that this gate complexity could be parallelized such that the circuit depth scales polylogarithmically in $n$. Thus, the overall gate complexity is

$$\widetilde{\mathcal{O}}\bigg(\frac{(n + C_f \log(n))(f(x_0) - f^*)}{\epsilon^{1.75}}\bigg).$$

The space complexity to represent the $d$-dimensional system on a grid with grid spacing $\mathcal{O}(\epsilon)$ is $\mathcal{O}(n \log(1/\epsilon))$.

In related work, [3] analyzes the complexity of escaping a saddle point when one has access to *noisy* queries to the value of the function $f$. Additionally, lower bounds on the $\epsilon$-dependence of quantum algorithms for this problem are given in [4].

**Existing resource estimates**

This problem has received relatively little attention, and no resource estimates have been performed.

**Caveats**

Reference [1] gives the query complexity of the quantum algorithm but does not perform a full end-to-end resource analysis. (However, it does numerically study the performance of the quantum algorithm in a couple of toy examples.) Additionally, many practical scenarios are said to obey a "cheap gradient principle" [5, 6], which says that computing the gradient is almost as easy as computing the function itself, and in these scenarios, no significant quantum speedup is available. Finally, in the setting of variational quantum algorithms, this does not avoid the issue of barren plateaus, which refers to the situation where a large portion of the parameter space has a gradient (and Hessian) that vanishes exponentially with $n$. These regions would be characterized as $\epsilon$-approximate local minima unless $\epsilon$ is made exponentially small in $n$.

**Comparable classical complexity and challenging instance sizes**

The best classical algorithm [7] for this problem makes

$$\widetilde{\mathcal{O}}\left(\frac{\log(n)(f(x_0) - f^*)}{\epsilon^{1.75}}\right)$$

queries to the *gradient* of $f$. Note that $\Omega(n)$ queries to the value of $f$ would be needed to construct a query to the gradient. (When the quantum algorithm in [1] was first discovered, the best classical algorithm required $\mathcal{O}(\log(n)^6)$ gradient queries [8, Theorem 3], and this was later improved.) Although the literature focuses mainly on the query complexity, examination of [7, Algorithm 1] indicates that an additional $\widetilde{\mathcal{O}}(n(f(x_0) - f^*)/\epsilon^{1.75})$ arithmetic operations would be required to process the results of the gradient queries and compute the next point to be queried (e.g., adding pairs of $n$-dimensional vectors).

**Speedup**

The quantum algorithm in [1] has the same query complexity as the classical algorithm in [7]; the difference is that the quantum algorithm makes (coherent) queries to an evaluation oracle, while the classical algorithm requires access to a gradient oracle. Thus, if classical gradient queries are just as cheap as evaluation queries (as is often the case), there is no speedup. If it were the case that gradient queries are not directly available, then the speedup in query complexity could be exponential. However, even in this case, the speedup in gate complexity can be at most polynomial, since both the classical and quantum algorithms have $\mathrm{poly}(n)$ gate complexity, and classical gradient queries can be constructed from $\widetilde{\mathcal{O}}(n)$ classical evaluation queries, which can be accomplished with at most $C_f$ classical gates. Indeed, the largest polynomial speedup in gate complexity occurs when $C_f = \mathcal{O}(n)$—in this case, the quantum algorithm has $\widetilde{\mathcal{O}}(n)$ gate complexity and the classical algorithm has $\widetilde{\mathcal{O}}(n) \cdot C_f = \mathcal{O}(n^2)$ gate complexity, a quadratic speedup.

**Outlook**

It is unclear whether the algorithm for finding local minima could lead to a practical speedup, as it depends highly on the (non)availability of an efficient classical procedure for implementing gradient oracles; a quantum speedup is possible only when such oracles are difficult to implement classically, and even so, the speedup in gate complexity would be modest. However, the algorithm represents a useful end-to-end problem where the quantum gradient estimation primitive can be applied. It is also notable that the quantum algorithm employs Hamiltonian simulation, a primitive not used in most other approaches to continuous optimization. Relatedly, [9, 10] propose a quantum subroutine called "quantum Hamiltonian descent" which is a genuinely quantum counterpart to classical gradient descent, via Hamiltonian simulation of an equation similar to Eq. (13). Unlike classical gradient descent, it can exploit quantum tunneling to avoid getting stuck in local minima; thus, it can potentially find *global* minima of nonconvex functions. Establishing concrete end-to-end problems where quantum approaches based on Hamiltonian simulation yield an advantage in nonconvex optimization is an interesting direction for future work.

# 6 Cryptanalysis

Computation and communication are secured by cryptography. For example, a user's data can be made private, along with messages that they send or receive, from malicious agents who interfere to try to learn the sensitive information. A set of algorithms collectively called a *cryptosystem* endows the security. The attempt to break security is known as *cryptanalysis*, which has its own set of algorithms. Historically, both cryptography and cryptanalysis considered classical, polynomial-time algorithms as the only realistic ones. The advent of quantum computation forces us to consider attacks via quantum algorithms. Generally, we want to know what is the best algorithm for cryptanalysis, in order to understand the effect on the cryptosystem in the worst case. Quantum attacks can void the security of widely used cryptosystems (see Section 6.1 on breaking cryptosystems). More broadly, quantum cryptanalysis can reduce a cryptosystem's security (see Section 6.2 on weakening cryptosystems), such that it becomes more expensive to implement in a secure manner. While the properties of quantum mechanics can also be used to devise more secure cryptosystems (e.g., quantum key distribution) [1, 2, 3], we consider this area of cryptography to be outside the scope of the present discussion on quantum algorithms.

*The authors are grateful to Matthew Campagna and Samuel Jaques for reviewing this section of the survey.*

**This application area contains:**

## 6.1 Breaking cryptosystems

**Overview**

Much of modern cryptography relies on computational assumptions.[15] A cryptosystem is a protocol to achieve some security goal, such as hiding information, ensuring integrity of information, or computing a function correctly. A cryptosystem is secure if, assuming a particular mathematical problem is hard to solve (or assuming the existence of certain functions, e.g., pseudorandom), an adversary cannot compromise the security goal. The earliest such cryptosystems used particular problems from number theory, and variants are widely deployed to this day [1]. These cryptosystems are in the class of public-key or asymmetric cryptography. Public-key cryptography uses key pairs: a private key known only to one user, and a public key that can be widely distributed to allow any user to perform tasks like encryption. In contrast, symmetric-key cryptography uses a single secret key that must be preshared between communicating parties.

Quantum computers use quantum algorithms to solve computational problems, and in some cases they provide a speedup over the best known classical techniques. When they are applied to the underlying computational task in a cryptosystem, a large speedup over classical methods can break the cryptosystem, in that an adversary efficiently learns the secret key or the encrypted information to a non-negligible degree. One of the first discovered and most famous applications of quantum computing is Shor's algorithm [2], which breaks or solves the integer factorization problem, and both the general discrete logarithm problem and the elliptic curve discrete logarithm problem. These problems are believed to be classically hard, and are the basis of security for the most common public-key cryptosystems, like Diffie–Hellman, Rivest–Shamir–Adleman (RSA), and elliptic curve cryptography (ECC). The applications of these public-key cryptosystems include encryption to hide the contents of a message, signatures that prevent tampering and impersonation, and key exchange to generate a key for symmetric-key cryptography [3]. In this section, we restrict our focus to two of the most widely used cryptosystems: RSA and ECC.

**Actual end-to-end problem(s) solved**

The RSA cryptosystem [4] relies on a user choosing a large number $N$ that is the product of two prime numbers; arithmetic is done modulo $N$. Denote by $n = \lceil \log_2(N) \rceil$ the number of bits specifying $N$. Along with the modulus $N$, two integers $e$ and $d$ are used as exponents, such that $(m^e)^d = m \bmod N$ for all values $0 \le m < N$. The pair $(N, e)$ is the public key, and the value $d$ is the private key. A message $m$ is encrypted as $m^e \bmod N$. Exponentiation with $d$ performs decryption, recovering $m$ as specified in the relation above. Some applications store the factors of $N$, which must also be kept private. The user generates $e$ and $d$ using their knowledge of the prime factors of $N$. First, a suitable $e$ is generated, and then $d$ is computed from the prime factors of $N$ and $e$ using some number-theoretic facts together with the extended Euclidean algorithm. However, if an adversary is able to find the factors of $N$ after the construction by the user, they can also solve for $d$ and thereby decrypt messages. The security of RSA is based on the observed difficulty of factoring large numbers like $N$, that is, the integer factorization problem.

ECC is based on a different problem, the elliptic curve discrete logarithm problem (ECDLP), which has the advantage of smaller key sizes for equivalent security levels, compared to RSA. Consequently, fewer resources (e.g., communication, complexity of encryption and decryption) are required to implement ECC. Elliptic curves are constructed over a finite field $K$, as the set of solutions to the equation

$$y^2 = x^3 + ax + b\,, \quad a, b \in K\,,$$

which specify points on an elliptic curve [5, 6]. The set of solutions $P = (x, y)$ forms an abelian group under a specially defined addition operation. Collectively, the set of parameters $K, a, b$ and a so-called base point $G$ (a solution to the equation of prime additive order $N$, i.e., $NG$ is the additive identity) specify the cryptosystem. A private key is a random integer $k$ satisfying $1 \leq k < N$, and a public key is the value $P = kG$, the result of adding $G$ to itself $k$ times. The assumption of hardness is in the following problem (ECDLP): *Given $P$ and $G$, where $P = kG$ for some secret value $k$, find $k$.* ECC is constructed from the observation that calculating $P$ from $k, G$ is efficient, whereas it is computationally infeasible for an adversary to compute $k$ from points $P$ and $G$.

---

[15]An example of a cryptosystem not requiring computational assumptions is the one-time pad.

**Dominant resource cost/complexity**

Shor's algorithm [2] solves the number-theoretic problem of order finding: given $n$-bit positive integer $N$ and $x$ coprime to $N$, find the smallest integer $r$ such that $x^r = 1 \bmod N$. Factoring was shown to reduce to order finding. In particular, there is an efficient, otherwise classical algorithm, of classical complexity $\mathcal{O}(n^3)$ [7], that uses order finding as a quantum subroutine. To describe the quantum algorithm for order finding, let the function $f$ denote modular exponentiation, that is, $f(e) = x^e \bmod N$, and note that $f$ is periodic with (unknown) period $r$. Also, let $L$ be a large integer such that an interval of length $L$ contains many periods, that is, $L \gg r$. It can be shown that $L \geq N^2$ is sufficient. There are three steps. First, an equal superposition over the numbers $\{0, \dots, L-1\}$ is formed and the function $f$ is computed into an ancilla register yielding the state $L^{-1/2} \sum_{e=0}^{L-1} |e\rangle |f(e)\rangle$. Second, a measurement is performed on the ancilla register, which, due to the periodicity of the function $f$, yields a state $(\lceil L/r \rceil)^{-1/2} \sum_{j=0}^{\lfloor L/r \rfloor} |rj + y\rangle$ for $0 \leq y < r$ a random and unknown integer.[16] Third, a quantum Fourier transform is performed. In the case that $L$ is a multiple of $r$, the result is

$$\frac{\sqrt{r}}{L} \sum_{j=0}^{L/r} \sum_{z=0}^{L-1} e^{2\pi i z(rj+y)/L} |z\rangle = \frac{1}{\sqrt{r}} \sum_{\ell=0}^{r-1} e^{2\pi i \ell y/r} |\ell L/r\rangle\,, \qquad (14)$$

where the equality follows since coefficients of $|z\rangle$ for which $z$ is not equal to $\ell L/r$ for some integer $\ell$ vanish due to destructive interference. Measurement of this state then produces an outcome $\ell L/r$ for a randomly sampled $\ell$. The value of $r$ can be classically computed by dividing the measurement outcome by $L$ and determining the value of the denominator of the rational number that results; repetition may be required since $\ell$ and $r$ could have common divisors. If $L/r$ is not an integer, the measurement outcome is (with high probability) an integer close to $\ell L/r$ for some integer $\ell$. One can deduce the rational number $\ell/r$ (which allows for the determination of $r$) from the estimate of $\ell L/r$ by writing it as a continued fractions expansion, with classical complexity $\mathcal{O}(n^3)$ [7].

This entire procedure can alternatively be viewed as quantum phase estimation applied to the unitary $U$ that sends $|y\rangle \mapsto |xy \bmod N\rangle$ for all $y$ relatively prime to $N$, performed with at least $2n$ bits of precision.

---

[16] If $r\lfloor L/r \rfloor + y \geq L$, then the $j = \lfloor L/r \rfloor$ term does not appear in the expression.

The number of qubits for order finding—and hence for Shor's factoring algorithm—is $\mathcal{O}(n)$, which stems from the number of bits specifying the problem: the first register has size $2n$, and the ancilla register holding the result $f(e)$ has size $n$. Naively, the number of gate operations is $\mathcal{O}(n^2)$ for the quantum Fourier transform and $\mathcal{O}(n^3)$ for implementing the coherent modular exponentiation $|e\rangle|0\rangle \mapsto |e\rangle|x^e \bmod N\rangle$. The $\mathcal{O}(n^3)$ arises from decomposing modular exponentiation into $\mathcal{O}(n)$ modular multiplications, one for each bit of $e$—using schoolbook multiplication, the gate cost per multiplication is $\mathcal{O}(n^2)$. Implementing this modular arithmetic with reversible circuits represents the bottleneck in the complexity. These circuits are closely related to those in classical computing that have been optimized. Still, improvements can lead to better resource counts [8]. The best scaling in theory is achieved with algorithms that have large prefactors in their complexity, making them impractical to implement except when $n$ is large: total gate complexity of $\mathcal{O}(n^2 \log(n))$ is possible asymptotically, using integer multiplication with $\mathcal{O}(n \log(n))$ scaling [9], although analyses optimized for $n \approx 2048$ use methods for which the total gate complexity scales as $\widetilde{\mathcal{O}}(n^3)$ [10]. Alternatively, optimization may be performed to, for example, increase qubit count and decrease gate count or gate depth. For example, an approximate version of the quantum Fourier transform is implemented with $\mathcal{O}(n \log(n))$ gates and allows factoring with $\mathcal{O}(\log(n))$-depth quantum circuits [11], at the cost of extra overhead in number of qubits and gates; allowing for $\mathcal{O}(\log^2(n))$-depth preserves the circuit size $\mathcal{O}(n^3)$.

A related approach proposed by Regev [12] for quantum factoring has quantum circuit size of only $\widetilde{\mathcal{O}}(n^{3/2})$ gates (assuming fast integer multiplication at cost $\widetilde{\mathcal{O}}(n)$), but the circuit has to be run $\mathcal{O}(n^{1/2})$ times. Thus, it achieves the same overall asymptotic gate complexity as Shor's algorithm. The idea of the algorithm is to extend period finding to higher dimensions and optimize resource counts. Unlike Shor's algorithm, which is proven to succeed, Regev's approach relies on a number-theoretic assumption, albeit a plausible one. The reduction in quantum circuit depth and natural parallelism of the approach may lead to more favorable resource counts in practice. In particular, a constant fraction of the runs can fail [13], which may allow it to be implemented fault tolerantly with less overhead. Initial work on optimizing the algorithm has established a tradeoff between the number of qubits and the number of gates. Linear qubit cost of $\mathcal{O}(n)$ is possible (although the constant prefactor is larger than that of Shor's algorithm) while still maintaining $\widetilde{\mathcal{O}}(n^2)$ total gate complexity [13, 14].

Essentially the same quantum algorithm of Shor is readily applied to elliptic curves, as well as the discrete logarithm problem (i.e., find $r$ such that $a^r = b$ for $a, b \in G$ where $G$ is a group), which is also used as a computationally hard problem for cryptography. These applications are all instances of the *hidden subgroup problem*: Find the generators for subgroup $K$ of a finite group $G$, given a quantum oracle performing $U|g\rangle|h\rangle = |g\rangle|h \oplus f(g)\rangle$, where $f : G \to X$ ($X$ is a finite set) is a function that is promised to be constant on the cosets of $K$ and take unique values on each coset. In the case of period finding, $G$ is the group $\mathbb{Z}/L\mathbb{Z}$ under addition, and the hidden subgroup is $K = \{0, r, 2r, \ldots, L - r\}$ (technically a subgroup only if $r$ divides $L$); one can verify that $f(g) = x^g \bmod N$ is constant on each coset of $K$. The procedure outlined above for period finding can be applied to other groups, where it is called "the standard method" [15], which requires generalizing the quantum Fourier transform to arbitrary groups. A simple example is Simon's problem [16]—indeed, historically speaking Simon's algorithm inspired Shor's [17]—where $G$ is the abelian group $(\mathbb{Z}/2\mathbb{Z})^n$ of bit strings of length $n$ under addition, $K = \{0, c\}$ for some hidden bit string $c$, and the generalized Fourier transform is simply the Hadamard transform $H^{\otimes n}$. For abelian groups, the hidden subgroup $K$ can be determined with polylog$(|G|$ queries to $f$, but the method does not work for nonabelian groups, such as the symmetric group and the dihedral group.

**Existing resource estimates**

The minimum recommended key size for RSA is 2048 bits [18]. Optimizations in the circuits [19, 20] and incorporation of hardware constraints [21] have led to decreasing but also more realistic resource estimates. For key size $n = 2048$, an optimized resource estimate was performed in [10], geared toward implementation on a device with nearest-neighbor connectivity in 2D, where logical qubits are encoded with the surface code. Their implementation used roughly $3n \approx 6000$ logical qubits and roughly $0.3n^3 \approx 3 \times 10^9$ non-Clifford gates (a mixture of Toffoli and $T$ gates). Accounting for magic state distillation and routing in 2D, it was shown how the computation could be laid out on a 2D grid of 14,000 logical qubits.

For ECC, the minimum recommended key size to ensure 128-bit security (quantifying the number of attacks needed to learn the encrypted information; see Section 6.2 on weakening cryptosystems for details), is $n = 256$ bits [18] (achieving the same level of security with RSA requires a key size of 3072 bits [22, 23]). For breaking 256-bit ECC, an early resource esimate concluded that around three times fewer logical qubits, and 100 times fewer Toffoli gates are required (compared to 3072-bit RSA) [23]. Similar to factoring, improvements have been made in logical circuit compilation [24] and how this translates into hardware implementations [25, 26, 27]. In [27], a method was given requiring $1.1 \times 10^8$ Toffoli gates and 6000 logical qubits. Additionally, by offloading some of the work to a brute-force classical computation and exploiting a simplification that arises when computing multiple ECC keys in parallel, the total Toffoli count per key was shown to approach $4.4 \times 10^7$ [27]. As a conclusion, breaking elliptic curve cryptography is easier than factoring for quantum computers in practice [28], relative to their practical difficulty on classical computers.

The physical resources required to implement these logical circuits fault tolerantly depends on many details of the hardware, including the error rate, the physical gate speed, and the available connectivity. In both cases (2048-bit RSA [10, 29] and 256-bit ECC [25, 26, 27]), given current hardware schemes restricted to nearest-neighbor 2D connectivity with logical qubits encoded into surface codes, the number of physical qubits is estimated to be on the order of 10 million and the computation runs for at least 3–10 hours (significantly longer than this for platforms with relatively slower physical gate speeds). For a discussion on how to convert between logical and physical resources, see Section 17 on fault-tolerant quantum computation. Optimization based on the particular architecture can give improvements to these estimates. For example, if one assumes a logarithmic number of nonlocal links, as could be envisaged in photonic implementations, the estimated runtime can be reduced to less than one minute per 256-bit ECC key [27]. The algorithms considered in the resource estimates above do not achieve the best known asymptotic scaling, which comes at the cost of large constant prefactors.

**Caveats**

While the popular cryptosystems based on number-theoretic problems are rendered insecure for public-key cryptography, there exist alternatives that are believed to be secure against quantum computers: for example, based on error correcting codes or lattices [3]. These alternative computational problems are believed to be hard for both classical and quantum computers. The National Institute of Standards and Technology (NIST) of the United States has provided standards and encouraged implementation [30]. The class of symmetric-key cryptography (see a standard text [1] for details) involves computations that do not have much structure, and also is not broken by quantum computers.[17] Instead, the number of bits of security is reduced.

Prior experimental demonstrations of Shor's algorithm have used knowledge of the answer in order to optimize the circuit and thus lead to sizes that are experimentally feasible on non-error-corrected devices. Meaningful demonstration should avoid such shortcuts [32], which are not available in realistic cryptographic scenarios.

### Comparable classical complexity and challenging instance sizes

The best known classical algorithm for factoring is the number field sieve, which has time complexity superpolynomial in number of bits $n$: namely, it scales as $\mathcal{O}(\exp(p \cdot n^{1/3} \log^{2/3}(n)))$, where $p > 1.9$. With a hybrid quantum-classical algorithm applying amplitude amplification on the number field sieve, $p = 1.387$ can be achieved using a number of qubits scaling only as $\mathcal{O}(n^{2/3})$ [33]. Classically, problems of size 795 bits have been factored, taking 76 computer core-years, which distributed in parallel over a cluster took 12 days; the same team then extended the record to 829 bits [22].

Several algorithms attacking elliptic curve cryptography have complexity $\mathcal{O}(2^{n/2})$ [34], leading to the recommended doubling of key size compared to bits of security. In practice, a problem of size 117 bits was solved [35]. The corresponding security is estimated [36] to be about 60 bits, compared to 70 bits for the RSA record above.

### Speedup

The number of gates to implement Shor's algorithm is $\widetilde{\mathcal{O}}(n^2)$ asymptotically using fast multiplication on large numbers [37]. More practically, without incurring the time overhead and additional storage space of fast multiplication, the scaling is $\mathcal{O}(n^3)$. Assuming classical and quantum gates are polynomially related in time complexity, the speedup for solving the factoring problem is superpolynomial, and the speedup for solving the ECDLP is exponential. However, there are no tight lower bounds on the classical complexity of factoring or ECDLP; it remains possible that more efficient classical algorithms could be discovered.

### NISQ implementations

The large circuit depth, complicated operations, and high number of qubits needed to implement Shor's algorithm make faithful NISQ implementation challenging. However, there have been several attempts to ease implementation at the expense of losing the guarantees of Shor's algorithm, in the hope that the output is still correct with some nonzero probability, which could be vanishing.

One approach [38] is to simplify several operations and make them approximate. The outcome is that the circuit depth is $\mathcal{O}(n^2)$, saving a factor of $n$ [20]. The depth is then about $10^8$ to factor a 1024-bit instance of RSA, so for relevant sizes, error correction is still required. Implementation of the approximate algorithm, including experimentally, allowed for the successful factorization of larger problem instances than had been possible before. This approximate version is not NISQ in the usual sense of involving noisy circuits, but rather introduces some uncontrolled approximation error in return for reducing the depth, for the possibility of a useful result. Another approach is to encode the factoring problem in a variational optimization circuit. Again, performance is not guaranteed; moreover, variational optimization applied to generic problems is expected to have, at best, a quadratic improvement compared to classical methods, leaving no hope for breaking cryptography. Classical simulation on small problem sizes shows that the algorithm can succeed [39], as does experimental implementation on a superconducting quantum processor [40]. We emphasize that, generally, these NISQ approaches have no evidence or arguments for scaling to cryptographically relevant system sizes.

---

[17]If the adversary can query the cryptosystem's algorithms in superposition, some symmetric-key cryptography can be broken using period finding to extract the key [31]. However, this capability of the adversary is not considered realistic.

## Outlook

The existence of Shor's algorithm implies common RSA and elliptic curve schemes are theoretically not secure, and resource estimates have made clear what scale of quantum hardware would break them. While such hardware does not exist currently, progress toward such a device can be used to inform the speed of transitioning to quantum-resistant encryption [41]. Currently, from a hardware perspective, the field of quantum computing is far from implementing algorithms that would break encryption schemes used in practice. The estimates above suggest that the resources required would be millions of physical qubits performing billions of Toffoli gates running on the timescale of hours or days. In contrast, the current state of the art is on the order of one hundred noisy physical qubits, with progress toward demonstration of a single logical qubit. Running fault-tolerant quantum computation requires extra overhead, such as magic state factories (see Section 26 on quantum error correction and Section 27 on lattice surgery). Thus, the gap between state-of-the-art hardware and the requirements for breaking cryptosystems is formidable. Moreover, a linear increase in key size will increase, for example, the number of Toffoli gates by a power of three, which can be substantial. Therefore, considering the experimental challenges, likely only the most sensitive data will be at risk first, rather than common transactions. Consequently, these highly confidential communications will likely adopt post-quantum cryptography first to avoid being broken. However, insecure protocols often linger in practice, so quantum computers can exploit any vulnerabilities in deployed systems that have not been addressed. For example, RSA keys of size 768 bits have been found in commercial devices (note that such key sizes can already be broken classically [22]). In addition, intercepted messages, encrypted with RSA or elliptic curves, can be stored now and decrypted later, once large-scale quantum computers become available.

The resilience of candidates for post-quantum cryptography is under active investigation. In particular, specialized quantum attacks [42, 43] can reduce the number of bits of security, weakening the cryptosystem. Relaxing to toy variants of relevant cryptosystems in order to find new attacks, quantum algorithms can provide polynomial-time solutions [44]. Classical algorithms have even broken certain candidate cryptosystems [45, 46]. Note that these attacks affect the feasibility of particular proposals, but there exist other post-quantum candidates that have no known weaknesses. With the completion of NIST's standardization process, approved post-quantum cryptography is rapidly deployed. For example, several popular messaging platforms [47, 48] recently adopted post-quantum key derivation hybridized with ECC, so that the scheme is secure as long as at least one of the underlying cryptosystems is secure.

A sensitive area that warrants additional discussion is cryptocurrency, since much of it relies on the compromised public-key cryptography based on abelian groups. Moreover, changing the cryptographic protocol of the currency requires that most of the users reach a consensus to do so, which can be challenging to coordinate, even if the technical hurdles of adopting post-quantum encryption are overcome. Cryptocurrency wallets that have revealed their public key (e.g., via a transaction reusing a public key assigned to that wallet previously) can be broken using Shor's algorithm. An attack is also possible during the short time window in which the key is revealed during a single transaction [49]. Different cryptocurrencies have different levels of susceptibility to these types of attacks [50, 51]. Nevertheless, the mining of cryptocurrency is not broken, but only weakened by quantum computers.

## 6.2  Weakening cryptosystems

**Overview**

The discovery of Shor's algorithm (see Section 6.1 on breaking cryptosystems) prompted interest in post-quantum cryptography, the study of cryptosystems assuming the presence of large-scale, working quantum computers [1]. While some existing systems retained confidence in their security, others that were broken by quantum algorithms were superseded by those that accomplish the same task, but are believed to maintain a high level of security against quantum attacks.

Even if a cryptosystem is not broken altogether, its degree of security can be weakened by quantum algorithms. The strength of a cryptosystem is typically quantified by the number of bits of security (also called the security parameter), that is, $n$ bits corresponds to guessing the desired information with probability $1/2^n$ and accessing what is being protected. When considering computational assumptions, a simplified definition of the security parameter $n$ is that cryptanalysis requires a computational cost of $2^n$ operations, captured by the best known attack. *Breaking* a cryptosystem means only an efficient number of operations (i.e., poly$(n)$) are needed, while an attack that *weakens* a cryptosystem still takes $2^m > \text{poly}(n)$ operations, for some $m < n$.

In contrast to public-key cryptosystems, symmetric-key cryptography was discovered earlier and has fewer capabilities. However, it relies less on the presumed hardness of underlying mathematical problems, and correspondingly has only been weakened by quantum cryptanalysis, as discussed in more detail below.

**Actual end-to-end problem(s) solved**

In symmetric-key cryptography, two communicating parties share the same key $K$, which is used both in encryption $Enc_K$ and decryption $Dec_K$. As usual, the cryptographic algorithm $(Enc_K, Dec_K)$ is known to everyone, including adversaries. The most significant break of a symmetric-key algorithm is an adversary learning the key, given $r$ pairs of plaintext (the message $m$) and corresponding ciphertext $c$ (its encryption).[18] Such a pair can be accessed by, for example, forcing a certain test message to be transmitted. Precisely, an input $K$ is sought for which the following function outputs 1:

$$f(K) = \left( (Enc_K(m_1) = c_1) \wedge \cdots \wedge (Enc_K(m_r) = c_r) \right),$$

that is, find a key such that all the messages encrypt correctly. A straightforward attack is to use brute force and test every key; in practice, sophisticated classical attacks do not perform better than this approach in asymptotic scaling.

**Dominant resource cost/complexity**

The main, generic quantum attack is to use amplitude amplification: given a classical algorithm with success probability $\mathcal{O}(2^{-n})$ of finding a solution, the probability is increased quadratically to $\mathcal{O}(2^{-n/2})$. Thus, applying amplitude amplification to the task of solving for the key, the security of cryptosystems goes from $n$ bits to $n/2$.

---

[18]There are more sophisticated attack models: for example, many communicating adversaries may have the goal of compromising one of multiple keys [2]. Correspondingly, there are other definitions of security, but this simple and powerful one generally is considered for quantum attacks.

The function queried in superposition must be efficient to evaluate with a quantum circuit, which is often the case in cryptography [1]. However, the operations are typically long sequences of Boolean arithmetic. As such, a universal gate set and fault-tolerant quantum computation are still required. To store the key, $\mathcal{O}(n)$ register qubits are needed, and many more ancilla qubits are used for the reversible arithmetic.

**Existing resource estimates**

Consider the Advanced Encryption Standard (AES) [3], a symmetric encryption algorithm that is widely used in cryptosystems, for example, for encrypting web traffic. At a high level, it mixes the plaintext and adds it to the key to obtain the ciphertext. An attack based on amplitude amplification needs around 3000–7000 logical qubits [4] for AES-$k$, where $k$ denotes key size in bits, and $k \in \{128, 192, 256\}$. For these sizes, the number of necessary problem instances $r$ is three to five. While the number of logical qubits roughly doubles going from AES-128 to AES-256, the number of $T$ gates goes from $2^{86} \approx 10^{25}$ to $2^{151} \approx 10^{45}$. More recent work [5] allows for higher qubit counts (by about 70%) in exchange for much lower Toffoli depth, such that their product is reduced. Such optimization resulted in about 99% reduction in this metric.

**Caveats**

Since the quantum attack only halves the exponent in the complexity, a simple fix is to double the key length, for example, by adopting AES-256 instead of AES-128. This modification results in increased, but usually tolerable, cost in implementation (i.e., complexity of encryption and communication resources). In addition, there exist cryptosystems with an information-theoretic security guarantee, assuming adversaries with unlimited computational power, which covers against quantum attacks [1].

Furthermore, it is important to note that to realize the full quadratic benefit of amplitude amplification, the $\mathcal{O}(2^{n/2})$ function queries must be performed in series. In contrast, classical brute-force attacks can exploit the parallelism available in high-performance classical computers, potentially increasing the value of $n$ for which a quantum approach would be advantageous over classical methods.

**Comparable classical complexity and challenging instance sizes**

Classical algorithmic attacks on AES have reduced the security by only a few bits [6]. More practical are side-channel attacks, which make use of physical byproducts, such as energy consumption. For example, when comparing bits between a key and another string, a flipped value can result in logic that increases energy consumption, compared to the same value where nothing happens. The two cases are distinguished and information about the key is learned. Currently, 128 bits of security is roughly the minimum recommended amount [7]. The use of parallelization forces the adoption of relatively large key sizes, compared to what is necessary for a single processor ($\sim 60$ bits).

**Speedup**

The basic speedup is quadratic: $\mathcal{O}(\sqrt{N})$ function evaluations compared to $\mathcal{O}(N)$ classically, where $N$ denotes the number of possibilities for the key; that is, $n = \lceil \log_2(N) \rceil$. However, the function queries in amplitude amplification cannot be parallelized. Then, the evaluation time of the function sets a bottleneck [1]. That is, the problem size is limited by the number of function evaluations $T$ that can be run in an acceptable period of time. For $\sqrt{N} > T$, employing $p$ parallel quantum processors, each executes $T = \sqrt{N/p}$ evaluations. Then, $p = \mathcal{O}(N/T^2)$ and the total number of evaluations is $pT = \mathcal{O}(N/T)$, whereas classically, the number of processors is $\mathcal{O}(N/T)$ and total evaluations is $\mathcal{O}(N)$. The advantage is a factor of $T$, which is the bottleneck, rather than the larger $\sqrt{N}$. However, the advantage can be overshadowed by faster or cheaper classical processing. That is, if classical computers evaluate the function $T$ times faster than quantum processors, there is no advantage in runtime with using the quantum device. Furthermore, this argument assumes the same cost of parallelization for classical and quantum, which is optimistic for quantum devices. An example of this effect is in mining cryptocurrency [8]: while a quantum computer needs quadratically fewer attempts to succeed, the development of fast, specialized, classical hardware negates the advantage. Essentially, for brute-force attacks, parallelization has the most significant impact in cryptanalysis.

## NISQ implementations

The key can be encoded as the ground state of a Hamiltonian, and then variational methods can be applied to solve for it. The scaling is expected to be the same as amplitude amplification. However, since the variational algorithm does not have a set time complexity, the solution may be found much slower or faster [9]. If the fluctuations are large enough, they can potentially pose a challenge to cryptography, which makes worst-case guarantees. However, there is no reason to expect that the success probability will scale favorably with key size and compromise security in practice. Another approach is to use amplitude amplification, but adapt it to near-term devices, so that the NISQ-optimized versions perform better in real experiments [10].

## Outlook

Here, we focused on the example of symmetric-key encryption. Nonetheless, the effect of amplitude amplification to halve the effective bits of security is generic for computational problems, assuming efficient construction of the oracle. From the cryptographic standpoint, this attack is mild and can be counteracted by doubling the number of bits of security in the scheme. In practice, the increase in key size can be unwieldy in certain applications, such as cryptocurrencies, but fundamental security is not threatened.

# 7 Solving differential equations

*The authors are grateful to Dong An, Di Fang, and Ashley Montanaro for reviewing this section of the survey.*

**Overview**

Many applications in engineering and science rely on solving differential equations. Accordingly, this constitutes a large fraction of research-and-development high-performance computing (HPC) workloads across a wide variety of industries. There have been many proposals to speed up differential equation solving using a quantum computer. At this point, the consensus is that we lack compelling evidence for practical quantum speedup on industry-relevant problems. However, the field is progressing rapidly, and this conclusion is subject to change.

Some of the main application areas that have been considered are:

- **Computational fluid dynamics** (CFD), usually involving simulation of the Navier–Stokes equation. The main industries relying on CFD simulations are automotive, aerospace, civil engineering, wind energy, weather/climate modeling, and defense. While most simulations focus on air or fluid flow on solid objects, other processes, such as foaming, are also important to model. Large CFD calculations are routinely in the petaflop regime and are run on millions of CPU cores. Specific quantum proposals include [1, 2, 3, 4, 5, 6, 7, 8, 9].

- **Geophysical modeling**, involving simulation of the wave equation. The main industries are oil and gas, hydroelectric, geophysics. Large seismic imaging simulations can easily be in the petaflop regime. Quantum proposals for simulating the wave equation include [10, 11, 12].

- A wide variety of engineering problems involving the **finite element method** (FEM) for studying structural properties of solid objects. The main industries are civil engineering, manufacturing (including automotive), aerospace, and defense. The simulations are typically slightly smaller in scale than CFD, though still often requiring large HPC clusters. Quantum FEM proposals include [13, 14, 15, 16].

- **Maxwell's equations and the heat equation** have applications in chip design and other electronic component design, as well as for navigation and radar technology. Specific quantum proposals include [17, 13, 18].

- **Plasma physics** simulations, involving the simulation of the Vlasov equation, are widespread in nuclear fusion research. Quantum approaches include [19, 20, 21].

- **Risk modeling**, involving the simulation of stochastic differential equations, is extensively used in finance (especially derivatives pricing), insurance, and energy markets. Specific quantum proposals include [22, 23, 24, 25, 26].

Differential equations can be categorized according to a number of properties: (i) *ordinary vs. partial* depending on the number of differential variables, (ii) *stochastic vs. deterministic* depending on whether the function is a random variable or not, (iii) *linear vs. nonlinear*. We will focus mainly on linear ordinary and partial differential equations, which have received the most attention in the quantum computing literature, and only comment briefly on stochastic and nonlinear differential equations.

In order to solve a differential equation numerically, one typically specifies a discretization scheme. Two important classes of discretization schemes are: (i) *grid-based* schemes, including finite difference methods (FDMs), as well as the finite volume method (FVM) and the FEM combined with various choices of support grids and preconditioning (see [27, 28] for an introduction). For example, in the finite difference framework, the continuous space is discretized on a uniform grid and the continuous operators are replaced by finite difference operations on neighboring grid points. Alternatively, (ii) one can discretize space by expansion in a functional basis (Fourier, Hermite, etc.), and solve the discrete problem in this basis. This second class of methods is often referred to as *spectral methods* [29]. There is often a tradeoff between error convergence and regularity requirements, with higher-order grid-based methods and spectral methods offering faster error convergence with the number of grid points or basis functions utilized, but requiring more stringent assumptions on the smoothness of the solution of the differential equation, which are not always satisfied in the applications listed above.

Given a discretization scheme, solving linear differential equations can be accomplished by solving linear systems of equations. In cases where one is interested in very high precision, requiring very fine discretization, the linear system of equations can be too large for straightforward numerical solutions on a classical computer. In particular, if one wants high-precision results integrated over time, and/or systems with many continuous variables, then the simulations can be challenging both in time and memory. Quantum algorithms aim to offer a speedup over classical methods by leveraging the existence of quantum linear system solvers, or more generally, primitives in quantum linear algebra, which enable quantum algorithms to perform efficient manipulations of vectors that are exponentially large in the number of qubits and elementary operations involved. However, at a technical level, various complications arise, including the difficulty of reading out useful information at the end of the algorithm, and assumptions about the differential equation that must be true for the methods to work. Ultimately, polynomial speedups for end-to-end problems appear to be possible, but for differential equations in a fixed number of spatial dimensions, exponential speedups for real-world applications are not generally attainable.

**Actual end-to-end problem(s) solved**

We are interested in solving a general linear partial differential equation (PDE) of the form

$$\mathcal{L}(u(x)) = f(x) \quad \text{for} \quad x \in \mathbb{C}^d\,, \tag{15}$$

where $\mathcal{L}$ is a linear differential operator acting on the function $u(x)$, and $f(x) \in \mathbb{C}$ is the inhomogeneous term (which is 0 for homogeneous PDEs). In addition to Eq. (15), we are given boundary conditions on $u(x)$ and its derivatives, which ideally are sufficient to ensure a unique solution—for example, Dirichlet boundary conditions refer to a specification of a function $b(x)$ and a requirement that $u(x) = b(x)$ for $x$ contained in some subset $\Omega \in \mathbb{C}^d$. As a canonical example of a linear PDE, consider the Poisson equation in $d$ dimensions, given by

$$\frac{\partial^2 u}{\partial x_1^2} + \cdots + \frac{\partial^2 u}{\partial x_d^2} = f(x)\,. \tag{16}$$

As another example, we consider a linear ordinary differential equation (ODE) of the form

$$\frac{\mathrm{d}\bar{u}(t)}{\mathrm{d}t} = A(t)\bar{u}(t) + \bar{b}(t)\,, \tag{17}$$

where we refer to the variable $t \in \mathbb{R}$ as time (although it could represent a different quantity), $\bar{u}(t)$ and $\bar{b}(t)$ are $N$-dimensional vectors, and $A(t)$ is an $N \times N$ matrix. Boundary conditions on $\bar{u}(t)$ are also specified, often in the form of an initial condition at $t = 0$, with one seeking the solution at some final time $T$; this is known as an *initial value problem*. Higher-order linear ODEs can always be transformed into first-order linear ODEs. Note that if $A(t)$ is anti-Hermitian and $\bar{b}(t) = 0$, then Eq. (17) becomes the time-dependent Schrödinger equation, which is solved directly with Hamiltonian simulation. Equation (17) could be viewed within the framework of Eq. (15) with $d = 1$ and $u$ a vector-valued rather than scalar-valued function. We separate these cases because the existing literature typically uses either Eq. (15) or Eq. (17) as its starting point, and the methods pursued in each case are distinct.

For nonlinear PDEs, the linear equations in Eq. (15) and Eq. (17) are replaced by nonlinear ones. For example, one can extend Eq. (17) to consider an ODE with a polynomial nonlinearity of the form

$$\frac{\mathrm{d}\bar{u}(t)}{\mathrm{d}t} = F_M(t)\bar{u}(t)^{\otimes M} + A(t)\bar{u}(t) + \bar{b}(t)\,, \tag{18}$$

where $F_M(t)$ is a tensor encoding the nonlinearity, although note that existing quantum algorithms have focused on the case where $F_M(t)$ and $A(t)$ are time independent. This class of differential equations includes important potential applications such as CFD, since the Navier–Stokes equation is nonlinear with a quadratic nonlinearity ($M = 2$).

What does it mean to "solve" the differential equation? In the most general sense, this refers to obtaining an expression for the solution $u(x)$ (for Eq. (15)) or $\bar{u}(t)$ (for Eq. (17)). While closed-form solutions can be derived for some simple differential equations, this is not possible in general, and the solution typically must be computed numerically. However, in specific applications, we often do not need complete information about the solution function $u(x)$ or $\bar{u}(t)$ to accomplish a certain goal. An end-to-end problem involving the solution of a differential equation boils down to estimating the value of some *property* of the solution, denoted by $\mathcal{P}[u] \in \mathbb{R}$, up to specified additive error $\epsilon$. For linear PDEs, a straightforward example is when the property $\mathcal{P}$ is simply the value of $u$ at a specific point $x^*$, that is, $\mathcal{P}[u] = u(x^*)$. More generally, we restrict to the case where $\mathcal{P}[u]$ is a linear functional of $u$, that is, $\mathcal{P}[u] = \langle r, u \rangle := \int_{x \in \Omega} \mathrm{d}x\, r(x)u(x)$ for some subset $\Omega \subset \mathbb{R}^d$ and function $r : \Omega \to \mathbb{R}$ for which $\langle r, r \rangle = 1$ [14]. For example, in [17], a quantum algorithm for solving Maxwell's equations based on the FEM was given, where the quantity of interest was not the electric field itself at any specific point, but rather the electromagnetic scattering cross section. In this case, the cross section was given by the square of a linear functional of $u$.

Properties of ODEs can be treated in the same framework, where we are interested in computing quantities of the form $\mathcal{P}[u] = \int_{t \in \Omega} \mathrm{d}t\, \bar{r}^{\mathsf{T}}(t)\bar{u}(t)$, which are linear functionals of the entries of $\bar{u}(t)$ over some subset $\Omega$ of the interval $[0, T]$. However, we note that for initial value problems, often of primary interest are properties at the final time $T$, in which case $\mathcal{P}[u]$ would reduce to an inner product $\bar{r}^{\mathsf{T}}\bar{u}(T)$. For example, in [30], the drag force on a ship hull was expressed as a linear functional of the solution to the lattice Boltzmann equation evolved forward in time.

**Dominant resource cost/complexity**

There are many distinct approaches to constructing a quantum algorithm for solving the end-to-end problem above. The exact complexity will of course depend on the method, but it will also depend on specific details related to how the differential equation and its boundary conditions are specified to the quantum algorithm (input model), as well as instance-specific factors such as how smooth the solution to the differential equation is. Here we overview some of the available choices and complexity considerations, focusing the bulk of the discussion on methods that leverage the quantum linear system solver (QLSS) as a primitive, as these have received the most attention in the literature.

**Discretization of linear PDEs:** Any numerical method must perform some form of discretization. First, we focus on linear PDEs such as Eq. (16) where there is no time variable. The choice of discretization will depend sensitively on the problem at hand. In the case of the Poisson equation in Eq. (16) with Dirichlet boundary conditions, quantum algorithms leveraging discretization schemes based on FDM, FEM, and spectral methods were discussed in [31], [14], and [32], respectively. The key goal is to minimize the number $N$ of grid points or basis functions while achieving discretization error $\mathcal{O}(\epsilon)$. Using low-order grid-based methods, a problem in $d$ spatial dimensions requires taking $N = (1/\epsilon)^{\Omega(d)}$ grid points, with some caveats on solution norm and regularity [14]. Alternative sparse-grid or spectral methods can improve the $1/\epsilon$ dependence to logarithmic, but still scale exponentially with $d$ [32]; however, these generally require stricter regularity requirements on the solution to the differential equation, which may not be satisfied in applications.

After appropriate discretization, the linear differential equation in Eq. (15) (along with its boundary conditions) reduces to a matrix equation:

$$L|u\rangle = |f\rangle. \tag{19}$$

This is the same linear equation that would be obtained for a classical method using the same discretization scheme. Information about the solution function $u(x)$ is encoded into the $N$ components of the vector $|u\rangle$.[19] Classical methods typically manipulate a full description of all components of the vector $|u\rangle$, whereas quantum methods can create the normalized quantum state $|u\rangle/\||u\rangle\|$ encoding these $N$ components into its amplitudes with $\mathcal{O}(\log(N))$ qubits.

If the linear PDE has a time variable, one option is to treat it equally as the other $d-1$ variables (see, e.g., [18]), but it is often treated separately. First, discretization of the other $d-1$ variables using $N$ total grid points or basis functions is performed as above, which reduces the linear PDE in Eq. (15) to an ODE with $N$ variables as in Eq. (17). Time is then discretized and propagated as discussed for ODEs below.

**Discretization of time in linear ODEs:** To solve the linear ODE on $N$ variables in Eq. (17)—whether it came about via discretization of a PDE or otherwise—the time interval $[0,T]$ is discretized into grid points, and the solution at one grid point is related to the solution at the prior grid point by a time-ordered exponential of the matrix $A$. If $A$ is time independent, this exponential can be approximated by a truncated Taylor series [33, 34, 35], and if $A$ is time dependent, it can be approximated by a truncated Dyson series [36]. The number of grid points needed scales linearly with $T$, and the series is truncated at order $\mathrm{polylog}(T/\epsilon)$. An alternative approach when $\bar{u}(t)$ is sufficiently smooth in time uses spectral methods, which approximate $\bar{u}(t)$ as a truncated series over a complete set of basis functions [37]. In any case, the relation between the solution at different time grid points or basis functions leads to a linear system of equations, now of size roughly $N' = \mathcal{O}(NT)$, but again of the form $L|u\rangle = |f\rangle$ as in Eq. (19).

---

[19]In this section, we adopt a convention where quantum states like $|u\rangle$ need not be normalized states. In fact, the norm, denoted by $\||u\rangle\|$ will be important for reasoning about the complexity.

Here the solution $|u\rangle$ is a "history state" meaning that it is given by a *superposition* of states $|t\rangle|\bar{u}(t)\rangle$ for different discrete values of $t$ across the entire interval $[0, T]$. Since one is often interested only in $\bar{u}(T)$, additional trivial time steps can be included at the end to boost the portion of the history state amplitude on the $t = T$ branch [38].

It is important to emphasize that classical methods for solving ODEs do not solve the same linear equation $L|u\rangle = |f\rangle$ arrived at by these methods. Rather, they typically handle time in a time-marching fashion where the value of $\bar{u}(t)$ at one time step is directly computed from its value at one or more previous time steps. In [39], a quantum time-marching strategy was proposed for time-dependent homogeneous linear ODEs, which generates a sequence of quantum states $|\bar{u}(0)\rangle, |\bar{u}(t_1)\rangle, |\bar{u}(t_2)\rangle, \ldots, |\bar{u}(T)\rangle$ (rather than a superposition of these states). This method avoids the need to solve a linear system, but it does utilize primitives from quantum linear algebra. Similarly, the methods of [40, 41, 42, 43, 44, 45] avoid the need to solve linear systems by mapping ODEs to equations that can be simulated with Hamiltonian simulation [40, 41, 42, 43], or linear combination of Hamiltonian simulation [44, 45].

**Solving the linear system:** Once the linear PDE or ODE has been reduced to a linear system $L|u\rangle = |f\rangle$, it can be solved on a quantum computer by applying the QLSS. The QLSS subroutine prepares a quantum state approximating the normalized solution vector $|u\rangle/\|\,|u\rangle\,\|$ up to some specified precision $\xi$ in $\ell_2$ norm, where $\|\,|u\rangle\,\| = \sqrt{\langle u|u\rangle}$ is the norm of the quantum state encoding the solution to the linear system. To do so, the QLSS assumes access to oracles that (coherently) query the matrix elements of $L$ and prepare the normalized state $|f\rangle/\|\,|f\rangle\,\|$. Optimal QLSSs [46, 47] (see also alternative near-optimal methods in [48, 49, 50, 51, 52, 35]) make $\mathcal{O}(s\kappa \log(1/\xi))$ queries to these oracles, where $\kappa$ is the condition number of the matrix $L$ (i.e., the ratio of the largest and smallest singular values), and $s$ is the maximum number of nonzero elements in any row or column of $L$ ("sparsity"). Additionally, one can compute an estimate for the norm $\|\,|u\rangle\,\|$ up to relative error $\xi$ using $\widetilde{\mathcal{O}}(s\kappa/\xi)$ queries (note the worse $\xi$-dependence) [47, 50]. For simplicity, we assume that to achieve $\epsilon$ overall error on the end-to-end problem, it will suffice to take $\xi = \mathcal{O}(\epsilon)$, although there can also be factors that depend on the choice of discretization and norms of the solution $u$ (see, e.g., [14]).

Henceforth, let $N'$ refer to the size of the linear system being solved, so $N' = N$ for the PDE example described above, and $N' = \mathcal{O}(NT)$ for the ODE example (with $N$ reserved for the size of the matrix $A$).

The oracles for querying the matrix elements of the $s$-sparse $N' \times N'$ matrix $L$ and for preparing the $N'$-dimensional state $|f\rangle/\|\,|f\rangle\,\|$ are assumed to have cost $\mathrm{polylog}(N')$. This is valid if the matrix elements of $L$ can be efficiently computed "on the fly," which is plausible when they are given by succinct expressions, for example, based on a simple finite difference formula. However, if entries of $L$ and $|f\rangle/\|\,|f\rangle\,\|$ depend on arbitrary, classically stored data related to, for instance, object geometries, boundary conditions, or grid point locations, then the assumption of $\mathrm{polylog}(N')$ cost per query requires access to a log-depth quantum random access memory (QRAM). This requirement is necessary in many practical applications of PDEs involving highly complex geometries in three spatial dimensions, such as CFD and seismic modeling. The assumption of log-depth QRAM brings significant caveats—for example, while the quantum circuits for implementing the QRAM operation can be parallelized to have depth polylog($N$), they cannot avoid having size poly($N$) for accessing a database with poly($N$) entries; see Section 17 on loading classical data for more information.

With these assumptions, the QLSS portion of the quantum algorithm can be performed exponentially faster in the parameter $N$, and with exponential saving in memory, compared to any classical method that manipulates vectors of size $N$, which includes Gaussian elimination and iterative methods like conjugate gradient. Specifically, the quantum complexity to obtain the state $|u\rangle/\|\,|u\rangle\,\|$ is given by

$$s\kappa \cdot \text{polylog}(N', 1/\epsilon)$$

and the cost to obtain $\|\,|u\rangle\,\|$ is $s\kappa\epsilon^{-1} \cdot \text{polylog}(N', 1/\epsilon)$. The number of qubits is $\mathcal{O}(\log(N'))$, although if a log-depth QRAM is necessary, this may require poly($N'$) ancilla qubits.

**Reading out estimates for properties of the solution to the differential equation:**
Preparing the $\log(N')$-qubit state $|u\rangle/\|\,|u\rangle\,\|$ does not immediately yield an estimate for the property $\mathcal{P}[u]$. Indeed, reading out useful information from $|u\rangle/\|\,|u\rangle\,\|$ represents a major bottleneck of the algorithm. Consider the case that $\mathcal{P}[u]$ corresponds to the value $u(x^*)$ at a specific point $x^*$ (for PDEs), or the amplitude $\langle x^*|\bar{u}(T)\rangle$ on one of the basis states (for ODEs). Then, the estimation of $\mathcal{P}[u]$ to precision $\epsilon$ is performed with amplitude estimation, which introduces multiplicative overhead $\mathcal{O}(\|\,|u\rangle\,\|/\epsilon)$ into the complexity. To read out all $N$ amplitudes of the state $|u\rangle$ in this fashion, a linear factor of $N$ would be reintroduced, although more advanced methods of pure state tomography can reduce this to $\sqrt{N}$ [53]. In the more general case that $\mathcal{P}[u]$ is a linear functional, the value of $\mathcal{P}[u]$ can be expressed (up to discretization error) as an overlap $\mathcal{P}[u] = \langle r|u\rangle$ between some preparable normalized state $|r\rangle$ and the solution vector $|u\rangle$ that solves $L|u\rangle = |f\rangle$. Thus, to compute $\mathcal{P}[u]$, one computes the overlap between $|r\rangle$ and $|u\rangle/\|\,|u\rangle\,\|$, and then multiplies by $\|\,|u\rangle\,\|$. Overlap estimation [54] is a straightforward application of amplitude estimation, and achieving precision $\epsilon/\|\,|u\rangle\,\|$ introduces $\mathcal{O}(\|\,|u\rangle\,\|/\epsilon)$ multiplicative overhead. Thus, the overall scaling of the complexity is

$$\frac{s\kappa \,\|\,|u\rangle\,\|}{\epsilon} \cdot \text{polylog}(N', 1/\epsilon) \,. \tag{20}$$

For initial value problems governed by the ODE in Eq. (17), one is often interested in properties $\mathcal{P}[u]$ that depend only on the solution $\bar{u}(T)$ at the final time $T$. When $|u\rangle = \sum_i |t_i\rangle |\bar{u}(t_i)\rangle$ is a history state encoding of the solution $\bar{u}(t)$ over the whole interval, the complexity of computing useful information will grow with the ratio

$$q = \frac{\sup_{t \in [0,1]} \|\,|\bar{u}(t)\rangle\,\|}{\|\,|\bar{u}(T)\rangle\,\|} \approx \mathcal{O}\left(\frac{\|\,|u\rangle\,\|}{\|\,|\bar{u}(T)\rangle\,\|}\right),$$

that is, the factor by which the norm of the solution has decayed compared to its maximum on the interval $[0,T]$ (the approximation is correct assuming $\sup_{t \in [0,T]} \|\,|\bar{u}(t)\rangle\,\|$ accounts for a constant fraction of the total norm $\|\,|u\rangle\,\|$ ). This arises from the fact that $|T\rangle|\bar{u}(T)\rangle$ contributes at most $\|\,|\bar{u}(T)\rangle\,\| \approx \mathcal{O}(\|\,|u\rangle\,\|/q)$ amplitude to the history state $|u\rangle$ and thus the additive precision $\epsilon$ will need to be on the order of $\|\,|u\rangle\,\|/q$, or smaller, to learn something useful about $\bar{u}(T)$. Since the complexity scales linearly with $\|\,|u\rangle\,\|/\epsilon$, the complexity grows with $q$. Unfortunately, $q$ can be large and growing with $T$ if the solution to the ODE is decaying. Furthermore,

the dependence of the complexity on $q$ is necessary since otherwise the algorithm would be able to perform postselection on low-probability events, a power ruled out by widely believed complexity-theoretic conjectures; see [33, 55].

The persisting polylog($N$) dependence in Eq. (20) suggests an exponential speedup in the parameter $N$ compared to classical methods, but this conclusion depends on the scaling of the parameters $s$, $\kappa$, and $\||u\rangle\|/\epsilon$ with $N$.

**Condition number:** The sparsity $s$ and condition number $\kappa$ depend on the differential equation and the choice of discretization, but can often be controlled. For example, for PDEs discretized by the FEM, in many instances we have $s = \mathcal{O}(1)$ and $\kappa \leq \mathcal{O}(N^{2/d})$ (see, e.g., [56, Theorem 9.7.1]). Additionally, preconditioning of linear systems to reduce $\kappa$ is an effective technique in classical iterative approaches to solving linear systems such as the conjugate gradient method, and several studies have examined the possibility of integrating these into the QLSS in some circumstances [17, 57, 58, 59]. In the best case scenario, these could reduce $\kappa$ to $\mathcal{O}(1)$.

In the setting of ODEs like Eq. (17), upper bounds on $\kappa$ can be derived. These bounds can have the form $\kappa \leq CT$, where $C$ is a factor that depends on the spectral properties of $A$.[20] The upper bounds on $\kappa$ also require an assumption that the ODE is "dissipative" [38, 33, 34, 36]; otherwise, the value of $C$ in the bound can grow exponentially with $T$ [34], consistent with the observation that the norm of the history state $|u\rangle$ can be exponentially larger than the norm of the initial state $|\bar{u}(0)\rangle$. A sufficient condition for the ODE to be dissipative is that $A$ is diagonalizable and the real parts of its eigenvalues are non-positive [33] so that the solutions are stable and do not grow with time, although this technical definition was relaxed slightly in [34] and now includes nondiagonalizable $A$. The requirement of dissipation is not as relevant for classical solvers based on time-marching strategies, which can renormalize growing solutions at each step and do not generally require solving linear systems.

The linear dependence on $T$ in the complexity cannot be improved in general since this factor appears in the complexity of optimal Hamiltonian simulation, which corresponds to the special case that $A$ is anti-Hermitian and $\bar{b} = 0$ [60]. However, it has been shown that many ODEs admit "fast forwarding" and the dependence on $T$ can be reduced. For example, for stable ODEs (when all eigenvalues of $A$ have a negative real part), a bound of the form $\kappa \leq \widetilde{\mathcal{O}}(\sqrt{T})$ was derived in [35]; see also [55].

**Final complexity:** For dissipative ODEs on $N$ variables, propagated to time $T$, the final complexity inherits a linear dependence on $CT$ via the condition number, and existing literature typically includes the factor $q$ defined above directly in the complexity statements. These statements are phrased to say that the state $|\bar{u}(T)\rangle$ can be obtained to error $\xi$ in complexity roughly $TsqC \cdot \text{polylog}(N, 1/\xi)$ [33, 34, 35, 36], which accounts for postselecting the time register to $t = T$ but not yet the complexity to read out a property of interest.[21] Defining $\epsilon' = \epsilon q/\||u\rangle\|$ to be the normalized precision parameter that should be small in order to learn something interesting, and taking $\xi = \mathcal{O}(\epsilon')$, the total end-to-end complexity including readout could then be expressed as

$$\frac{TsqC}{\epsilon'} \cdot \text{polylog}(N, q/\| \, |u\rangle \, \|\epsilon') \,. \tag{21}$$

---

[20]Generally, the bound on $\kappa$ scales with the number of grid points in time. The linear-in-$T$ bound is achieved when using the Dyson series method with high-order truncation [36].

[21]We briefly mention the complexity of alternative approaches to ODEs that avoid solving linear systems. The quantum time-marching method of [39] has a different complexity form, but has similar features, growing with time (in this case, as $T^2$) and depending on an "amplification ratio" $Q > q$, but offering other potential benefits, such as minimal regularity requirements (even allowing $A(t)$ to have jump discontinuities) and needing fewer queries to the initial condition $\bar{u}(0)$. Meanwhile, the linear combination of Hamiltonian simulation method [45] shares the feature of needing minimal queries to $\bar{u}(0)$ while matching the complexity stated above.

For PDEs in $d$ dimensions, we recall that $N$ and $\epsilon$ are not independent parameters: in general, we are interested in simulating the PDE to a fixed precision, and adapt $N$ to reach the desired precision. As discussed, depending on the discretization method, $N$ scales either as $(1/\epsilon)^{\mathcal{O}(d)}$ or $(\text{polylog}(1/\epsilon))^d$, but either way, we have that $\text{polylog}(N) \leq d^{\mathcal{O}(1)} \cdot \text{polylog}(1/\epsilon)$. For PDEs with a time variable and initial conditions specified at $t = 0$, where $d - 1$ dimensions are discretized and time is integrated via mapping to an ODE, we substitute this value of $N$ into Eq. (21). This gives a total complexity of

$$d^{\mathcal{O}(1)} \cdot \widetilde{\mathcal{O}}\big(TCq/\epsilon'\big)$$

for reading out a property of the (renormalized) solution to the PDE at time $T$ to precision $\epsilon'$.

For static PDEs where all $d$ dimensions are discretized via a grid-based method like the FEM, we instead substitute $\kappa = \mathcal{O}(N^{2/d})$, $s = \mathcal{O}(1)$, and $N = (1/\epsilon)^{\mathcal{O}(d)}$ into Eq. (20), yielding overall end-to-end complexity

$$\frac{\| \, |u\rangle \, \| \, d^{\mathcal{O}(1)}}{\epsilon^{1+\mathcal{O}(1)}}$$

to compute a global property of the PDE, given conditions on the boundary. If preconditioning improves $\kappa$ to $\mathcal{O}(1)$, the $\epsilon$ dependence is improved to essentially linear in $1/\epsilon$—see [14] for a more careful analysis specific to the FEM that arrives at similar expressions, but better accounts for impact of solution norm and smoothness.

Generally, the main conclusion is that—irrespective of the discretization scheme—the quantum complexity is polynomial in the desired precision $1/\epsilon$, although for $d$-dimensional PDEs the complexity may scale as $\text{poly}(d) \cdot \text{poly}(1/\epsilon)$ rather than $\text{poly}((1/\epsilon)^d)$. Thus, for fixed dimension $d$ there is potentially room for a polynomial-in-$1/\epsilon$ quantum speedup, the size of which grows exponentially in the dimension $d$. Ultimately, the necessity of the $\mathcal{O}(1/\epsilon)$ scaling is traced back to the fact that the quantum solver produces a quantum state encoding the normalized solution to the differential equation, potentially exponentially faster than leading classical methods such as conjugate gradient, but the exponential speedup is lost in the readout step, where one must learn an observable of interest to error $\epsilon$. Moreover, this conclusion holds not just for "bad" observables (like full state tomography), but for any observable, due to the $\Omega(1/\epsilon)$ cost of quantum readout.

**Nonlinear differential equations:** The immediate issue with applying the above methods to nonlinear differential equations such as the nonlinear ODE of Eq. (18) is that discretization no longer leads to a system of linear equations. Early work on quantum algorithms for nonlinear ODEs handled this issue by dividing time into short time steps and preparing the quantum state encoding the solution at one time step using multiple copies of the solution at the previous time step [61]. Since quantum states cannot be cloned, the complexity of this strategy necessarily grows exponentially in the number of time steps. More recent quantum algorithms for nonlinear differential equations instead proceed by first linearizing the differential equation and then applying the methods sketched above [62, 63, 64, 65, 34, 66]. Specifically, the most heavily studied approach has been Carleman linearization, where one exactly maps a nonlinear

ODE with polynomial nonlinearity such as Eq. (18) to a linear ODE on an infinite-dimensional variable ($\bar{u}, \bar{u}^{\otimes 2}, \bar{u}^{\otimes 3}, \ldots$), and then truncates to form an approximate finite-dimensional linear ODE. Quantum algorithms based on this method were first studied in [63] and further developed in [65, 34, 66, 67], and it has also been analyzed in the context of specific differential equations such as reaction-diffusion equations [65, 66, 68], the Navier–Stokes equation (via the lattice Boltzmann equation) [1, 30], and differential equations related to training classical neural networks [69]. The complexity of the quantum algorithm has a similar scaling as that for linear ODEs quoted in Eq. (21), growing linearly in $T$, $q$, and $1/\epsilon'$. However, this complexity scaling requires an additional assumption: a quantity $R$ capturing the nonlinearity-to-dissipation ratio of the differential equation must satisfy $R < 1$ for the errors to be controlled (see [63, 65, 66, 70]), and it is not always clear when this condition holds.

For example, in the case of the Navier–Stokes equation, the size of the nonlinearity—and hence the value of $R$—grows with the "Reynolds number" of the fluid, and the condition $R < 1$ would be violated when simulating high-Reynolds-number turbulent flows. Turbulent flows are potentially handled by applying the Carleman linearization method to the lattice Boltzmann equation rather than the Navier–Stokes equation directly [1, 30], in which case the size of the nonlinearity does not scale with the Reynolds number. Indeed, generally speaking, for this approach to nonlinear differential equations, there is a delicate interplay between the size of the input state at time $t = 0$, the form of the nonlinearity, and the amount of dissipation in the linear term; see [66] for a discussion.

Separate from these approaches, methods have been proposed that map nonlinear classical dynamics to linear phase-space dynamics that can be simulated with Hamiltonian simulation [71, 21, 13, 72, 43].

**Comments on the complexity of alternative methods and problems:** We briefly comment on two further classes of applications involving PDEs, but which typically have very different characteristics. The first is stochastic differential equations (SDEs), which are simulated extensively in computational finance and more generally in risk modeling. There, one typically samples trajectories of the SDE (via Monte Carlo methods), and evaluates observables stochastically. Quantum-accelerated Monte Carlo has been worked on extensively (see Section 8.2 on pricing financial options). Where classical methods require sampling $\mathcal{O}(1/\epsilon^2)$ trajectories to obtain an $\epsilon$-estimate of a certain quantity, the quantum method can create a superposition of trajectories and read out the relevant amplitude at $\mathcal{O}(1/\epsilon)$ cost—a quadratic quantum advantage. On the classical side, a key advantage of these methods is that they avoid an exponential scaling in the number of continuous dimensions $d$ (i.e., the "curse of dimensionality"), unlike the "Eulerian" approaches discussed above that discretize the $d$-dimensional space into $N \geq \exp(\Omega(d))$ grid points or basis functions. Thus, they are relatively effective when $d$ is large, and less preferred when $d$ is small due the fact that their $\epsilon$ dependence cannot be better than $\mathcal{O}(1/\epsilon^2)$. In fact, in some cases, classical and quantum trajectory-based methods can be applied not only to SDEs but also to (deterministic) PDEs, and thus they compete against QLSS-based PDE and ODE methods we have discussed. For example, the quantum and classical complexity of a Monte Carlo approach for the heat equation was studied in [18] and compared against alternative approaches—for the end-to-end problem considered, the classical Monte Carlo approach outperformed all classical alternatives when $d > 4$, and the quantum-accelerated Monte Carlo approach outperformed all (quantum or classical) alternatives when $d > 2$. For SDEs, an alternative to Monte Carlo is to map the SDE to a Fokker–Planck equation via the Itô calculus and solve the Fokker–Planck PDE. This has been proposed in [73]. However, for most SDEs of interest in risk analysis, Monte Carlo simulation converges in a number of samples scaling linearly in the number of variables, leaving very little room for a quantum speedup in these applications given our current understanding.

The last class of problems to be mentioned are multi-particle Schrödinger equations. They are (i) high dimensional, (ii) complex, and (iii) require high-precision solutions for practical applications. Hence, they match all of the criteria under which a quantum advantage might be expected. The second-quantized approach to solving the full configuration interaction molecular Schrödinger equation is a specific case of the spectral method, although here one must solve an eigenvalue equation rather than a linear system. Unsurprisingly, this case has already gathered a lot of attention (see Section 2 on quantum chemistry).

**Existing resource estimates**

An explicit resource estimate for linear PDEs was reported in [74] for solving Maxwell's equations to estimate an electromagnetic scattering cross section in 2D. Following the asymptotic analysis of [17], it employed an FEM-based discretization scheme to form a linear system of size $N = 3 \times 10^8$, targeting accuracy $\epsilon = 0.01$. The estimates did not incorporate preconditioning methods and assumed a value for the condition number $\kappa \approx 10^4$, ultimately finding that $10^{29}$ $T$ gates would be needed to complete the computation. However, this work predated asymptotic and practical advancements to the complexity of the QLSS [46, 47], and modern estimates for the same problem would likely lead to more reasonable resource counts. Note also that much of the art in classical PDE solvers is to find appropriate preconditioning schemes to control the condition number. In [17], it was shown that one common class of preconditioners works within the framework of the quantum algorithm, but it is as yet unclear if this is the case more generally.

For ODEs, [35] gave a detailed performance analysis of the Taylor series truncation approach developed in [33, 34] applied to general time-independent dissipative ODEs. They gave explicit upper bounds on the condition number $\kappa$ of the linear system in terms of the total evolution time $T$ and the "log-norm" of the matrix $A$. They considered the task of outputting the history state $|u\rangle$ or the final-time state $|\bar{u}(T)\rangle$. By combining the bound on $\kappa$ with explicit upper bounds from [75] on the query complexity of the QLSS, they determined an upper bound on the number of times the algorithm needs to query a block-encoding of the ODE matrix $A$ to accomplish this task. The estimated number of queries per unit time varied from 10 to $10^5$ over the parameter regime considered, and these figures would be reduced with subsequent improvements to the QLSS complexity, such as those reported in [47].

In [30], the query bounds of [35] were applied to the specific end-to-end CFD problem of computing the drag force on a ship hull in the incompressible (and potentially turbulent) parameter regime, by solving the nonlinear lattice Boltzmann equation (linearized via Carleman linearization). They considered a simplified version of the problem where the ship hull is modeled as a sphere. They estimated that the quantum algorithm would need $10^{20}$–$10^{24}$ $T$ gates and roughly $10^3$–$10^5$ logical qubits (depending on the value of the Reynolds number) to compute the drag force on the sphere. Classically, the lattice Boltzmann method is a high-accuracy method and would be computationally intractable for this instance in the high-Reynolds-number regime. In practice, classical methods resort to lower accuracy methods, which for this instance can be completed within several minutes on a laptop. Computing the drag force on actual ship hulls with the quantum method is expected to be significantly more resource intensive compared to the flow-past-a-sphere instance due to the need to coherently load the boundary conditions describing the ship hull each time the block-encoding is queried.

**Caveats**

A key caveat is that many analyses in the literature do not consider the full end-to-end problem that needs to be solved for applications. Often, these works only consider the cost of preparing the quantum state $|u\rangle$ that encodes the solution to the differential equation into its amplitudes, and they report this cost in terms of the number of queries to oracles of the input data. Thus, these works study the task of solving differential equations as a primitive—similar to Hamiltonian simulation, that is, simulation of the time-dependent Schrödinger ODE—rather than as an end-to-end application. As discussed above, reading out useful information to precision $\epsilon$ introduces a $\Omega(1/\epsilon)$ multiplicative overhead and dramatically changes the complexity scaling, precluding exponential end-to-end speedups. Furthermore, whereas a full classical description of the solution could be computed just once, and subsequently many properties read out from that description, the state $|u\rangle$ is consumed during readout, and the number of times $|u\rangle$ needs to be prepared grows with the number of properties one wants to learn. In some cases, one may only seek to learn a few observables, but in other cases, extracting the desired information might require near full tomography of the quantum state $|u\rangle$, which in certain situations removes all quantum advantage [18].

The readout caveat can potentially be avoided if a small number of *samples* from the state $|u\rangle$ measured in the computational basis, as opposed to properties $\mathcal{P}[u]$ as defined above, would be useful for the end-to-end application. However, in such cases one must also be careful to compare against classical methods for the same task, where quantum-inspired methods can be competitive [76]. In [42], it was shown that BQP-hard problems can be encoded into an ODE describing coupled oscillators and solved by sampling from $|u\rangle$, but this situation would be unlikely to arise naturally in applications. In [69], it was suggested that samples from $|u\rangle$ encoding the solution to certain nonlinear differential equations could be useful for training neural networks.

Another caveat is that complexity statements often report only the number of times the algorithm queries oracles for the input data. In applications where the input data encoding complex boundary conditions or object geometries is not efficiently computable but rather stored in a large classical database, one must assume access to a log-depth QRAM in order to implement these oracles efficiently, an assumption that has its own caveats.

For simulating time evolution of ODEs, it is important to emphasize the dependence of the complexity in Eq. (21) on the parameter $g$, which for dissipative systems grows with time, potentially exponentially, as the size of the solution decays. Furthermore, for nonlinear differential equations, we reiterate that existing quantum algorithms are often based on the assumption that the nonlinearity-to-dissipation ratio is sufficiently small, which may not be satisfied in practical instances. Generally speaking, strong nonlinearities can cause the solution to develop discontinuities, and linearization schemes might break down for problems of interest if the solutions lack sufficient regularity, as can be the case for simulations of turbulence in CFD.

Finally, we note that due to the large number of methods available to classical solvers of differential equations, an important caveat is that any claim of quantum advantage must be sure to compare against the best possible classical method, and consider the possibility that this classical method might benefit from parallelization.

**Comparable classical complexity and challenging instance sizes**

Classical algorithms for linear PDEs can compute a classical description of the solution $u$ by solving the same linear equation $L|u\rangle = |f\rangle$ solved by the quantum algorithm. For an $s$-sparse $N \times N$ linear system, the complexity of an exact Gaussian elimination approach is $\mathcal{O}(N^\omega)$, where $\omega < 2.37$ is the matrix multiplication exponent. However, in practice, approximate iterative methods such as the conjugate gradient method are more common. The complexity of conjugate gradient scales as $\widetilde{\mathcal{O}}(Ns\sqrt{\kappa} \log(1/\epsilon))$ when $L$ is positive semidefinite [77, Chapter 10.2]. For a discretization scheme like the FEM where $N = (1/\epsilon)^{\Omega(d)}$, and using the aforementioned bound $\kappa \leq \mathcal{O}(N^{2/d})$, the complexity of the conjugate gradient approach is $s(1/\epsilon)^{\Omega(d)}$, which has exponential dependence on the spatial dimension $d$ but for fixed $d$ scales as $\text{poly}(1/\epsilon)$. Additionally, in practice, the conjugate gradient method benefits from preconditioning techniques to reduce $\kappa$, and from parallel implementation on graphics processing units (GPUs). For a sense of scale, [78] used the preconditioned conjugate gradient method within a finite element analysis to compute the thermal conductivity and elasticity of certain 3D cast iron samples imaged with micro-computed tomography (a task chosen mainly to benchmark their method). Among other reported results, their implementation solved the end-to-end problem, which required solving several linear systems, with $N \geq 10^6$ in less than 1 second, and $N \geq 4 \times 10^8$ in less than 30 minutes using a single GPU with 8 gigabytes of RAM.

For linear and nonlinear initial value problems, classical methods could apply conjugate gradient or other linear system solvers to the same linear equation $L|u\rangle = |f\rangle$ that the quantum algorithm constructs to solve the ODE. The complexity of this approach would have a linear-in-$N$ dependence, but since the solution would be a full classical description of the history state, it would not need to also pay the factor of $q$ arising from postselecting on the $t = T$ branch of the history state, and it would avoid the $\mathcal{O}(1/\epsilon)$ readout cost.

However, most classical methods do not follow this route and instead integrate the ODE with a time-marching method, where a description of the length-$N$ vector is propagated forward in time, for which there are many options [79, 80]. Similar to the quantum complexity, nearly linear-in-$T$ scaling is achieved by high-order methods, so long as $A(t)$ is smooth up to corresponding order, or by spectral methods [29]. In the time-independent or smooth case, one would achieve $NTC' \cdot \text{polylog}(T, 1/\epsilon)$ complexity, where $C'$ is some constant depending on the spectral properties of $A$, similar to $C$. As in the quantum case, care must be taken to choose $\epsilon$ appropriately when solutions are exponentially growing or decaying.

We also mention that there has recently been work on using machine learning methods to classically simulate nonlinear PDEs, especially for CFD [81, 82]. These methods are generally very fast, but they are heuristic, so they are suitable in some instances but not when high-confidence, high-accuracy solutions are required.

**Speedup**

For linear PDEs in $d$ dimensions, solved via discretization with $N = (1/\epsilon)^{\Omega(d)}$ grid points and inverting the corresponding linear system, the speedup is a reduction from time roughly $\epsilon^{-\Omega(d)}$ classically to $d^{\mathcal{O}(1)}\epsilon^{-1}\epsilon^{-\mathcal{O}(1)}$ quantumly, here omitting dependencies on $\||u\rangle\|$ and $\log(\epsilon^{-1})$. The $\mathcal{O}(1)$ powers depend on the details of the discretization and the efficacy of preconditioning. Other discretization schemes may give rise to slightly distinct complexity forms, but in any case, the conclusion is that for fixed dimension $d$, the speedup is at best *polynomial*, a point that has been made in more detail in, for example, [14, 18].

The speedup can be exponential in the parameter $d$. However, in many engineering applications, the number of dimensions is fixed to be fewer than four (three for space, one for time), limiting the advantage quantum methods can obtain. Furthermore, for PDEs with large $d$, trajectory-based classical strategies can avoid the exponential-in-$d$ complexity scaling, and in cases where these methods apply, the best possible speedup is typically a quadratic reduction from $\mathcal{O}(1/\epsilon^2)$ to $\mathcal{O}(1/\epsilon)$; see, for example, [18] and Section 8.2 on option pricing.

For integrating ODEs of $N$ variables forward in time, there can be an exponential speedup in the parameter $N$. However, since $N$ and $\epsilon$ are typically related by $N \leq (1/\epsilon)^{\Omega(d)}$ (e.g., when the ODE arises via discretization of a PDE), the $\mathcal{O}(1/\epsilon)$ cost of readout will contribute a much larger factor than the polylog($N$) cost of the QLSS, and the best possible speedup is again polynomial. An exponential speedup could be possible if $N \geq \exp(\Omega(1/\epsilon))$, or if samples from the state $|u\rangle$ were directly useful within the end-to-end application, but this assessment would also require that the solution-decay-factor $q$ appearing in the quantum complexity does not cancel the speedup.

In general, these methods do offer the possibility of an exponential saving in space, since the quantum methods can represent the vector using a logarithmic number of qubits. Nevertheless, the overall take-home message is that quantum algorithms can potentially outperform classical algorithms, but major gains are only to be expected when the number of spatial dimensions is large, or if there is otherwise a reason that the linear systems involved are much larger than the precision demanded in the output. This intuition is corroborated by the analysis of quantum computing algorithms for ab initio chemistry, where the number of dimensions scales with the number of electrons.

**NISQ implementations**

Various proposals at NISQ implementations of PDE solvers have been made; see [83] and references therein. The idea is to start from some discretization of the PDE $L|\psi(\theta)\rangle = |b\rangle$, where $|\psi(\theta)\rangle$ is an appropriately chosen variational circuit, and to optimize the parameters $\theta$ of the circuit. This is an example of a variational quantum algorithm. Another proposal applies a variational approach to nonlinear PDEs [84]. Note that even if these methods find parameters to generate a good approximation of the solution, they would still pay the $\mathcal{O}(1/\epsilon)$ cost to read out properties. Thus, they offer at best a polynomial speedup over classical methods. It is difficult to imagine that sufficient size and precision can be reached in the NISQ regime to be competitive with the best classical solvers.

**Outlook**

While the simulation of differential equations is one of the most important large-scale computational tasks, constituting a sizable fraction of HPC workloads in industry, at present the benefit of quantum solvers for real-world problems appears limited to relatively modest polynomial speedups. Extensive work on quantum algorithms for solving differential equations has developed methods with rigorous analyses and likely close-to-optimal complexities; the challenge that remains is how these methods can integrate into an end-to-end application pipeline, in such a way as to reduce the cost. To find a high-value application related to solving differential equations (beyond potentially ab initio chemistry), one would likely need to find a situation that meets some or all of the following criteria: (i) involves a very large number of spatial dimensions, (ii) has simple geometry or initial conditions in order to avoid the need for a QRAM input model,

(iii) requires high precision, ruling out heuristic classical approximate methods, (iv) requires learning a relatively small number of properties of (or ideally requires only samples from) the solution vector. There remains the possibility for substantial improvements in memory usage, but this is not currently a bottleneck in classical PDE solving.

# 8   Finance

The financial services industry is among those beginning to explore the potential future benefits of quantum computing. Finance has the distinct feature that more powerful and more accurate simulations can lead to direct competitive advantage, in a way that is harder to identify in other industries. In this application area, researchers strive to find quantum speedups for use cases of interest to financial services. A number of use cases have been proposed as candidates for quantum solutions, such as:

- **Derivative pricing** (such as options [1, 2], and collateralized debt obligations (CDOs) [3]): Derivatives are financial instruments that are built upon an underlying asset (or assets) that can depend on the value of the asset in potentially complicated ways. In the derivative pricing problem, one needs to determine a fair price of the financial instrument, which is the price that would be received by the seller or paid by the purchaser when an asset or liability is transferred between market participants in an orderly transaction. Typically, one needs to compute the expected value of the fair purchase price of underlying assets at some later date when pricing a derivative. A similar and related problem is known as *computing the Greeks* [4]. The Greeks of a financial derivative are quantities that determine the sensitivity of the derivative to various parameters in the problem. For example, the Greeks of an option are given by the derivative of the price of the option with respect to some parameter, for example, $\Delta := \partial V / \partial X$, where $V$ is the price of the option and $X$ is the price of the underlying asset.

- **Credit valuation adjustments (CVAs)** [5]: CVA is the problem of determining the fair price of a derivative, portfolio, or other financial instrument while taking into account the purchaser's (potentially poor) credit rating, and the risk of default. CVA is typically given by the difference between the default risk-free portfolio and the value of the portfolio taking into account the possibility of default.

- **Value at risk (VaR)** [6]: Many forms of risk analysis can be considered, with VaR being a common example. VaR measures the total value a financial instrument (such as a portfolio) might lose over a predefined time interval within a fixed confidence interval. For example, the VaR of a portfolio might indicate that, with 95% probability, the portfolio will not lose more than \$$Y$. A similar technique works as well for the related conditional value at risk (CVaR) problem.

- **Portfolio optimization** [7]: The goal of portfolio optimization is to determine the optimal allocation of funds into a universe of investable assets such that the resulting portfolio maximizes returns and minimizes risk, while also respecting other constraints.

While there are are many more use cases and several approaches for generating quantum speedups, broadly speaking, many use cases stem from one of two paths to quantum improvements: quantum enhancements to Monte Carlo methods (for simulating stochastic processes), and constrained optimization. In the first case, the approach generally involves encoding a relevant, problem-specific function into a quantum state, and then using quantum amplitude estimation to sample from the distribution quadratically fewer times than classical Monte Carlo methods [8]. In the second case, a financial use case is reduced to a constrained optimization problem, and a quantum algorithm for optimization is used to solve the problem.





Among the use cases studied in these two areas, option pricing and portfolio optimization often serve as archetypal examples of Monte Carlo and constrained optimization problems, respectively, and their associated quantum algorithms have the most follow-up work. Moreover, these two classes of problems comprise a considerable fraction of the classical compute used in the financial services industry. For these reasons, we will focus on these two use cases in this section, though the approaches, caveats, and complexities typically translate to other relevant use cases.

In addition to the use cases described above, other areas of interest to the financial services industry include post-quantum cryptography, as well as quantum-secure networking and quantum key distribution. However, many of these topics or their proposed quantum implementations are outside the scope of this discussion. Quantum machine learning is yet another popular use case within quantum approaches to finance, but oftentimes these results are quantum approaches to standard machine learning problems, which are then applied to a financial application. As such, we will also not study machine learning in this finance-specific section and instead refer interested readers to any of the excellent review articles on quantum finance (e.g., [9, 10]) for more details.

*The authors are grateful to Nikitas Stamatopoulos for reviewing this section of the survey, to Patrick Rebentrost for reviewing Section 8.1, and to Ashley Montanaro for reviewing Section 8.2.*

## This application area contains:

## 8.1 Portfolio optimization

**Overview**

Given a set of possible assets into which one can invest, the problem of portfolio optimization (PO) involves finding the optimal allocation of funds in these assets so as to maximize returns while minimizing risk. The Markowitz model, as it is commonly called, is widely used in the financial industry, owing to its simplicity and broad applicability. Sophisticated constraints, transaction cost functions, and modifications to the problem can be used to model realistic, modern PO problems. Numerically solving these optimization problems is a routine part of existing workflows in financial services operations. Several quantum approaches to solving the PO problem have been proposed, each with their own advantages and drawbacks.

**Actual end-to-end problem(s) solved**

Consider a set of $n$ investable assets with a fixed total budget. Define $w_i \in \mathbb{R}$ to be the fraction of the total budget that is invested into asset $i$. Thus, the $n$-dimensional vector $w$ defines a portfolio. Let $r$ be a known $n$-dimensional vector denoting the expected return for each of the available assets, that is, the percentage by which the value of each asset is expected to grow over some defined time period. Let $\Sigma \in \mathbb{R}^{n \times n}$ be the covariance matrix governing the random (and possibly correlated) fluctuations in the asset returns away from their mean $r$. In practice, the input parameters $\Sigma$ and $r$ can be inferred from historical stock price data, or through more sophisticated analyses. The covariance matrix can be used to define a portfolio's "risk" $w^{\mathsf{T}}\Sigma w$, which is precisely the variance in the returns it generates, assuming the underlying model is accurate. Denote the all-ones vector by $\mathbf{1}$, and for any pair of vectors $u, v$ let $\langle u, v \rangle$ denote the standard inner product between $u$ and $v$. The goal of the Markowitz formulation of PO is to find the optimal portfolio (i.e., vector of weights $w$) that either:

- maximizes the expected return subject to a fixed risk parameter $\sigma_0^2$

$$
\begin{aligned}
\max_{w} \ & \langle w, r \rangle \\
\text{s.t.} \quad & w^{\mathsf{T}}\Sigma w = \sigma_0^2 \\
& \langle \mathbf{1}, w \rangle = 1
\end{aligned}
\tag{22}
$$

- minimizes risk subject to a fixed return parameter $r_0$

$$
\begin{aligned}
\min_{w} \ & w^{\mathsf{T}}\Sigma w \\
\text{s.t.} \quad & \langle w, r \rangle = r_0 \\
& \langle \mathbf{1}, w \rangle = 1
\end{aligned}
\tag{23}
$$

- maximizes return and minimizes risk with a tradeoff determined by a parameter known as the "risk aversion parameter" $\lambda$:

$$
\begin{aligned}
\max_{w} \ & \langle w, r \rangle - \lambda w^{\mathsf{T}}\Sigma w \\
\text{s.t.} \quad & \langle \mathbf{1}, w \rangle = 1
\end{aligned}
\tag{24}
$$





or the alternative for the square root of risk (standard deviation rather than variance)

$$\max_w \; \langle w, r \rangle - q\sqrt{w^\intercal \Sigma w}$$
$$\text{s.t.} \quad \langle \mathbf{1}, w \rangle = 1, \tag{25}$$

where $q$ plays the same role as $\lambda$.

Typically, it is satisfactory to find a vector that optimizes the objective function up to additive error $\epsilon$, for some prespecified value of $\epsilon$.

When solving the above Markowitz model formulations of PO, the absence of inequality constraints leads to simpler optimization problems that can be solved with efficient classical approaches. For example, the optimization problem in Eq. (23) is a simple quadratic program without complicated constraints, for which one can derive a closed-form expression for $w$ using Lagrange multipliers [1]. More general PO problems that include practically relevant constraints (such as the simple "long-only" constraint $w_i \geq 0$, which prohibits "short" positions in which $w_i$ can be less than zero) cannot generically be solved analytically, and one needs to employ more sophisticated numerical solvers. Real-world PO problems include a number of possible constraints (see [2] for a discussion), including, but not limited to:

- Long only—$w_j \geq 0$ for all $j$.

- Investment bands—$w_j \in [w_j^{\min}, w_j^{\max}]$.

- Turnover constraints—$|\Delta w_j| \leq U_j$ for some fixed fraction $U_j$, where $\Delta w_j$ represents the change in holdings of asset $w_j$ from one portfolio to the next.

- Cardinality constraints—minimum, maximum, or exact number of nonzero assets in the portfolio.

- Sector constraints—specified minimum and/or maximum allocations to groups of assets (e.g., the energy or healthcare sectors).

- Transaction costs—typically represented as a function of $|\Delta w_j|$, and often added as a term in the objective function rather than as a constraint.

- Market impact—the effect on the price of an asset that a market participant has when buying or selling the asset. Related to liquidity, market impact can be seen as a type of transaction cost that arises when a transaction causes the price of the asset to move.

The PO can also be formulated in an "online" manner, where, for example, asset performance data arrives one day at a time, and one has the opportunity to update the portfolio at the end of each day [3].

As is often the case with optimization problems, the problem formulation *strongly* affects the solution strategy and the problem "hardness." If the PO problem is unconstrained and continuous (i.e., each $w_i$ is a real number), then the problem is relatively easy. If convex inequality constraints, such as the long-only or turnover constraints, are imposed, then the problem is harder but can still be tackled by relatively efficient methods for convex optimization. By contrast, if one discretizes the problem (so that $w$ now represents an integer number of asset shares or lots being traded), or if one applies some of the constraints above (such as integer-valued constraints like cardinality), then the problem becomes nonconvex and considerably harder to





solve. In general, with discrete constraints, the problem can be formulated as an instance of an integer program (IP) (if all variables are discrete) or a mixed-integer program (MIP) (if some variables are discrete and others are continuous), which are NP-hard and therefore intractable to solve in polynomial-time (in $n$) under widely believed assumptions. Alternatively, given the IP formulation of the problem as a starting point, one can encode the integer variables in a binary representation, thereby allowing the problem to be formulated as a quadratic unconstrained binary optimization (QUBO) instance [4]. These formulations allow quantum algorithms for combinatorial optimization to be employed; for example, the MIP formulation can be solved with a branch-and-bound approach [5], and the QUBO formulation can be solved via Grover-type methods, or heuristically through (NISQ-friendly) quantum annealing approaches (e.g., [6]).

**Dominant resource cost/complexity**

An early approach to solving this optimization problem using a quantum algorithm was presented in [7], in which the Markowitz problem is written as minimizing risk with fixed return (Eq. (23)), and without other complicated constraints. This simple optimization problem boils down to an equality constrained convex program; it can be solved by introducing Lagrange multipliers and solving a linear system (represented by a matrix $G$) involving the input data $r$ and $\Sigma$ [7]. The approach of [7] is to use a quantum linear system solver (QLSS) and prepare the quantum state $|w\rangle$ whose amplitudes are proportional to the optimal weights $w_i$. The complexity to do so to error $\epsilon$ is $\widetilde{\mathcal{O}}(\kappa\zeta\log(1/\epsilon))$, where $\kappa$ is the condition number of the matrix $G$ being inverted and $\zeta = \|G\|_F/\|G\|$ is the ratio of its Frobenius norm to its spectral norm. The $\widetilde{\mathcal{O}}$ suppresses logarithmic factors, including a factor coming from applying unitaries that block-encode the matrix $G$ in polylog($n$) depth, essentially equivalent to the assumption that log-depth quantum random access memory (QRAM) is available. It is a priori unclear what the value of $\kappa$ and $\zeta$ would be for actual PO instances and whether they depend on $n$, but the explicit logarithmic dependence of this complexity on $n$ is appealing. However, a drawback of this approach is that it produces the quantum state $|w\rangle$ rather than an estimate for the optimal portfolio $w$. Learning the $n$ entries of $w$ to precision $\epsilon$ in 2-norm incurs multiplicative overhead of $\widetilde{\mathcal{O}}(n/\epsilon)$ using quantum pure state tomography [8] for total time complexity $\widetilde{\mathcal{O}}(n\kappa\zeta/\epsilon)$.[22]

When convex linear inequality constraints, such as long-only or turnover constraints, are included, the above approach will not work. However, a more sophisticated method can be applied, which first maps the PO instance to a convex program (specifically, a second-order cone program (SOCP)) and then makes use of interior point methods to solve the program. These interior point methods can be quantized, forming quantum interior point methods (QIPMs) [10, 11]. The QIPM is an iterative method, where each iteration involves solving a linear equation with a QLSS and classically reading out the solution with tomography. Thus, the procedure within each iteration is similar to the procedure above for solving the unconstrained PO problem, but the linear system to be solved is different (and changes with each iteration). A preliminary study of the effectiveness of this approach for PO was given by [12], followed by a more extensive study in [13]. The QIPM produces an $\epsilon$-optimal classical estimate for $w$, and has

---

[22]Reference [7] suggests several possible nonstandard problems that can be solved with $|w\rangle$ without actually learning the entries of $w$, such as sampling values of $i$ with large $|w_i|$, and estimating overlaps $\langle\tilde{w}, w\rangle$ with hypothesized portfolios $\tilde{w}$. In general, inner products $\langle u, w\rangle$ of arbitrary normalized vectors $u$ with $w$ can be learned to precision $\epsilon$ using overlap estimation [9] (an application of amplitude estimation), incurring multiplicative overhead of $\mathcal{O}(1/\epsilon)$, but no explicit linear-in-$n$ dependence. However, the practical utility of such tasks within the existing workflows of financial institutions is unclear.





time complexity $\widetilde{\mathcal{O}}(n^{1.5}\zeta\kappa\xi^{-1}\log(1/\epsilon))$, where $\kappa$ and $\zeta$ are the maximum condition number and Frobenius-to-spectral-norm ratio for the matrices that must be inverted over the course of the algorithm, respectively, and $\xi$ is the precision to which tomography must be performed. Note that in principle $\xi$ can stay constant even as the overall precision estimate $\epsilon \to 0$ [13].

With the addition of discrete constraints, PO is instead formulated as a nonconvex MIP. MIPs are typically solved with a branch-and-bound approach (for a summary in a financial context, see, e.g., [14, Chapter 11]). Key to this approach is the ability to solve convex relaxations of the MIP (where the discrete constraints are dropped) in $\mathrm{poly}(n)$ time (perhaps via classical or quantum interior point methods for SOCPs, as above). To impose the discrete constraints, a tree is constructed and explored, where generating the children of a given node in the tree requires solving one of these relaxations. Thus, the number of convex relaxations that must be solved is proportional to the tree size $T$, which is generally exponentially large in $n$. Reference [5] (extending prior work of [15]) showed that a quantum algorithm can produce the same output while exploring quadratically fewer nodes, solving roughly $\widetilde{\mathcal{O}}(\sqrt{T})$ convex relaxations (but doing so coherently, which could introduce overheads), for a total complexity of $\widetilde{\mathcal{O}}(\sqrt{T})\cdot\mathrm{poly}(n)$. The value of $T$ is instance dependent and requires empirical estimation: a preliminary numerical analysis of the value of $T$ for a certain ensemble of PO instances up to $n = 56$ found that $T \sim 2^{0.14n}$ to $2^{0.20n}$ [5].

The algorithm for online PO of [3], which leverages the multiplicative weights update method, has time complexity scaling as $\widetilde{\mathcal{O}}(\sqrt{n})$, a potentially quadratic speedup compared to the analogous classical algorithm. However, the quantum algorithm has worse dependence on the number of time steps.

The assessment of the number of qubits used by these algorithms requires a nuanced discussion of loading classical data. A key feature of all of the approaches above is that they require (repeatedly) accessing the classical data representing the historical stock information (i.e., the returns $r$ and the covariance matrix $\Sigma$) in the quantum algorithm. The size of this data is typically $\mathcal{O}(n^2)$. Loading can be performed using block-encoding dense matrices of classical data and QRAM, which achieves $\mathcal{O}(\log(n))$ depth (time), at the expense of $\mathcal{O}(n^2)$ space. Here, several caveats are inherited from the QRAM primitive. Moreover, for practical values of $n$, this $\mathcal{O}(n^2)$ space cost could be prohibitively large, although it is possible this space cost could manifest as a dedicated QRAM hardware element of the device, rather than as part of the main processor. If log-depth QRAM of sufficient size is not desired or not available, the data could instead be loaded with only $\mathcal{O}(\log(n))$ space and in $\mathcal{O}(n^2)$ time, but this overhead in time would likely preclude the possibility of quantum speedup at least in the first two cases, where the formulation is convex and classical $\mathrm{poly}(n)$-time algorithms exist.

**Existing resource estimates**

A detailed, end-to-end resource analysis of the PO problem using QIPMs was performed in [13]. The authors followed the approach of [12] and performed a careful accounting of all quantum resources, including constant prefactors. The authors found that one needs $800n^2$ logical qubits, a $T$-depth of

$$(2\times 10^8)\kappa\zeta n^{1.5}\xi^{-2}\log_2(n)\log_2(\epsilon^{-1})\log_2(\kappa n^{14/27}\xi^{-1}),$$

and a $T$-count of

$$(7\times 10^{11})\kappa\zeta n^{3.5}\xi^{-2}\log_2(n)\log_2(\epsilon^{-1})\log_2(\kappa\zeta\xi^{-1}),$$





where $\kappa$ is the maximum condition number encountered in the algorithm, $\zeta$ is the maximum Frobenius-to-spectral-norm ratio, and $\xi$ is the minimum tomographic precision required. The $\xi^{-2}$ dependence can asymptotically be improved to $\xi^{-1}$ at the expense of a more sophisticated protocol for tomography [8]. Furthermore, these estimates used explicit bounds on the complexity of the QLSS from [16] which have since been improved upon in [17, 18]. Using the updated bounds from [18] would immediately reduce the $T$-count and $T$-depth estimates by at least three orders of magnitude. Note also that this calculation incorporated optimized circuits for block-encoding dense matrices of classical data with $\mathcal{O}(\log(n))$ $T$-depth but $\mathcal{O}(n^2)$ $T$-count [19], leading to the large discrepancy between those two quantities. The authors performed numerical simulations of PO instances to determine the instance-specific quantities. Using numerically determined values for $\kappa\zeta$ and $\xi$, and using realistic values of $\epsilon = 10^{-7}$ and $n = 100$, these resource counts imply that one would need $8 \times 10^6$ logical qubits, $2 \times 10^{24}$ $T$-depth, and $8 \times 10^{29}$ $T$-count. These logical estimates for the number of non-Clifford gates could in principle be turned into estimates for the number of physical qubits and runtime on actual hardware, using the methods discussed in the section on fault-tolerant quantum computation. However, the authors of [13] did not do so, in part because the logical costs were sufficiently high that the qualitative conclusion about the practicality of the algorithm was already clear.

**Caveats**

The quantum algorithms for PO discussed above inherit many of the caveats of their underlying primitives, namely, QLSS, tomography, and classical data loading. One salient caveat is that the QLSS-based approaches depend on a number of instance-specific parameters $\kappa, \zeta, \xi$, which are difficult to predict without running numerical simulations. The asymptotic speedup is subject to assumptions about the scaling of these parameters. Additionally, for a speedup to be possible, log-depth QRAM must be available on large datasets, which, while theoretically possible, presents practical challenges.

The branch-and-bound approach does *not* require log-depth QRAM to achieve its nearly quadratic speedup since the runtime will be dominated by the exponential tree-size factor (although it would help to have fast QRAM to reduce by $\text{poly}(n)$ factors the time needed to solve the convex relaxations at each step). However, a caveat to that approach is that to obtain the quadratic speedup, the convex relaxations of the MIP (which would be SOCPs), would need to be solved coherently. In principle, this is always possible, but it would likely require a substantial amount of coherent classical arithmetic and additional $\text{poly}(n)$ overheads in time and space.

**Comparable classical complexity and challenging instance sizes**

Convex formulations of the PO problem are typically solved classically via mapping to SOCPs. Optimized software packages can solve these SOCPs efficiently, and many are based on interior point methods. These interior point methods have theoretical runtime complexity of roughly $\widetilde{\mathcal{O}}(n^{\omega+0.5}\log(1/\epsilon))$, where $\omega \approx 2.373$ is the matrix multiplication exponent, although for practical instance sizes, the effective value of $\omega$ is typically closer to 3. Note that the example PO problem with 100 assets solved in [13] and described above can typically be solved within seconds on a laptop using traditional classical methods. Problem sizes found in the financial services industry can include as many as tens of thousands of assets.





In the IP or MIP formulation of PO, classical solutions will have complexity exponential in $n$. As a point of reference, the numerical experiments reported in [5] classically solved hundreds of PO instances up to size $n = 56$ (and likely could have gone significantly higher).

**Speedup**

Recall that the QIPMs used to solve the SOCP for constrained PO are virtually identical to their classical counterpart; they differ by their use of a quantum subroutine to solve linear systems. Thus, any speedup obtained by the quantum approach to solving the SOCP will necessarily come from speedups from the QLSS plus tomography approach to solving a linear system. The approach for unconstrained PO was also based on the same primitives. The performance of the quantum method is often compared against classical Gaussian elimination. However, since the quantum approach necessarily produces an approximate solver (due to tomography), another valid comparison to make is against approximate classical solvers, such as the conjugate gradient method [20] or the randomized Kaczmarz method [21]. In the case of the randomized Kaczmarz method, the classical complexity for solving an $L \times L$ linear system to precision $\xi$ scales as $\mathcal{O}(L\kappa^2\zeta^2 \log(\xi^{-1}))$ (where $\kappa$ is the condition number and $\zeta$ the Frobenius-to-spectral-norm ratio) compared to $\mathcal{O}(L^3)$ for Gaussian elimination (asymptotically $\mathcal{O}(L^\omega)$). Thus, the quantum method provides the greatest speedup when $\kappa\zeta \propto L$ and $\xi = \mathcal{O}(1)$, in which case the QIPM for constrained PO runtime scales as $\widetilde{\mathcal{O}}(n^{2.5})$, whereas the classical runtime scales as $\widetilde{\mathcal{O}}(n^{3.5})$, where $n$ is the number of stocks in the portfolio (see [13, Table XI] for a more complete discussion). For unconstrained PO, which only requires one linear system, the comparison would be $\widetilde{\mathcal{O}}(n^2)$ vs. $\widetilde{\mathcal{O}}(n^3)$. In either case, the speedup is subquadratic. Moreover, the numerical simulations in [13] were not consistent with these optimistic assumptions on $\kappa\zeta$ and $\xi$, suggesting, rather, that the QIPM would have minimal if any speedup over classical IPMs, albeit based on small instance sizes up to $n = 120$.

The speedup for the branch-and-bound approach to the MIP formulation of PO is quadratic (up to log factors), although, as mentioned, in contrast to the convex formulations, both the quantum and classical algorithms generally have runtime exponential in $n$.

**NISQ implementations**

Several alternative approaches to PO using quantum solutions have been proposed.

- Hybrid-HHL [22]. This work generalizes the algorithm of [7], described above, by employing midcircuit measurements and conditional logic to obtain a NISQ version of the QLSS that readily solves the PO problem.

- Variational approaches based on the quantum approximate optimization algorithm (QAOA) [23, 24, 25] . These approaches typically use the quadratic objective function from Eq. (24), but instead consider $w_i \in \{0, 1\}$ as binary variables indicating whether or not an asset is part of the portfolio (a substantial deviation from the normal formulation). Constraints are dealt with by adding penalties to the objective function. Alternatively, constraints can be enforced by choosing clever versions of the ansatz [26] or by making measurements to project into the feasible space [24].

- Quantum annealing approaches: [4, 27, 28, 29, 6]. As in the previous case, these approaches require the problem to be formulated as a binary optimization problem. However, in this





case, they typically start with the IP formulation and encode integers in binary through one of several possible encodings [4] (thus, the number of binary variables will be greater than $n$). Constraints in the PO problem can also be included in the objective function using a variety of tricks, resulting in the desired QUBO, which can then be solved using a quantum annealer.

**Outlook**

The QIPM approach (and QLSS-based techniques more generally) for continuous formulations of PO have the potential to offer polynomial (but subquadratic) speedups for the PO problem. However, these speedups are subject to conjectures about the scaling of certain instance-specific parameters and preliminary empirical estimates are not suggestive of a maximal speedup. In any regard, the resource estimates of [13] illustrate that the non-Clifford resources required to implement the QIPM for this use case are prohibitive, even at problem sizes that are trivial to solve with classical computers. An asymptotic quantum advantage for this problem could exist for sufficiently large sets of assets, but without drastic improvements to the quantum algorithm and the underlying primitives (e.g., QRAM, QLSS), it is unlikely this approach will be fruitful. Even if such improvements are made, the algorithm only provides a polynomial speedup that is subquadratic, at best, greatly limiting the upside potential of this approach.

The branch-and-bound approach for discrete formulations has the possibility of a larger quadratic speedup, but, as has been observed (see, e.g., [30, 31]) in the context of Grover-like quadratic speedups in combinatorial optimization, it is unclear whether the quadratic speedup is sufficient to overcome the inherently slower quantum clock speeds and overheads due to fault-tolerant quantum computation for practical instance sizes.

## 8.2   Monte Carlo methods: Option pricing

**Overview**

Many financial instruments require an estimate of the average of some function of a stochastic variable within a window of time. To compute this average, one can use Monte Carlo methods to perform many simulations of the stochastic process over the time window, evaluate the function (which can potentially depend on the path taken by the stochastic variable during the entire window), and numerically estimate the average. While the setup and details of the problems may vary from one use case to another, the underlying methods are often quite similar. As an archetypal example of this problem, we will focus on the problem of pricing derivatives, such as options, but we remark that many of these results can be carried over to other use cases, such as computing the Greeks, credit valuation adjustments, and value at risk.

Derivatives are financial instruments that, roughly speaking, allow the parties involved to benefit when an asset (such as a stock) increases or decreases in value, but without having to hold the asset itself. One type of derivative—called an "option"—is a contract that permits the holder to either purchase ("call option") or sell ("put option") an underlying asset at a fixed, predetermined price (the "strike price") at or prior to some predetermined time in the future ("the exercise window"). The seller of the option is obligated to either sell or buy the asset, should the holder choose to exercise the option.

How, then, should one decide on a price for the option (i.e., the amount the holder must pay for the contract, not the strike price)? The well-known Black–Scholes (or Black–Scholes–Merton) model provides one approach to pricing options, making a few assumptions about the underlying assets and the rules of the contract. More complicated options can be considered that include, for example, multiple assets in the contract (e.g., basket options), multiple possible exercise windows (e.g., Bermudan or American options), etc.

Typically, options are priced by running Monte Carlo sampling on the value of the underlying asset(s) and determining the expected profit or loss from a given position, which can be translated into a price that the purchaser must pay. Options with a larger potential downside for the seller should cost a larger amount to purchase. For more information on options and Monte Carlo methods in the context of computational finance, see [1, 2].

**Actual end-to-end problem(s) solved**

The task is to price an option based on an underlying asset. The price of the asset is a random variable $X$ that follows a known (or assumed) stochastic process that models the market for the underlying asset. The option has a known payoff function $f(X)$ (e.g., the difference between the price of the asset at each time step minus the strike price over the trajectory, or zero, whichever is larger). For options that depend on more than one underlying asset or on asset prices at multiple distinct points in time, the random variable $X$ would represent a vector of data containing all information needed to compute the payoff. Given these inputs, the end-to-end problem is to compute an estimate of the expected payoff $\mathbb{E}_X(f(X))$ that lies within a certain error tolerance $\epsilon$ with high probability. This quantity is then used to determine the fair price of the option, which we take to be the expected value of the derivative at the contract's expiration date, discounted to the pricing date.

Using the assumed stochastic model for the price of the asset, one can develop a stochastic differential equation for the average payoff of the option. In limited cases, one can compute the average payoff analytically, as in the case of the famous Black–Scholes formula for the price of





European call options, for which the 1997 Nobel Prize in Economics was awarded. The Black–Scholes differential equation for the price of an asset at time $t$ can be derived by assuming the price of the underlying stock follows a geometric Brownian motion

$$\mathrm{d}X_t = X_t\alpha\mathrm{d}t + X_t\sigma\mathrm{d}W_t\,,$$

where $X_t$ is the price of the underlying asset at time $t$, $\alpha$ is a parameter known as the "drift" of the asset, $\sigma$ is the volatility (the standard deviation of the underlying returns), and $\mathrm{d}W_t$ is an increment of an accompanying Brownian motion $W_t$. Using Itô's lemma, one can derive a differential equation for the price of the option at time $t$ and, in limited cases (with several assumptions), one can solve the differential equation analytically. In practice, however, different types of contract have more complex definitions and fewer assumptions and, as a consequence, the differential equation cannot be solved analytically. Quantum approaches to numerically solving the stochastic differential equation have been proposed, including finite difference methods [3], Hamiltonian simulation [4], and quantum random walks [5]. For more detail on quantum approaches to solving differential equations, see Section 7 on solving differential equations. In many real-world derivative pricing use cases, the underlying differential equation becomes intractable. Thus, the most common classical method of computing the average payoff of an option is not through solving the stochastic differential equation, but rather through Monte Carlo sampling the random process $X$ directly. To do so, one generates a large number of price trajectories over the chosen time range, and the average payoff is computed numerically. In what follows, we will focus on quantum approaches to Monte Carlo estimation—also known as quantum-accelerated[23] Monte Carlo methods—which was pioneered in [6] and subsequently applied to several problems in finance (e.g., [7, 8, 9, 10, 11, 12]). However, we remark that other approaches to solving this problem that do not make use of Monte Carlo methods have also been proposed (e.g., [13]), and that this is an area of active research.

To compute different quantities, such as value at risk or credit valuation adjustments, similar approaches are often employed: simulate the underlying stochastic evolution several times and estimate the desired quantity numerically. The function to be computed may be quite different, but the approach is often the same.

**Dominant resource cost/complexity**

In [7, 9], the quantum speedup of Monte Carlo estimation from [6] is applied to solve the option pricing problem. We briefly explain the method and its dominant cost. First of all, this requires discretizing the set of values the random variable $X$ can take, which we index by the label $x$. Let $N$ denote the number of values and $n = \lceil \log_2(N) \rceil$ denote the number of qubits needed to hold the state $|x\rangle$. The first step is to load the probabilities for the future prices of the asset into the amplitudes of a quantum state, that is, the state

$$\sum_x \sqrt{p_x}|x\rangle\,,$$

where $p_x$ is the probability that $x$ is observed in the corresponding classical Monte Carlo simulation.

---

[23] Not to be confused with quantum Monte Carlo methods, which are classical algorithms for simulating certain quantum systems.





Second, a subroutine is employed that computes information about the payoff function into an ancilla register using coherent arithmetic. More precisely, the angle $\theta_x$ is computed (rounded to some finite number of bits of precision), where $\sin(\theta_x) = \sqrt{f(x)}$. (For simplicity, here we assume $0 \leq f(x) \leq 1$ for all $x$, but we revisit this point later.) This yields

$$\sum_x \sqrt{p_x} |x\rangle |\theta_x\rangle \,.$$

Third, the amplitude $\sqrt{f(x)}$ is loaded into the amplitude of an ancilla register by applying the map $|\theta\rangle|0\rangle \mapsto |\theta\rangle(\sin(\theta)|0\rangle + \cos(\theta)|1\rangle)$. This gives

$$\left( \sum_x \sqrt{p_x f(x)} |x\rangle |\theta_x\rangle \right)|0\rangle + \left( \sum_x \sqrt{p_x(1 - f(x))} |x\rangle |\theta_x\rangle \right)|1\rangle \,.$$

The probability of measuring the final ancilla in $|0\rangle$ is precisely $\mathbb{E}_X(f(X))$. Thus, the final step is to apply quantum amplitude estimation, which requires $\mathcal{O}(\epsilon^{-1})$ calls to the unitary that produces the state above to obtain an estimate to error $\epsilon$.

If $0 \leq f(x) \leq 1$ does not hold, the above approach needs to be modified, for example, by shifting and rescaling $f$ over a sequence of intervals of increasing length, as discussed in [6, 7]. Roughly speaking, to make sure that $f(x)$ falls within the interval $[0,1]$, at least for a large fraction of the randomly chosen values of $x$, we should expect the function $f$ will need to be scaled down by a factor on the order of the standard deviation $\sigma = \sqrt{\mathbb{E}_X(f(X)^2) - (\mathbb{E}_X f(x))^2}$. Thus, to achieve error $\epsilon$, quantum amplitude estimation must be performed to precision $\epsilon/\sigma$ instead of $\epsilon$.

There are three components to the algorithm that each contribute to the resource cost:

- Loading the distribution with amplitudes $\sqrt{p_x}$. The gate complexity of this step is roughly the same as the time complexity of classically drawing a Monte Carlo sample, although for certain distributions it could be faster (e.g., a quadratic quantum speedup can be obtained if $p_x$ is the stationary distribution of a Markov process [14]). Alternatively, if a functional form for $p_x$ is known, the methods of [15] could be used to approximately prepare the state—note that the Grover–Rudolph approach to state preparation [16] is incompatible with a quantum speedup in the context of Monte Carlo estimation [17]. Finally, [9] proposes using a quantum generative adversarial network (qGAN), a variational quantum algorithm, which could reduce the resources but requires a training phase.

- Coherent arithmetic to compute the rotation angle $\theta_x$. This depends on the complexity of the function $f$, but can generally be accomplished in comparable gate complexity as classical arithmetic, that is, $\mathrm{poly}(n)$. In [18], it was shown how the payoff can instead be put directly into the amplitude, without ever computing $\theta_x$, using quantum signal processing methods [15].

- Quantum amplitude estimation to precision $\epsilon/\sigma$, which requires $\mathcal{O}(\sigma/\epsilon)$ repetitions of the above two costs to achieve an $\epsilon$-estimate on the quantity $\mathbb{E}_X f(X)$.

Overall, from [19, Theorem 1.1] the complexity is

$$\frac{\sigma}{\epsilon} \cdot \mathrm{poly}(n) \,, \tag{26}$$





with the $\text{poly}(n)$ factor generally on the same order as the time required to draw and process a single classical Monte Carlo sample. This complexity does not require one to have an upper bound on $\sigma$, and it improves over the original work of [6] (which in turn is based on the algorithm of [20] for the uniform distribution) and follow-up work [21] by removing additional $\log(\sigma/\varepsilon)$ factors originating from the need to rescale the (potentially unbounded) random variable, to account for the contribution of its tails. In fact, the method can also work even for random variables with infinite variance [22]. However, in practice, the more advanced techniques and analyses of [19, 22] may not be necessary, as the underlying assets are typically modeled with distributions such as Gaussians where the tails are well behaved and the variance of the relevant random variable is controlled.

The general approach to Monte Carlo estimation sketched above has been extended and optimized in various ways; see, for example, [23, 24, 25, 26, 27]. For instance, in [26], the authors study a quantum algorithm for the optimal stopping problem by developing a quantum version of least-squares Monte Carlo. The algorithm finds a quadratic speedup over related classical methods, thereby demonstrating that American-style options—which are more complex than European-style options because they allow the holder to exercise the option and buy/sell the underlying asset at any point in the exercise window, rather than just the end—also maintain a quadratic speedup over classical Monte Carlo.

### Existing resource estimates

Detailed resource estimations for benchmark option pricing problems (known as autocallable and target accrual redemption forward, or TARF) were studied in [28]. The authors studied real-world use cases and problem sizes that are simple enough to analyze, yet complex enough to capture desirable features (such as path dependence and multiple underlyings), making them relevant to current financial institutions. For a basket autocallable with 3 underlying assets, 5 payment days, and a knock-in put option with 20 barrier dates, the authors found that one would need about 8000 logical qubits, a $T$-depth of $5.4 \times 10^7$, and a $T$-count of about $1.2 \times 10^{10}$, using the most efficient methods they studied. For a TARF with 1 underlying and 26 payment dates, one needs about $1.2 \times 10^4$ logical qubits, a $T$-depth of about $8.2 \times 10^7$, and a $T$-count of about $9.8 \times 10^9$. A follow-up analysis [18] involving a quantum signal processing approach subsequently reduced these estimates to $4.7 \times 10^3$ logical qubits, $4.5 \times 10^7$ $T$-depth, and $2.4 \times 10^9$ $T$-count. For comparison, classical Monte Carlo methods are roughly estimated to require 1–10 seconds for $4 \times 10^4$ samples to achieve the same accuracy on these examples.

Similar analyses were performed in [11] for the computation of the Greeks, which are quantities that measure the sensitivity of a derivative to various parameters. To compute the Greeks of an option, one needs to compute the derivative of the payoff function with respect to, for example, the price of the underlying. To do this on a quantum computer, one needs to be able to estimate both the expectation of the payoff function and have a way of computing gradients. The authors apply several quantum methods of computing gradients in order to calculate the Greeks, in addition to the quantum approaches to Monte Carlo methods used. Using a quantum gradient method to compute Greeks of an option, the authors estimate that one would need about $1.2 \times 10^4$ logical qubits and a $T$-depth of around $10^8$.





**Caveats**

There are many types of options and derivatives that may not be accurately captured by these simple models. Some payoff functions are path dependent, and hence one cannot simply use the asset value at some fixed time to compute the cost, but rather the cost depends on the trajectory the random variable takes in each Monte Carlo sample.

Moreover, classical approaches to Monte Carlo sampling often allow for massive parallelization, as each simulation of the underlying asset can be done independently. By contrast, quantum algorithms for this problem require a *serial* approach, as the subroutines in the quantum algorithm must be run one after another without measurement and restart if the quadratic advantage is to be realized. When the slower clock speed found in quantum devices is also taken into account, the requirements for a quantum speedup over classical methods become more stringent, as much larger problem sizes are required to achieve practical advantage. For further reading, see [29, Section 2.3], for example.

It is worth noting that in certain cases the number of classical samples needed to achieve error $\epsilon$ can be reduced from the naive $\mathcal{O}(\sigma^2/\epsilon^2)$, cutting into the quadratic quantum speedup. In particular, quasi–Monte Carlo methods, which sample possible trajectories of the underlying assets nonrandomly, can achieve a nearly quadratic speedup compared to traditional classical Monte Carlo methods, but gain an exponential dependence on the number of underlying assets ("curse of dimensionality") which limits their use; see [2, Chapter 5]. The number of samples can also potentially be reduced classically via multilevel Monte Carlo methods [30], although a quantum algorithm for multilevel Monte Carlo also exists [31]. In general, when and how these various methods work is delicate and must be evaluated on a case-by-case basis.

**Comparable classical complexity and challenging instance sizes**

Classical approaches to option pricing comprise some of the largest computational costs incurred by financial institutions. In the traditional approach to solving the option pricing problem, Monte Carlo sampling is required to simulate the evolution of the underlying asset over the time horizon of the option, and it can be slow to converge. In particular, denote the expectation value of $f(X)$ by $V := \mathbb{E}_X(f(X))$, and the variance of $f(X)$ by $\sigma^2$. Classical Monte Carlo methods compute an estimate $\hat{V}$ for $V$ formed by averaging $f(X)$ for $M$ independent samples of $X$. By Chebyshev's inequality,

$$\Pr(|V - \hat{V}| \geq \epsilon) \leq \frac{\sigma^2}{M\epsilon^2}.$$

Thus, classically one needs $M \sim \mathcal{O}(\sigma^2/\epsilon^2)$ samples to find an estimate $\hat{V}$ within a 99% confidence interval [6].

In typical industrial scenarios, options can be priced to sufficient operational precision after roughly a few seconds of runtime, sampling as many as tens of thousands of Monte Carlo trajectories.

Alternatively, a tensor network–based classical approach to option pricing was proposed by [32] that could lead to significant advantages over traditional classical methods in some cases.

**Speedup**

The classical algorithm requires $M = \mathcal{O}(\sigma^2/\epsilon^2)$ samples whereas the quantum algorithm requires only $\widetilde{\mathcal{O}}(\sqrt{M}) = \widetilde{\mathcal{O}}(\sigma/\epsilon)$ samples. The gate cost of a sample is roughly the same classically and





quantumly, and thus the speedup is (nearly) quadratic, inherited from the quadratic speedup of quantum amplitude estimation.

**Outlook**

In [28], the authors place an upper bound on the resources required for pricing options on quantum computers, and they provide a goalpost for quantum hardware development to be able to outperform classical Monte Carlo methods. In particular, the authors estimate that a quantum device would need to be able to execute about $10^7$ layers of $T$ gates per second. Moreover, the code distance for fault-tolerant implementation would need to be chosen large enough to support $10^{10}$ total error-free logical operations. These requirements translate to a logical clock rate of about 50 MHz that would be needed in order to compete with current classical Monte Carlo methods. This clock speed is orders of magnitude faster than what is foreseeably possible given the current status of physical hardware and currently known methods for performing logical gates in the surface code.

While the resource requirements for pricing of derivatives are quite stringent, this is nevertheless an area of active research. For example, a new "analog" quantum representation of stochastic processes was developed in [33] that can compute $\epsilon$-accurate estimates of time averages (over $T$ time steps) of certain functions of stochastic processes in time polylog$(T) \cdot \epsilon^{-c}$, where $3/2 < c < 2$, an exponential speedup over classical methods in the parameter $T$. The analog nature of their method leads to additional caveats, and finding concrete applications of this method remains an interesting open question.

# 9  Machine learning with classical data

There has been significant recent interest in exploring the interplay between quantum computing and machine learning. Quantum resources and quantum algorithms have been studied in all major parts of the traditional machine learning pipeline: (i) the dataset, (ii) data processing and analysis, (iii) the machine learning model leading to a hypothesis family, and (iv) the learning algorithm (see [1, 2, 3] for reviews). In this section, we predominantly focus on quantum approaches for the latter three categories—that is, here we mostly consider quantum algorithms applied to classical data. These approaches include algorithms hinging on the quantum linear system solver (or quantum linear algebra more generally) as the source for possible quantum speedup over classical learning algorithms. These also include *quantum neural networks* (using the framework of variational quantum algorithms) and *quantum kernels*, where the classical machine learning model is replaced with a quantum model. Additionally, in this section, we discuss quantum algorithms that aim to speed up data analysis tasks, namely, *tensor principal component analysis (TPCA)* and *topological data analysis*.

Quantum machine learning is an active area of research. As such, we expect the conclusions made in this section to evolve over time, as new results are discovered. At present, our evaluation suggests that few of the considered quantum machine learning algorithms show any promise of quantum advantage in the immediate future. This conclusion stems from a number of factors, including issues of loading classical data into the quantum device and extracting classical data via tomography, and the success of classical "dequantized" algorithms [4]. More specialized tasks such as tensor PCA and topological data analysis may provide larger polynomial speedups (i.e., better than quadratic) in some regimes, but their application scope is less broad. Finally, other techniques such as quantum neural networks and quantum kernel methods contain heuristic elements which make it challenging to perform concrete analytic end-to-end resource estimates [5].


*The authors are grateful to Ewin Tang for reviewing Section 9.1, to Eric Anschuetz for reviewing Section 9.2, to Matthew Hastings and Robin Kothari for reviewing Section 9.3, to Vedran Dunjko for reviewing Sections 9.4 and 9.5, and to Marco Cerezo for reviewing Section 9.5.*


**This application area contains:**

## 9.1    Quantum machine learning via quantum linear algebra

**Overview**

Linear algebra in high-dimensional spaces with tensor product structure is the workhorse of quantum computation as well as of much of machine learning (ML). It is therefore unsurprising that efforts have been made to find quantum algorithms for various learning tasks, including but not restricted to cluster-finding [1], principal component analysis [2], least-squares fitting [3, 4], recommendation systems [5], binary classification [6], and Gaussian process regression [7]. One of the main computational bottlenecks in all of these tasks is the manipulation of large matrices. Significant speedup for this class of problems has been argued for via quantum linear algebra, as exemplified by the quantum linear system solver (QLSS). The main question marks for viability are (i) can quantum linear algebra be fully dequantized [8] for ML tasks, (ii) can the classical training data be loaded efficiently into a quantum random access memory (QRAM), and (iii) do the quantum ML algorithms that avoid the above-mentioned pitfalls address relevant machine learning problems? Our current understanding suggests that significant quantum advantage would require an exceptional confluence of (i)–(iii) that has not yet been found in the specific applications analyzed to date, though modest speedups are plausible.

**ML applications**

The structure of this section differs from other sections in this survey, due to the disparate nature of many of the quantum machine learning proposals and the fact that they are often heuristic. Rather than cover every proposal, we explore three specific applications. Each example explains which end-to-end problem is being solved and roughly how the proposed quantum algorithm solves that problem, arriving at its dominant complexity. In each case, the quantum algorithm assumes access to fast coherent data access (log-depth QRAM) and leverages quantum primitives for solving linear systems (and linear algebra more generally). Under certain conditions, these primitives can be exponentially faster than classical methods that manipulate all the entries of vectors in the exponentially large vector space. However, for these examples, it is crucial to carefully define the end-to-end problem, as exponential advantages can be lost at the readout step, where the answer to a machine learning question must be retrieved from the quantum state encoding the solution to the linear algebra problem. In the three examples below, this is accomplished with some form of amplitude or overlap estimation, a primitive that brings a multiplicative $\mathcal{O}(1/\epsilon)$ factor into the overall complexity when seeking precision $\epsilon$. This $\mathcal{O}(1/\epsilon)$ readout cost could be avoided if it were the case that $\mathcal{O}(1)$ *samples* from the output state prepared by quantum linear algebra were sufficient for solving the end-to-end problem—situations where this may arise include training large neural networks by solving nonlinear differential equations [9] and determining an optimal set of random features in kernel-based supervised learning [10]—but we do not cover these examples in detail here.

Furthermore, even if these quantum algorithms are exponentially faster than classical algorithms that manipulate the full state vector, in some cases this speedup has been "dequantized" via classical algorithms that merely sample from the entries of the vector. Specifically, for some end-to-end problems, there exist classical "quantum-inspired" algorithms [8, 11, 12] that solve the problem in time only polynomially slower than the quantum algorithm assuming an analogous data-input model. Namely, the assumption that the quantum algorithm has fast QRAM access to the classical data is analogous to the assumption that the classical algorithm has fast "sample-and-query" (SQ) access to the data—SQ access allows the classical algorithm to sample





an entry from the database with probability proportional to its value squared, or to compute the value of any specific entry of the database. The reason that it is fair to compare quantum algorithms relying on QRAM access with classical algorithms relying on SQ access is that both utilize a certain tree-like data structure to enable fast implementation at the circuit level. A large, one-time cost (scaling linearly in the total size of the classical dataset) may be required to "load" the data structure with the classical data, but the data structure is dynamic in the sense that if a single entry in the database is added or changed, updating the data structure has low cost (scaling logarithmically in the total size of the classical dataset). Once the data structure has been set up, one can implement QRAM access (resp. SQ access) using a quantum (resp. classical) circuit with depth only logarithmic in the size of the database. We will not cover the particulars of the quantum-inspired algorithms in more detail, but we note that most of the machine learning tasks based on linear algebra for which quantum algorithms have been proposed have also been dequantized in some capacity [11].

However, it is worth emphasizing that in some cases it remains possible that there could be an exponential quantum advantage if the quantum algorithm is able to exploit additional structure in the matrices involved, such as sparsity, that the classical algorithm cannot. The three examples below roughly illustrate the spectrum of possibilities: some tasks are fully dequantized, whereas others, to the best of our current knowledge, could still support exponential advantages if certain conditions are met.

## Example 1: Gaussian process regression

**Actual end-to-end problem:** Gaussian process regression (GPR) is a nonparametric, Bayesian method for regression. GPR is closely related to kernel methods [13], as well as to other regression models, including linear regression [14]. Our presentation of the problem follows that of [14, Chapter 2] and [15]. Given training data $\{x_j, y_j\}_{j=1}^{M}$, with inputs $x_j \in \mathbb{R}^N$ and noisy outputs $y_j \in \mathbb{R}$, the goal is to model the underlying function $f(x)$ generating the output $y$

$$y = f(x) + \epsilon_{\text{noise}},$$

where $\epsilon_{\text{noise}}$ is drawn from i.i.d. Gaussian noise with variance $\sigma^2$. Modeling $f(x)$ as a Gaussian process means that for inputs $\{x_j\}_{j=1}^{M}$, the outputs $\{f(x_j)\}_{j=1}^{M}$ are treated as random variables with a joint multivariate Gaussian distribution, in such a way that any subset of these values are jointly normally distributed in a manner consistent with the global distribution. While this multivariate Gaussian distribution governing $\{f(x_j)\}_{j=1}^{M}$ will generally be correlated for different $j$, the additional additive error $\epsilon_{\text{noise}}$ on our observations $y_j$ is independent from the choice of $f(x_j)$ and uncorrelated from point to point. The Gaussian process is specified by the distribution $\mathcal{N}(m, K)$ where $m$ is the length-$M$ vector obtained by evaluating a "mean function" $m(x)$ at the points $\{x_j\}_{j=1}^{M}$, and $K$ is an $M \times M$ covariance kernel matrix obtained by evaluating a covariance kernel function $k(x, x')$ at $x, x' \in \{x_j\}_{j=1}^{M}$—$\mathcal{N}$ then denotes the multivariate Gaussian distribution with the corresponding mean and covariance. The functional form of the mean and covariance kernel are specified by the user and determine the properties of the Gaussian process, such as its smoothness.[24] These functions typically contain a small number of hyperparameters which can be optimized using the training data. A commonly used covariance kernel function is the squared exponential covariance function $k(x, x') = \exp\left(-\frac{1}{2\ell^2}\|x - x'\|^2\right)$

---

[24]This can be visualized by sampling a function from the distribution, which means sampling a value of $f(x_j)$ from the distribution for each $x_j$, and plotting the values of $f(x_j)$ as a curve.





where $\ell$ is a hyperparameter controlling the length scale of the Gaussian process, and $\|\cdot\|$ denotes the standard Euclidean norm in the case of vector arguments and the spectral norm in the case of matrix arguments.

Given choices for $m(x)$ and $k(x, x')$ and the observed data $\{x_j, y_j\}_{j=1}^M$, our task is to predict the value $f(x_*)$ of a new test point $x_*$. Because the Gaussian process assumes that all $M + 1$ values $\{f(x_1), \ldots, f(x_M), f(x_*)\}$ have a jointly Gaussian distribution, it is possible to condition upon the observed data to obtain the distribution for $f(x_*)$ which is $p(f_*|x_*, \{x_j, y_j\}) \sim \mathcal{N}(\bar{f}_*, \mathbb{V}[f_*])$. Our goal is to compute $\bar{f}_*$, the mean (linear predictor) of the distribution for $f(x_*)$, as well as the variance $\mathbb{V}[f_*]$, which gives uncertainty on the prediction. Computing the underlying multivariate Gaussian distribution can be bypassed by exploiting the closure of Gaussians under linear operations, in particular, conditioning. This re-expresses the problem as linear algebra with the kernel matrix. Assuming the common choice of $m(x) = 0$ and defining the length-$M$ vector $k_* \in \mathbb{R}^M$ to have its $j$-th entry given by $k(x_*, x_j)$, we obtain

$$\bar{f}_* = k_*^{\mathsf{T}}[K + \sigma^2 I]^{-1} y$$

$$\mathbb{V}[f_*] = k(x_*, x_*) - k_*^{\mathsf{T}}[K + \sigma^2 I]^{-1} k_*$$

which characterize the prediction for the test point. The advantages of GPR are a small number of hyperparameters, model interpretability, and that it naturally returns uncertainty estimates for the predictions. Its main disadvantage is the computational cost.

**Dominant resource cost/complexity:** In classical implementations, the cost is dominated by performing the inversion $[K + \sigma^2 I]^{-1}$, typically via a Cholesky decomposition, resulting in a complexity of $\mathcal{O}(M^3)$ (see [14, Chapter 8] and [16] for approximations to reduce the classical cost). In [7], a quantum algorithm was proposed that leverages the QLSS to perform this inversion more efficiently. The quantum computer uses the classical data to infer the linear predictor and variance for a test point $x_*$, and this process must be repeated for the computation of each new test point output. We analyze the complexity of computing $\bar{f}_*$, with a simple extension for $\mathbb{V}[f_*]$. Given classically observed/precomputed values of $y$ and $k_*$, the quantum algorithm uses state preparation from classical data (based on QRAM) to prepare quantum states representing $|y\rangle$ and $|k_*\rangle$,[25] each with a gate depth of $\mathcal{O}(\log(M))$ (though using $\mathcal{O}(M)$ gates overall). The algorithm also uses a block-encoding of classical data (also using QRAM) for $A := [K + \sigma^2 I]$, with a normalization factor of $\alpha = \|K + \sigma^2 I\|_F$, where $\|\cdot\|_F$ denotes the Frobenius norm.[26] The state-of-the-art QLSS has complexity $\mathcal{O}(\alpha \kappa \|A\|^{-1} \log(1/\epsilon))$ calls to an $\alpha$-normalized block-encoding of matrix $A$ with condition number $\kappa$ (see Eq. (61)). In this case, the minimum singular value of $A$ is at least $\sigma^2$, so $\kappa/\|A\| \leq \sigma^{-2}$. The QLSS produces the normalized state $|A^{-1}y\rangle$, and a similar approach yields an estimate for the norm $\|A^{-1}y\|$ to relative error $\epsilon$ at cost $\widetilde{\mathcal{O}}(\alpha \kappa \|A\|^{-1} \epsilon^{-1})$ [18, 19]. Given unitary circuits performing these tasks, we can estimate the quantity $\bar{f}_* = \langle k_*|A^{-1}y\rangle \cdot \|k^*\| \cdot \|A^{-1}y\|$ to precision $\epsilon$ using overlap estimation with gate depth upper bounded by

$$\widetilde{\mathcal{O}}\left(\log(M) \cdot \|K + \sigma^2 I\|_F \sigma^{-2} \cdot \frac{\|k_*\| \|[K + \sigma^2 I]^{-1} y\|}{\epsilon}\right),$$

---

[25] For any vector $v$, the notation $|v\rangle$ denotes the normalized quantum state whose amplitudes in the computational basis are proportional to the entries of $v$, for example, $|y\rangle = \frac{1}{\|y\|}\sum_j y_j|j\rangle$.

[26] It may be more efficient to load in the $\{x_j\}$ values and then coherently evaluate the kernel entries using quantum arithmetic. Some ideas in this direction are explored in [17]. One might also consider block-encoding $K$ and $\sigma^2 I$ separately and combining them with linear combination of unitaries.





where the three factors come from state preparation (i.e., QRAM), QLSS, and overlap estimation, respectively. Using QRAM for state preparation as described above would use $\mathcal{O}(M^2)$ ancilla qubits. Note that classical "quantum-inspired" methods for solving linear systems, based on SQ access, also have poly($\|A\|_F, \kappa, \epsilon^{-1}, \log(M)$) complexity [11, 20, 12], and thus the quantum algorithm as stated above offers at most a polynomial speedup in the case of dense matrices.

On the other hand, [7] considers the case where the vectors and kernels are sparse[27] and uses this to reduce the cost of the quantum algorithm and of QRAM. In this case, using block-encodings of sparse matrices, the factor $\|A\|_F$ in the complexity expression is replaced by a factor $s\|A\|_{\max}$, where $s$ is the sparsity of the matrix $A$ and $\|A\|_{\max}$ is the maximum magnitude of any entry of $A$—log-depth QRAM with $\Omega(M)$ ancilla qubits would still be necessary to implement the sparse access oracle to the $sM$ arbitrary nonzero entries of $A$ in depth $\mathcal{O}(\log(M))$. The upshot is that in the sparse case, because the algorithm assumes the kernel is not low rank, this algorithm is not dequantized by SQ access [11] and may still offer an exponential speedup over quantum-inspired methods. However, we note that the assumption of sparsity in $[K + \sigma^2 I]$ may also enable the use of more efficient classical algorithms for computing the inverse (see Section 18 on QLSSs). Moreover, we must include the classical precomputation of evaluating the entries of this matrix. A related, and similarly efficient, quantum algorithm is proposed in [15] for optimizing the hyperparameters of the GPR kernel by maximizing the marginal likelihood of the observed data given the model.

## Example 2: Support vector machines

**Actual end-to-end problem:** The task for the support vector machine (SVM) is to classify an $N$-dimensional vector $x_*$ into one of two classes ($y_* = \pm 1$), given $M$ labeled data points of the form $\{(x_j, y_j) \colon x_j \in \mathbb{R}^N, y_j = \pm 1\}_{j=1,\ldots,M}$ used for training. The training phase solves a continuous optimization problem to find a maximum-margin hyperplane, described by normal vector $w \in \mathbb{R}^M$ and offset $b \in \mathbb{R}$, which separates the training data. That is, data points with $y_j = 1$ lie on one side of the plane, and data points with $y_j = -1$ lie on the other side. Once trained, the classification of $x_*$ is inferred via the formula

$$y_* = \text{sign}(b + \langle w, x_* \rangle), \tag{27}$$

where $\langle \cdot, \cdot \rangle$ denotes the standard inner product between vectors.

In the "hard-margin" version of the problem where all training points must be classified correctly (assuming it is possible to do so, i.e., the data is linearly separable), the solution $(w, b)$ is given by

$$\underset{(w,b)}{\text{argmin}} \|w\|^2, \qquad \text{subject to:} \qquad (\langle w, x_j \rangle + b)y_j \geq 1 \qquad \forall j. \tag{28}$$

In the "soft-margin" version of the problem, the hyperplane need not correctly classify all training points. The relation $(\langle w, x_j \rangle + b)y_j \geq 1$ is relaxed to $(\langle w, x_j \rangle + b)y_j \geq 1 - \xi_j$, with $\xi_j \geq 0$. Now, $(w, b)$ are determined by

$$\underset{(w,b,\xi)}{\text{argmin}} \|w\|^2 + \gamma\|\xi\|_1, \quad \text{subject to:} \quad (\langle w, x_j \rangle + b)y_j \geq 1 - \xi_j \quad \forall j, \tag{29}$$

where $\|\cdot\|_1$ denotes the vector 1-norm, and $\gamma$ is a user-specified hyperparameter related to how much to penalize points that lie within the margin. Both Eqs. (28) and (29) are convex programs,

---

[27]For the squared exponential covariance function mentioned above, the kernel matrix will not be sparse, but [7] notes several applications of GPR where sparsity is well justified.





in particular, quadratic programs, which can also be rewritten as second-order cone programs [21]. Another feature of these formulations is that the solution vectors $w$ and $\xi$ are usually sparse; the $j$-th entry is only nonzero for values of $j$ where $x_j$ lies on or within the margin near the hyperplane—these $x_j$ are called the "support vectors."

In [22], a "least-squares" version of the SVM problem was proposed, which has no inequality constraints:[28]

$$\underset{(w,b,\xi)}{\text{argmin}} \|w\|^2 + \frac{\gamma}{M}\|\xi\|^2, \quad \text{subject to:} \quad (\langle w, x_j \rangle + b)y_j = 1 - \xi_j \quad \forall j. \tag{30}$$

This is an equality-constrained least-squares problem, which is simpler than a quadratic program and can be solved using Lagrange multipliers and inverting a linear system. Specifically, one introduces a vector $\beta \in \mathbb{R}^M$ and solves the $(M+1) \times (M+1)$ linear system $Au = v$, where

$$A = \begin{pmatrix} 0 & \mathbf{1}^\intercal/\sqrt{M} \\ \mathbf{1}/\sqrt{M} & K/M + \gamma^{-1}I \end{pmatrix}, \qquad u = \begin{pmatrix} b \\ \beta \end{pmatrix}, \qquad v = \frac{1}{\sqrt{M}} \begin{pmatrix} 0 \\ y \end{pmatrix}, \tag{31}$$

with $K$ the kernel matrix for which $K_{ij} = \langle x_i, x_j \rangle$, $\mathbf{1}$ the all-ones vector, and $I$ the identity matrix. The vector $w$ is inferred from $\beta$ via the formula $w = \sum_j \beta_j x_j/\sqrt{M}$.

However, unlike the first two formulations, the least-squares formulation does not generally have sparse solution vectors $(w, b)$ (see [23]). Additionally, its solution can be qualitatively different, due to the fact that correctly classified data points can lead to negative $\xi_j$ that apply penalties to the objective function through the appearance of $\|\xi\|^2$.

**Dominant resource cost/complexity:** The hard-margin and soft-margin formulations of SVM are quadratic programs, which can be mapped to second-order cone programs and solved with quantum interior point methods (QIPMs). This solution was proposed in [21], and, assuming access to log-depth QRAM it can find $\epsilon$-accurate estimates for the solution $(w, b)$ in time scaling as $\tilde{\mathcal{O}}(M^{0.5}(M+N)\kappa_{\text{IPM}}\zeta \log(1/\epsilon)/\xi')$, where $\kappa_{\text{IPM}}$, $\zeta$, and $\xi'$ are instance-specific parameters related to the QIPM. This compares to $\mathcal{O}(M^{0.5}(M+N)^3 \log(1/\epsilon))$ for naively implemented classical interior point methods. In [21], numerical simulations on random SVM instances were performed to compute these instance-specific parameters, and the results were consistent with a small polynomial speedup. However, the resource estimate of [24] for a related problem suggests a practical advantage may be difficult to realize with this approach.

The least-squares formulation can be solved directly with the QLSS, as pursued in [6]. This can be compared to classically solving the linear system via Gaussian elimination, with cost $\mathcal{O}(M^3)$. The QLSS requires the ability to prepare the state $|v\rangle$, which can be accomplished in $\mathcal{O}(\log(M))$ depth through methods for preparation of states from classical data, although requiring $\mathcal{O}(M)$ total gates and ancilla qubits. One also needs a block-encoding of the matrix $A$. One method is through block-encodings from classical data, which requires classical precomputation of the $\mathcal{O}(M^2)$ entries of $K$ (incurring classical cost $\mathcal{O}(M^2N)$) and producing a block-encoding with normalization factor $\alpha = \|A\|_F$ (Frobenius norm). Henceforth, we assume that $\|x_j\| \leq 1$ for all $j$, which can always be achieved by scaling down the training data (inducing a scaling up of $w$ and $\sqrt{\gamma}$ by an equal factor). This implies $\|K/M\|_F \leq 1$ and hence $\|A\|_F \leq \sqrt{2} + 1 + \sqrt{M}\gamma^{-1}$. A better block-encoding can be obtained by block-encoding $K/M$ via the method for Gram

---

[28]Our definition of the least-squares SVM is equivalent to the normal presentation found in [22, 6]; however, we choose slightly different conventions for normalization of certain parameters, such as $\gamma$, with respect to $M$. The goal of our choices is to make the final complexity expression free of any explicit $M$ dependence.





matrices[29] and $\gamma^{-1}I$ via the trivial method, and then combining these with the rest of $A$ via linear combination of block-encodings. This avoids the need to classically calculate the inner products $\langle x_i, x_j \rangle$, and has a better normalization $\alpha \leq \sqrt{2} + 1 + \gamma^{-1}$.

Given these constructions, the QLSS outputs the state $|u\rangle = (b|0\rangle + \sum_{j=1}^{M} \beta_j |j\rangle) / \sqrt{b^2 + \|\beta\|^2}$; the cost is $\widetilde{\mathcal{O}}(\alpha \kappa_A / \|A\|)$ queries to the block-encoding of $A$, where $\kappa_A$ is the condition number of $A$. We may assert that $\|A\| \geq 1$. This follows by noting that the lower-right block of $A$ (as defined in Eq. (31)) is positive semidefinite, and that 1 is an eigenvalue of $A$ when the lower-right block is set to zero. The condition number should be upper bounded by an $M$-independent function of $\gamma$ due to the appearance of the regularizing $\gamma^{-1}I$.

Reading out all $M + 1$ entries of $|u\rangle$ via tomography would multiply the cost by $\Omega(M)$. However, in [6], it was observed that to classify a test point $x_*$ via Eq. (27), one can use overlap estimation rather than classically learning the solution vector. In our notation and normalization, this can be carried out as follows. Let $|x_j\rangle = \sum_{i=1}^{M} x_{ji} |i\rangle / \|x_j\|$, with $x_{ji}$ denoting the $i$-th entry of the vector $x_j$. Starting with $|u\rangle$, we prepare $|x_j\rangle$ into an ancilla register, using methods for controlled state preparation from classical data, forming

$$|\tilde{u}\rangle = \frac{b|0\rangle|0\rangle + \sum_{j=1}^{M} \beta_j |j\rangle \left( \|x_j\| |x_j\rangle + \sqrt{1 - \|x_j\|^2} |M+1\rangle \right)}{\sqrt{b^2 + \|\beta\|^2}}.$$

One also creates a reference state $|\tilde{x}_*\rangle$ encoding $x_*$, defined as

$$|\tilde{x}_*\rangle = \frac{1}{\sqrt{2}} |0\rangle|0\rangle + \frac{1}{\sqrt{2M}} \sum_{j=1}^{M} |j\rangle \left( \|x_*\| |x_*\rangle + \sqrt{1 - \|x_*\|^2} |M+2\rangle \right).$$

The right-hand side of Eq. (27) is then given by $\sqrt{2}\sqrt{b^2 + \|\beta\|^2} \langle \tilde{u} | \tilde{x}_* \rangle$. Thus, the overlap $\langle \tilde{u} | \tilde{x}_* \rangle$ must be estimated to precision $\epsilon = 1 / \sqrt{2(b^2 + \|\beta\|^2)}$ in order to distinguish $\pm 1$ and classify $x_*$. Additionally, the norm $\|u\| = \sqrt{b^2 + \|\beta\|^2}$ must be calculated; this can separately be done to relative error $\epsilon'$ at cost $\widetilde{\mathcal{O}}(\alpha \kappa_A / \epsilon')$ (see Section 18 on QLSSs). We may also note that as $u = A^{-1}v$ and $\|v\| = 1$, we have $\|u\| \leq \kappa_A / \|A\|$. Thus, the overall circuit depth required to classify $x_*$ is

$$\widetilde{\mathcal{O}}\left( \frac{\alpha \kappa_A^2}{\|A\|^2} \right).$$

There is no explicit $\mathrm{poly}(N, M)$ dependence. However, for certain datasets and parameter choices, such dependence could be hidden in $\kappa_A$ or $\alpha$, making an apples-to-apples comparison with Gaussian elimination less clear.

Furthermore, this task has been dequantized under the assumption of SQ access [26, 11, 12]. In time scaling as $\mathrm{poly}(\|A\|_F, \epsilon^{-1}, \log(NM))$, one can classically sample from the solution vector $|u\rangle$ to error $\epsilon$, and furthermore, one can estimate inner products $\langle \tilde{u} | \tilde{v} \rangle$ in time $\mathcal{O}(1/\epsilon^2)$

---

[29] We sketch a possible instantiation of this method here. Define $|x_i\rangle = \|x_i\|^{-1} \sum_{k=1}^{M} x_{ik} |k\rangle$ where $x_{ik}$ is the $k$-th entry of $x_i$. Suppose $M = 2^m$ is a power of 2. Following the setup in block-encodings and [25, Lemma 47], we must define sets of $M$ orthonormal states $\{|\psi_i\rangle\}$ and $\{|\phi_j\rangle\}$. We choose $|\psi_i\rangle = (\|x_i\| |x_i\rangle + \sqrt{1 - \|x_i\|^2} |M+1\rangle)(H^{\otimes m} |i\rangle) |0^m\rangle$, where $H$ denotes the Hadamard transform. We choose $|\phi_j\rangle = (\|x_j\| |x_j\rangle + \sqrt{1 - \|x_j\|^2} |M+2\rangle) |0^m\rangle (H^{\otimes m} |j\rangle)$. These states can be prepared in $\mathcal{O}(\log(M))$ depth using $\mathcal{O}(M)$ total gates and ancilla qubits with methods for controlled state preparation from classical data. It can be verified that these sets are orthonormal, and that $\langle \psi_i | \phi_j \rangle = \langle x_i, x_j \rangle / M$. Hence, the Gram matrix construction yields a block-encoding of $K/M$ with normalization factor 1.





[27].[30] However, the cost can be reduced through a trick that is analogous to how the quantum algorithm can block-encode the $\gamma^{-1}I$ part of $A$ separately to avoid the dependence on a large $\|A\|_F$. In particular, [11, Corollary 6.18] gives a classical complexity that would be polynomially related to the quantum complexity above under appropriate matching of parameters, but the power of this polynomial speedup could still be significant. In any case, such a speedup crucially requires log-depth QRAM access to the training data, which requires total gate complexity $\Omega(NM)$ and $\mathcal{O}(NM)$ ancilla qubits.

**Example 3: Supervised cluster assignment**

**Actual end-to-end problem:**  Suppose we are given access to a vector $x \in \mathbb{C}^N$ and a set of $M$ samples $\{y_j \in \mathbb{C}^N\}_{j=1,\ldots,M}$. We want to estimate the distance between $x$ and the centroid of the set $\{y_j\}$ to judge whether $x$ was drawn from the same set as $\{y_j\}$. If we have multiple sets $\{y_j\}$, we can infer that $x$ belongs to the one for which the distance is shortest; as a result, this is also called the "nearest-centroid problem." Specifically, the computational task is to estimate $\|x - \frac{1}{M}Y\mathbf{1}\|$ to additive constant error $\epsilon$ with probability $1 - \delta$, where $Y \in \mathbb{C}^{N \times M}$ is the matrix whose columns are $y_j$, and $\mathbf{1}$ is the vector of $M$ ones—the vector $Y\mathbf{1}/M$ is the centroid of the set.

**Dominant resource cost/complexity:**  Naively computing the centroid incurs classical cost $\mathcal{O}(NM)$. In [1], a quantum solution to this problem was proposed. Let $\bar{x} = x/\|x\|$ and let $\bar{Y}$ be normalized so that all columns have unit norm. Define the $N \times (M + 1)$ matrix $R$ and length-$(M + 1)$ vector $w$ as follows:

$$R = \begin{pmatrix} \bar{x} & \bar{Y}/\sqrt{M} \end{pmatrix}, \qquad w = \begin{pmatrix} \|x\| \\ -1_Y/\sqrt{M} \end{pmatrix},$$

where $1_Y$ is the length-$M$ vector containing the norms of the columns of $Y$, defined such that $\bar{Y}1_Y = Y\mathbf{1}$. Then, $Rw = x - \frac{1}{M}Y\mathbf{1}$. Using methods for block-encoding and state preparation from classical data, one constructs $\mathcal{O}(\log(NM))$-depth circuits that block-encode $R$ (with normalization factor $\|R\|_F = 2$) and prepare the state $|w\rangle$. If we apply the block-encoding of $R$ to $|w\rangle$ and measure the block-encoding ancillas, the probability that we obtain $|0\rangle$ is precisely $(\|Rw\|/2\|w\|)^2$. Thus, using amplitude estimation, one learns $\|Rw\|$ to precision $\epsilon$ with probability at least $1 - \delta$ at cost $\mathcal{O}(\|w\|\log(1/\delta)/\epsilon)$ calls to the log-depth block-encoding and state preparation routines.

The advantage over naive classical methods essentially boils down to the assumption of efficient classical data loading for a specific dataset. Subsequently, this quantum algorithm was dequantized, and it was understood that a similar feat is possible classically in the SQ access model [8]. Specifically, the classical algorithm runs in time $\widetilde{\mathcal{O}}(\|w\|^2 \log(1/\delta)/\epsilon^2)$, reducing the exponential speedup to merely quadratic.

---

[30]The method of doing so is succinct to describe (see, e.g., [28]). First, one uses sample access to the vector $\tilde{u}$ to generate an index $i$ at random, with probability $|\tilde{u}_i|^2/\|\tilde{u}\|^2$. Then, one uses query access to $\tilde{u}$ and $\tilde{v}$ to compute the quantity $\mathcal{R} = (\tilde{v}_i\|\tilde{u}\|)/(\tilde{u}_i\|\tilde{v}\|)$. The expectation value of $\mathcal{R}$ is precisely $\langle\tilde{u}|\tilde{v}\rangle$, and the variance is upper bounded by 1. Thus, an estimate of $\langle\tilde{u}|\tilde{v}\rangle$ to $\epsilon$ precision is obtained by averaging $\mathcal{O}(1/\epsilon^2)$ samples of the random variable $\mathcal{R}$.





**Caveats**

The overwhelming caveat in these and other proposals is access to the classical data in quantum superposition. These quantum machine learning algorithms assume that we can load a vector of $N$ entries or a matrix of $N^2$ entries in polylog($N$) time. While efficient quantum data structures, that is, QRAM, have been proposed for this task, they introduce a number of caveats. In order to coherently load $N$ pieces of data in $\log(N)$ time, QRAM uses a number of ancilla qubits, arranged in a tree structure. To load data of size $N$, the QRAM data structure requires $\mathcal{O}(N)$ qubits, which is exponentially larger than the $\mathcal{O}(\log(N))$ data qubits used in the algorithms above. This spatial complexity does not yet include the overheads of quantum error correction and fault-tolerant computation, in particular the large spatial resources required to distill magic states in parallel. As such, we do not yet know if it is possible to build a QRAM that can load the data sufficiently quickly, while maintaining moderate spatial resources.

In addition, achieving speedups by efficiently representing the data as a quantum state may suggest that classical methods based on tensor networks could achieve similar performance, in some settings. Taking this line of reasoning to the extreme, a number of efficient classical algorithms have been developed by "dequantizing" the quantum algorithms. That is, by assuming an analogous access model (the SQ access model) to the training data, as well as some assumptions on sparsity and/or rank of the inputs, there exist approximate classical sampling algorithms with polynomial overhead as compared to the quantum algorithms [8, 27]. This means that any apparent exponential speedup must be an artifact of the data loading/data access assumptions.

A further caveat is inherited from the QLSS subroutine, which is that the complexity is large when the matrices involved are ill conditioned. This caveat is somewhat mitigated in the Gaussian process regression and support vector machine examples above, where the matrix to be inverted is regularized by adding the identity matrix.

**End-to-end resource analysis**

To the best of our knowledge, full end-to-end resource estimation has not been performed for any specific quantum machine learning tasks.

**Outlook**

Much of the promise of quantum speedup for classical machine learning based on linear algebra hinges on the extent to which quantum algorithms can be dequantized. At present, the results of [8] seem to prohibit an exponential speedup for many of the problems proposed, but there is still the possibility of a large polynomial speedup. The most recent asymptotic scaling analysis [11] for dequantization methods still allows for a power 4 speedup in the Frobenius norm of the "data matrix" and a power 11 speedup in the polynomial approximation degree (see [29] for more details). However, the classical algorithms are steadily improving, and their scaling might be further reduced.

It is also worth noting that the classical probabilistic algorithms based on the SQ access model are not currently used in practice. This could be due to a number of reasons, including the poor polynomial scaling, the fact that the access model might not be well suited to many practical scenarios, or simply because the method is new and has not been tested in practice (see [30, 31] for some work in this direction).

On the other hand, some machine learning tasks based on quantum linear algebra are not known to be dequantized, such as Gaussian process regression under the assumption that the





kernel matrix is sparse. In particular, avoiding dequantization and achieving an exponential quantum speedup appears to require that the matrices involved are simultaneously sparse, high rank, and well conditioned.[31] In this situation, quantum algorithm can leverage block-encodings for which the normalization factor is equal to the sparsity, rather than general block-encodings of classical data for which the normalization factor is the Frobenius norm. The complexity of quantum-inspired classical algorithms based on SQ access will still grow polynomially with the Frobenius norm even when the matrices are sparse,[32] although other classical algorithms may be able to exploit the sparsity more directly. Perhaps unsurprisingly, the prototypical matrices that satisfy these criteria are sparse unitary matrices, such as those naturally implemented by a local quantum gate. For unitary matrices, the condition number is 1, and the Frobenius norm is equal to the square root of the Hilbert space dimension—exponentially large in the system size. (As a simple example, consider the identity matrix on $n$ qubits.) A central question is whether situations like this occur in interesting end-to-end machine learning problems. Even if they do, an exponential speedup is not guaranteed. An additional hurdle arises in the quantum readout step, which incurs a cost scaling as the inverse in the precision target. To avoid exponential overhead, the end-to-end problem must not require exponentially small precision.

## Further reading

For further reading on specific machine learning tasks where quantum algorithms have been proposed, we refer the reader to [32, 33, 34]. We have not covered dequantization techniques in great detail; for an accessible summary and perspective, see [28], and for a more detailed overview, see [35].

---

[31] Dequantization can also be avoided even when the matrices involved are dense, provided that they are given by a product of a small number of sparse matrices. For example, it was described in [10] how an exponential speedup may be possible for a certain ML task, where the matrix to be inverted is neither sparse nor low rank; rather, it is related to a sparse matrix via the discrete Fourier transform (a dense unitary matrix). A block-encoding for the relevant matrix is constructed by leveraging the quantum Fourier transform (QFT). Note that the QFT can be decomposed into a product of sparse matrices corresponding to the local unitary gates in the quantum circuit for QFT.

[32] For example, consider the complexity of algorithms for pseudoinversion of an $s$-sparse matrix $A$. Let the rank of $A$ be $r$, which may be as large as the matrix dimension, and suppose all nonzero singular values of the matrix lie in the interval $[1/\kappa, 1]$. This implies that $\sqrt{r}/\kappa \leq \|A\|_F \leq \sqrt{r}$, and thus $\kappa \|A\|_F \geq \sqrt{r}$. As a consequence, the complexity of quantum-inspired algorithms with SQ access in [11, 20, 12]—scaling as $\mathrm{poly}(\|A\|_F, \kappa, 1/\epsilon)$—will necessarily grow as a polynomial of the matrix rank, even for fixed sparsity. On the other hand, the complexity of the quantum algorithm with QRAM that applies the QLSS and amplitude estimation—scaling as $\mathrm{poly}(s\|A\|_{\max}, \kappa, 1/\epsilon)$—is independent of the matrix rank for well-conditioned $A$ (i.e., $\kappa = \mathcal{O}(1)$).

## 9.2   Quantum machine learning via energy-based models

### Overview

An important class of models in machine learning is *energy-based models*, which are heavily inspired by statistical mechanics. The goal of energy-based models is to train a physical model (i.e., tune the interaction strengths between a set of particles) such that the model closely matches the training set when the model is in thermal equilibrium (made more precise below). Energy-based models are an example of *generative* models since, once they are trained, they can then be used to form new examples that are similar to the training set by sampling from the model's thermal distribution.

Due to their deep connection to physics, energy-based models are prime candidates for various forms of quantization. However, one challenge faced by quantum approaches is that the statistical mechanical nature of the learning problem also often lends itself to efficient, approximate classical methods. As a result, the best quantum algorithms may also be heuristic in nature, which prevents an end-to-end complexity analysis. While energy-based models are less widely used than deep neural networks today, they were an important conceptual development in machine learning [1] and continue to foster interest due to their sound theoretical basis, and their connection to statistical mechanics.

There are a number of proposals for generalizing energy-based models to quantum machine learning. The starting point is a graph where the vertices are divided into *visible* $\{v\}$ and *hidden* $\{h\}$ nodes. When each node is assigned a value in some discrete or continuous set, this constitutes a "configuration" $(h, v)$ of the model. A training set $\mathcal{D}$ is provided as input, containing a list of configurations of the visible vertices. The hidden nodes are not part of the training set, but including them is essential for the model to be able to capture latent variables in the data.

A graphical model is then built on the vertices—each vertex is a physical system (such as a spin-1/2 particle) and edges between vertices represent physical interactions. The model is described by an energy functional $H(h, v)$, which assigns an energy value to each possible configuration $(h, v)$ of the vertices. For example, in Boltzmann machines (BMs), the vertices are assigned binary variables, and the interactions are Ising interactions. The model can be used to generate samples (e.g., via Markov chain Monte Carlo methods) from the thermal distribution (also known as the Boltzmann distribution or the Gibbs distribution) at unit temperature, that is, the distribution where each configuration $(h, v)$ is sampled with probability proportional to $e^{-H(h,v)}$. In unsupervised learning tasks, provided a set of training samples of configurations of the visible units $v$, the goal is to tune the interaction weights of the model such that the model's thermal distribution best matches the distribution that generated the training set.

Quantum algorithms can potentially be helpful for training classical graphical models. One can also generalize the model itself by allowing the physical systems on each vertex to be quantum, and interactions between systems to be noncommuting.

### Actual end-to-end problem(s) solved

**Classical graphical models:**   Let $G = (V, E)$ denote a graph with vertices $V$ and edges $E$. For classical models, each vertex $j$ is assigned a binary variable $z_j = \pm 1$. The variables are split into visible and hidden nodes, $z \in \{v\} \cup \{h\}$. For classical BMs, the energy functional is





often taken to be quadratic[33] with weights $\{b_i, w_{ij}\}$:

$$H(z) = \sum_{i \in V} b_i z_i + \sum_{(i,j) \in E} w_{ij} z_i z_j. \tag{32}$$

Note that interactions can occur between any pair of nodes (hidden or visible). In the special case of a restricted Boltzmann machine (RBM), each edge must pair up a hidden node with a visible node (i.e., the graph is bipartite). This restriction makes the model less expressive than graphs with edges between hidden nodes, but it leads to simplifications for certain training approaches.

The thermal distribution corresponding to the energy functional (at unit temperature) associates each configuration $v$ of visible nodes with a probability $p(v)$ such that

$$p(v) = \sum_h p(h, v), \qquad p(h, v) = \frac{\mathrm{e}^{-H(h,v)}}{\mathcal{Z}}, \qquad \mathcal{Z} = \sum_{h,v} \mathrm{e}^{-H(h,v)},$$

where $\mathcal{Z}$, the partition function, is the normalization to ensure probabilities sum to 1. Even though hidden nodes are integrated out in the calculation of $p(v)$, they impact the distribution of $p(v)$ through their interactions with the visible nodes.

Given a training set $\mathcal{D} = \{v_1, v_2, \ldots, v_{|\mathcal{D}|}\}$ of sample configurations of the visible nodes, the goal of the training phase is to modify the weights $\theta \in \{b_i\} \cup \{w_{ij}\}$ such that samples from the thermal distribution of the model most closely match the training samples. Ideally, this is done by finding the set of weights that maximizes the likelihood of observing the samples, that is, $\prod_{v \in \mathcal{D}} p(v)$, or, equivalently, minimizing the (normalized) log-likelihood loss function, defined as

$$L(b, w) = -\frac{1}{|\mathcal{D}|} \sum_{v \in \mathcal{D}} \log(p(v)). \tag{33}$$

The loss function can be minimized using some variant of gradient descent, which requires the evaluation of the derivatives $\partial_\theta L$ for $\theta \in \{b_i\} \cup \{w_{ij}\}$. For the energy functional above, these derivatives can be readily calculated from ensemble averages (see, e.g., [3]). For example,

$$\frac{\partial L}{\partial w_{ij}} = \langle z_i z_j \rangle_{v \in \mathcal{D}} - \langle z_i z_j \rangle, \tag{34}$$

where $\langle \cdot \rangle$ denotes an average over samples from the thermal distribution $p(h, v)$, while $\langle \cdot \rangle_{v \in \mathcal{D}}$ denotes an average where $v$ is drawn at random from the training set $\mathcal{D}$, and $h$ is sampled from the thermal distribution conditioned on that choice of $v$. Without any further restrictions, the gradients will typically be difficult to evaluate (or even to estimate accurately). An exact computation requires computing a sum over the exponential number of configurations of the vertices.

In some cases, good estimates of the gradients can be obtained by repeatedly drawing samples from the thermal distribution and computing averages. Samples can be generated with

---

[33]This quadratic energy functional is related to the Sherrington–Kirkpatrick (SK) model [2] with an external field, which is a model for spin glasses in the statistical mechanics literature. For the SK model, the couplings $w_{ij}$ are chosen randomly for each pair of nodes, and it is typically computationally hard to find configurations with optimal energy (see Section 4.2 on beyond quadratic speedups in combinatorial optimization for additional information).





Markov chain Monte Carlo (MCMC) methods such as the Metropolis–Hastings algorithm or simulated annealing; however, the time required to sample from a distribution close to the thermal distribution depends on the mixing time of the Markov chain, which is generally unknown and can also be exponential in the graph size. Additionally, many samples need to be generated to produce a robust average, with precision $\epsilon$ requiring $\mathcal{O}(1/\epsilon^2)$ samples. Approximate classical methods, such as contrastive divergence [4], avoid this issue by initializing the Markov chain at one of the training samples and deliberately taking a small number of steps—this does not exactly correspond to optimizing the log-likelihood but in some cases has empirical success [5]. Indeed, here we see the benefit of restricting to bipartite graphs in RBMs: since there are no edges between hidden nodes, the Gibbs distribution over the hidden nodes is independent from node to node, conditioned on a fixed setting of the visible nodes. This enables a simple exact calculation for the first term of Eq. (34), and it is also key to the success of estimating the second term with contrastive divergence, where the hidden layer and the visible layer are conditionally sampled in alternating fashion.

Once the model has been trained, new samples can also be generated via the same MCMC methods. The end-to-end tasks are (i) training the model, and then, (ii) generating samples from the trained model to accomplish some larger machine learning goal.

**Quantum graphical models:** A separate end-to-end problem is found by generalizing the model itself to be quantum. For example, one can start with a classical BM and promote the binary variables to qubits. The energy functional is promoted to a quantum Hamiltonian and augmented with a transverse field, which does not commute with the Ising interactions. The result is a quantum Boltzmann machine (QBM), described by a transverse-field Ising (TFI) Hamiltonian [6] (cf. Eq. (4)):

$$H_{\mathrm{QBM}} = -\sum_{i \in V}(\kappa_i X_i + b_i Z_i) - \sum_{(i,j) \in E} w_{ij} Z_i Z_j \,, \tag{35}$$

where $X_i$ and $Z_i$ are the Pauli-$X$ and Pauli-$Z$ operators on qubit $i$, and $b_i, \kappa_i, w_{ij}$ are real variational parameters of the model. The ground or Gibbs state of $H_{\mathrm{QBM}}$ can be prepared in a variety of ways, including the adiabatic algorithm, Gibbs sampling, or as a variational quantum algorithm. These states can be measured (in the $Z$ basis or in the $X$ basis), yielding samples of the variables $v, h$ drawn from different distributions than the thermal distribution for the classical BM. Alternatively, one can trace out the hidden nodes, viewing the leftover quantum state on the visible nodes as the output of the QBM; this may be suitable if the input data is also quantum. As in the classical case, the training phase for a QBM consists of varying the weights via gradient descent to maximize a likelihood function. However, the noncommutativity of the Hamiltonian leads to complications: the gradients of the loss function are no longer directly given by sample expectation values as was the case in Eq. (34), although workarounds have been proposed [6, 7, 8, 9, 10]. For example, in the case of classical input data, sample expectation values can be used to optimize function that is not equal to the loss function, but can be shown to be an upper bound on it using the Golden–Thompson inequality [6]. Alternatively, in the case of quantum input data, and assuming the QBM has no hidden nodes, the relevant loss metric is the relative entropy and its gradients can again be related to sample averages [7]—this scenario is closely related to the Hamiltonian learning problem. In any case, the end-to-end problem is to train these models and generate samples.





**Dominant resource cost/complexity**

**Complexity of classical graphical models:** Recall that for classical BMs, one wishes to produce samples from the thermal distribution corresponding to the energy functional in Eq. (32), that is, Gibbs sampling (of diagonal Hamiltonians), either to assist in training the model or, if it has already been trained, to make inferences or generate new data. Specifically, given $H(h,v)$, one wishes to draw samples of $(h,v)$ with probability proportional to $\mathrm{e}^{-H(h,v)}$, either with $v$ free or with $v$ fixed (sometimes referred to as "clamped") to a particular value from the training set $\mathcal{D}$. Classically, one approach is simulated annealing or other MCMC algorithms. Quantumly, one can take one of several analogous approaches, including "quantum simulated annealing" [11] and quantum annealing, discussed as follows.

For quantum simulated annealing, one prepares the coherent Gibbs state $\sum_{v,h}\sqrt{p(h,v)}|v,h\rangle$, and a quadratic speedup is obtained over classical simulated annealing. The method is to construct a Hamiltonian whose ground state is the coherent Gibbs state at temperature $T$ (i.e., for which probabilities $p(h,v)$ are proportional to $\mathrm{e}^{-H(h,v)/T}$), and follow an adiabatic path from $T = \infty$ to $T = 1$. Following the path is accomplished by repeatedly performing quantum phase estimation (QPE) to project onto the ground state of the Hamiltonian at a given temperature. As is typical for the adiabatic algorithm, the cost of this procedure is dominated by the inverse of the spectral gap—this is the precision required for QPE to succeed. Specifically, for a graphical model with $|V|$ vertices, the runtime will be poly$(|V|)/\Delta$, where $\Delta$ is the minimum spectral gap. Importantly, $\Delta$ can be related to the maximum mixing time $t_{\mathrm{mix}}$ of the simulated annealing Markov chain, as $1/\Delta = \mathcal{O}(\sqrt{t_{\mathrm{mix}}})$, which leads to the quadratic speedup, although it is possible that $\Delta$ is exponentially small in $|V|$.

An alternative method for preparing (and sampling from) the coherent Gibbs state was proposed in [3]. There, one begins in an easy-to-prepare coherent mean-field state approximating the coherent Gibbs state. Then, one performs rejection sampling with amplitude amplification to gain a quadratic speedup over the analogous classical method. Additionally, it was proposed to use amplitude estimation to gain a quadratic improvement in the number of samples needed to achieve precision $\epsilon$, from $\mathcal{O}(1/\epsilon^2)$ to $\mathcal{O}(1/\epsilon)$, mirroring later analyses that work for general quantum-accelerated Monte Carlo methods [12]. If these $\mathcal{O}(1/\epsilon)$ quantum samples are each for the same training sample $v \in \mathcal{D}$, this is straightforward; however, if the samples are drawn randomly from $v \in \mathcal{D}$, achieving the quadratic speedup from amplitude estimation requires accessing the data in $\mathcal{D}$ coherently and quickly. Such data access is provided by the quantum random access memory (QRAM) primitive, for which the circuit *depth* can be logarithmic in the size of the training data, at the expense of a number of ancilla qubits (and total gates) that is linear in the size of the training data.

For quantum annealing, the idea is to add a uniform transverse field, as in the QBM of Eq. (35) with $\kappa_i = \kappa_j$ for all $i, j$. The transverse field is initially strong, and slowly turned off. This is similar to the adiabatic algorithm, but differs in that it is specifically carried out at finite ambient temperature. Thus, the system-bath interaction of the device naturally drives the state to the Gibbs state, which coincides with the classical thermal distribution once the transverse field is turned off. This is a heuristic method; it is efficient but there are few success guarantees. The hope is that the inclusion of an initial transverse field induces nonclassical fluctuations that help the system avoid becoming trapped in local minima as the transverse field is turned off.

Overall, computing the gradient of the loss function with respect to one parameter, up to precision $\epsilon$, will require complexity $\mathcal{O}(S/\epsilon)$, where $S$ is the complexity of sampling from the Gibbs state. The above assumes log-depth QRAM to be able to estimate the $\langle z_i z_j \rangle_{v \in \mathcal{D}}$ term





of Eq. (34). The complexity of $S$ will be $\text{poly}(|V|)\sqrt{t_{\text{mix}}}$ if a quantum simulated annealing approach is used, or some hard-to-analyze quantity if the quantum annealing approach is used. If the number of training samples is small, one can also sequentially compute the sum over $v \in \mathcal{D}$ and avoid the assumption of log-depth QRAM, leading to complexity $\mathcal{O}(S|\mathcal{D}|/\epsilon')$ (where $\epsilon' \geq \epsilon$ may be order-1). This must be carried out for all $|E| + |V|$ weights in the model, although these could be simultaneously estimated to precision $\epsilon$ at cost $\tilde{\mathcal{O}}(\sqrt{|E| + |V|}/\epsilon)$ samples, using methods from [13], which leverage the quantum gradient estimation primitive. It is not clear what value of $\epsilon$ is required in practice. Reference [3] takes $\epsilon \sim 1/\sqrt{|\mathcal{D}|}$, to match the natural uncertainty coming from a finite number of training samples. In this case, the overall complexity is dominated by

$$\tilde{\mathcal{O}}\Big( S \cdot \sqrt{|V| + |E|} \cdot \sqrt{|\mathcal{D}|} \Big) \tag{36}$$

assuming log-depth QRAM, and

$$\tilde{\mathcal{O}}\Big( S \cdot \sqrt{|V| + |E|} \cdot |\mathcal{D}| \Big) \tag{37}$$

without log-depth QRAM (the precision for each training sample can be taken as $\epsilon' = \mathcal{O}(1)$). The linear dependence on $|\mathcal{D}|$ could potentially be mitigated by first classically computing a "core set" $\mathcal{D}'$ satisfying the requirements that $|\mathcal{D}'| \ll |\mathcal{D}|$ and that replacing $\mathcal{D}$ with $\mathcal{D}'$ causes minimal change to the loss function in Eq. (33) [14].

**Complexity of quantum graphical models:** For QBMs, the dominant cost comes from producing samples from the quantum Gibbs state for the system in Eq. (35), that is, the state $\rho \propto \mathrm{e}^{-H_{\text{QBM}}}$, which can be accomplished through methods for Gibbs sampling. Rigorous methods for Gibbs sampling may scale exponentially in the size of the graph, without further assumptions. Such scaling would likely not be tolerable in practice. However, Monte Carlo–style methods for Gibbs sampling, which follow a similar approach as MCMC, but in an inherently quantum way, may be more effective in this case. These could have $\text{poly}(|V|)$ scaling for some parameter settings, but must also have exponential scaling in the worst case, as sampling low-energy Ising-model configurations is known to be NP-hard.

One can also heuristically apply quantum annealing, beginning from a large transverse field and reducing its strength slowly to some final nonzero value. However, some hardware platforms may only admit global control over the transverse field, preventing one from tuning the transverse-field strengths $\kappa_i$ individually. In any of these approaches, it is difficult to make any rigorous statements about the Gibbs sampling complexity.

### Existing resource estimates

There are no logical resource estimates for quantum annealing. However, [15, 16] discuss in detail how to embed the fully connected architecture of a RBM into the 2D lattice architecture available on planar quantum annealers. Reference [16] reports an embedding ratio scaling which is roughly quadratic—that is, a graphical model with $|V|$ vertices requires $\mathcal{O}(|V|^2)$ qubits to accommodate the architectural limitations of the device. A proper resource estimation has not been performed for the fault-tolerant algorithm of [3].





**Caveats**

There are two main caveats to quantum approaches to training classical models, which apply to both the annealing and to the fault-tolerant setting. First, classical heuristic algorithms, such as greedy methods or contrastive divergence, often perform well in practice and are the method of choice for existing classical analyses. These methods are also often highly parallelizable. If the quantum algorithm offers a speedup over a slower, exact classical method, this may not be relevant if the faster approximate classical methods are already sufficient. Second, the situations where one might hope for the heuristic quantum annealing approach to perform better might not be relevant problems, for instance, in highly regular lattice-based problems.

A caveat of the QBM is that the gradients of the loss function are not exactly related to sample averages, and imperfect workarounds, such as those proposed in [6], must be pursued. Like many other situations in machine learning, the resulting end-to-end solution is heuristic and evidence of its efficacy requires empirical demonstration.

**Comparable classical complexity and challenging instance sizes**

For classical models, an exact computation of the gradients would scale exponentially in the size of the graph, as $\mathcal{O}(|\mathcal{D}|2^{|V|})$ for the gradient of a single parameter. Approximate methods based on simulated annealing or other MCMC methods would scale as $\mathcal{O}(S_c/\epsilon^2)$, where $S_c$ is the classical sample time, scaling as $S_c = \mathrm{poly}(|V|)t_{\mathrm{mix}}$. On the other hand, these methods can also be implemented heuristically at reduced cost (e.g., contrastive divergence, where one does not wait for the chain to mix), and they can also be implemented on parallel architectures. For instance, in [17], an architecture was proposed to train deep BMs. Experiments demonstrated that heuristic training methods could be carried out for graphs of size 1 million in 100 seconds on field-programmable gate arrays available in 2010. Much larger sizes would be accessible to a scaled-up version of the same architecture on modern hardware. It is unlikely that any exact method, quantum or classical, could match this efficiency.

For the quantum models, the classical complexity of sampling from the Gibbs state of the model would be exponential in the graph size $|V|$. Thus, training these models would likely not be pursued classically.

**Speedup**

For the classical models, the speedup can be quadratic in most of the parameters: producing a sample can in some cases be sped up quadratically, and the number of samples required to achieve a certain precision also enjoys a quadratic speedup (e.g., $t_{\mathrm{mix}}$ to $\sqrt{t_{\mathrm{mix}}}$ and $\mathcal{O}(1/\epsilon^2)$ to $\mathcal{O}(1/\epsilon)$). The methods that give these provable quadratic speedups are based on primitives such as amplitude amplification, where superquadratic speedups are not possible without exploiting additional structure. Larger superpolynomial speedups are only possible under optimistic assumptions about the success of heuristic quantum annealing approaches at producing samples faster than classical approaches.

For the quantum models, the speedup is technically exponential, assuming that for the models considered, quantum algorithms for Gibbs sampling scale efficiently while approximate classical methods (e.g., tensor networks) scale exponentially. Indeed, it was shown in [18] that certain QBMs are BQP-complete, in the sense that its ground state can be efficiently prepared on a quantum computer, and any quantum computation (including those with exponential speedup)





can be encoded into its ground state. However, this construction is artificial, and it has yet to be demonstrated that there are specific real-world machine learning tasks where these models offer a speedup over the best available classical machine learning model for the same task.

**Outlook**

While energy-based models are naturally in a form that can readily be extended to the quantum domain, there still lacks decisive evidence of quantum advantage for a specific end-to-end classical machine learning problem. There remains some uncertainty on the outlook of these approaches due to the centrality of heuristic quantum approaches. One may hold out hope that these heuristics could outperform classical heuristics in some specific settings, but the success of classical heuristics and effectiveness of approximate classical approaches present a formidable barrier to achieving any quantum advantage in this area.

**Further reading**

We refer the reader to [5] for more information on quantum approaches to energy-based models.

## 9.3 Tensor PCA

**Overview**

Inference problems play an important role in machine learning. One of the most widespread methods is principal component analysis (PCA), a technique that extracts the most significant information from a stream of potentially noisy data. In the special case where the data is generated from a rank-1 vector plus Gaussian noise—the spiked matrix model—it is known that there is a phase transition in the signal-to-noise ratio [1]: above the transition point, the principal component can be recovered efficiently, while below the transition point, the principal component cannot be recovered at all. In the tensor extension of the problem, there are two transitions. One information theoretical, below which the principal component cannot be recovered, and another computational, below which the principal component can be recovered, but only inefficiently, and above which it can be recovered efficiently. Thus, the tensor PCA problem offers a much richer mathematical setting, which has connections to optimization and spin glass theory; however, it is yet unclear if the tensor PCA framework has natural practical applications. A quantum algorithm [2] for tensor PCA was proposed which has provable runtime guarantees for the spiked tensor model; it offers a potentially *quartic* speedup over its classical counterpart and also efficiently recovers the signal from the noise at a smaller signal-to-noise ratio than other classical methods. This algorithm was further developed in [3] and extended to also give a quartic speedup for a related discrete optimization problem called "planted noisy $k$XOR," which is argued to have possible relevance in cryptography.

**Actual end-to-end problem(s) solved**

Consider the spiked tensor problem. Let $v \in \mathbb{R}^N$ (or $\in \mathbb{C}^N$)[34] be an unknown signal vector, and let $p \in \mathbb{N}$ be a positive integer. Construct the tensor

$$T = \lambda v^{\otimes p} + V,$$

where $V$ is a random tensor in $\mathbb{R}^{N^p}$ (or $\mathbb{C}^{N^p}$), with each entry drawn from a normal distribution with mean 0 and variance 1. The vector $v$ is assumed to have norm $\sum_j v_j^* v_j = \sqrt{N}$ and can be identified with a quantum state. The quantity $\lambda$ is the signal-to-noise ratio.

The main question we are interested in is for what values of $\lambda$ can we detect or reconstruct $v$ from (full) access to $T$, and how efficiently can this be done? In [4], it was shown that the maximum likelihood solution $w^{\mathrm{ML}}$ to the objective function

$$w^{\mathrm{ML}} = \underset{w \in \mathbb{C}^n}{\mathrm{argmax}} \langle T, w^{\otimes p} \rangle$$

will have high correlation with $v$ as long as $\lambda \gg N^{(1-p)/2}$, where $\langle \cdot, \cdot \rangle$ denotes the standard dot product after writing the $N^p$ entries of the tensor as a vector. However, the best known *efficient* classical algorithm [5] requires $\lambda \gg N^{-p/4}$ to recover an approximation of $v$. Using the spectral method, that is, mapping the tensor $T$ to a $N^{p/2} \times N^{p/2}$ matrix and extracting the maximal eigenvalue, recovery can be done in time complexity $\mathcal{O}(N^p)$, ignoring logarithmic prefactors.

Hastings [2] proposes classical and quantum algorithms to solve the spiked tensor model by first mapping $T$ to a bosonic quantum Hamiltonian with $N$ modes, $n_{\mathrm{bos}}$ bosons, and $p$-body

---

[34]Reference [2] provides reductions between real and complex cases.





interactions, where $n_{\mathrm{bos}}$ is a tunable integer parameter satisfying $n_{\mathrm{bos}} > p/2$

$$H_{\mathrm{PCA}}(T) = \frac{1}{2} \left( \sum_{\mu_1, \ldots, \mu_p = 1}^{N} T_{\mu_1, \ldots, \mu_p} \left( \prod_{i=1}^{p/2} a_{\mu_i}^{\dagger} \right) \left( \prod_{j=1+p/2}^{p} a_{\mu_j} \right) + \mathrm{h.c.} \right), \quad (38)$$

where h.c. stands for Hermitian conjugate. Here, the operators $a_\mu$ and $a_\mu^\dagger$ are annihilation and creation operators of a boson in mode $\mu$, and we restrict to the sector for which $\sum_\mu a_\mu^\dagger a_\mu = n_{\mathrm{bos}}$. As $n_{\mathrm{bos}}$ increases, the number of particles increases and the complexity of the algorithm grows, but the threshold for $\lambda$ above which recovery is possible also decreases.

Hastings shows that the vector $v$ can be efficiently recovered from a vector in the large energy subspace of $H_{\mathrm{PCA}}(T)$ when the largest eigenvalue of $H_{\mathrm{PCA}}(T)$ is at least a constant factor larger than $E_{\max}$, where $E_{\max}$ corresponds to the case where there is no signal. It is shown that, roughly,

$$E_{\max} \sim n_{\mathrm{bos}}^{p/4+1/2} N^{p/4},$$

$$E_0 \approx \lambda(p/2)! \binom{n_{\mathrm{bos}}}{p/2} N^{p/2} \approx \lambda n_{\mathrm{bos}}^{p/2} N^{p/2},$$

where $E_0$ is the maximum eigenvalue of $H_{\mathrm{PCA}}(T)$. Thus, if $\lambda \gg N^{-p/4}$, there will be a gap between $E_0$ and $E_{\max}$, and this gap grows as $n_{\mathrm{bos}}$ increases. This enables the quantum algorithm to recover the signal even for signal strength as weak as $\lambda \gg N^{-p/4}$.

Hastings considers the case where $p$ is constant and $N$ grows, and assumes that $n_{\mathrm{bos}} = \mathcal{O}(N^\theta)$ for some $p$-dependent constant $\theta > 0$ chosen sufficiently small. In fact, ultimately, it is determined that in the recovery regime $\lambda \gg N^{-p/4}$, the parameter $n_{\mathrm{bos}}$ need only scale as $\mathrm{polylog}(N)(N^{-p/4}/\lambda)^{4/(p-2)}$. In any case, terms in the complexity $\mathcal{O}(N^p)$ are dominated by terms $\mathcal{O}(N^{n_{\mathrm{bos}}})$.

The idea of mapping the order-$p$ tensor $T$ to an order-2 tensor (i.e., a matrix) $H_{\mathrm{PCA}}$, and then solving the planted inference problem by (classically) extracting spectral information from $H_{\mathrm{PCA}}$, was also independently discovered in [5], where it is called the "Kikuchi method." There, a tunable parameter $\ell$ plays the role of $n_{\mathrm{bos}}$, although the two formulations offer different intuitions to motivate the method; their relationship is discussed in detail in [3].

**Dominant resource cost/complexity**

Hastings shows that the dominant eigenvector of $H_{\mathrm{PCA}}$ can be classically extracted in $\widetilde{\mathcal{O}}(N^{n_{\mathrm{bos}}})$ time via the power method, where the tilde indicates that we ignore polylogarithmic factors.

He proposes three quantum algorithms for the same problem. The first runs quantum phase estimation on a random state. Since the random state will have squared overlap $\Omega(N^{-n_{\mathrm{bos}}})$ with the high-energy subspace, the expected number of repetitions of phase estimation is $\mathcal{O}(N^{n_{\mathrm{bos}}})$. The second algorithm proposes to further use amplitude amplification, reducing the complexity to $\mathcal{O}(N^{n_{\mathrm{bos}}/2})$. The third algorithm further improves the complexity by choosing a specific initial high-energy state, and showing that the overlap with the state scales as $\Omega(N^{-n_{\mathrm{bos}}/2})$, which combined with amplitude amplification, leads to a $\mathcal{O}(N^{n_{\mathrm{bos}}/4})$ complexity. As discussed above, the estimates assume that factors of $\mathcal{O}(N^p)$ can be ignored, since they are negligible with respect to the query complexity of $N^{\mathcal{O}(n_{\mathrm{bos}})}$.





This constitutes a quartic speedup over the classical spectral algorithm acting on $H_{PCA}$ for the same choice of $n_{bos}$ that is also presented in [2]. Since the ansatz state is a product state, it can be prepared efficiently.

Hastings further argues that the Hamiltonian simulation of $H_{PCA}$ within the phase estimation subroutine can be accomplished by viewing $H_{PCA}$ as a sparse or local matrix. Specifically, we can view $H_{PCA}$ as a second-quantized Hamiltonian with $N$ registers storing a $\mathcal{O}(\log(n_{bos}))$-bit number representing the occupancy of each mode. The total number of qubits needed is $\mathcal{O}(N \log(n_{bos}))$. Each of the $\mathcal{O}(N^p)$ terms in Eq. (38) corresponds to a single nonzero entry of one row of $H_{PCA}$ in this basis, so the $\mathcal{O}(n_{bos}^N) \times \mathcal{O}(n_{bos}^N)$ matrix $H_{PCA}$ is $\mathcal{O}(N^p)$-sparse. Alternatively, a more compact representation would be to view $H_{PCA}$ in a first-quantized picture, allocating $n_{bos}$ registers each storing a $\mathcal{O}(\log(N))$ bit number corresponding to which mode each of the $n_{bos}$ bosons are in—the total number of qubits needed is $\mathcal{O}(n_{bos} \log(N))$. In order to enforce permutational symmetry, each term in Eq. (38) would decompose into $(p/2)! \binom{n_{bos}}{p/2}$ separate terms that act locally on $p/2$ of the registers. This is closer to the approach taken in the quantum algorithm of [3].

Either way, we can efficiently perform Hamiltonian simulation (e.g., by first constructing a block-encoding of the sparse or local Hamiltonian) to perform quantum phase estimation at total gate complexity $poly(N, n_{bos})$ (here interpreting $p = \mathcal{O}(1)$), which is negligible compared to $N^{n_{bos}}$.

**Caveats**

The spiked tensor model does not immediately appear to be related to any practical problems. Additionally, efficient recovery requires that the signal-to-noise ratio be rather high, which may not occur in real-world settings, and when it does, it is not clear that formulating the problem as a tensor PCA problem will be the most efficient path forward. Relatedly, while the runtime of the algorithm scales subexponentially in $N$, for large values of $n_{bos}$, its $N$ dependence of $\mathcal{O}(N^{n_{bos}})$ may still lead to a practically intractable algorithm.

**Comparable classical complexity and challenging instance sizes**

The algorithms proposed in [2] (see also [5, 3]) improve on other spectral methods for the spiked tensor model, whenever $n_{bos} > p/2$ for sufficiently large $p$. The threshold for which the new algorithms beat the older ones decreases as $n_{bos}$ increases, although the complexity of the algorithm increases with $n_{bos}$.

**Speedup**

The quartic speedup over the classical power method is achieved by combining a quadratic speedup from amplitude amplification with a quadratic speedup related to choosing a clever initial state for phase estimation. The existence of the clever initial state is essential for the beyond-quadratic speedup, and it has been related [3] to the BQP-hardness of the guided local/sparse Hamiltonian problem [6].

As discussed above, there is no readout issue, as the vector $v$ can be efficiently recovered from the single particle density matrix obtained from the eigenvector of $H_{PCA}(T)$. The quantum algorithm has $\mathcal{O}(N \log(n_{bos}))$ space complexity, which is an exponential improvement over the classical spectral algorithm presented in [2] for the same problem. Furthermore, the quartic





speedup in time and exponential speedup in space is possible even in the absence of a large-scale quantum random access memory (QRAM), since the overall runtime of the algorithm is much larger than the size of the input dataset ($\mathcal{O}(N^p)$), and linear (rather than logarithmic) data access cost is tolerable.

In [3], the quartic speedup was extended to apply generally to instances where the "Kikuchi method" is used; improvements to the classical algorithm within this framework would imply commensurate improvements to the quantum algorithm to maintain the quartic relationship.

## Outlook

The quartic speedup is very compelling, as beyond-quadratic speedups are rare in quantum algorithms. It is also appealing that the speedup also does not necessarily rely upon an assumption of large-scale QRAM. However, it is currently unclear how applicable this quartic speedup can be in real-world situations.

## Further reading

We refer the reader to [3] for a discussion on the intuition and scope of the quartic speedup and additional technical details. We also note that [7] has studied the quantum approximate optimization algorithm (QAOA) applied to the spiked tensor model, although the result is not directly comparable as it is only shown to succeed for larger values of $\lambda$, where additional classical algorithms are also successful.

## 9.4 Topological data analysis

**Overview**

In topological data analysis (TDA), we aim to compute the dominant topological features (connected components and $k$-dimensional holes), known as Betti numbers, of $N$ data points sampled from an underlying topological manifold or of a graph with $N$ vertices. These features may be of independent interest (e.g., the number of connected components in the matter distribution in the universe) or can be used as generic features to compare datasets. We refer to [1] for a recent survey of applications of TDA.

Quantum algorithms for TDA leverage the ability of a register of qubits to efficiently represent a quantum state that stores all cliques in the clique complex built on the topological manifold. The textbook classical algorithm exactly computes the Betti numbers, but its complexity scales polynomially with the number of cliques in the complex, which may grow combinatorially with $(N, k)$. In contrast, quantum algorithms naturally estimate Betti numbers normalized by the number of cliques in the complex, and their complexity scales polynomially in $(N, k)$ for clique-dense complexes. If the error is rescaled so as to be a constant additive error estimate for the Betti number, currently known quantum algorithms provide a quadratic speedup over the best classical algorithms for the equivalent problem. For relative-error estimates of the Betti number, quartic and superpolynomial speedups have been shown for certain families of graphs. In addition, a number of complexity-theoretic results have been shown that provide evidence that estimating normalized Betti numbers is efficient for quantum computers, but classically hard. Nevertheless, finding practical applications for relative error estimates of high-dimensional Betti numbers is an open research question.

**Actual end-to-end problem(s) solved**

We construct a simplicial complex built from $N$ data points sampled from an underlying manifold. The simplicial complex is a higher-dimensional generalization of a graph, constructed by connecting data points within a given distance of each other. A simplicial complex constructed in this way is known as a clique complex or a Vietoris–Rips complex. We can consider a sequence, known as a filtration, of complexes constructed by connecting points at increasingly large distances (the distance is referred to as the length scale of the clique complex). We denote the number of $k$-simplices in the complex at length scale $i$ as $|S_k^i|$.

We then compute the Betti numbers $\beta_k^i$ (the number of $k$-dimensional holes at a given length scale $i$) or the persistent Betti numbers $\beta_k^{i,j}$ (the number of $k$-dimensional holes present at both scale $i$ and scale $j$) of the simplicial complex. The persistent Betti numbers $\beta_k^{i,j}$ are used to infer the dominant topological features, considered to be those with the longest persistence as the length scale is increased. The births and deaths of features are typically plotted on a "persistence diagram." Different datasets can be compared by using stable distance measures between their diagrams, or by vectorizing the diagrams and using kernel methods or neural networks. For graphs, there is only a single length scale $i$, and so $\beta_k^i$ is the quantity of interest. For statements common to both $\beta_k^{i,j}$ and $\beta_k^i$, we will use the notation $\beta_k^*$. Typical classical applications consider low values of $k$, motivated primarily by computational cost and interpretability of the resulting topological features.





**Dominant resource cost/complexity**

Quantum algorithms naturally estimate $\beta_k^* / |S_k^i|$ to additive error $\epsilon$ [2, 3, 4, 5, 6]. For a complex built from $N$ data points, we can either use $N$ qubits to encode the simplicial complex, or $\mathcal{O}(k \log(N))$ qubits when $k \ll N$. Quantum algorithms have two subroutines:

(i) Preparing a state that encodes the simplices present in the simplicial complex. This reduces to finding $k$-simplices present in the complex at the given length scale (using either classical rejection sampling or Grover's algorithm / amplitude amplification). Using Grover's algorithm this scales as $\left( \binom{N}{k+1} / |S_k^i| \right)^{1/2}$. More efficient clique-finding methods can be used for special classes of graphs [7].

(ii) Projecting onto the eigenspace of an operator that encodes the topological features of the complex in the amplitude of the quantum state (using either quantum phase estimation or quantum singular value transformation). This introduces a dependence on the gap(s) $\Lambda$ of the operator(s) used to encode the topology.

The most efficient approaches use amplitude estimation to compute the normalized (persistent) Betti number. The most expensive subroutines within the quantum algorithms are the membership oracles that determine if a given simplex is present in the complex, the cost of which we denote by $m_k$. In the classically challenging clique-dense regime, the overall cost of the most efficient known quantum algorithms for computing $\beta_k^* / |S_k^i|$ to error $\epsilon$ is approximately

$$\widetilde{\mathcal{O}} \left( \frac{m_k}{\epsilon} \sqrt{\frac{\beta_k^*}{|S_k^i|}} \left( \sqrt{\frac{\binom{N}{k+1}}{|S_k^i|}} + \frac{\text{poly}(N, k)}{\Lambda} \right) \right).$$

When we choose $\epsilon = \Delta / |S_k^i|$ (i.e., computing $\beta_k^*$ to constant additive error $\Delta$) the complexity is

$$\widetilde{\mathcal{O}} \left( \frac{m_k \sqrt{\beta_k^*}}{\Delta} \left( \sqrt{\binom{N}{k+1}} + \frac{\sqrt{|S_k^i|} \cdot \text{poly}(N, k)}{\Lambda} \right) \right).$$

It is clear that regardless of how $|S_k^i|$ and $\beta_k^*$ scale, the runtime is polynomial in $\binom{N}{k+1}$.

When we choose $\epsilon = r \beta_k^* / |S_k^i|$ (i.e., computing $\beta_k^*$ to relative error $r$) the complexity is

$$\widetilde{\mathcal{O}} \left( \frac{m_k}{r} \left( \sqrt{\frac{\binom{N}{k+1}}{\beta_k^*}} + \sqrt{\frac{|S_k^i|}{\beta_k^*}} \cdot \frac{\text{poly}(N, k)}{\Lambda} \right) \right).$$

For clique-dense complexes with large Betti numbers, where $\beta_k^*$ and $|S_k^i|$ are only polynomially smaller than $\binom{N}{k+1}$, the runtime is polynomial in $N$ and $k$.

**Existing resource estimates**

In [5], the gate depth (and non-Clifford gate depth) of all subroutines (including explicit implementations of the membership oracles) was established for computing $\beta_k^{i,j}$ and $\beta_k^i$. However, that reference did not consider a final compilation to $T$/Toffoli gates for concrete problems of interest.





In [6], the Toffoli complexity of estimating $\beta_k^i$ was determined. The Toffoli complexity for estimating $\beta_k^i$ to relative error for a family of graphs with large $\beta_k^i$ was determined for $k = 4, 8, 16, 32$ and $N \leq 10^4$. The resulting Toffoli counts ranged from $10^8$ ($N = 100, k = 4$) to $10^{17}$ ($N = 10^4, k = 32$), using $\mathcal{O}(N)$ logical qubits.

### Caveats

Quantum algorithms naturally solve a different problem than the textbook classical algorithm solves. Namely, the quantum algorithm estimates $\beta_k^*/|S_k^i|$ to error $\epsilon$, with runtime $\text{poly}(\epsilon^{-1})$, for a single length scale (pair of length scales). The algorithm must be repeated for all length scales to compute the persistence diagram. In contrast, the textbook classical algorithm for dimension $k$ and length scale $j$ exactly computes the full persistence diagram for all $\beta_{k' \leq k}^{i,j' \leq j}$.

Quantum algorithms (and classical algorithms based on the power method discussed below) depend on the eigenvalue gap(s) $\Lambda$ of the operator(s) that encode the topology. The scaling of these gaps has not been studied for typical applications.

Finally, typical classical applications consider dimension $k \leq 3$, and applications of high-dimensional Betti numbers are not yet known.

### Comparable classical complexity and challenging instance sizes

While classical algorithms are technically efficient for constant dimension $k$, they are limited in practice. For a number of benchmark calculations on systems with up to $10^9$ simplices, we refer to [8].

The textbook classical algorithm, which for dimension $k$ and length scale $j$ exactly computes the full persistence diagram for all $\beta_{k' \leq k}^{i,j' \leq j}$ (rather than a single Betti number), has a worst-case scaling of $\mathcal{O}(|S_{k,k+1}^j|^\omega)$ where $\omega \leq 2.37$ is the exponent of matrix multiplication, and we have defined a shorthand notation $|S_{k,k+1}^j| := |S_k^j| + |S_{k+1}^j|$ [9]. In practice, the textbook classical algorithm is observed to scale as $\mathcal{O}(|S_{k,k+1}^j|)$ due to sparsity in the complex [9]. Moreover, for problems that do not naturally have this sparse structure, well-studied classical heuristics can be applied to sparsify the complex [10].

Classical algorithms based on the power method [11] can achieve worst-case scaling of $\mathcal{O}(|S_k^i|)$ for computing individual Betti numbers $\beta_k^i$ (an improvement over the worst-case scaling of the textbook algorithm above). The classical power method scales approximately as

$$\widetilde{\mathcal{O}}\left( \frac{|S_k^i|(Nk\beta_k^i + (\beta_k^i)^2)\lambda_{\max}}{\Lambda} \log\left(\frac{1}{\epsilon}\right) \right)$$

to compute $\beta_k^i$ to additive error $\epsilon$, where $\lambda_{\max}$ is a bound on the largest eigenvalue of the operator encoding the topology. The power method has recently been extended to compute persistent Betti numbers, with a similar complexity [5]. Although the power method for persistent Betti numbers is more efficient than the worst-case performance of the textbook classical algorithm described above, it must be repeated for each pair of length scales to compute the persistence diagram, which is a disadvantage in practice.

Recently, randomized classical algorithms have been proposed for estimating $\beta_k^i/|S_k^i|$ to additive error [6, 12]. The algorithm of [12] scales as

$$\left( \frac{N}{\lambda_{\max}} \right)^{\mathcal{O}\left( \frac{1}{\sqrt{\Lambda}} \log\left(\frac{1}{\epsilon}\right) \right)} \cdot \text{poly}(N)$$





assuming that we can efficiently sample and check $k$-simplices. When $k = \Omega(N)$, the algorithm runs in polynomial time for clique complexes with constant gap $\Lambda$ and error $\epsilon = \Omega(1/\mathrm{poly}(N))$ (or $\epsilon$ constant and $\Lambda = \Omega(1/\log^2(N))$).

**Speedup**

As discussed above, quantum algorithms naturally compute $\beta_k^* / |S_k^i|$ to additive error $\epsilon$, with runtime $\mathrm{poly}(\epsilon^{-1})$. A number of complexity-theoretic results have been shown for this problem. Reference [7] showed that estimating the kernel dimension of general Hamiltonians is DQC1-hard.[35] In [14], it was shown that estimating normalized quasi-Betti numbers (which accounts for miscounting low-lying but nonzero singular values) of general cohomology groups is also DQC1-hard. The hardness of estimating normalized (persistent) Betti numbers of a clique complex, subject to a gap assumption of $\Lambda = \Omega(1/\mathrm{poly}(N))$—which is the problem solved by existing quantum algorithms—has not been established (see [14, Section 1.1]).

Reference [15] showed that determining if the Betti number of a (clique-dense) clique complex is nonzero is NP-hard in general. This was superseded by the results of [16, 17] which showed that this problem is QMA$_1$-hard.

We can consider the speedup of quantum algorithms for TDA in two regimes, constant additive error and relative error:

- For constant additive error, the most natural comparison is between the quantum algorithm and classical algorithms based on the power method. Quantum algorithms are able to achieve a quadratic speedup over the classical power method in the clique-dense regime [11, 5]. An exponential speedup is not possible for general graphs, due to the aforementioned NP- and QMA-hardness results [15, 16, 17].

- For the task of computing $\beta_k^i$ to relative error, graph families have been identified for which the quantum algorithm provides superpolynomial [6] or quartic [6, 15] speedups over the classical power method. Recently introduced randomized classical algorithms [6, 12] may scale efficiently for this same task of estimating $\beta_k^i$ to relative error. For example, when $k = \Omega(N)$ the algorithm of [12] runs in polynomial time for clique complexes with constant gap $\Lambda$ and error $\epsilon = \Omega(1/\mathrm{poly}(N))$ or $\epsilon$ constant and $\Lambda = \Omega(1/\log^2(N))$. These are more restrictive conditions than quantum algorithms (which can simultaneously have both $\Lambda, \epsilon = \Omega(1/\mathrm{poly}(N))$). These features will not occur for all graphs.

**NISQ implementations**

In [18], a NISQ-amenable compilation of the quantum algorithm described above was proposed, trading deep quantum circuits for many repetitions of shallower circuits, which comes at the cost of worsening the asymptotic scaling of the algorithm (see the table in [5] for a quantitative comparison). A proof-of-principle experiment was performed demonstrating this method [18]. In [14], it was shown that the TDA problem can be mapped to a fermionic Hamiltonian, and it was proposed to use the variational quantum eigensolver to find the ground states of this Hamiltonian (the degeneracy of which gives $\beta_k^i$). It is unclear what ansatz circuits one should

---

[35]DQC1 is a complexity class that is physically motivated by the "one clean qubit model" [13]. This model has a single pure state qubit which can be initialized, manipulated, and measured freely, as well as $N - 1$ maximally mixed qubits.





use to make this approach advantageous compared to classical algorithms, as naive (e.g., random) trial states would have exponentially small overlap with the target states.

**Outlook**

The complexity-theoretic results that provide evidence for the classical hardness and quantum tractability of estimating normalized (persistent) Betti numbers suggest that quantum algorithms for TDA may be an interesting area to search for new quantum speedups.

Nevertheless, it is important to emphasize that current quantum algorithms do not provide more than quadratic speedups for the practical problem solved in current TDA applications, and complexity-theoretic results suggest that exponential speedups will not be possible for general graphs for this task.

As such, an important open problem is to identify applications for the task naturally solved by quantum computers (providing relative error estimates for clique-dense graphs with large Betti numbers). If new applications can be identified for datasets that are both clique-dense and have large high-dimensional (persistent) Betti numbers (that are practically interesting to compute to relative error), then quantum algorithms may be of practical relevance.

## 9.5   Quantum neural networks and quantum kernel methods

**Overview**

In this section, we discuss two collections of proposals to use a quantum computer as a platform to execute machine learning models, often known as *quantum neural networks* and *quantum kernel methods*. Some early ideas in this space were motivated by the constraints of near-term, "NISQ" [1] devices. Despite this, not all subsequent proposals are necessarily implementable on NISQ devices. Moreover, the proposals need not be restricted to running on NISQ devices, but could also be run on devices with explicit quantum error correction. For simplicity, we present concrete examples based on supervised machine learning tasks. However, outside of these examples, we keep our discussion more general and note that the techniques are also applicable to other settings, such as unsupervised learning and generative modeling.

Given access to some data, our goal is to obtain a function or distribution that emulates certain properties of the data, which we will call a *hypothesis*. This is obtained by first defining a *hypothesis set* or *model family*, and using a learning algorithm to output a hypothesis from this set. For example, in supervised learning, we have data $x_i \in X$ that have respective labels $y_i \in Y$. The goal is then to find a hypothesis function $h : X \to Y$ that approximates the "true" unknown underlying labeling function, such that it correctly labels previously unseen data with high probability. Note that we have left the exact descriptions of the sets $X$ and $Y$ ambiguous. They could, for instance, correspond to sets of numbers or vectors. More generally, this description encompasses the possibility of operating on quantum data such that each $x_i$ or $y_i$ corresponds to a quantum state.

Quantum neural networks and quantum kernel methods use a quantum computer to assist in constructing the model family, in place of a classical model such as a neural network. Specifically, here we prepare some quantum state(s) encoding the data and measure some observable(s) to ultimately construct model predictions. We first elaborate on both quantum neural networks and quantum kernel methods.

**Quantum neural networks**

**Actual end-to-end problem(s) solved:**   Given data $x$, we consider a model constructed from a parameterized quantum circuit:

$$h_{\boldsymbol{\theta}}(x) = f(\mathrm{tr}[\rho(x, \boldsymbol{\theta})O]) , \qquad (39)$$

where $\rho(x, \boldsymbol{\theta})$ is a quantum state (output of some parameterized quantum circuit) that encodes both the data $x$ as well as a set of adjustable parameters $\boldsymbol{\theta}$, $O$ is some chosen measurement observable, and $f$ is some function that can be enacted as classical postprocessing on the measurement result (we remark that $O$ itself can also be trainable [2], but we do not explicitly indicate this in the notation for simplicity of exposition). As a basic example, if $x$ corresponds to a classical vector, $\rho(x, \boldsymbol{\theta})$ could correspond to initializing in the $|0\rangle\langle 0|$ state and applying some data-encoding gates $U(x)$ followed by parameterized gates $V(\boldsymbol{\theta})$. Alternatively, the data itself could be a quantum state, and a more general operation in the form of a parameterized channel $\mathcal{V}(\boldsymbol{\theta})$ could be applied. There is also no a priori reason why data encoding and trainable gates need to be applied each once in separate steps rather than in a mixed or repeated fashion.

The model is optimized via a learning algorithm which aims to find the optimal parameters $\boldsymbol{\theta}^*$ by minimizing a loss function. For instance, in supervised learning, given some labeled





training dataset $T = \{(x_i, y_i)\}$, a suitable choice of loss should compare how close each $h_{\boldsymbol{\theta}}(x_i)$ is to the true label $y_i$ for all data in $T$. The quality of the model can then be assessed on a set of previously unseen data outside of $T$. It is important to pause here and reflect that an optimized loss does not guarantee good performance on unseen data. This is referred to in the literature as the gap between *empirical* and *total* risk, or simply the generalization gap/error. Conversely, a small generalization error alone is not sufficient to guarantee good performance (one should then also ask for a good loss on training data).

We remark that the setting we presented has substantial overlap with the setting of variational quantum algorithms (VQAs)—indeed, a quantum neural network can be thought of as a VQA that incorporates data—thus, the same challenges and considerations that apply to VQAs also apply here. There will additionally be extra considerations due to the role of the data.

**Dominant resource cost/complexity:** The encoding of data $x$ and parameters $\boldsymbol{\theta}$ in Eq. (39) should be sufficiently expressive that it (i) leads to good performance on data and (ii) is (at minimum) not efficiently simulable classically [3], if one is to seek quantum advantage. These requirements set some criteria for minimum circuit complexity.

The learning algorithm to find optimal parameters is usually performed by classical heuristics, such as gradient descent, and can have significant time overhead, requiring evaluation of Eq. (39) at many parameter values (see Section 20 on VQAs for more details).

The size of the training dataset required can also have direct implications for runtime, with a larger amount of training data typically taking a longer time to process. Reference [4] proves that good generalization can be achieved with the size of the training data $|T|$ growing in tandem with the number of adjustable parameters $M$. Specifically, it is shown that the generalization error with high probability scales as $\mathcal{O}(\sqrt{M \log(M)/|T|})$. Thus, only a mild amount of data is required for good generalization. We stress again that this alone does not say anything about the ability for quantum neural networks to obtain low training error.

**Scope for advantage:** Quantum neural networks could achieve advantage in a number of ways, for example, by improving on runtime or by using less training data. In supervised learning settings, generalization performance is a separate consideration and an additional domain for possible advantage. Machine learning with quantum neural networks has yielded some promising performance empirically and encouraging theoretical guarantees exist for certain stages of the full pipeline in restricted settings [5, 6, 4, 7, 8] (loss minimization can remain a challenge [9, 3], see Section 20 on VQAs again for more details). Nevertheless, there are currently no practical use cases with full end-to-end performance guarantees in the same way that we have for other quantum algorithms. However, due to the heuristic nature of classical machine learning, one may debate whether such a guarantee is possible, or even if seeking theoretical quantum advantage in the traditional algorithmic sense is the most appropriate goal [10].

## Quantum kernel methods

**Actual end-to-end problem(s) solved:** Quantum kernel methods are a quantum instance of a class of techniques known as kernel methods, of which support vector machines are a prominent example. We first briefly review the general framework. Given a dataset





$T = \{x_i\} \subset X$, the model can be written as

$$h_{\boldsymbol{\alpha}}(x) = \sum_{i:\, x_i \in T} \alpha_i k(x, x_i)\,, \tag{40}$$

where $\boldsymbol{\alpha} = (\alpha_1, \alpha_2, \dots)$ is a vector of parameters to be optimized, and $k(x, x')\colon X \times X \to \mathbb{R}$ is a measure of similarity known as the kernel function. This model has several theoretical motivations:

- The matrix with entries $K_{ij} = k(x_i, x_j)$ is usually defined to be symmetric positive semidefinite for any choice of $\{x_1, \dots, x_m\} \subseteq X$ and $k(x_i, x_j)$. By Mercer's theorem, it is thus an inner product of feature vectors $\phi(x_i), \phi(x_j)$ which embed the data $x_i$ and $x_j$ in a (potentially high-dimensional) Hilbert space. Linear statistical methods can be used to learn a linear function in this high-dimensional space, only using the information of the inner products $k(x_i, x_j)$ and never having to explicitly evaluate $\phi(x_i)$ and $\phi(x_j)$, which can be much harder to compute.

- The Representer Theorem [11] states that the optimal model over the dataset $T$ (optimal for $T$, though not necessarily for expanded datasets) can be expressed as a linear combination of kernel values evaluated over $T$—that is, the optimal model exactly takes the form in Eq. (40). This is known as the kernel trick.

- Further, if the loss function is convex, then the dual optimization program to find the optimal parameters $\boldsymbol{\alpha}^*$ is also convex [12].

A key question that remains is then how to choose a kernel function. Quantum kernel methods embed data in quantum states, and thus evaluate $k(x_i, x_j)$ on a quantum computer. Similar to quantum neural networks or any other quantum model, the quantum kernel should be hard to simulate classically [3]. As an example, we present two common choices of quantum kernel (see [13] for a more general discussion).

- The fidelity quantum kernel

$$k_F(x, x') = \mathrm{tr}[\rho(x)\rho(x')]\,, \tag{41}$$

which can be evaluated either with a SWAP test or, given classical data with unitary embeddings, it can be evaluated with the overlap circuit $|\langle 0|U(x')^\dagger U(x)|0\rangle|^2$.

- The fidelity kernel can run into issues for high-dimensional systems (increasing qubit count), as the inner product in Eq. (41) can be very small for $x \neq x'$. This motivated the proposal of a family of projected quantum kernels [14], of which one example is the Gaussian projected quantum kernel

$$k_P(x, x') = \exp\left(-\gamma \sum_{\ell=1}^{n} \big\|\rho_\ell(x) - \rho_\ell(x')\big\|_2^2\right)\,, \tag{42}$$

where $\rho_\ell(x)$ is the reduced density matrix of the $n$-qubit state $\rho(x)$ on qubit $\ell$, and $\gamma$ is a hyperparameter.





**Dominant resource cost/complexity:** During the optimization of the dual program to find the optimal parameters $\boldsymbol{\alpha}^*$, $\mathcal{O}(|T|^2)$ expectation values corresponding to the kernel values in Eq. (40) need to be accurately evaluated, as well as when computing $h_{\boldsymbol{\alpha}^*}(x)$ for a new data point $x$ with the optimized model. This can lead to a significant overhead in applications with large datasets. Alternatively, the primal optimization problem has reduced complexity in the dataset size, but greatly exacerbated dependence on the error [15]. The gate complexity is wholly dependent on the choice of data encoding leading to the kernel function. As the kernel function should be classically nonsimulable, this sets some minimum requirements in terms of circuit complexity. However, in the absence of standardized techniques for data encoding, it is hard to make more precise statements.

**Scope for advantage:** In [16], the authors demonstrate that using a particular constructed dataset and data embedding, concrete quantum advantage can be obtained for a constructed machine learning problem based on the discrete logarithm problem. The original work was based on the fidelity kernel, but a similar advantage can also be more simply obtained for the projected quantum kernel [14] and adapted beyond kernel methods to the reinforcement learning setting [17]. Beyond this, concrete advantage (up to similar computational assumptions) can be shown more generally for any learning problem where the underlying (unknown, to be learned) labeling function constitutes a BQP-hard family [18]. While great strides have been made in understanding the complexity of quantum kernel methods [19, 14], at present there do not yet exist examples of explicit end-to-end theoretical guarantees of advantage for classical data relevant for a real-world problem. As with quantum neural networks, it may be debated whether or not this is a reasonable question for theoretical research efforts.

### Caveats

One consideration we have not discussed so far is how to encode classical data into a quantum circuit, which is a significant aspect of constructing the quantum model. There are many possibilities, such as amplitude encoding or encoding data into rotation angles of single-qubit rotations (see, e.g., [20, 21, 22, 23]). While certain strategies are popular, there is no universal strategy. In general, it is unclear what is the best choice for a given problem at hand, and thus selecting the data-encoding strategy can itself be a heuristic process. The same question extends to the choice of quantum neural network or quantum kernel. While certain choices may perform well in specific problem instances, there is at present a lack of strong evidence why such approaches may be advantageous over their classical counterparts in general.

While optimization of parameterized quantum circuits is predominantly a concern for quantum neural networks, the search for good quantum kernels has also motivated proposals of trainable kernels [22, 24, 25] where a parameterized quantum circuit is used to construct the quantum kernel (note that this is distinct from the "classical" optimization of $\boldsymbol{\alpha}$ in Eq. (40)). In the case that the parameter optimization process is performed using heuristics, it is subject to the same challenges and considerations that arise with VQAs (see Section 20 for more details).

Finite statistics is an important consideration for both settings. Where there is optimization of parameterized quantum circuits, one must take care to avoid the barren plateau phenomenon [9] (again, see Section 20 for more details and further references). Analogous effects can also occur in the kernel setting [26], which can arise even purely due to the data-encoding circuit [14, 27].





**Outlook**

The use of classical machine learning models is generally heuristic, guided by empirical evidence or sometimes physical intuition. Despite this, classical machine learning has found remarkable success in solving many practical problems of interest. The quantum techniques outlined in this section also broadly follow this approach (although theoretical progress has also been substantial in certain areas), and there is no a priori reason why they cannot also be useful. Nevertheless, it can be challenging to make concrete predictions for quantum advantage, particularly for learning problems with classical data (see [16, 18] as some exceptions). For practical problems this is exacerbated by our limited analytic understanding for end-to-end applications, even in the fully classical setting. Indeed, it may ultimately be challenging to have the same complete end-to-end theoretical analysis that other quantum algorithms enjoy, aside from a few select examples [10]. Within the realm of quantum data, there appears to be ripe potential for concrete provable advantage [28, 29, 30], however, this is beyond the scope of this section.

**Further reading**

We refer the reader to [12, 22] for pedagogical expositions of quantum kernel methods, to [31, 32] for comprehensive reviews of quantum neural networks, and to [33] for a review of quantum machine learning models at large, including an exposition of machine learning with quantum data.

# Quantum algorithmic primitives

To deliver an advantage over classical approaches, end-to-end quantum solutions must exploit known quantum phenomena capable of providing a quantum speedup. The disparate collection of known quantum applications is built from a common group of *quantum algorithmic primitives*, which are the source of quantum advantage. Algorithmic primitives are typically not suited for directly solving an end-to-end problem, due to their reliance on unspecified oracles or because their input and/or output does not exactly match that of the end-to-end problem (e.g., some primitives output a quantum state rather than classical data, and thus they have no direct classical analog). Nevertheless, it can be very fruitful to think of algorithms as compositions of different algorithmic primitives, both for higher-level intuitive overview and for independently studying and optimizing the primitives themselves.

This part surveys a variety of quantum algorithmic primitives. For each, we sketch the basic idea of what they do and how they work, as well as discussing example use cases and important caveats. We generally assume that these primitives will need to be implemented in fault-tolerant fashion when they are used within an end-to-end solution for a given application, but we comment on NISQ implementations in passing.

**This part contains:**













# 10  Quantum linear algebra

At a high level of abstraction, quantum computers compose unitary matrices, and do so with classically unparalleled efficiency. This hints at quantum speedups for linear algebra tasks. However, often one needs to work with large non-unitary matrices; thus, for performing general linear algebra tasks we often wish to embed certain non-unitary matrices into unitary matrices represented by efficient quantum circuits, and then apply them to quantum states, take their sums or products, or implement more general matrix functions. These tasks are collectively referred to as "quantum linear algebra," the building blocks of which are discussed in this section.

The techniques described in this section evolved over the past decades and converged to the presented unified framework within several distinct research threads. Block-encodings emerged as a natural approach for embedding non-unitary matrices into quantum circuits, inspired by approaches based on purification, dilation (e.g., Stinespring representation [1] or Stinespring dilation [2]), and postselection. Quantum signal processing (QSP) was discovered as a byproduct of the characterization of simple single-qubit pulse sequences used in nuclear magnetic resonance [3], for synthesizing polynomial transformations applicable to a "signal parameter" encoded as a matrix element of a single-qubit rotation matrix. Meanwhile, it was extensively studied how matrix functions could be synthesized using the linear combinations of unitaries technique on matrix exponentials implemented by Hamiltonian simulation [4, 5, 6], or Chebyshev polynomials of operators implemented via quantum walk techniques [7, 8, 9]. Such matrix exponentials or Chebyshev polynomials can be implemented, for example, via qubitization of a block-encoded operator. In parallel to progress on advanced amplitude amplification [10, 11] techniques, it was recognized [12, 13] that QSP can be "lifted" for applying polynomial transformations to the eigenvalues of quantum walk operators (such as those implemented by qubitization), and thus for implementing a rich family of matrix functions, immediately yielding an optimal algorithm for time-independent Hamiltonian simulation. The concepts of qubitization and QSP were later generalized and unified into the framework of quantum singular value transformation [14], providing generalizations and more efficient implementations of a number of existing quantum algorithms and leading to the discovery of several new algorithms.

*The authors are grateful to Lin Lin for reviewing this section of the survey.*

**This primitive area contains:**

## 10.1   Block-encodings

**Rough overview (in words)**

In a quantum algorithm, the quantum gates that are applied to quantum states are necessarily unitary operators. However, one often needs to apply a linear transformation to some encoded data that is not represented by a unitary operator, and furthermore, one generally needs coherent access to these non-unitary transformations. How can we encode such a non-unitary transformation within a unitary operator? Block-encoding is one method of providing exactly this kind of coherent access to generic linear operators. Block-encoding works by embedding the desired linear operator as a suitably normalized block within a larger unitary matrix, such that the full encoding is a unitary operator, and the desired linear operator is given by restricting the unitary to an easily recognizable subspace. To be useful for quantum algorithms, this block-encoding unitary must also be realized by some specific quantum circuit acting on the main register and additional ancilla qubits.

Block-encodings are ubiquitous within quantum algorithms, but they have both benefits and drawbacks. They are easy to work with, since one can efficiently perform manipulations of block-encodings, such as taking products or convex combinations. On the other hand, this improved working efficiency comes at the cost of having more limited access. For example, if a matrix is stored in classical random access memory, the matrix entries can be explicitly accessed with a single query to the memory, whereas if one only has access to a block-encoding of the matrix, estimating a matrix entry to precision $\varepsilon$ requires $\mathcal{O}(1/\varepsilon)$ uses of the block-encoding unitary in general (by utilizing an amplitude estimation subroutine).

Block-encodings also provide a layer of abstraction that assists in the design and analysis of quantum algorithms. One can simply assume access to a block-encoding and count the number times it is applied. To run the algorithm, it is necessary to choose a method for implementing the block-encoding. There are many ways of constructing block-encodings that could be suited to the structure of the input. For instance, there are efficient block-encoding strategies for density matrices, positive operator-valued measures (POVMs), Gram matrices, sparse matrices, matrices that are stored in quantum data structures, structured matrices, and operators given as a linear combination of unitaries (with a known implementation). We discuss these constructions below. For unstructured, dense matrices, the strategy for Gram matrices can be instantiated using state preparation and quantum random access memory (QRAM) as subroutines. For more details on a particular block-encoding scheme for loading matrices of classical data, see Section 17.3 on block-encoding dense matrices of classical data.

**Rough overview (in math)**

Our goal is to build a unitary operator that gives coherent access to an $M \times M$ matrix $A$ (we will later relax the assumption that $A$ is square), with normalization $\alpha \geq \|A\|$, where $\|A\|$ denotes the spectral norm of $A$. As the name suggests, block-encoding is a way of encoding the matrix $A$ as a block in a larger unitary matrix

$$U_A = \begin{array}{c} \\ |0^a\rangle \\ |0^a\rangle_\perp \end{array} \begin{array}{cc} |0^a\rangle & |0^a\rangle_\perp \\ \begin{pmatrix} A/\alpha & \cdot \\ \cdot & \cdot \end{pmatrix} \end{array},$$





where the labels $|0^a\rangle$ and $|0^a\rangle_\perp$ indicate which portion of the vector space each block corresponds to—specifically, whether the first $a$ qubits are equal to or orthogonal to the state $|0^a\rangle$, respectively. Three of the four blocks are unspecified and can take on any values such that $U_A$ is unitary. More precisely, we say that the unitary $U_A$ is an $(\alpha, a, \epsilon)$-block-encoding of the matrix $A \in \mathbb{C}^{M \times M}$ if

$$\|A - \alpha(\langle 0^a| \otimes I)U_A(|0^a\rangle \otimes I)\| \leq \epsilon, \tag{43}$$

where $a \in \mathbb{N}$ is the number of ancilla qubits used for embedding the block-encoded operator, and $\alpha, \epsilon \in \mathbb{R}_+$ define the normalization and error, respectively. Note that $\alpha \geq \|A\| - \epsilon$ is necessary for $U_A$ to be unitary. The definition above can be extended for general matrices, though additional embedding or padding may be needed (e.g., to make the matrix square).

Once a block-encoding is constructed, it can be used in a quantum algorithm to apply the matrix $A$ to a quantum state by applying the unitary $U_A$ to the larger quantum system. The application of the block-encoding can be thought of as a probabilistic application of $A$—applying $U_A$ to $|0^a\rangle|\psi\rangle$ and postselecting on the first register being in the state $|0^a\rangle$ gives an output state proportional to $A|\psi\rangle$ in the second register.

There are several ways of implementing block-encodings based on the choice of matrix $A$ [1, Section 4.2].[36]

- Unitary matrices are $(1, 0, 0)$-block-encodings of themselves. Controlled unitaries (e.g., CNOT) are essentially $(1, 1, 0)$-block-encodings of the controlled operation.

- Given an $s$-qubit density matrix $\rho$ and an $(a + s)$-qubit unitary $G$ that prepares a *purification* of $\rho$ as $G|0^a\rangle|0^s\rangle = |\rho\rangle$ (s.t. $\mathrm{tr}_a|\rho\rangle\langle\rho| = \rho$, where $\mathrm{tr}_a$ denotes trace over the first register), then the operator [2]

$$(G^\dagger \otimes I_s)(I_a \otimes \mathrm{SWAP}_s)(G \otimes I_s)$$

  is a $(1, a + s, 0)$-block-encoding of the density matrix $\rho$, where $I_x$ denotes the identity operator on a register with $x$ qubits, and $\mathrm{SWAP}_s$ denotes the operation that swaps two $s$-qubit registers [1, Lemma 45].

- Similarly, one can construct block-encodings of POVM operators, given access to a unitary that implements the POVM [3]. Specifically, if $U$ is a unitary that implements the POVM $M$ to precision $\epsilon$, such that for all $s$-qubit density operators $\rho$ we have

$$\left| \mathrm{tr}(\rho M) - \mathrm{tr}\left[ U(|0\rangle\langle 0|^{\otimes a} \otimes \rho)U^\dagger(|0\rangle\langle 0| \otimes I_{a+s-1}) \right] \right| \leq \epsilon,$$

  then $(I_1 \otimes U^\dagger)(\mathrm{CNOT} \otimes I_{a+s-1})(I_1 \otimes U)$ is a $(1, 1 + a, \epsilon)$-block-encoding of $M$ [1, Lemma 46].

- One can also implement a block-encoding of a Gram matrix using a pair of state preparation unitaries $U_L$ and $U_R$. In particular, the product

$$U_A = U_L^\dagger U_R$$

  is a $(1, a, 0)$-block-encoding of the Gram matrix $A$ whose entries are $A_{ij} = \langle \psi_i | \phi_j \rangle$, where [1, Lemma 47]

$$U_L|0^a\rangle|i\rangle = |\psi_i\rangle, \qquad U_R|0^a\rangle|j\rangle = |\phi_j\rangle.$$

---

[36]References to locations in [1] typically refer to the longer arXiv version, rather than the STOC version.





- One can generalize the above strategy from Gram matrices to arbitrary matrices to produce $(\alpha, a, \epsilon)$-block-encodings of general matrices $A$, where again $\alpha \geq \|A\|$. See Section 17.3 on block-encoding dense matrices of classical data for details.

- Sparse matrices: Given a matrix $A \in \mathbb{C}^{2^w \times 2^w}$ that is $s_r$-row sparse and $s_c$-column sparse (meaning each row and column has at most $s_r$ and $s_c$ nonzero entries, respectively), then, defining $\|A\|_{\max} = \max_{i,j} |A_{ij}|$, one can create a $(\sqrt{s_r s_c} \|A\|_{\max}, w + 3, \epsilon)$-block-encoding of $A$ using oracles $O_r$, $O_c$, and $O_A$, defined as follows [1, Lemma 48]

$$O_r : |i\rangle|k\rangle \mapsto |i\rangle|r_{ik}\rangle, \qquad\qquad \forall i \in [2^w] - 1, k \in [s_r],$$
$$O_c : |\ell\rangle|j\rangle \mapsto |c_{\ell j}\rangle|j\rangle, \qquad\qquad \forall \ell \in [s_c], j \in [2^w] - 1,$$
$$O_A : |i\rangle|j\rangle|0^b\rangle \mapsto |i\rangle|j\rangle|A_{ij}\rangle, \qquad\qquad \forall i, j \in [2^w] - 1.$$

  In the above, $r_{ij}$ is the index of the $j$-th nonzero entry in the $i$-th row of $A$ (or $j + 2^w$ if there are less than $i$ nonzero entries), $c_{ij}$ is the index of the $i$-th nonzero entry in the $j$-th column of $A$ (or $i + 2^w$ if there are less than $j$ nonzero entries), and $|A_{ij}\rangle$ is a $b$-bit binary encoding of the matrix element $A_{ij}$. To build the block-encoding, one needs one query to each of $O_r$ and $O_c$, and two queries of $O_A$. This input model is known as the sparse access model. If, in addition to being sparse, the matrix also enjoys some additional *structure*, for example, there are only a few distinct values that the matrix elements can take, the complexity can be further improved [4, 5]. Finally, note that the sparsity dependence can be essentially quadratically improved—reducing the block-encoding normalization factor from $\sqrt{s_r s_c} \|A\|_{\max}$ to $(\max(s_r, s_c))^{(1+o(1))/2} \|A\|_{1 \to 2}$, where $\|A\|_{1 \to 2} = \max_v \|Av\|_2 / \|v\|_1$—using advanced Hamiltonian simulation techniques [6, Theorem 2] combined with taking the logarithm of unitaries [1, Corollary 71], however, the resulting subroutine may be impractical and comes with a worse precision dependence.

- For matrices given as a linear combination of unitary operators (LCU), we can block-encode the matrix using the LCU technique [7]. We provide a full description in §Linear combinations of Section 10.2, and only give a brief outline here. For $A = \sum_{i=1}^{L} c_i V_i$ with $V_i$ unitary, we define the oracles PREPARE (acting on $\lceil \log_2(L) \rceil$ ancilla qubits) and SELECT (acting on the ancilla and register qubits), and implement a $(\sum_i |c_i|, \lceil \log_2(L) \rceil, 0)$-block-encoding of $A$, using $U := \text{PREPARE}^\dagger \cdot \text{SELECT} \cdot \text{PREPARE}$. The Hamiltonians of physical systems can often be written as a linear combination of a moderate number of Pauli operators, leading to a prevalence of this technique in quantum algorithms for chemistry [8, 9] and condensed matter physics [8, 10, 11].

In addition to the definition of block-encoding in Eq. (43), one can also define an asymmetric version as follows

$$\left\| A - \alpha(\langle 0^a | \otimes I) U_A (|0^b\rangle \otimes I) \right\| \leq \epsilon,$$

where $a$ may not equal $b$. In this case, $U_A$ can be considered to be an $(\alpha, (a, b), \epsilon)$- or an $(\alpha, \max(a, b), \epsilon)$-block-encoding of $A$. This can be useful for block-encoding a non-square matrix.

**Dominant resource cost (gates/qubits)**

The complexity of block-encoding an operator depends on the type of data or operator being encoded and any underlying assumptions. For instance, unitaries are naturally block-encodings





of themselves, and hence their resource requirements depend entirely on their circuit-level implementation without any additional overhead for being a "block-encoding." By contrast, approaches that make use of state preparation and QRAM to implement the block-encoding tend to have larger complexities, as those two subroutines typically dominate the resource requirements. For example, the best known circuits that implement block-encodings of matrices of classical data for general, dense $N \times N$ matrices use $\mathcal{O}(N \log(1/\epsilon))$ qubits to achieve minimum $T$-gate count (which also scales as $\mathcal{O}(N \log(1/\epsilon))$), or a larger $\mathcal{O}(N^2)$ number of qubits to achieve minimum $T$-gate depth (which scales as $\mathcal{O}(\log(N) + \log(1/\epsilon))$ [12]. In the sparse access model, one can use $\mathcal{O}(w + \log^{2.5}(s_r s_c/\epsilon))$ one- and two-qubit gates, and $\mathcal{O}(b + \log^{2.5}(s_r s_c/\epsilon))$ ancilla qubits [1], in addition to the calls to the matrix entry $O_A$ and sparse access oracles $O_r$ and $O_c$, which must be implemented either by computing matrix entries "on-the-fly" or by using a primitive (see [13] for asymptotic resource statements for general sparse matrices with varying ancilla availability). Assuming appropriate binary representations of the numbers $A_{ij}$, the exponents of the above logarithms can be reduced to 1 using the techniques of [14] (see also [8, Section III.D] and [15, Supplementary Material VII.A.2]).

The value of block-encodings is not that it is always cheap to implement them (as it depends on the relevant cost metric and the data access model); rather, the concept of block-encodings is powerful because it allows a practitioner of quantum algorithms to study and optimize the block-encoding construction independently of how it is used within the larger algorithm.

**Caveats**

The definition of block-encoding requires $\|A\|/\alpha \leq 1$. If $\|A\|/\alpha = 1$, then the block-encoding achieves an optimal normalization factor $\alpha$. However, note that often the above constructions lead to suboptimal normalization factors in the sense that $\alpha \gg \|A\|$. In practical applications, this suboptimality usually leads to a corresponding increase in the overall complexity.

For a given desired block-encoding, there can be several independent, yet equally valid implementations, and one can sometimes optimize for various resources when building the block-encoding. For example, many block-encoding strategies require a step in which some classical data is loaded into QRAM, but there are several ways of performing this data-loading step.

When using a block-encoding as part of a larger quantum algorithm, it is important to ensure that the overhead introduced by implementing a block-encoding will not outweigh any potential quantum speedups, as block-encoding can be very resource intensive.

The use of $|0\rangle^{\otimes a}$ as the "signal" state is just one convention—we can use any "signal" state, given a unitary to prepare it [2]. One can also consider a more general definition known as "projected unitary encodings" which allows using an arbitrary subspace, rather than just a state-indexed block [1].

**Example use cases**

Block-encodings are ubiquitous in quantum algorithms, and they prevail in quantum algorithms that need coherent access to some linear operator or a method of implementing a non-unitary transformation on quantum data. Some specific examples:

- We can manipulate block-encoded operators—for example, take convex or linear combinations, products, tensor products, and other transformations of an input operator.





- The combination of qubitization with quantum signal processing, or quantum singular value transformation can be used to realize algorithms by applying polynomial transformations to block-encoded matrices. Prominent examples are Hamiltonian simulation via qubitization, and matrix (pseudo) inversion [1, Theorem 41] that can be used for solving large linear systems of equations [16] or more generally least-squares regression problems [17].

- Block-encoding can be used to provide coherent access to classical data in a quantum algorithm; for example, loading classical data into a quantum interior point method for portfolio optimization [18].

**Further reading**

Reference [17] provides an instructive overview of the concept of block-encoding and showcases its power in several applications related to (generalized) regression problems. Meanwhile, [1] is a comprehensive collection of technical results about block-encodings and quantum linear algebra more generally.

## 10.2 Manipulating block-encodings

**Rough overview (in words)**

Given one or more block-encodings, we often want to form a single block-encoding of a product, tensor product, or linear combination of the individual block-encoded operators. This can be achieved as outlined below, using additional ancilla qubits.

**Rough overview (in math) and resource cost**

We will consider the case of two operators $A$ and $B$, with straightforward generalizations to additional operators [1]. We are given an $(\alpha, a, \epsilon_a)$-block-encoding $U_A$ of $A$, and a $(\beta, b, \epsilon_b)$-block-encoding $U_B$ of $B$. Operators $A$ and $B$ act on system qubits $s$.

**Products:** The operation $U_{AB} := (I_b \otimes U_A)(U_B \otimes I_a)$ is an $(\alpha\beta, a+b, \alpha\epsilon_b + \beta\epsilon_a)$-block-encoding of $AB$ [1, Lemma 53], where $I_x$ denotes the identity operator on $x$ qubits (see Fig. 1). For example, if $a = b$, this construction uses twice as many ancilla qubits for block-encoding the product compared to the block-encoding of the individual matrices. In fact, we can assume without loss of generality that $a = b$ (by taking the maximum of the two) and improve the construction using the circuit in Fig. 2, which uses $a + 1$ ancilla qubits instead of $2a$. This idea has been generalized to encompass products of $L$ block-encodings using only $a + \lceil \log_2(L) \rceil + 1$ ancillas (rather than $aL$), where it is known as the "compression gadget"; see [2, Lemma 13] and [3, Lemma 3].

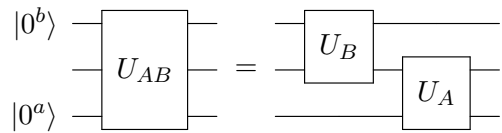

Figure 1: Implementing the block-encoding $U_{AB}$ of $AB$ that acts on $s$ qubits. The cost is $a + b$ ancilla qubits, and one call to each of $U_A$, $U_B$.

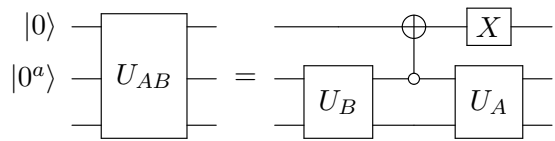

Figure 2: Implementing the block-encoding $U_{AB}$ of $AB$ for the case where both $U_A$ and $U_B$ act on $a$ ancilla qubits. The controlled gate is an $a$-controlled generalized Toffoli gate.

**Tensor products:** The operation $U_{A\otimes B} := (U_A \otimes U_B)$ is an $(\alpha\beta, a + b, \alpha\epsilon_b + \beta\epsilon_a)$-block-encoding of the operator $A \otimes B$, as depicted in Fig. 3.





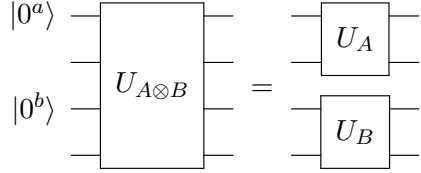

Figure 3: Implementing the block-encoding $U_{A \otimes B}$ of $A \otimes B$ that acts on $2s$ qubits. The cost is $a + b$ ancilla qubits, and one call to each of $U_A$, $U_B$.

**Linear combinations:** Linear combinations of block-encodings can be viewed as a generalization of the linear combination of unitaries (LCU) trick [4]. We wish to implement a block-encoding of $\sum_{i=0}^{L-1} c_i A_i$, where $c_i \in \mathbb{R}$ (the LCU trick can also be extended to complex coefficients) and define $\lambda := \sum_{i=0}^{L-1} |c_i|$. We consider $L$ block-encodings $U_i$ that are $(1, m, \epsilon_i)$-block-encodings of $A_i$. We note that in cases where the block-encodings have different $\alpha_i$ or $m_i$ values, the former can be absorbed into the $c_i$ values and the latter can be taken as $m = \max_i m_i$.

We first define an operator PREPARE by the following action on $|0^{\lceil \log_2(L) \rceil}\rangle$

$$\text{PREPARE}|0^{\lceil \log_2(L) \rceil}\rangle = \frac{1}{\sqrt{\lambda}} \sum_j \sqrt{|c_j|}|j\rangle$$

that prepares a weighted superposition on an ancilla register, such that the amplitudes are proportional to the square roots of the absolute values of the desired coefficients. We also define[37]

$$\text{SELECT} = \sum_{j=0}^{L-1} |j\rangle\langle j| \otimes \text{sign}(c_j)U_j.$$

We then have the following result

$$\left(\langle 0^{\lceil \log_2(L) \rceil}| \otimes I\right) \text{PREPARE}^\dagger \cdot \text{SELECT} \cdot \text{PREPARE}\left(|0^{\lceil \log_2(L) \rceil}\rangle \otimes I\right)$$
$$= \frac{1}{\lambda} \sum_{i=0}^{L-1} c_i U_i, \tag{44}$$

that is, $U_{\text{LC}} := \text{PREPARE}^\dagger \cdot \text{SELECT} \cdot \text{PREPARE}$ is a $(\lambda, \lceil \log_2(L) \rceil, 0)$-block-encoding of the LCU $\sum_i c_i U_i$, as depicted in Fig. 4. This is the standard LCU trick [4], and it does not require $U_i$ to be block-encodings (or we can view them as $(1, 0, 0)$-block-encodings of themselves). This technique can be used in Hamiltonian simulation, or to instantiate a block-encoding.

If, as specified above, $U_i$ are block-encodings of $\tilde{A}_i$ (which approximate $A_i$), we also have the following result

$$\left\|\left(\sum_{i=0}^{L-1} c_i A_i\right) - \lambda\left(\langle 0^{m + \lceil \log_2(L) \rceil}| \otimes I\right) U_{\text{LC}}\left(|0^{m + \lceil \log_2(L) \rceil}\rangle \otimes I\right)\right\| \le \sum_{i=0}^{L-1} |c_i|\epsilon_i.$$

Hence, $U_{\text{LC}}$ is a $(\lambda, \lceil \log_2(L) \rceil + m, \lambda \max_i \epsilon_i)$-block-encoding of $\sum_{i=0}^{L-1} c_i A_i$.

---

[37]To be precise, for $j \notin \{0, 1, \ldots, L-1\}$ we define $\text{sign}(c_j)U_j := I$.





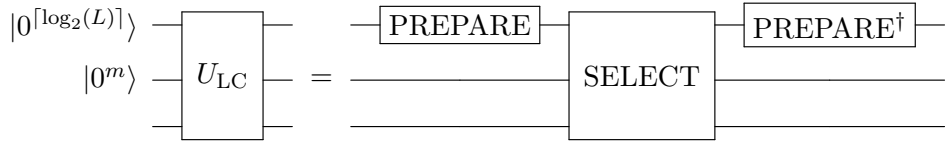

Figure 4: Implementing the block-encoding $U_{\text{LC}}$ of $\sum_i c_i A_i$ that acts on $s$ qubits. We require $\lceil \log_2(L) \rceil + m$ ancilla qubits. The regular LCU circuit is obtained by omitting the register $|0^m\rangle$ and the requirement that $U_i$ are block-encodings. The gate complexity of PREPARE depends on the coefficients $c_i$ but is $\Theta(L)$ in the worst case (using no additional ancilla qubits) [5]. We can also define PREPARE that leads to entanglement with a garbage register PREPARE$|0^{\lceil \log_2(L) \rceil}\rangle|0^g\rangle = \lambda^{-0.5} \sum_i \sqrt{|c_i|}|i\rangle|G_i\rangle$, which can be seen to satisfy the relations required to implement the linear combination, Eq. (44). It can sometimes (e.g., [6]) be cheaper to implement this garbage-entangled PREPARE; see Section 17.2 on preparing states from classical data. The cost of SELECT depends on the form of $U_i$, but in the worst case requires $\Theta(L)$ primitive gates and $\Theta(L)$ calls to $|0\rangle\langle 0| \otimes I + |1\rangle\langle 1| \otimes U_i$ [7, 6], although this can be improved in some relevant special cases (e.g., [8]). When $U_i$ are multiqubit Pauli operators, [9] provides a depth-optimized implementation of SELECT that achieves $\mathcal{O}(\log(Ln))$ depth using $\Theta(Ln)$ total gates and total ancilla qubits.

### Caveats

Performing linear algebraic manipulations of block-encodings using these primitives can quickly increase the ancilla count of the algorithm and worsen the normalization factor of the block-encoding. Amplifying a subnormalized block-encoding is possible, but costly, requiring an amount of time scaling roughly linearly in the amplification factor; see [10, 1]. Given a single block-encoded operator $A$, the above primitives can be used to implement a block-encoding of a polynomial in $A$. However, this can be achieved with much lower overhead using quantum singular value transformation (QSVT).

### Example use cases

- Linear combination of block-encodings are used to obtain mixed-parity functions in QSVT required for Hamiltonian simulation.

- LCU trick is used for Hamiltonian simulation via Taylor series and to instantiate block-encodings of chemistry or condensed matter physics Hamiltonians (see, e.g., [6, 8]).

### Further reading

- References [11, Section 3.3] and [12, Section 7.3] contain a comprehensive discussion of manipulating block-encodings, including proofs of many of the results stated above.

## 10.3 Quantum signal processing

**Rough overview (in words)**

Quantum signal processing (QSP) [1] describes a method for nonlinear transformations of a signal parameter encoded in a single-qubit gate, using a structured sequence that interleaves the "signal gate" with fixed parameterized "modulation" gates. The technique was originally motivated by the desire to characterize pulse sequences used in nuclear magnetic resonance [1]. Remarkably, it has been shown [1, 2] that there is a rich family of polynomial transformations that are in one-to-one correspondence with appropriate modulation sequences; moreover, given such a polynomial, one can efficiently compute the corresponding modulation parameters.

Even more remarkably, this analysis holds not just for single-qubit "signal gates" but can be extended for multiqubit operators that *act* like single-qubit rotations when restricted to appropriate 2D subspaces [3]. This insight enables the implementation of block-encodings of polynomials of Hermitian or normal matrices when used in conjunction with qubitization. The two-step process of qubitization and QSP can be unified and generalized through quantum singular value transformation (QSVT).

**Rough overview (in math)**

We follow the "Wx convention" of QSP [4, 5]. We define the single-qubit signal operator

$$W(x) := \begin{pmatrix} x & i\sqrt{1-x^2} \\ i\sqrt{1-x^2} & x \end{pmatrix} = e^{i \arccos(x) X}$$

which is a single-qubit $X$ rotation. We can verify that

$$W(x)^2 = \begin{pmatrix} 2x^2 - 1 & \cdot \\ \cdot & \cdot \end{pmatrix}$$

$$W(x)^3 = \begin{pmatrix} 4x^3 - 3x & \cdot \\ \cdot & \cdot \end{pmatrix}$$

$$\vdots$$

$$W(x)^n = \begin{pmatrix} T_n(x) & \cdot \\ \cdot & \cdot \end{pmatrix},$$

where $T_n(x)$ is the $n$-th Chebyshev polynomial of the first kind, showcasing that even a simple sequence of the signal unitaries can implement a rich family of polynomials of the signal $x$.

More complex behavior is obtained by interleaving $W(x)$ with parameterized single-qubit $Z$ rotations $e^{i\phi_j Z}$. We define a QSP sequence as

$$U_{\mathrm{QSP}}(\Phi) := e^{i\phi_0 Z} \prod_{j=1}^{d} W(x) e^{i\phi_j Z},$$

where $\Phi$ denotes the vector of angles $(\phi_0, \phi_1, \ldots, \phi_d)$. The QSP sequence implements the following unitary

$$U_{\mathrm{QSP}}(\Phi) = \begin{pmatrix} P(x) & iQ(x)\sqrt{1-x^2} \\ iQ^*(x)\sqrt{1-x^2} & P^*(x) \end{pmatrix}, \tag{45}$$





where $P(x), Q(x)$ are complex polynomials obeying a number of constraints (see below), and $P^*(x)$, $Q^*(x)$ denote their complex conjugates. Note also that the relationship between the sequence $\Phi$ and the corresponding polynomial $P(x)$ can be understood through nonlinear Fourier analysis [6].

**Dominant resource cost (gates/qubits)**

A QSP circuit that implements a degree-$d$ polynomial in the signal parameter requires $d$ uses of $W(x)$ and $d+1$ fixed-angle $Z$ rotations. There are efficient classical algorithms to determine the angles for a given target polynomial, either using high-precision arithmetic with $\sim d \log(d)$ bits of precision [2] (or more [4]—though this can be mitigated using heuristic techniques [7, 8]) or in some regimes using more efficient optimization-based or iterative algorithms [9, 10, 11, 12, 6, 13]. In particular, this line of work has culminated in [13] with an algorithm for finding the angles that is provably numerically stable, that is, it requires only polylog($d/\epsilon$) bits of precision to achieve $\epsilon$ error. An alternative method computes the angles directly if both $P(x)$ and $Q(x)$ are known, and offers a way to compute $Q(x)$ if only $P(x)$ is known [14, 15]. Although these procedures are efficient in theory, in practice it may still be nontrivial to find the angles. Nevertheless, researchers reportedly computed angle sequences corresponding to various degree $d \approx 10^7$ polynomials [14, 15].

**Caveats**

As discussed above, not all polynomials can be implemented by a QSP sequence. Implementable polynomials must obey a number of constraints, which can be somewhat restrictive. For the standard QSP circuit $U_{\mathrm{QSP}}(\Phi)$ given above, the achievable polynomials pairs $P(x), Q(x) \in \mathbb{C}$ can be characterized by the following three conditions

- $\mathrm{Deg}(P) \leq d$ and $\mathrm{Deg}(Q) \leq d-1$,

- $\mathrm{Parity}(P) = \mathrm{Parity}(d)$ and $\mathrm{Parity}(Q) = \mathrm{Parity}(d-1)$,

- $\forall\, x \in [-1, 1] : |P(x)|^2 + (1-x^2)|Q(x)|^2 = 1$ (required for Eq. (45) to be unitary).

This last requirement can be particularly limiting. A useful way to circumvent this for real functions is to encode the polynomial in the matrix element $\langle +|U_{\mathrm{QSP}}(\Phi)|+\rangle$ rather than in $\langle 0|U_{\mathrm{QSP}}(\Phi)|0\rangle$, where $|+\rangle = (|0\rangle + |1\rangle)/\sqrt{2}$. This matrix element evaluates to

$$\langle +|U_{\mathrm{QSP}}(\Phi)|+\rangle = \mathrm{Re}[P(x)] + i\sqrt{1-x^2}\,\mathrm{Re}[Q(x)]\,.$$

Given a real target polynomial $f(x)$ with parity equal to $\mathrm{Parity}(d)$, we can guarantee that the matrix element evaluates to $f(x)$ by choosing $\mathrm{Re}[P(x)] = f(x)$ and $\mathrm{Re}[Q(x)] = 0$. The third condition above then reduces to $1 - f(x)^2 = |\mathrm{Im}[P(x)]|^2 + (1-x^2)|\mathrm{Im}[Q(x)]|^2$. By [4, Lemma 6], there exist choices for $\mathrm{Im}[P(x)]$ and $\mathrm{Im}[Q(x)]$ that satisfy this identity as well as the first two conditions above, provided $|f(x)| \leq 1\ \forall\, x \in [-1, 1]$. In summary, we may implement any real polynomial $f(x)$ satisfying the requirements [4, Corollary 10]:

- $\mathrm{Deg}(f) = d$,

- $\mathrm{Parity}(f) = \mathrm{Parity}(d)$,





- $\forall \, x \in [-1, 1] : |f(x)| \leq 1$.

There are several related conventions considered in the literature for the explicit form of the single-qubit operators used in QSP; a thorough discussion is given in [5, Appendix A]. One common form that links closely to qubitization and QSVT is the reflection convention, which replaces $W(x)$ by the reflection

$$R(x) = \begin{pmatrix} x & \sqrt{1-x^2} \\ \sqrt{1-x^2} & -x \end{pmatrix}, \tag{46}$$

and adjusts the parameters $\{\phi_j\}$ accordingly [4].

### Example use cases

- Functions of Hermitian or normal matrices, in conjunction with qubitization, including for Hamiltonian simulation.

- Functions of general matrices via QSVT.

- Reference [16] applied QSP to beyond-Heisenberg-limit calibration of two-qubit gates in a superconducting system.

### Further reading

- A pedagogical discussion of QSP [5].

- Detailed proofs of the key results of QSP [1, 4].

- Lecture notes on QSP [17, Section 7.6].

## 10.4 Qubitization

**Rough overview (in words)**

Qubitization is a method for using a block-encoding $U_A$ of a Hermitian operator $A$ to manipulate $A$, for example, implement $A^2$, or, more generally, some function $f(A)$ [1]. However, the eigenvalues of $U_A$ are typically unrelated to those of $A$, and plain repeated applications of $U_A$ do not in general produce the desired behavior. Qubitization converts the block-encoding $U_A$ into a unitary operator $W$ (sometimes called a qubiterate or a qubitized quantum walk operator) having the following guaranteed advantageous properties:

- $W$ is a block-encoding of the operator $A$.

- The spectrum of $W$ is directly related to the spectrum of $A$.

- Repeated application of $W$ leads to structured behavior that can be cleanly analyzed.

This combination of features means that qubitization can be used for applying polynomial transformations to the spectrum of $A$. For example, repeated application of $W$ implements Chebyshev polynomials of $A$, while other polynomials can also be implemented by using quantum signal processing [2, 1, 3].

The key observation is that a qubitization unitary $W$ has eigenvalues and eigenvectors that relate in a nice way to those of $A$. Thus, one can also perform quantum phase estimation on $W$ to learn these quantities [4, 5], providing a potentially cheaper alternative to such tasks compared to approaches based on explicit Hamiltonian simulation for implementing $U = \mathrm{e}^{\mathrm{i}At}$.

**Rough overview (in math)**

We are given a $(1, m, 0)$-block-encoding $U_A$ of Hermitian $A$, such that

$$A = (\langle 0^m| \otimes I)U_A(|0^m\rangle \otimes I) \Longleftrightarrow U_A = \begin{pmatrix} A & \cdot \\ \cdot & \cdot \end{pmatrix}.$$

First, we will assume $U_A$ is also Hermitian (implying $U_A^2 = I$, where $I$ is the identity matrix). Let $A$ have spectral decomposition $A = \sum_\lambda \lambda |\lambda\rangle\langle\lambda|$. An application of $U_A$ to an eigenstate $|\lambda\rangle$ of $A$ gives

$$U_A|0^m\rangle|\lambda\rangle = \lambda|0^m\rangle|\lambda\rangle + \sqrt{1 - \lambda^2}|\perp_{0^m,\lambda}\rangle,$$

where $|\perp_{0^m,\lambda}\rangle$ is a state perpendicular to $|0^m\rangle$.[38] Noting $U_A^2 = I$ reveals that the 2D subspace $S_\lambda$ spanned by $\{|0^m\rangle|\lambda\rangle, |\perp_{0^m,\lambda}\rangle\}$ is invariant under the action of $U_A$. $U_A$ restricted onto $S_\lambda$ can be described by the matrix

$$
\begin{array}{c}
\phantom{|\perp_{0^m,\lambda}\rangle} \\
|0^m\rangle|\lambda\rangle \\
|\perp_{0^m,\lambda}\rangle
\end{array}
\begin{array}{c}
\begin{array}{cc} |0^m\rangle|\lambda\rangle & |\perp_{0^m,\lambda}\rangle \end{array} \\
\begin{pmatrix} \lambda & \sqrt{1 - \lambda^2} \\ \sqrt{1 - \lambda^2} & -\lambda \end{pmatrix},
\end{array}
$$

which is a 2D reflection with eigenvalues $\pm 1$. Clearly, repeated application of (self-inverse) $U_A$ can have limited effect on any input state. Qubitization uses a reflection $Z_{|0^m\rangle} = (2|0^m\rangle\langle 0^m| - I)$

---

[38]If $\lambda = \pm 1$, then there is no need for $|\perp_{0^m,\lambda}\rangle$, and the subspace $S_\lambda$ becomes 1D.





to transform $U_A$ into a Grover-like operator $W = Z_{|0^m\rangle}U_A$ which has the following matrix when restricted onto the invariant subspace $S_\lambda$ in the $\{|0^m\rangle|\lambda\rangle, |\perp_{0^m,\lambda}\rangle\}$ basis

$$[W]_{\{|0^m\rangle|\lambda\rangle, |\perp_{0^m,\lambda}\rangle\}} = \begin{pmatrix} 1 & 0 \\ 0 & -1 \end{pmatrix} \begin{pmatrix} \lambda & \sqrt{1-\lambda^2} \\ \sqrt{1-\lambda^2} & -\lambda \end{pmatrix}$$

$$= \begin{pmatrix} \lambda & \sqrt{1-\lambda^2} \\ -\sqrt{1-\lambda^2} & \lambda \end{pmatrix},$$

showing that $W$ is still a $(1, m, 0)$-block-encoding of $A$. This has the form of a $Y$-axis rotation

$$[W]_{\{|0^m\rangle|\lambda\rangle, |\perp_{0^m,\lambda}\rangle\}} = \begin{pmatrix} \cos(\theta_\lambda) & \sin(\theta_\lambda) \\ -\sin(\theta_\lambda) & \cos(\theta_\lambda) \end{pmatrix},$$

where $\theta_\lambda = \arccos(\lambda)$. Therefore, $W$ has eigenvalues $e^{\pm i \arccos(\lambda)}$ with respective eigenvectors $(|0^m\rangle|\lambda\rangle \pm i|\perp_{0^m,\lambda}\rangle)/\sqrt{2}$, which can be accessed using quantum phase estimation.

Furthermore, we can see that on the span of the subspaces $S_\lambda$ repeated application of $W$ acts as

$$W^d = \bigoplus_\lambda \begin{pmatrix} \cos(d\theta_\lambda) & \sin(d\theta_\lambda) \\ -\sin(d\theta_\lambda) & \cos(d\theta_\lambda) \end{pmatrix}$$

$$= \bigoplus_\lambda \begin{pmatrix} T_d(\lambda) & \sqrt{1-\lambda^2}U_{d-1}(\lambda) \\ -\sqrt{1-\lambda^2}U_{d-1}(\lambda) & T_d(\lambda) \end{pmatrix}$$

$$= \begin{pmatrix} T_d(A) & \cdot \\ \cdot & \cdot \end{pmatrix},$$

where $T_d(\cdot)$ and $U_d(\cdot)$ are degree-$d$ Chebyshev polynomials of the first and second kind, respectively. Therefore, $W^d$ applies the polynomial transformation $T_d$ to each eigenvalue of $A$, thereby implementing $T_d(A)$.

**Dominant resource cost (gates/qubits)**

The resource cost of qubitization is inherited from the cost of the block-encoding. Given a Hermitian $(\alpha, m, 0)$-block-encoding $U_A$, the qubitization operator $W$ is a (non-Hermitian) $(\alpha, m, 0)$-block-encoding, and it uses no additional qubits. The operation $Z_{|0^m\rangle} = (2|0^m\rangle\langle 0^m| - I)$ can be implemented (up to a global phase) with an $m$-qubit (anti)controlled $-Z$ gate, equivalent to an $m$-qubit Toffoli up to single-qubit gates. An example qubitization circuit is shown below in Fig. 5 for $m = 3$. Implementing a block-encoding of a degree-$d$ Chebyshev polynomial applied to $A$ requires $d$ calls to $U_A$ and $Z_{|0^m\rangle}$.

If the block-encoding $U_A$ is not Hermitian, qubitization can be achieved using the construction of [1, Lemma 10] that uses one additional qubit, one call to controlled $U_A$, and one call to controlled $U_A^\dagger$ to implement the Hermitian block-encoding

$$U_A' := ((HX) \otimes I)(|0\rangle\langle 0| \otimes U_A + |1\rangle\langle 1| \otimes U_A^\dagger)(H \otimes I). \tag{47}$$

An alternative to qubitization is based on quantum singular value transformation (QSVT) that uses the sequence $Z_{|0^m\rangle}U_A^\dagger Z_{|0^m\rangle}U_A$, analogous to the earlier $W^2$, acting as

$$\begin{pmatrix} \lambda & \sqrt{1-\lambda^2} \\ -\sqrt{1-\lambda^2} & \lambda \end{pmatrix}^2$$





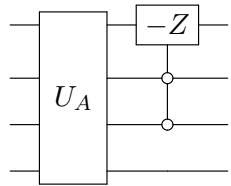

Figure 5: An example qubitization circuit using the Hermitian $(1, 3, 0)$-block-encoding $U_A$.

on a 2D subspace analogous to $S_\lambda$. The approach can be extended to odd-degree polynomials with a single additional application of $Z_{|0^m\rangle} U_A$ [3]. The advantage of this approach is that it does not require $U_A$ to be Hermitian, thus there is no need for an additional qubit or calls to controlled $U_A^{\pm 1}$. This approach may be referred to as "quantum eigenvalue transformation" [6, 7] as this is a special case of QSVT applied to Hermitian $A$.

**Caveats**

The original formulation of qubitization [1] discussed above requires a Hermitian or normal block-encoded matrix $A$. The concept can be extended to general (non-square) matrices via QSVT, providing a significant generalization, however, in some cases quantum signal processing and its generalized versions [8, 9] can exploit additional structure that comes, for example, from the extra symmetries of Hermitian block-encodings, leading to potential constant-factor savings. Consider, for example, Hamiltonian simulation, where QSVT separately implements $\sin(tH)$ and $\cos(tH)$ using a block-encoding $U_H$ of the Hamiltonian $H$, and applies a three-step oblivious amplification procedure on top of linear combination of unitaries to implement $\exp(itH)$ [3]. Meanwhile, quantum signal processing implements $\exp(itH)$ directly but requires an additional ancilla qubit and controlled access to a Hermitian block-encoding $U_H'$, which, when implemented via Eq. (47), uses both controlled $U_H$ and $U_H^\dagger$, resulting in a factor of $\sim 4$ overhead. Altogether these considerations suggest that the QSVT-based approach might have a slightly better constant-factor overhead, particularly when controlled $U_H$ is significantly more costly to implement than $U_H$. If $U_H$ is already Hermitian, then quantum signal processing can have an improved complexity.

**Example use cases**

- Some quantum algorithms in quantum chemistry that compute energies perform phase estimation on a qubitization operator $W$ implemented via calls to a block-encoding of the electronic structure Hamiltonian. This avoids the approximation error incurred when performing phase estimation on $e^{iHt}$, implemented via Hamiltonian simulation [4, 5].

- Qubitization acts as a precursor to QSVT, which extends the concept to general matrices and unifies it with quantum signal processing.

**Further reading**

- Original introduction of qubitization [1] and QSVT [3].





- A pedagogical overview of quantum signal processing, its lifting to QSVT, and their applications [10].

- Reference [6, Chapters 7 & 8] provides an accessible derivation of qubitization and QSVT.

## 10.5   Quantum singular value transformation

**Rough overview (in words)**

Quantum singular value transformation (QSVT) can be viewed as both a unification and generalization of qubitization and quantum signal processing. Given a block-encoding $U_A$ of a general matrix $A$, QSVT enables the transformation of the singular values of $A$ by a polynomial $f(\cdot)$. In QSVT, there is a one-to-one correspondence between the desired polynomial transformation and its quantum circuit implementation whose parameters can be found by efficient classical algorithms.

It transpires that a number of existing quantum algorithms have simple and optimal (or near-optimal) implementations via the QSVT framework, including Hamiltonian simulation [1, 2, 3], amplitude amplification and estimation [3, 4], quantum linear systems solving [3, 5], Gibbs sampling [3], algorithms for topological data analysis [6, 7, 8], and quantum phase estimation [5, 9].

**Rough overview (in math)**

We are given a $(1, m, 0)$-block-encoding $U_A$ of operator $A$ (for simplicity, we will restrict our presentation to square matrices $A$, noting there is a straightforward generalization to non-square $A$ [3]), such that

$$A = (\langle 0^m| \otimes I)U_A(|0^m\rangle \otimes I),$$

where $|0^m\rangle$ denotes $|0\rangle^{\otimes m}$. The matrix $A$ has a singular value decomposition (SVD)

$$A = \sum_i \sigma_i |w_i\rangle\langle v_i|.$$

QSVT provides a method for implementing

$$f^{(SV)}(A) := \begin{cases} \sum_i f(\sigma_i)|w_i\rangle\langle v_i| & \text{if } f \text{ is odd, and} \\ \sum_i f(\sigma_i)|v_i\rangle\langle v_i| & \text{if } f \text{ is even,} \end{cases}$$

for certain definite-parity polynomials $f \colon [-1, 1] \to \mathbb{C}$, such that $|f(x)| \leq 1 \; \forall \; x \in [-1, 1]$. Crucially, QSVT does not require us to know the SVD in advance; the transformation is carried out automatically by following an SVD-agnostic procedure outlined below. Note that $f^{(SV)}(A)$ only coincides with the matrix function $f(A)$ for Hermitian $A$ (see §Caveats, below). In the Hermitian case, we can also obtain block-encodings of mixed-parity or complex functions by taking linear combinations of block-encodings—see [10] for examples.

By considering $U_A|0^m\rangle|v_i\rangle$ and $U_A^\dagger|0^m\rangle|w_i\rangle$ one can show that (see [11] for a step-by-step derivation) $U_A$ and $U_A^\dagger$ act as linear maps between the 2D subspaces $S_i := \mathrm{Span}\{|0^m\rangle|v_i\rangle, |\perp_i\rangle\}$ and $S_i' := \mathrm{Span}\{|0^m\rangle|w_i\rangle, |\perp_i'\rangle\}$, with $U_A$ being a transition matrix between these bases given by

$$\begin{array}{cc} & \begin{array}{cc} |0^m\rangle|v_i\rangle & \quad |\perp_i\rangle \end{array} \\ \begin{array}{c} |0^m\rangle|w_i\rangle \\ |\perp_i'\rangle \end{array} & \begin{pmatrix} \sigma_i & \sqrt{1-\sigma_i^2} \\ \sqrt{1-\sigma_i^2} & -\sigma_i \end{pmatrix}, \end{array} \tag{48}$$





where both $|\perp_i\rangle, |\perp'_i\rangle$ are orthogonal to $|0^m\rangle$ (but not necessarily to each other).[39] The 2D subspace $S_i$ is invariant under the operation $W := Z_{|0^m\rangle} U_A^\dagger Z_{|0^m\rangle} U_A$ (where $Z_{|0^m\rangle} = (2|0^m\rangle\langle 0^m| - I)$). The operation $W$, restricted onto the 2D subspace $S_i$, is written as

$$
\begin{pmatrix}
\sigma_i & \sqrt{1-\sigma_i^2} \\
-\sqrt{1-\sigma_i^2} & \sigma_i
\end{pmatrix}^2 .
$$

An additional application of $Z_{|0^m\rangle} U_A$ maps back into the $S'_i$ subspace. By analogy with qubitization, repeated application of $W$ applies a Chebyshev polynomial to each of the singular values of $A$. In analogy with quantum signal processing, by lifting the $Z_{|0^m\rangle}$ reflection operation to a (controlled) rotation $e^{i\phi_j Z_{|0^m\rangle}}$ we can impose polynomial transformations of the singular values of $A$, which then induce the claimed polynomial transformation of $A$. It is typically convenient to use an additional ancilla qubit to implement $e^{i\phi_j Z_{|0^m\rangle}}$.

We define a QSVT circuit as the unitary sequence

$$
U_\Phi := \begin{cases}
e^{i\phi_1 Z_{|0^m\rangle}} U_A \prod_{j=1}^{(d-1)/2} \left( e^{i\phi_{2j} Z_{|0^m\rangle}} U_A^\dagger e^{i\phi_{2j+1} Z_{|0^m\rangle}} U_A \right) & \text{if } d \text{ is odd, and} \\
\prod_{j=1}^{d/2} \left( e^{i\phi_{2j-1} Z_{|0^m\rangle}} U_A^\dagger e^{i\phi_{2j} Z_{|0^m\rangle}} U_A \right) & \text{if } d \text{ is even,}
\end{cases}
$$

where $\Phi = (\phi_1, \phi_2, \ldots, \phi_d)$. We have that

$$
(\langle 0^m| \otimes I) U_\Phi (|0^m\rangle \otimes I) = P^{(SV)}(A) = \begin{cases}
\sum_i P(\sigma_i)|w_i\rangle\langle v_i|, & \text{for odd } d, \text{ and} \\
\sum_i P(\sigma_i)|v_i\rangle\langle v_i|, & \text{for even } d,
\end{cases}
$$

that is, the unitary $U_\Phi$ is a block-encoding of $P^{(SV)}(A)$, were $P$ is the same polynomial that appears in quantum signal processing because the 2D matrix of Eq. (48) has the same form as the analogous 2D matrix in Eq. (46). We note that the constraints on the polynomials typically preclude direct implementation of the desired function as outlined above. By exploiting that $-\Phi$ implements $P^*$, we can use the circuit shown in Fig. 6 to implement a block-encoding of

$$
P_\Re(A) = (\langle +| \otimes \langle 0^m| \otimes I)(|0\rangle\langle 0| \otimes U_\Phi + |1\rangle\langle 1| \otimes U_{-\Phi})(|+\rangle \otimes |0^m\rangle \otimes I)
$$

for any definite-parity polynomial $P_\Re : [-1, 1] \to [-1, 1]$ by appropriately choosing $\Phi$ to implement a complex polynomial that fulfills the QSP conditions and then taking linear combinations of $U_\Phi, U_{-\Phi}$ to give a block-encoding of $P_\Re(A)$ [3, 5, 10].

**Dominant resource cost (gates/qubits)**

Given a degree-$d$ even-parity polynomial $f : [-1, 1] \to [-1, 1]$ and a $(1, m, 0)$-block-encoding $U_A$ of $A$, one can implement a block-encoding of $f(A)$ using $d/2$ calls to $U_A$, $d/2$ calls to $U_A^\dagger$, $2d$ $m$-controlled Toffoli gates, and $d$ single-qubit $Z$ rotations (as shown in Fig. 6). Implementing a degree $d+1$ odd polynomial additionally requires another call to $U_A$, another two $m$-controlled Toffoli gates, and another single-qubit $Z$ rotation. The QSVT circuit implements a $(1, m+1, 0)$-block-encoding of $f(A)$.

---

[39]If $\sigma_i = 1$, then there is no need for $|\perp_i\rangle, |\perp'_i\rangle$, and the subspaces $S_i$, $S'_i$ become 1D.





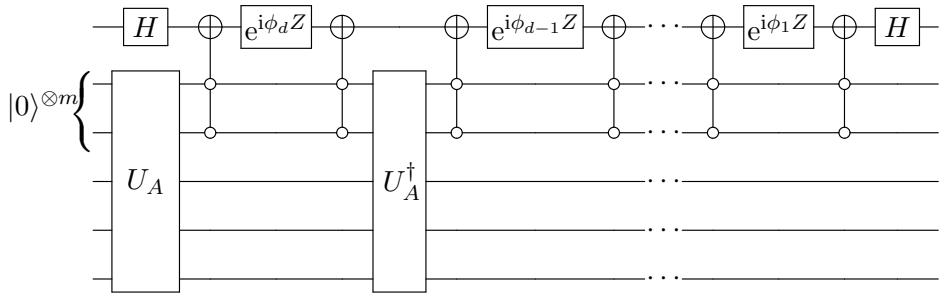

Figure 6: The QSVT circuit $U_\Phi$ which transforms a block-encoding $U_A$ of $A$ into a block-encoding of $f(A)$ for definite-parity $f : [-1, 1] \to [-1, 1]$ polynomial of degree $d$. As discussed in the main text, the angles $\{\phi_i\}$ can be calculated using efficient classical algorithms.

If $U_A$ is imperfect (i.e., it is a $(1, m, \epsilon)$-block-encoding of $A$), then [3, Lemma 22] shows that the error in $f(A)$ is bounded by $4d\sqrt{\epsilon}$; that is, QSVT implements a $(1, m + 1, 4d\sqrt{\epsilon})$-block-encoding of $f(A)$. Moreover, if the norm of $A$ is bounded away from 1, for example, $\|A\| \leq 1/2$, then the perturbation bound can be improved to $\mathcal{O}(d\epsilon)$ [3, Lemma 23].

Given an initial state $|\psi\rangle$, the success probability of implementing $f(A)|\psi\rangle$ is given by $|\langle\psi|f(A)^\dagger f(A)|\psi\rangle|^2$.

#### Caveats

Since the output must be subnormalized to ensure the existence of a unitary block-encoding of $f(A)$, $f$ must satisfy $|f(x)| \leq 1 \; \forall \; x \in [-1, 1]$.

As noted above, $f^{(SV)}(A)$ is only guaranteed to coincide with the matrix function $f(A)$ for Hermitian $A$. As an example, choosing $f(x) = x^2$ we have $f^{(SV)}(A) = \sum_i \sigma_i^2 |v_i\rangle\langle v_i| = A^\dagger A$ whereas $A^2 = \sum_{i,j} \sigma_i \sigma_j |w_i\rangle\langle v_i|w_j\rangle\langle v_j|$. As discussed above, for the Hermitian case we can implement a block-encoding of a mixed-parity function $f$ by taking linear combinations of block-encodings of its even and odd parts. However, in the general case when $|w_i\rangle$ and $|v_i\rangle$ do not coincide, it does not seem to be possible to remove the parity constraint, as the odd $\sum_i f_{\text{odd}}(\sigma_i)|w_i\rangle\langle v_i|$ and even $\sum_i f_{\text{even}}(\sigma_i)|v_i\rangle\langle v_i|$ singular value transforms potentially map to different subspaces.

As discussed in Section 10.3 on quantum signal processing, computing the angle sequence $\Phi$ can be a nontrivial classical task. Several approaches for accomplishing this task have been studied [12, 3, 13, 14, 10, 15, 16, 17, 18, 19, 20, 21], and researchers have reported computing angle sequences for polynomials up to $d \approx 10^7$ [20, 21].

As noted above, if $f(A)$ has small singular values, then preparing a quantum state $f(A)|\psi\rangle$ might require many repeated uses of its block-encoding, thus the normalization factor of $f$ plays a crucial role in efficiency.

In many applications, one seeks to apply a function that is not a polynomial (e.g., $e^x$, $e^{ix}$, $\text{erf}(x)$). In such cases, one needs to first approximate the desired function by a polynomial (incurring an approximation error $\epsilon$) in order to apply QSVT.





**Example use cases**

- **Linear equation solving**: Apply a polynomial approximation of $1/x$ to a block-encoding of $A^\dagger$ to get an approximate block-encoding of the pseudoinverse $A^+$.

- **Hamiltonian simulation**: Apply polynomial approximations of $\sin(x)$ and $\cos(x)$ to a block-encoding of a Hamiltonian $H$ and combine them with **linear combination of unitaries** and **amplitude amplification** to obtain a block-encoding of $e^{iHt}$.

- **Fixed-point amplitude amplification** [22]: Construct a polynomial that maps values in the domain $[a_{\min}, 1]$ to the range $[1-\delta, 1]$, and apply this polynomial to a state-preparation unitary that prepares the desired state with amplitude $a$. The result is amplification of the amplitude to at least $1-\delta$ as long as $a > a_{\min}$.

- For additional applications, see [3, 9, 5, 23, 24].

**Further reading**

- The QSVT framework was introduced in [3] and is also discussed in detail in [25].

- A pedagogical tutorial of the QSVT framework is given in [5, 11].

- A streamlined derivation of QSVT is presented in [26].

# 11 Hamiltonian simulation

The task of Hamiltonian simulation is to approximately compile the evolution under a Hamiltonian $H(t)$, for time $t$, into a sequence of quantum gates. For a time-independent Hamiltonian, solving the Schrödinger equation (setting $\hbar = 1$) yields a time evolution operator $U(t) = e^{-iHt}$. In this section, we will discuss the equivalent operator $U(t) = e^{iHt}$, which is the more common definition in an algorithmic setting. We will assume $t \geq 0$, without loss of generality. The Hamiltonian of interest can arise from physical systems (e.g., quantum chemistry, condensed matter systems, or quantum field theories) but may also be constructed for other applications, such as differential equation simulation. Quantum simulation does not give full access to the amplitudes of the wavefunction during the simulation, unlike classical approaches based on exact diagonalization (or similar methods). Instead, we are only able to measure observables with respect to the time-evolved state, or use the state as an input to other quantum subroutines. Nevertheless, there are no known efficient classical methods that achieve this for general local or sparse Hamiltonians, suggesting an exponential quantum speedup. In fact, as a quantum computation can be expressed as a time evolution under a sequence of local (time-dependent) Hamiltonians, quantum simulation (i.e., time evolution and measurement of a given observable) is a BQP-complete problem.

Hamiltonian simulation algorithms require access to the Hamiltonian. There are three commonly used input models. The Pauli input model assumes that the Hamiltonian is given classically as a sum of products of Pauli operators, for example, $H = \sum_l h_l H_l$, where $h_l$ are coefficients and $H_l$ are multiqubit Pauli products. The $d$-sparse access model assumes that the Hamiltonian is a sparse matrix with at most $d$ nonzero elements per row or column. We require that the locations of the nonzero elements and their values are efficient to compute classically. The density matrix access model assumes that the Hamiltonian corresponds to a density matrix, which we are either provided access to [1], or given a unitary that prepares a purification of the density matrix [2]. All of these input models can be used to prepare block-encodings of the Hamiltonian, which provides a standard-form access model that generalizes the above input models. Block-encodings are the input model of choice for some algorithms for Hamiltonian simulation (e.g., qubitization with quantum signal processing) [2].

Hamiltonian simulation can be used as a subroutine in a range of algorithms, including quantum phase estimation, quantum linear system solvers, Gibbs state preparation, and the quantum adiabatic algorithm. We remark that some of these algorithms are implicitly using Hamiltonian simulation to provide coherent, unitary access to the Hamiltonian. This can be particularly useful if few ancilla qubits are available, which may inhibit the use of some approaches to coherently access the Hamiltonian (e.g., block-encodings based on linear combinations of unitaries) but does not prevent the use of Hamiltonian simulation based on product formulas.

**In this section, we consider four commonly studied algorithms for Hamiltonian simulation:**







Each algorithm has its own advantages and disadvantages, as described at a high level in Table 6. Specific optimizations of each algorithm may be available for a given Hamiltonian. One can also consider hybridized methods combining two or more of the algorithms [3, 4, 5, 6, 7, 8]. There are also other methods for Hamiltonian simulation, such as quantum walks [9, 10, 11] or density matrix–based Hamiltonian simulation [1, 12], which we do not discuss due to their less widespread use as algorithmic primitives for the applications discussed elsewhere in this survey.

| | Product formula (order $k$) | qDRIFT | Taylor and Dyson series | QSP/QSVT |
|---|---|---|---|---|
| # Qubits | $\mathcal{O}(n)$ | $\mathcal{O}(n)$ | $\mathcal{O}(n + \log(L))$ | $\mathcal{O}(n + \log(L))$ |
| Access model | Pauli<br>Sparse | Pauli<br>Sparse | Pauli<br>Sparse<br>Block-encoding | Block-encoding |
| Scaling | $\mathcal{O}\left(5^{2k}nL\|H\|_1 t(\|H\|_1 t\epsilon^{-1})^{\frac{1}{2k}}\right)^a$ | $\mathcal{O}(n\|H\|_1^2 t^2\epsilon^{-1})$ | $\tilde{\mathcal{O}}(\|H\|_1 tnL\log(\epsilon^{-1}))$ | $\mathcal{O}(nL(\|H\|_1 t + \log(\epsilon^{-1}))^b$ |
| Pros | Commutator scaling.<br>Simple implementation.<br>Empirical performance.<br>Minimal ancilla qubits. | $L$-independent scaling.<br>No ancilla qubits. | $\log(1/\epsilon)$ scaling.<br>Time-dependent simulations. | Optimal scaling with $t, \epsilon$.<br>Few ancilla qubits for algorithm. |
| Cons | Scaling with $t, \epsilon$ at low orders.<br>Exponential prefactor (in order $k$). | Scaling with $t, \epsilon$. | Many ancilla qubits if<br>using noncompressed variant [3]. | Time-dependent simulation.<br>Ancilla/gate cost of block-encoding. |

Table 6: High-level comparison of Hamiltonian simulation techniques. Quantitative comparisons assume a Pauli input model (which can easily be used to prepare a block-encoding of the Hamiltonian). For the stated complexity, we consider evolution $U(t) = e^{iHt}$ for time $t$ under a time-independent Hamiltonian $H$ on $n$ qubits, given as a sum of $L$ Pauli products $H = \sum_{j=1}^{L} h_j P_j$. The evolution is approximate to error $\epsilon$ in the spectral norm (diamond norm for qDRIFT). We define $\|H\|_1 = \sum_{j=1}^{L} |h_j|$. The qubit requirement for the Taylor and Dyson series method omits additional additive factors that scale logarithmically with the norm and/or derivative of the Hamiltonian. In specific applications it may be possible to reduce the number of qubits and/or gate complexity further by exploiting knowledge of the system, such as symmetries, commutation structure, or energy scales. For example, the factor of $n$ present in the above complexities may be reduced by exploiting locality in the Pauli product terms of the Hamiltonian.

---

[a] The factor of $n$ can be reduced to $w$ when each Pauli term $P_j$ acts nontrivially on at most $w$ sites. The factor $\|H\|_1^{1+1/2k}$ can be reduced by exploiting commutativity of the various $P_j$.

[b] The factor $nL$ derives from an upper bound on the gate complexity of block-encoding, and it can often be significantly improved by exploiting structure in $h_j$ and $H_j$.


The authors are grateful to Yuan Su for reviewing this section of the survey.

## 11.1  Product formulas

**Rough overview (in words)**

Product formulas (or Trotter–Suzuki formulas/Trotterization) [1] are the most commonly used approach for Hamiltonian simulation, and are applicable to Hamiltonians in the Pauli access model and the sparse access model (see below for definitions of these models). Product formulas divide the evolution into a repeating sequence of short unitary evolutions under subterms of the Hamiltonian. These subterm evolutions have a known decomposition into elementary quantum gates. The error in product formulas depends on the commutators between different terms in the decomposition; if all of the terms in the Hamiltonian commute, product formulas are exact.

Product formula approaches have also been extended to treat time-dependent Hamiltonians [2, 3, 4, 5, 6]. In the following discussion, we will restrict our focus to the time-independent case, noting that the time-dependent approaches are executed in the same way, but have a slightly more complex error analysis.

**Rough overview (in math)**

Given a Hamiltonian $H$, desired evolution time $t$, and error $\epsilon$, return a circuit $U(t)$ made of elementary gates such that

$$\|U(t) - \mathrm{e}^{\mathrm{i}Ht}\| \le \epsilon.$$

In the above, we use the operator norm $\|\cdot\|$ (the maximal singular value) to quantify the quality of approximation, which controls the error for arbitrary input states (in trace distance) and for observables. This worst-case metric is mathematically convenient, but, as discussed below, tighter bounds may be obtained by using error metrics more closely aligned with the specification of the problem.

A product formula generates $U(t)$ through a product of easy-to-implement evolutions under terms in the Hamiltonian. For a Hamiltonian decomposition $H = \sum_{j=1}^{L} H_j$ with $L$ terms, the first-order product formula with $r$ steps is

$$S_1(t) = \left( \prod_{j=1}^{L} \mathrm{e}^{\mathrm{i}H_j t/r} \right)^r.$$

The error in the first-order product formula is upper bounded as [7]

$$\|S_1(t) - \mathrm{e}^{\mathrm{i}Ht}\| \le \frac{t^2}{2r} \sum_{i=1}^{L} \left\| \sum_{j=i+1}^{L} [H_i, H_j] \right\| \le \frac{\|H\|_1^2 t^2}{2r},$$

where $\|H\|_1 = \sum_{j=1}^{L} \|H_j\|$. Higher-order formulas can be defined recursively, and are referred to as $(2k)$th-order product formulas. The error in a recursively defined $(2k)$th-order product formula is bounded by [7]

$$\|S_{2k}(t) - \mathrm{e}^{\mathrm{i}Ht}\| = \mathcal{O}\left( \frac{\|H\|_1^{2k+1} t^{2k+1}}{r^{2k}} \right).$$

Product formulas can be applied to $d$-sparse Hamiltonians (at most $d$ nonzero elements per row/column) with efficiently row computable nonzero elements [8]. Access to the nonzero





elements of the Hamiltonian is provided via oracles $O_f$ and $O_H$. The oracle $O_f$ returns the column index ($j$) of the $k \in \{1, \ldots, d\}$th nonzero element in row $i$. The oracle $O_H$ returns the value of the matrix element $H_{ij}$.

$$O_f : O_f|k\rangle|i\rangle|0\rangle = |k\rangle|i\rangle|j\rangle$$
$$O_H : O_H|i\rangle|j\rangle|0\rangle = |i\rangle|j\rangle|H_{ij}\rangle.$$

Using graph-coloring algorithms, a $d$-sparse Hamiltonian $H$ can be efficiently decomposed into a sum of efficiently simulable sparser Hamiltonians [9, 10]. Illustrating this idea with the approach of [9], we can decompose $H = \sum_{j=1}^{6d^2} H_j$, where each $H_j$ is 1-sparse. The nonzero elements of a given $H_j$ can be computed coherently using $\mathcal{O}(\log^*(n))$ queries to $O_f, O_H$, where $\log^*$ is the iterated logarithm.[40] Time evolution under a 1-sparse Hamiltonian can be implemented efficiently using the approach of [8]. To simulate $e^{iHt}$ using, for example, a first-order product formula, we sequentially apply each $e^{iH_jt}$ using the methods outlined above.

As a special case of the $d$-sparse access model, one can consider Hamiltonians given as a linear combination of $L$ Pauli terms $H = \sum_{j=1}^{L} H_j = \sum_{j=1}^{L} \alpha_j P_j$, as each Pauli tensor product is already a 1-sparse matrix (so in this case, $d \leq L$). Time evolution under each Pauli term (or in some cases, groups of Pauli terms) can be simulated efficiently (see, e.g., [1, 11]), considerably simplifying the $d$-sparse construction by removing the need for oracles $O_f$ and $O_H$.

**Dominant resource cost (gates/qubits)**

For an $n$-qubit Hamiltonian, product formulas act on $n$ qubits. In the Pauli access model, no additional ancilla qubits are required. In the sparse access model, ancilla qubits may be required to implement the oracles $O_f$ and $O_H$ and to implement time evolution under 1-sparse Hamiltonians $H_j$.

The gate complexity is obtained by choosing the number of Trotter steps $r$ sufficiently large to obtain an error $\epsilon$ and multiplying by the complexity of implementing each step of the product formula. It is necessary to balance the improved asymptotic scaling with $t$ and $\epsilon$ of higher-order Trotter formulas against the exponentially growing prefactor of the higher-order formulas. In practical simulations of chemistry, condensed matter systems, or quantum field theories, a low-order formula (2nd–6th) typically minimizes the gate count.

A recursively defined $(2k)$th-order product formula (i.e., the first-order formula is given by $k = 1/2$, and is the base case) for simulating a $d$-sparse Hamiltonian for time $t$ to accuracy $\epsilon$ requires [10]

$$\mathcal{O}\left(5^{2k}d^2(d + \log^* n)\|H\|t\left(\frac{d\|H\|t}{\epsilon}\right)^{1/2k}\right)$$

calls to the oracles $O_f$ and $O_H$.

A recursively defined $(2k)$th-order product formula for simulating an $L$-term Hamiltonian in the Pauli access model for time $t$ to accuracy $\epsilon$ requires [7]

$$\mathcal{O}\left(5^{2k}nLt\left(\frac{t\alpha_{\mathrm{comm},k}}{\epsilon}\right)^{1/2k}\right) \tag{49}$$

---

[40] For practical purposes, the iterated logarithm is essentially constant, since $\log^*(n) \leq 5$ for all $n \leq 2^{65536}$.





elementary single- and two-qubit gates, where

$$\alpha_{\mathrm{comm},k} = \sum_{i_1,i_2,\dots,i_{2k+1}} \|[H_{i_{2k+1}},\dots[H_{i_2},H_{i_1}]]\|.$$

The dependence on $\alpha_{\mathrm{comm},k}$ can be tightened and calculated for lower-order formulas (see [7] for full calculations). The dependence on $n$ can be reduced to $w$ for local Hamiltonians with Pauli terms that each act on at most $w$ qubits.

**Caveats**

The error bounds of product formulas in the Pauli access model have been the object of significant investigation. Evaluating the tightest spectral norm bounds requires computing a large number of commutators between the terms in the Hamiltonian, which can be computationally intensive. Numerical simulations have shown that the commutator bounds can be loose by several orders of magnitude for chemical [12, 13] or spin [14] systems.

The spectral norm is the worst-case metric; it is an active area of research to find error metrics better suited to the problem at hand. For example, one may consider the *average*-case error over random input states [15, 16] by the normalized Frobenius norm $\|U(t) - e^{iHt}\|_F/\sqrt{2^n}$. Recently, in [15] it was shown that the average-case error can be much smaller than the worst-case error for systems with large connectivity. More directly, one can also compute the Trotter error associated with input states from the low-energy [17, 18] or low-particle-number subspace [19, 20].

The gate counts of product formula approaches can also be reduced by grouping together mutually commuting terms such that they can be implemented using fewer gates than would be required to implement all the terms individually [21, 22, 23]. One can also reduce the number of Trotter steps required by randomizing the ordering of the terms [24, 25, 6] (although this must be balanced against any compilation benefits that may be obtained from a fixed ordering).

**Example use cases**

- Physical systems simulation: quantum chemistry, condensed matter systems, quantum field theories.

- Algorithms: quantum phase estimation, quantum linear system solvers, Gibbs state preparation, quantum adiabatic algorithm.

**Further reading**

- A rigorous derivation of the error in product formulas [7].

- A comparison of product formula methods with other approaches to Hamiltonian simulation for a concrete problem of interest [14].

- Video lectures on product formulas for Hamiltonians in the Pauli access model and product formulas for $d$-sparse Hamiltonians.

## 11.2 qDRIFT

**Rough overview (in words)**

The *quantum stochastic drift protocol* [1], abbreviated as qDRIFT, operates in the Pauli access model[41] and approximates the Hamiltonian simulation channel (as opposed to the unitary) by randomly sampling a term from the Hamiltonian (according to the coefficient magnitudes), and then evolving under the chosen term. This process is repeated for a number of steps. Because it approximates the channel, rather than the unitary, it can be more difficult to use qDRIFT as a coherent subroutine in other algorithms (see §Caveats below).

The error in qDRIFT depends on the 1-norm of Hamiltonian coefficients. One main advantage of qDRIFT is that it does not explicitly depend on the number of terms in the Hamiltonian and has small constant overheads, making it well suited to systems with rapidly decaying interaction strengths, dominated by a few large terms. However, qDRIFT's time and error dependence are asymptotically worse than other methods, which seems to originate from the algorithm's randomized nature [5]. qDRIFT can also be extended to time-dependent Hamiltonian simulation with a Hamiltonian $H(t)$, where the gate count of the algorithm scales as $\int_0^t \|H(t')\| dt'$, rather than as $t \max_{t'} \|H(t')\|$ like other Hamiltonian simulation algorithms [6]. We will restrict our discussion below to the time-independent case.

**Rough overview (in math)**

Given a Hamiltonian in the Pauli decomposition $H = \sum_i h_i H_i$ (with $\|H_i\| = 1$), qDRIFT provides a stochastic channel $\mathcal{N}$ that, when applied for $N$ steps, approximates the Hamiltonian simulation channel

$$\|\mathcal{N}^N - e^{iHt}(\cdot)e^{-iHt}\|_\diamond \le \epsilon$$

to within diamond-norm error $\epsilon$.

qDRIFT proceeds by randomly sampling terms according to their relative importance

$$X_k \overset{i.i.d.}{\sim} \quad \frac{h_i H_i}{p_i}, \quad \text{where} \quad p_i = \frac{|h_i|}{\|H\|_1}$$

and $\|H\|_1 := \sum_i |h_i|$ is the sum of the strengths. Each step of qDRIFT then evolves the randomly sampled term $X_k$ for a short period of time $t/N$, where $N$ is a free parameter determining the number of qDRIFT steps, which controls the error in the simulation. The result is the following quantum channel:

$$\mathcal{N}[\rho] = \mathbb{E}[e^{i(t/N)X_k}\rho e^{-i(t/N)X_k}].$$

As discussed above, this channel is repeated for $N$ steps, in order to approximate the Hamiltonian simulation channel.

**Dominant resource cost (gates/qubits)**

For an $n$-qubit Hamiltonian, qDRIFT acts on $n$ register qubits, and no additional ancilla qubits are required.

---

[41]qDRIFT was originally formulated, and is typically presented, for the Pauli access model [1], but the algorithm appears compatible with the $d$-sparse access model by applying it to the $d$-sparse decompositions in [2, 3] [4].





In order to simulate the Hamiltonian evolution channel to within diamond-norm error $\epsilon$, we require

$$N = \mathcal{O}\left(\frac{\|H\|_1^2 t^2}{\epsilon}\right)$$

steps of qDRIFT [1, 5]. While the diamond-norm is a different error metric to the spectral norm used in other articles in this section, both metrics provide upper bounds on the error in an observable measured with respect to the time-evolved state [1]. For unitary channels, the diamond norm is effectively equal to the spectral norm (see, e.g., discussion in [7], up to constant factors).

The gate complexity is the number of steps multiplied by the individual costs of the elementary evolution $e^{i(t/N)X_k}$, which scales linearly with the locality of the Pauli operator $X_k$. When using qDRIFT to time evolve a state (e.g., for the purpose of measuring an observable), it is important to average the results over a sufficient number of independently sampled qDRIFT circuits [1].

**Caveats**

The qDRIFT algorithm has a quadratic dependence on time and a linear dependence on the inverse error $1/\epsilon$, while other Hamiltonian simulation methods can achieve linear time dependence and logarithmic inverse error dependence. A higher-order variant of qDRIFT was recently developed that improves the error dependence, but it is only suitable for estimating the expectation value of observables with respect to the time-evolved state, rather than approximating the unitary channel itself [8]. It is currently unclear how to design higher-order variants of qDRIFT that improve the time dependence, which appears to result from the randomized nature of the algorithm [5].

As discussed above, qDRIFT approximates the time evolution channel, rather than the unitary $e^{iHt}$. As a result, it can be difficult to incorporate as a subroutine in algorithms that seek to manipulate the unitary directly—for example, measuring $\mathrm{Tr}(U(t)\rho)$. Tasks of this form feature in some approaches for phase estimation [9], motivating alternate, qDRIFT-inspired approaches, in order to exploit qDRIFT-like benefits [10].

**Example use cases**

- Physical systems simulation: quantum chemistry, condensed matter systems, quantum field theories.

- Algorithms: quantum phase estimation, quantum linear system solvers, Gibbs state preparation, quantum adiabatic algorithm.

- Hybridization with other quantum simulation methods [11, 12, 13].

- Using importance sampling to incorporate variable gate costs for simulating different terms $X_k$ [14].

## 11.3 Taylor and Dyson series (linear combination of unitaries)

**Rough overview (in words)**

Taylor and Dyson series approaches for Hamiltonian simulation expand the time evolution operator as a Taylor series (time independent) [1] or Dyson series (time dependent) [2, 3] and use the linear combination of unitaries (LCU) primitive to apply the terms in the expansion, followed by (robust, oblivious) amplitude amplification to boost the success probability close to unity. These methods are close to being asymptotically optimal, achieving linear scaling in time and logarithmic dependence on the error. However, they use a large number of ancilla qubits, compared to other Hamiltonian simulation algorithms.

**Rough overview (in math)**

We focus on the time-independent case and follow the presentation in [1]. Given a Hamiltonian $H$, desired evolution time $t$, and error $\epsilon$, return a circuit $U(t)$ made of elementary gates such that

$$\|U(t) - \mathrm{e}^{\mathrm{i}Ht}\| \le \epsilon.$$

In the above, we use the operator norm (the maximal singular value) to quantify the worst-case error in the simulation.

The total evolution time $t$ is divided into $r$ segments. In each segment, we evolve under an approximation of $\mathrm{e}^{\mathrm{i}Ht/r}$. The Hamiltonian is decomposed into a linear combination of unitary operations $H = \sum_{l=1}^{L} \alpha_l H_l$, where we choose $\alpha_l$ real and positive by shifting phases into $H_l$, and $\|H_l\| = 1$. This decomposition appears naturally when the Hamiltonian is given as a linear combination of Pauli products. We approximate $\mathrm{e}^{\mathrm{i}Ht/r}$ using a Taylor expansion truncated to degree $K$

$$\mathrm{e}^{\mathrm{i}Ht/r} \approx U(t/r) := \sum_{k=0}^{K} \frac{1}{k!}(\mathrm{i}Ht/r)^k$$

$$= \sum_{k=0}^{K} \sum_{l_1,\ldots,l_k=1}^{L} \frac{(\mathrm{i}t/r)^k}{k!}\alpha_{l_1}\ldots\alpha_{l_k} H_{l_1}\ldots H_{l_k}.$$

Each segment $U(t/r)$ is implemented using robust oblivious amplitude amplification. Amplitude amplification is necessary because truncating the Taylor series at degree $K$ makes $U(t/r)$ non-unitary. However, textbook amplitude amplification necessitates reflecting around the initial state (as well as the "good" state), which would be problematic since Hamiltonian simulation requires synthesizing a unitary that works simultaneously for all input states. This issue can be circumvented using oblivious amplitude amplification: we are given a unitary $V$ such that for any state $|\psi\rangle$, we have $V|0^m\rangle|\psi\rangle = a|0^m\rangle U|\psi\rangle + b|0_\perp^m \phi\rangle$, for a unitary operator $U$, and the goal is to amplify the state $|0^m\rangle U|\psi\rangle$ to be obtained with probability 1 (we can recognize $V$ as an $(a, m, 0)$ unitary block-encoding of $U$). A further problem is that the above operator $U(t/r)$ is non-unitary, and so deviates from the formulation of oblivious amplitude amplification [4]. The proven "robustness" property of oblivious amplitude amplification [1] ensures that the error induced by treating $U(t/r)$ as a probabilistically implemented unitary does not accumulate.





The value of $K$ controls the error in the simulation and can be chosen as

$$K = \mathcal{O}\left(\frac{\log(\|H\|_1 t/\epsilon)}{\log\log(\|H\|_1 t/\epsilon)}\right),$$

where we define $\|H\|_1 := \sum_{l=1}^{L} \alpha_l$. The total time evolution is divided into $r = \lceil \|H\|_1 t/\ln(2)\rceil$ segments, each of duration $\ln(2)/\|H\|_1$, which ensures that a single application of robust oblivious amplitude amplification boosts the success probability of the segment to unity.

Within each segment, we apply $U(t/r)$ using the LCU primitive. This technique can be applied to Hamiltonians given in both the Pauli and $d$-sparse access models. For the Pauli access model, the Hamiltonian is already in the form of a linear combination of unitary operators. For the $d$-sparse case, we can use graph coloring algorithms [5, 6] to decompose the $d$-sparse Hamiltonian into a linear combination of unitaries, where each unitary is 1-sparse and self-inverse.

**Dominant resource cost (gates/qubits)**

In addition to the $n$-qubit data register, the Taylor series approach requires a number of ancilla registers to implement the LCU technique. In the original formulation [1], a register with $K$ qubits is used to control the degree of the Taylor expansion, storing the value as $|k\rangle = |1^{\otimes k}0^{\otimes(K-k)}\rangle$. An additional $K$ registers, each containing $\lceil \log_2(L)\rceil$ qubits, are used to index the possible values of each of the possible $H_{l_k}$. Hence, the overall space complexity of the original formulation [1] is $\mathcal{O}(n + K\log(L)) = \mathcal{O}(n + \log(\|H\|_1 t/\epsilon)\log(L))$. In [7] it was shown how to reduce the space complexity to $\mathcal{O}(n + \log(K) + \log(L))$ using quantum counter circuits.

Additional ancilla qubits may be required to implement the LCU gadget (e.g., in the sparse access model) or for the reflections used in robust oblivious amplitude amplification.

As discussed above, implementing each segment requires one use of robust oblivious amplitude amplification, which makes two calls to the LCU circuit and one call to its inverse. The method incurs approximation errors from truncating the Taylor series at degree $K$ and from the use of robust oblivious amplitude amplification. The resulting error per segment is bounded by $(e\ln(2)/(K+1))^{K+1}$.

The cost of the LCU circuit depends on the Hamiltonian access model. For the case of the Pauli access model, the LCU circuit requires two calls to a PREPARE operation that prepares the ancilla registers with the correct coefficients. In the compressed formulation [7], this requires $\mathcal{O}(L + K)$ gates (compared to $\mathcal{O}(LK)$ gates in the original formulation [1]). The LCU circuit also requires one call to a SELECT oracle, which can be implemented using $K$ controlled select$(H)$ operations. Each of these $K$ operations can be implemented using $\mathcal{O}(Ln)$ elementary gates [8, 9] (using quantum read-only memory). The overall gate complexity in the Pauli access model is thus

$$\mathcal{O}\left(\frac{\|H\|_1 tLn \log(\|H\|_1 t/\epsilon)}{\log\log(\|H\|_1 t/\epsilon)}\right) = \widetilde{\mathcal{O}}\left(\|H\|_1 tLn \log\left(\frac{1}{\epsilon}\right)\right).$$

Using the LCU approach applied to a 1-sparse decomposition of a $d$-sparse Hamiltonian, the overall complexity is [1]

$$\mathcal{O}\left(\frac{d^2\|H\|_{\max} tn \log^2(d^2|H|_{\max}t/\epsilon)}{\log\log(d^2\|H\|_{\max}t/\epsilon)}\right) = \widetilde{\mathcal{O}}\left(d^2\|H\|_{\max} tn \log^2\left(\frac{1}{\epsilon}\right)\right),$$





where $\|H\|_{\max} = \max_{i,j} |\langle i|H|j\rangle|$. Using the amplification technique of [10], which utilizes quantum singular value transformation (QSVT), the dependence on $d$ and $\|H\|_{\max}$ can be improved in some cases.

The extension to time-dependent Hamiltonians, through the use of a Dyson series, requires an additional "clock" register to store the time value and introduces a logarithmic dependence on the time derivative of the Hamiltonian [2, 3, 7].

**Caveats**

Concrete resource estimates for physical systems of interest have observed that the Taylor series approach may require more ancilla qubits and gates than product formulas or quantum signal processing approaches for Hamiltonian simulation [8], although these qubit counts would be improved by the subsequent compression approach to the algorithm [7]. The gate complexity of the algorithm can be reduced by exploiting anticommutativity in the Hamiltonian [11], adding a corrective operation [12], or pruning terms with small magnitudes from the expansion [13].

**Example use cases**

- Physical systems simulation: quantum chemistry (see [14, 15, 16, 7]), condensed matter systems, quantum field theories.

- Algorithms: quantum phase estimation, quantum linear system solvers, Gibbs state preparation, quantum adiabatic algorithm.

- Hamiltonian simulation in the interaction picture [7].

**Further reading**

- A comparison of several Hamiltonian simulation algorithms, including Taylor series [8].

- Derivations of the compressed variants of Hamiltonian simulation via Taylor/Dyson series [7, Appendices B & D].

- Video lectures on Hamiltonian simulation with Taylor series.

## 11.4   Quantum signal processing / quantum singular value transformation

**Rough overview (in words)**

Quantum signal processing (QSP) and quantum singular value transformation (QSVT) are techniques for applying polynomial transformations to block-encoded operators. These techniques can be used to implement Hamiltonian simulation, given a block-encoding of the Hamiltonian. Both approaches have optimal scaling with $t$ and $\epsilon$ for time-independent Hamiltonians.

QSP was initially developed for the $d$-sparse access model [1]. Through the introduction of block-encodings and qubitization, it was made applicable in a standard form to Hamiltonians in a Pauli access model, $d$-sparse access model, or given as density matrices (where we are given access to a unitary that prepares a purification of the density matrix) [2]. QSVT was later developed as a more general and direct route to the results of QSP [3].

Hamiltonian simulation via QSP or QSVT is less well suited to time-dependent Hamiltonians, as the need to Trotterize the time-dependent evolution breaks the optimal dependence on the parameters.

**Rough overview (in math)**

Access to the Hamiltonian $H$ is provided by an $(\alpha, m, 0)$-block-encoding $U_H$ (the case of approximate block-encodings can be treated using [3, Lemma 22]) such that

$$(\langle 0^m | \otimes I) U_H (|0^m\rangle \otimes I) = H/\alpha.$$

The Hamiltonian has a spectral decomposition of $\sum_\lambda \lambda |\lambda\rangle\langle\lambda|$. We seek to use $U_H$ to implement an operator $U(t)$ approximating

$$\left\| U(t) - \sum_\lambda e^{i\lambda t} |\lambda\rangle\langle\lambda| \right\| \leq \epsilon.$$

Qubitization converts $U_H$ into a more structured unitary $W$ (which is also a block-encoding of the Hamiltonian). The eigenvalues of $W$ are $e^{\pm i \arccos(\lambda/\alpha)}$, directly related to those of $H$. QSP then enables polynomial transformations to be applied to these eigenvalues, which defines the application of the polynomial to $W$. This concept can be generalized via QSVT, which effectively unifies the qubitization and QSP step.

In both cases, our goal is to implement a block-encoding of $U(t) \approx \sum_\lambda e^{i\lambda t} |\lambda\rangle\langle\lambda|$, which defines Hamiltonian simulation. In QSVT we separately implement polynomials approximating $\cos(\lambda t)$ and $i \sin(\lambda t)$, combine them using a linear combination of block-encodings, and boost the success probability using three-step oblivious amplitude amplification. Further details can be found in [3, 4]. Meanwhile, QSP implements $\exp(itH)$ directly but requires an additional ancilla qubit and controlled access to a Hermitian block-encoding $U_H'$, which, when implemented via Eq. (47), uses both controlled $U_H$ and $U_H^\dagger$ resulting in a factor of $\sim 4$ overhead [2]. Altogether, these considerations suggest that the QSVT-based approach might have a slightly better complexity, particularly when controlled $U_H$ is significantly more costly to implement than $U_H$. If $U_H$ is already Hermitian, then QSP can have a lower complexity.

**Dominant resource cost (gates/qubits)**

Using either QSP or QSVT, block-encoding a degree-$k$ polynomial $f(H)$ is performed using $\mathcal{O}(k)$ calls to the block-encoding $U_H$ [2, 3]. Hence, the degree of the polynomial approximating





$e^{iHt}$ determines the complexity of Hamiltonian simulation using these techniques. As noted in [3, Corollary 60], we can rigorously bound the resources for Hamiltonian simulation via QSVT for all values of $t$ as using

$$\mathcal{O}\left(\alpha t + \frac{\log(1/\epsilon)}{\log(e + \log(1/\epsilon)/\alpha t)}\right)$$

calls to the $(\alpha, m, 0)$-block-encoding $U_H$. This query complexity is optimal [5, 3], although the block-encoding can hide additional complexities, in practice. In some cases, the dependence on norm parameters can be improved by exploiting details of the simulated system; see [6, 7].

For a Pauli access model, the block-encoding is implemented using the linear combination of unitaries (LCU) primitives PREPARE and SELECT. For a Hamiltonian with $L$ terms $\alpha = \|H\|_1$, $m = \mathcal{O}(\log(L))$, and two additional qubits are required for QSVT. The overall gate complexity depends on the exact implementation of PREPARE and SELECT, which can often be tailored to the Hamiltonian of interest. In the worst case, PREPARE uses $\Theta(L)$ gates, and SELECT uses $\Theta(nL)$ gates (although these can be significantly improved by exploiting structure in the Hamiltonian; see, e.g., [8, 9]). Thus, the overall worst-case gate complexity is

$$\mathcal{O}\left(nL\left(\|H\|_1 t + \frac{\log(1/\epsilon)}{\log(e + \log(1/\epsilon)/\|H\|_1 t)}\right)\right).$$

For a $d$-sparse access model, $\alpha = d\|H\|_{\max}$, where $\|H\|_{\max} = \max_{i,j} |\langle i|H|j\rangle|$, $m = \mathcal{O}(\log(d))$, and two additional qubits are required for QSVT. The overall gate complexity depends on the cost of sparse access to elements of $H$. Assuming a circuit for sparse access with constant gate complexity, the overall gate complexity is

$$\mathcal{O}\left(d\|H\|_{\max} t + \frac{\log(1/\epsilon)}{\log(e + \log(1/\epsilon)/d\|H\|_{\max} t)}\right).$$

Using the QSVT-based amplification technique of [6], the dependence on $d\|H\|_{\max}$ can be improved in some cases.

The density matrix access model seeks to perform time evolution under $e^{i\rho t}$, given access to either multiple copies of $\rho$ or a unitary $U_\rho$ that prepares a purification of $\rho$. Given $U_\rho$, we can prepare a block-encoding of $\rho$ [2] (see Section 10.1 on block-encodings for details) with $\alpha = 1$. If the gate complexity of $U_\rho$ is $C(U_\rho)$, then the overall gate complexity is

$$\mathcal{O}\left(C(U_\rho)\left(t + \frac{\log(1/\epsilon)}{\log(e + \log(1/\epsilon)/t)}\right)\right).$$

### Caveats

The method was found to perform competitively with Trotterization (and better than Taylor series) in concrete resource estimates for simulating spin-chain Hamiltonians [10]. While that work had difficulty calculating the QSP phase factors, this issue has since been addressed with the development of classical algorithms for finding the phase factors (e.g., [11, 12, 13, 14], and successors). Nevertheless, this contributes a classical preprocessing cost to the algorithm.

It is currently unclear how to perform optimal time-dependent Hamiltonian simulation with these methods, without resorting to Trotterization. Some initial investigations have shown promising results using clock Hamiltonian constructions [15] or for time-periodic Hamiltonians [16, 17].





**Example use cases**

- Physical systems simulation: quantum chemistry, condensed matter systems (see [10]), quantum field theories, differential equations in plasma physics (see [18]).

- Algorithms: quantum phase estimation, quantum linear system solvers, Gibbs state preparation.

**Further reading**

- Pedagogical overviews [4, 19].

- Comparison of several Hamiltonian simulation algorithms [10].

# 12 Quantum Fourier transform

*The authors are grateful to Ronald de Wolf for reviewing this section of the survey.*

**Rough overview (in words)**

The quantum Fourier transform (QFT) is a quantum version of the discrete Fourier transform (DFT) and takes quantum states to their Fourier-transformed version.

**Rough overview (in math)**

The QFT is a quantum circuit that takes pure $N$-dimensional quantum states $|x\rangle = \sum_{i=0}^{N-1} x_i |i\rangle$ to pure quantum states $|y\rangle = \sum_{i=0}^{N-1} y_i |i\rangle$ with the Fourier-transformed amplitudes

$$y_k = \frac{1}{\sqrt{N}} \sum_{l=0}^{N-1} x_l \exp(2\pi \mathrm{i} k l / N) \quad \text{for } k = 0, \ldots, N-1. \tag{50}$$

**Dominant resource cost (gates/qubits)**

The space cost is $\mathcal{O}(\log(N))$ qubits and the quantum complexity of the textbook algorithm is $\mathcal{O}(\log^2(N))$. In terms of Hadamard gates, swap gates, and controlled phase shift gates $|0\rangle\langle 0| \otimes I + |1\rangle\langle 1| \otimes R_\ell$ with

$$R_\ell = \begin{pmatrix} 1 & 0 \\ 0 & \exp\left(2\pi \mathrm{i} 2^{-\ell}\right) \end{pmatrix},$$

the quantum circuit is given in Fig. 7 (see also [1, Fig. 5.1]), where $N = 2^n$. The swap gates

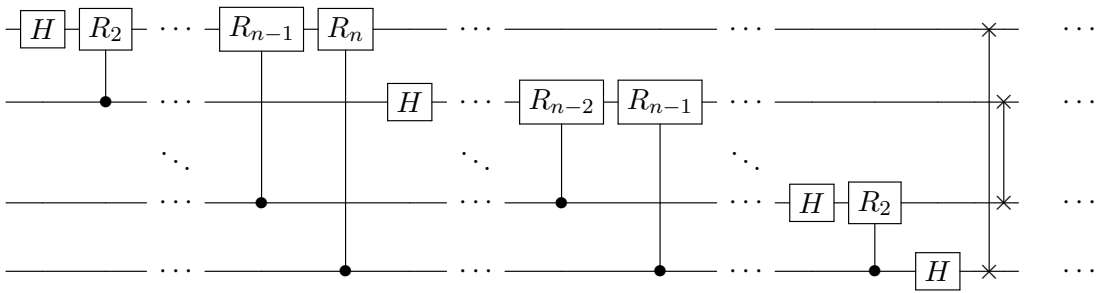

Figure 7: Quantum circuit implementation of QFT.

at the end of the circuit are required to reverse the order of the output qubits. The complexity can be improved to

$$\mathcal{O}\left(\log(N) \log\left(\log(N)\epsilon^{-1}\right) + \log^2\left(\epsilon^{-1}\right)\right)$$

when only asking for $\epsilon$-approximate solutions [2]. Finite constants and compilation cost for fault-tolerant quantum architectures are also discussed in the literature. For example, [3] gives an





implementation with $\mathcal{O}(\log(N) \log \log(N))$ $T$ gates and estimates finite $T$-gate costs for different instance sizes.

**Caveats**

- The QFT does not achieve the same task as the classical DFT that takes vectors $(x_0, \ldots, x_{N-1}) \in \mathbb{C}^N$ to vectors $(y_0, \ldots, y_{N-1}) \in \mathbb{C}^N$ with $y_k$ defined as in Eq. (50). The DFT can be implemented via the fast Fourier transform in classical complexity $\mathcal{O}(N \log(N))$, which is exponentially more costly than the quantum complexity $\mathcal{O}(\log^2(N))$ of the QFT. However, for the QFT to achieve the same task as the DFT, pure state quantum tomography would be required to read out and learn the Fourier-transformed amplitudes, which destroys any quantum speedup for the DFT.

- When QFT is employed in use cases, for example, for factoring, one has to be careful in finite-size instances when counting resources [4], and for this a semiclassical version of the QFT can be more quantum resource efficient [5].

- The QFT admits an efficient representation as a matrix product operator (a type of tensor network), meaning that the approximation improves exponentially in the bond dimension [6]. This suggests that quantum algorithms relying on the QFT for speedup must involve highly entangled input or intermediate states, in order to beat state-of-the-art tensor network methods.

**Example use cases**

- Even though the QFT does not speedup the DFT, QFT is used as a subroutine in more involved quantum routines with large quantum speedup. Examples include quantum algorithms for the discrete logarithm problem, the hidden subgroup problem, the factoring problem, to name a few. The QFT can be seen as the crucial quantum ingredient that allows for a superpolynomial end-to-end quantum speedup for these problems. We discuss this in the context of quantum cryptanalysis in Section 6.

- The QFT appears in the standard circuit for quantum phase estimation, where it is used to convert accrued phase estimation into a binary value that can be read out.

- The QFT is used for switching between the position and momentum bases in grid-based simulations of quantum chemistry [7] or quantum field theories [8].

**Further reading**

- Textbook reference [1, Section 5.1].

- The quantum Fourier transform can be generalized to other groups. The version presented above is for the group $\mathbb{Z}/(2^n \mathbb{Z})$. Its implementation for other abelian groups as well as nonabelian groups is discussed in [9] and the references therein.

# 13  Quantum phase estimation

*The authors are grateful to Patrick Rall and Ronald de Wolf for reviewing this section of the survey.*

**Rough overview (in words)**

The quantum phase estimation (QPE) subroutine produces an estimate of an eigenvalue of a unitary operator. It is a cornerstone of quantum algorithms primitives and has numerous applications. For example, Shor's algorithm for factoring can be viewed as an application of QPE together with modular exponentiation. Similarly, when combined with Hamiltonian simulation, QPE can produce an estimate for an eigenvalue of a Hamiltonian (given an appropriate initial state), an important problem in areas such as quantum chemistry. Generally, since quantum computations enact unitary operators, quantum phase estimation is an essential algorithmic tool for accessing information about these operators, specifically, information about their periodicities, and the properties of their eigenstates.

As one of the oldest quantum primitives discovered [1, 2], QPE has played a significant historical role in the development of quantum algorithms. In a typical use case, QPE is used as a first step to compute an estimate of the eigenvalue of the unitary into an ancilla register. Then, the ancilla register is used as a control for subsequent operations. However, in some applications, such as Gibbs sampling and solving the quantum linear system problem, this procedure must be applied *coherently* to a superposition of eigenstates with different eigenvalues, and the estimate of the eigenvalue must be uncomputed at the end. As discussed below, coherent usage of the QPE primitive in this manner must be handled with care, due to several identified caveats. While QPE still provides essential intuition for how these applications work, in some cases, modern techniques leveraging quantum signal processing and the quantum singular value transformation [3] lead to a cleaner and more direct analysis than QPE.

In the discussion below, we begin with the textbook presentation of QPE [2, 4], and expound on the aforementioned caveats. We also present example use cases, noting the instances where QPE was originally a key ingredient but no longer features directly in state-of-the-art solutions.

**Rough overview (in math)**

Let $U$ be a unitary with eigendecomposition $U = \sum_j e^{i2\pi\phi_j}|\psi_j\rangle\langle\psi_j|$. Given as input the state $|\psi_j\rangle$, the QPE subroutine produces an estimate $\hat{\phi}_j$ for $\phi_j$. The algorithm requires the ability to apply controlled $U^{2^p}$ for non-negative integers $p$. If $\phi_j$ is an exact multiple of $2^{-P}$, then an exact estimate of $\phi_j$ can be learned with certainty using only $p \in \{0, 1, \ldots, P - 1\}$. In general, an estimate $\hat{\phi}_j$ of $\phi_j$ satisfying $|\phi_j - \hat{\phi}_j| \leq \epsilon$ can be learned with high probability by taking the maximum value of $2^p$ on the order of $1/\epsilon$. The algorithm also requires application of an inverse quantum Fourier transform to orchestrate the constructive interference near the estimate for $\phi_j$. The quantum circuit for the standard approach to QPE is shown in Fig. 8.

Phase estimation can also be applied coherently onto a superposition of eigenstates. Suppose that the input state is $|\psi\rangle = \sum_j \alpha_j|\psi_j\rangle$. By linearity, if each phase $\phi_j$ is a multiple of $2^{-P}$ and





phase estimation is run with sufficient resolution, then QPE enacts the following unitary

$$|\psi\rangle|0\rangle \mapsto \sum_j \alpha_j |\psi_j\rangle |\phi_j\rangle, \tag{51}$$

where $|\phi_j\rangle$ holds a $P$-bit binary representation of $\phi_j$. If the auxiliary register is measured—here assuming for simplicity that the eigenvalues $\phi_j$ are nondegenerate—then with probability $|\alpha_j|^2$ (consistent with the Born rule) the estimate $\phi_j$ is obtained and the state collapses to the corresponding eigenstate $|\psi_j\rangle$.[42] If the phases $\phi_j$ are not multiples of $2^{-P}$, an approximate version of this operation can still be accomplished as long as the precision is sufficiently small to resolve the eigenvalues, subject to some caveats (discussed below).

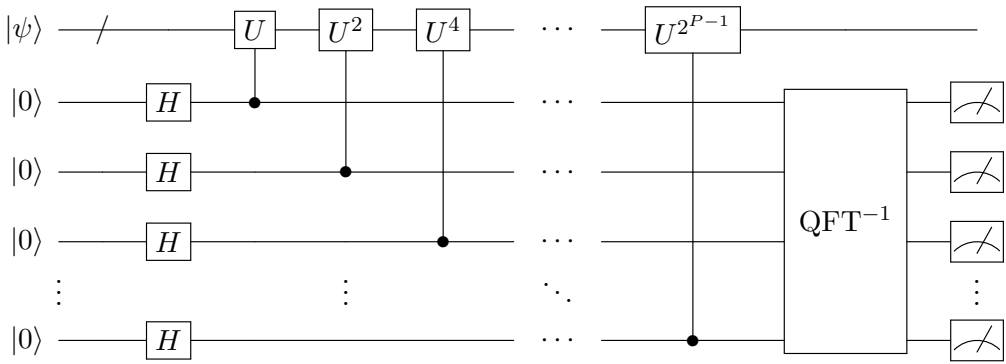

Figure 8: Quantum circuit implementation of QPE. The measurement outcomes on the $P$ ancilla qubits give a $P$-bit estimate of the phase $\phi_j$ (correct up to error $\mathcal{O}(2^{-P})$) with high probability.

### Dominant resource cost (gates/qubits)

The QPE subroutine is typically dominated by calls to the controlled unitary $U$. If resolution $\epsilon$ is desired, one must perform controlled $U^{2^p}$ operations for $p \in \{0, 1, \ldots, \lceil \log_2(1/\epsilon) \rceil + \mathcal{O}(1)\}$; thus, the number of calls to a controlled $U$ oracle will be $\mathcal{O}(1/\epsilon)$. This dependence on $\epsilon$ is optimal; the $\mathcal{O}(1/\epsilon)$ scaling is known as the *Heisenberg limit*.

In the context of estimating the eigenenergy of a Hamiltonian $H$, one can choose $U = \mathrm{e}^{\mathrm{i}H}$, and then implement controlled $U^t$, that is, controlled $\mathrm{e}^{\mathrm{i}Ht}$, with Hamiltonian simulation. In this case, given the ability to prepare an eigenstate of $H$, an $\epsilon$-approximation of the eigenvalue requires values of $t$ up to $\mathcal{O}(1/\epsilon)$.[43] However, one must also factor in the error in the Hamiltonian simulation. In a typical setting, access to the $n$-qubit Hamiltonian is given through a linear combination of $L$ unitaries, for instance, Pauli matrices. Let $\|H\|_1$ denote the sum of the coefficients in the combination. Then, methods for Hamiltonian simulation based on quantum signal

---

[42] Alternatively, if $\phi_j$ is known ahead of time (to sufficient precision), QPE can be wrapped inside of amplitude amplification and the state $|\psi_j\rangle$ can be prepared using $\mathcal{O}(|\alpha_j|^{-1})$ applications of the QPE circuit, rather than $\mathcal{O}(|\alpha_j|^{-2})$. Note that amplitude amplification can be understood through the QSVT [3] formalism, and in many applications, such as projecting onto the ground state of a Hamiltonian [5], one can achieve this sort of scaling directly without explicitly relying on QPE.

[43] The fact that learning energies to greater precision requires a proportionally greater amount of time $t$ is a manifestation of the energy-time Heisenberg uncertainty principle, and forms the origin of the term "Heisenberg limit."





processing can approximate $e^{iHt}$ to error $\mathcal{O}(\epsilon)$ with $\mathcal{O}(nL(\|H\|_1 t + \log(1/\epsilon)))$ gate complexity, whereas methods based on product formulas incur cost $\mathcal{O}(nL(\|H\|_1 t)^{1+1/2k}\epsilon^{-1/2k})$ for $(2k)$th-order product formulas, although the actual cost can be lower after accounting for structure in the Hamiltonian terms. Balancing the error from phase estimation against the error from Hamiltonian simulation can cause sub-Heisenberg-limited performance, such as in the case of the product formula approach. The overhead associated with imperfect Hamiltonian simulation can be avoided by applying QPE to different functions of $H$; for example, a promising choice is the qubitization operator, which acts in a similar way to $U = e^{i \arccos(H/\alpha)}$, where $\alpha$ is the normalization factor of the qubitization operator. The reason this is advantageous is that the qubitization operator can be implemented *exactly* given access to a block-encoding of $H$ [6, 7, 8]. In general, we require the unitary $U$ on which phase estimation is performed to be a known, classically invertible function of the Hamiltonian $U = f(H)$. The complexity of QPE depends on the desired uncertainty in the energy eigenvalue, which can be related to the uncertainty in the measured eigenphase via the magnitude of the derivative of the function, $\|f'(\cdot)\|$.

The number of qubits for QPE is simply the size of the register needed to hold the input state $|\psi_j\rangle$ plus the size of the register needed to hold the estimate $\hat{\phi}_j$ (i.e., roughly $\lceil \log_2(1/\epsilon) \rceil$ bits). Additionally, QPE requires an inverse quantum Fourier transform (QFT), which (using the textbook QFT implementation) adds only $\mathcal{O}(\log^2(1/\epsilon))$ additional gates to the protocol.

Another version of QPE [9] achieves the same task with only a single ancilla qubit, but, as a result, learns only one bit of the output at a time. Additionally, it requires an exact eigenstate as input. The latter problem can be avoided using a statistical approach [10, 11].

#### Caveats

The main caveats of QPE are related to the fact that eigenphases are not always exact integer multiples of $2^{-P}$, resulting in noncertain outcomes of QPE, which can lead to complications in certain applications.

- Fat tails and boosting of success probability: Whenever the phases $\phi_j$ are not exact integer multiples of $2^{-P}$ for some integer $P$, phase estimation will not return the answer $\phi_j$ with certainty. Rather, there will be a distribution of possible estimates $\hat{\phi}_j$ that is peaked near $\phi_j$. If one chooses $P = \lceil \log_2(1/\epsilon) \rceil + \mathcal{O}(1)$, then most of the probability mass of this distribution lies within $\epsilon$ of $\phi_j$. As $P$ is increased further, the distribution becomes more sharply peaked near $\phi_j$, and if an $\epsilon$-accurate estimate with $1 - \delta$ probability is desired, one must take $P = \lceil \log_2(1/\epsilon) \rceil + \mathcal{O}(\log(1/\delta))$, corresponding to a multiplicative $\mathcal{O}(1/\delta)$ overhead in the query complexity to $U$ and $\mathcal{O}(\log(1/\delta))$ additional ancilla qubits. This poor $\delta$ dependence is due to "fat tails" on the distribution of estimates of $\hat{\phi}_j$. One way to avoid this overhead is to take the median of estimates obtained from $\mathcal{O}(\log(1/\delta))$ repetitions of QPE [12, Lemma 1]. A downside of this approach is that it may be difficult to implement coherently on a superposition of eigenstates, in the sense of Eq. (51), since computing the median would require a coherent quantum sorting network. An alternative way to circumvent the fat tails problem is to modify the QPE protocol to have a nonuniform superposition in the register that controls applications of $U$; a judicious choice of superposition leads the distribution over estimates $\hat{\phi}_j$ to be a Kaiser window (see [13, Appendix D] and [14]) or discrete prolate spheroidal sequence (DPSS) function [15], which minimizes the probability of deviating from $\phi_j$ by more than $\epsilon$. See also [16], where a Gaussian profile is used to suppress the tails. Boosting the success probability to $1 - \delta$ in





this fashion incurs multiplicative $\mathcal{O}(\log(1/\delta))$ cost, rather than $\mathcal{O}(1/\delta)$. The overall cost in queries to $U$ by these methods matches a lower bound of $\Omega(\epsilon^{-1}\log(1/\delta))$ shown in [17].

- Performing coherent QPE: When $\phi_j$ are noninteger multiples of $2^{-P}$, the coherent operation in Eq. (51) cannot be straightforwardly performed with exact fidelity. This is because for each value of $j$, the second register will be in a superposition of many values of $\hat{\phi}_j$ (most but not all of the amplitude will lie on estimates close to $\phi_j$). To restore coherence, one might try coherently rounding the estimate $\hat{\phi}_j$ onto a coarser net of grid points (and then uncomputing the original estimate $\hat{\phi}_j$); however, there will always be edge cases where $\phi_j$ falls very near the midpoint between two grid points and rounding destroys some of the coherence in the input. This is true even as the precision of QPE is taken to zero ($\epsilon \to 0$). See [18] for a discussion. One possible way to mitigate this issue is presented in the "consistent phase estimation" protocol of [19, Section 5.2], where a random shift is applied to the grid points to avoid this situation for any particular eigenphase with high probability. However, this does not generically work simultaneously for all eigenphases. In [18], it is shown that performing the map of Eq. (51) is impossible without a "rounding promise" on the set of eigenphases $\{\phi_j\}$.

- Biased estimator: A further consequence of the noncertainty of the QPE output is that the estimate $\hat{\phi}_j$ is *biased*; that is, its expectation value is not exactly equal to $\phi_j$. This issue can also be fixed with a random shift idea, yielding an unbiased (and symmetrically distributed) version of QPE [20, 21].

**Example use cases**

- In quantum chemistry and condensed matter physics, QPE can be used to measure the eigenvalues (and especially the ground state energy) of the Hamiltonian $H$, which gives knowledge about reaction mechanisms, stable configurations, and other equilibrium properties. For QPE to succeed, a trial state $|\psi\rangle$ with substantial overlap with the eigenstate of interest must be input to QPE, which is challenging in the general case. The problem of ground state preparation has garnered intense study, and state-of-the-art techniques do not always follow the textbook method that relies on the QFT, presented above. For example, quantum signal processing can be leveraged directly to filter out unwanted eigenstates [22, 5], effecting a similar outcome as QPE.

- In Shor's algorithm, given a composite integer $N$ and a (randomly chosen) base $x < N$, QPE is used to determine the order of $x$, that is, the minimum integer $r$ for which $x^r \equiv 1$ mod $N$, which is in turn used to infer the prime factors of $N$. Here, the unitary $U$ is the modular multiplication unitary that sends $|y\rangle \mapsto |xy \mod N\rangle$.

- In amplitude estimation [23], given a unitary $U$ that prepares a state $U|\psi_0\rangle = a|\psi_g\rangle + b|\psi_b\rangle$, QPE is used to estimate $|a|$ or $|a|^2$. More advanced approaches to amplitude estimation not relying on QPE have since been developed. These leverage Grover's algorithm, or more generally quantum signal processing, without using the QFT. While these do not surpass the QPE-based method in asymptotic complexity, they potentially offer other benefits, such as improved practical performance and versatility. See [24] and references therein.

- In the Monte Carlo–style quantum algorithms for Gibbs sampling of quantum (i.e., non-diagonal in the computational basis) Hamiltonians, roughly speaking, the quantum state





undergoes a random walk on the eigenbasis of the Hamiltonian. Steps of this random walk are accepted or rejected according to how much the energy changes at each step. The QPE subroutine is used to simultaneously (approximately) project onto the eigenbasis of the Hamiltonian and to produce an estimate of the energy, used to determine whether the step should be accepted or rejected. Early studies [25, 26, 27] of this approach were hampered by the caveats related to rejecting quantum states and imperfect energy estimates, but recent works [28, 16, 29] circumvent these problems (by randomizing the grid points or completely abandoning phase estimation).

- To follow the ground state $|\psi_0(s)\rangle$ of a Hamiltonian $H(s)$ as some parameter $s$ is varied from 0 to 1, one can run the adiabatic algorithm. Alternatively, one can consider a discretization of steps $s_t \in \{s_0, \ldots, s_T\}$, where $0 = s_0 < s_1 < s_2 < \cdots < s_{T-1} < s_T = 1$, and run QPE on $H(s_t)$ in succession, each time causing a measurement in the eigenbasis of $H(s_t)$. Due to the quantum Zeno effect, as long as sufficiently small steps are taken, each projection will be onto the ground space with high probability (see, e.g., [30]). Larger steps can be tolerated if one boosts the probability that each step succeeds with amplitude amplification [31]. This approach is similar to the idea in Hastings' short-path algorithm [32, 33], which solves combinatorial optimization problems. However, note that modern implementations along these lines would likely elect to perform the ground state projection via eigenstate filtering [22] or related QSVT-based methods, rather than QPE.

- While state-of-the-art quantum linear system solvers (QLSSs) do not explicitly use QPE, the original QLSS by Harrow, Hassidim, and Lloyd [34] uses QPE to coherently measure the eigenvalues of a matrix $A$ into an auxiliary register. These eigenvalue estimates are subsequently inverted with coherent arithmetic in order to produce the state $A^{-1}|b\rangle$ corresponding to the solution to the system $Ax = b$. Achieving optimal asymptotic performance requires additional ingredients beyond QPE, and is best understood through the language of block-encodings and quantum linear algebra. This framework allows for manipulation of eigenvalues without explicitly reading them into an ancilla register with QPE.

- In certain machine learning tasks related to linear algebra, such as principal component analysis [35] and recommendation systems [36], quantum algorithms have been proposed that leverage QPE to access the information about the eigenvectors and eigenvalues. As explained in [18], these have not always fully accounted for the caveat related to coherent QPE, although typically these caveats can be circumvented using the framework of quantum linear algebra [37, 3].

**Further reading**

- The standard circuit and analysis of QPE appears in Nielsen and Chuang [4]. See also [2].

- Many variants of the QPE algorithm have been explored, which can be superior to the standard version in certain settings. See, for example, [18, 10] for additional references and informative overviews of various methods, along with their advantages and drawbacks.

- Reference [38] contains a pedagogical overview of QPE including some of its variants and applications.

# 14 Amplitude amplification and estimation

Quantum amplitude amplification and estimation provide means to boost or extract the amplitude of a marked quantum state that is produced in superposition with orthogonal states by a unitary matrix. They are among the most widely used quantum primitives, providing quadratic speedups over classical algorithms in many settings.

**This primitive area contains:**



*The authors are grateful to Patrick Rall for reviewing this section of the survey.*





## 14.1   Amplitude amplification

**Rough overview (in words)**

Given a quantum subroutine that succeeds with a probability less than one, amplitude amplification can be used to boost the success probability to 1 by making repeated calls to the subroutine and to a unitary that determines if the subroutine has succeeded. Amplitude amplification can be viewed as a generalization of Grover's search algorithm [1] and offers a quadratic speedup compared to classical methods in many instances.

**Rough overview (in math)**

We are given an initial state $|\psi_0\rangle$, a target ("good") state $|\psi_g\rangle$ that we can mark (i.e., the ability to reflect about the state), and a unitary $U$ (and its inverse $U^\dagger$) such that

$$U|\psi_0\rangle = |\psi\rangle = a|\psi_g\rangle + b|\psi_b\rangle \,,$$

where $|\psi_b\rangle$ is a ("bad") state orthogonal to the target state. In other words, $|a|^2$ is the probability of success of applying $U$ and measuring $|\psi_g\rangle$. In addition, we are given the ability to implement the reflection operator around the initial state $R_{\psi_0} = I - 2|\psi_0\rangle\langle\psi_0|$ and an operation that, when restricted to the subspace spanned by $\{|\psi_g\rangle, |\psi_b\rangle\}$, acts as the reflection around the target state $R_{\psi_g} = I - 2|\psi_g\rangle\langle\psi_g|$.

   Then, amplitude amplification allows us to boost the success probability to 1 through repeated calls to an operator $W = -UR_{\psi_0}U^\dagger R_{\psi_g}$, from the initial state $U|\psi_0\rangle = |\psi\rangle$. The standard analysis [2] proceeds by letting $a = \sin(\theta)$ and $b = \cos(\theta)$, and showing that the 2D subspace spanned by $|\psi_g\rangle, |\psi_b\rangle$ is invariant under $W$, which acts as a rotation operator such that $|\psi_g\rangle\langle\psi_g|W^m|\psi\rangle = \sin((2m+1)\theta)|\psi_g\rangle$.

   The algorithm can also be viewed through the lens of quantum singular value transformation (QSVT) whereby $U$ provides a generalized block-encoding (known as a projected unitary encoding) of the amplitude $a$. We can see this from $|\psi_g\rangle\langle\psi_g|U|\psi_0\rangle\langle\psi_0| = a|\psi_g\rangle\langle\psi_0|$. We choose to apply a polynomial $f(\cdot)$ satisfying the quantum signal processing conditions and $f(a) = 1$ to the block-encoded amplitude [3, Theorem 27 & 28]. For example, the textbook version of amplitude amplification is recovered by setting the QSVT rotation angles to $\pm\pi/2$.[44] This QSVT circuit applies a degree $2m + 1$ Chebyshev polynomial of the first kind $T_{2m+1}$ to the amplitude $a$, such that $|\psi_g\rangle\langle\psi_g|W^m|\psi\rangle = T_{2m+1}(a)|\psi_g\rangle = (-1)^m \sin((2m+1)\theta)|\psi_g\rangle$ for $a = \sin(\theta)$.

**Dominant resource cost (gates/qubits)**

The number of calls to $W$ is

$$m = \frac{\pi}{4\arcsin(a)} - \frac{1}{2} = \mathcal{O}\big(a^{-1}\big)$$

for small $a$. Each call to $W$ requires a call to each of $U, U^\dagger, R_{\psi_0}, R_{\psi_g}$. Often we have $|\psi_0\rangle = |0^{n+k}\rangle$, and $U$ acts on $n$ register qubits and $k$ ancilla qubits such that $U|0^{n+k}\rangle = a|\psi_g\rangle_n|0^k\rangle_k + b|\bot\rangle_{n,k}$, where $|\bot\rangle_{n,k}$ denotes a state orthogonal to $|0^k\rangle$ on the ancilla register. In this case the reflection operators are simple to implement using multicontrolled Toffoli gates.

---

[44]These rotation angles enable a gate compilation that removes the need for the QSVT ancilla qubit.





**Caveats**

The textbook version of amplitude amplification assumes that the success amplitude $a$ exactly equals $\sin(\pi/(4m+2))$ for an integer $m$. If this is not the case (e.g., when $a = 1/\sqrt{2}$), we can introduce a new qubit in $|0\rangle$ and apply an $R_y(2\phi)$ gate (i.e., a rotation about $Y$ by angle $2\phi$) to reduce the success probability (now defined by measuring $|\psi_g\rangle|0\rangle$) to $a\cos(\phi) = \sin(\pi/(4m'+2))$ for an integer $m'$.

In cases where we can only *lower bound* the success amplitude $a \geq a_0$, it is common to use fixed-point amplitude amplification [4]. This is best understood through QSVT [3, Theorem 27], where the reflection operators are replaced by parameterized phase operators $e^{i\theta|\psi_g\rangle\langle\psi_g|}$ and $e^{i\phi|\psi_0\rangle\langle\psi_0|}$.[45] The QSVT rotation angles are chosen to implement a polynomial that maps *all* amplitudes taking value at least $a_0$ to at least $(1 - \epsilon)$. The fixed-point amplitude amplification circuit uses a QSVT circuit that makes $\mathcal{O}(a_0^{-1}\log(\epsilon^{-1}))$ calls to $U, U^\dagger, e^{i\theta|\psi_g\rangle\langle\psi_g|}$, and $e^{i\phi|\psi_0\rangle\langle\psi_0|}$.

**Example use cases**

- Combinatorial optimization.

- Convex optimization via "minimum finding" subroutine (see [7, Appendix C]).

- Weakening cryptosystems.

- Tensor principal component analysis.

- Hamiltonian simulation using linear combinations of unitaries.

**Further reading**

- Both amplitude amplification and Grover search can be viewed through the lens of quantum walks on suitably constructed graphs. The quantum walks also take the form of a product of two reflections and more generally can be understood as quantizing a Markov chain describing a classical random walk [8]. We refer the interested reader to [9, 10, 11, 12].

- Oblivious amplitude amplification: Amplitude amplification can be extended to the case of oblivious amplitude amplification (OAA) [13]. The original formulation considered a setting where one is given unitary $U$ such that for any state $|\psi\rangle$, we have

$$U|0^m\rangle|\psi\rangle = a|0^m\rangle V|\psi\rangle + b|0^m_\perp\phi\rangle$$

for a unitary operator $V$. The goal is to amplify the probability for the state $|0^m\rangle V|\psi\rangle$ to 1. This is achieved through $\mathcal{O}(a^{-1})$ applications of an operator $W = U(I - 2|0^m\rangle\langle0^m|)U^\dagger(I - 2|0^m\rangle\langle0^m|)$ applied to $U|0^m\rangle|\psi\rangle$. We see that $W$ does not require reflections around the initial state $|\psi\rangle$. We can recognize $U$ as an $m$-qubit block-encoding of the operator $aV$, which can be transformed to a block-encoding of $V$ using QSVT.[46] The OAA subroutine

---

[45] It is shown in [5, Section 8.5] how these phase operators can be constructed using the corresponding controlled reflection operator. If only the uncontrolled reflection is available, a control can be added using, for example, [6, Fig. 5].

[46] We note that in this interpretation, one may be concerned that the phase information of the unitary $V$ is lost by transforming the singular values. This turns out not to be problematic, as the phase information of $V$ can be considered stored in the basis transformation matrices present in the singular value decomposition, rather than in the diagonal singular values matrix. This is taken care of automatically using QSVT. Phases are preserved when using an odd polynomial.





is used in the context of Hamiltonian simulation via Taylor series, where it would be problematic to have to reflect around the initial state during amplification.[47] It is also used in [16] (applied to isometries) for simulation of open quantum systems. OAA requires the block-encoded operator being amplified to preserve state norms (i.e., it must be an isometry), as this ensures that the success probability of the operation is independent of the state to which it is applied, which in turn enables amplification without reflection around the initial state.

It is also possible to amplify a block-encoding of a non-isometric operator $A$ using QSVT; see [3, Theorem 30] and [17]. Assume $\|A\| = 1$; given the ability to implement a block-encoding $U$ of $\sqrt{p}A$, we can use oblivious amplitude amplification to implement a block-encoding of $A$ using $\mathcal{O}(1/\sqrt{p})$ calls to $U, U^\dagger$. Note that for a general normalized state $|\psi\rangle$, it holds $\|A|\psi\rangle\| \leq 1$, with equality only achieved when $|\psi\rangle$ is the singular vector corresponding to the largest singular value of $A$. As a result, to boost the success probability of outputting $A|\psi\rangle/\|A|\psi\rangle\|$ to unity for a general input state requires using regular amplitude amplification, involving reflections around the initial state.

- While we are unaware of a standard reference for the use of an additional ancilla qubit to account for cases where the success amplitude $a \neq \sin(\pi/(4m + 2))$ for integer $m$, discussed above in §Caveats, it is explained more fully in these video lectures and also in [18, Appendix B].

---

[47]More precisely, a robust version of OAA is used which is applicable to an operator that is $\epsilon$ close to being unitary [14, 15].

## 14.2   Amplitude estimation

**Rough overview (in words)**

Given a quantum subroutine that succeeds with unknown success probability, amplitude estimation provides quadratic speedup over classical methods for estimating the success probability. Because many quantities of interest can be encoded in an amplitude or probability, amplitude estimation can be used as a widely applicable tool for obtaining Monte Carlo estimates with complexity $\mathcal{O}(1/\epsilon)$, instead of the $\mathcal{O}(1/\epsilon^2)$ achieved by classical estimation.

**Rough overview (in math)**

We are given an initial state $|\psi_0\rangle$, a target ("good") state $|\psi_g\rangle$, and a unitary $U$ and its inverse $U^\dagger$ such that

$$U|\psi_0\rangle = |\psi\rangle = a|\psi_g\rangle + b|\psi_b\rangle,$$

where $|\psi_b\rangle$ is a ("bad") state orthogonal to the target state. We assume that we can mark the target state $|\psi_g\rangle$ (i.e., the ability to reflect about the state). Thus, $p = |a|^2$ is the success probability of applying $U$ and measuring $|\psi_g\rangle$.[48] We are given the ability to implement the reflection operator around the initial state $R_{\psi_0} = I - 2|\psi_0\rangle\langle\psi_0|$ and an operation that, when restricted to the subspace spanned by $\{|\psi_g\rangle, |\psi_b\rangle\}$, acts as the reflection around the target state $R_{\psi_g} = I - 2|\psi_g\rangle\langle\psi_g|$. We can then estimate the success probability by performing quantum phase estimation on an operator $W = -UR_{\psi_0}U^\dagger R_{\psi_g}$, from the initial state $U|\psi_0\rangle = |\psi\rangle$. The standard analysis [1] proceeds by letting $|a| = \sin(\theta)$ and $|b| = \cos(\theta)$ (thus, the phases of $a$ and $b$ are absorbed into $|\psi_g\rangle$ and $|\psi_b\rangle$ and are not determined by the following procedure) and showing that the 2D subspace spanned by $\{|\psi_g\rangle, |\psi_b\rangle\}$ is invariant under $W$, where it acts as a rotation operator

$$W = \begin{pmatrix} \cos(2\theta) & \sin(2\theta) \\ -\sin(2\theta) & \cos(2\theta) \end{pmatrix}.$$

This operator has eigenvalues $e^{\pm 2i\theta}$, and we can estimate $\theta$ to additive error $\epsilon$ through quantum phase estimation. The estimate for $\theta$ can be converted into an estimate for $|a|$, or for the success probability $p = |a|^2$, which is often the quantity of interest.

**Dominant resource cost (gates/qubits)**

The classical approach for learning the probability $p$ to precision $\epsilon$ has complexity scaling as $M = \mathcal{O}(1/\epsilon^2)$, where the basic idea is to perform $M$ incoherent repetitions of applying $U$ and measuring in the $|\psi_g\rangle, |\psi_b\rangle$ basis, and then tally the measurement outcomes and construct the frequentist (or maximum likelihood) estimate of $p$. Amplitude estimation provides a quadratic speedup, learning the probability (and amplitude) with complexity scaling as $M = \mathcal{O}(1/\epsilon)$. The textbook variant has a constant success probability, which can be boosted to $1 - \delta$ with $\mathcal{O}(\log(1/\delta))$ overhead through standard methods (e.g., probability amplification by majority voting).

---

[48]Note that the original paper introducing amplitude estimation [1] uses the variable $a$ to denote the success probability. While the algorithm is referred to as amplitude estimation, it is often the success probability that we wish to compute, and the complexity of the algorithm is often presented accordingly.





More precisely, following the analysis of [1] one can see that to learn $|a|$ to error $\epsilon$ it suffices to utilize $M$ controlled applications of the walk operator $W$ where $M$ satisfies[49]

$$\epsilon \geq \frac{\pi\sqrt{1-|a|^2}}{M} + \frac{|a|\pi^2}{2M^2} \,. \tag{52}$$

The algorithm succeeds with probability at least $8/\pi^2$. For $|a| \approx 1 - \mathcal{O}(\epsilon)$, a further quadratic improvement is obtained (i.e., $M = \mathcal{O}(1/\sqrt{\epsilon})$ suffices).

To learn the success probability $p = |a|^2$ to error $\epsilon$ it suffices to utilize $M$ controlled applications of the walk operator $W$ where $M$ satisfies [1]

$$\epsilon \geq \frac{2\pi\sqrt{p(1-p)}}{M} + \frac{\pi^2}{M^2} \,. \tag{53}$$

The algorithm once again succeeds with probability at least $8/\pi^2$. Similar to above, if $p \approx \mathcal{O}(\epsilon)$ or $p \approx 1 - \mathcal{O}(\epsilon)$, then it suffices to take $M = \mathcal{O}(1/\sqrt{\epsilon})$.[50]

The overall gate complexity of an application involving amplitude estimation is given by $M$ times the gate complexity of implementing a controlled application of $W$.

A common setting is the case where $|\psi_0\rangle = |0^{n+k}\rangle$, and $U$ acts on $n$ register qubits and $k$ ancilla qubits such that $U|0^{n+k}\rangle = a|\psi_g\rangle|0^k\rangle_k + b|\psi_b\rangle|0_k^\perp\rangle_k$. In this case, the reflection operators are simple to implement, and $W$ can be controlled by making these reflections controlled (adding another control qubit to a multicontrolled Z gate). We require $\log(M)$ ancilla qubits for phase estimation (which can be reduced using modern variants, see below and [2]).

### Caveats

The textbook version of amplitude estimation described above produces biased estimates of $|a|$ and $p$. This is partly inherited from the biased nature of textbook quantum phase estimation. However, even if unbiased variants of phase estimation are used, the amplitude and probability estimates are not immediately unbiased, as they are obtained by applying nonlinear functions to the estimate of the phase. Unbiased variants of amplitude [2][51] and probability estimation [3, 4] have been developed to address this.

The variant of amplitude estimation described above is also "destructive" in the sense that the output state is collapsed into a state $\frac{1}{\sqrt{2}}(|\psi_g\rangle \pm i|\psi_b\rangle) \neq |\psi_0\rangle, |\psi\rangle$. A nondestructive variant may be desired if the initial state is expensive to prepare and we require coherent or incoherent repetitions of amplitude estimation. Nondestructive variants have been developed in [5, 4, 2].

---

[49]Specifically, Lemma 7 of [1] shows that if $\theta = \arcsin(|a|)$ and $\tilde{\theta} = \arcsin(|\tilde{a}|)$, then $|\theta - \tilde{\theta}| \leq \eta$ implies $|a^2 - \tilde{a}^2| \leq 2\eta\sqrt{a^2(1-a^2)} + \eta^2$. This is easily adapted to show that it also implies $|a - \tilde{a}| \leq \eta\sqrt{1-a^2} + a\eta^2/2$. They show that with probability at least $8/\pi^2$, $\theta$ is learned up to additive error at most $\eta = \pi/M$ with $M$ calls to $W$, which together with the above expressions implies Eqs. (52) and (53).

[50]We can compare to the classical approach of estimating $p$ by flipping a $p$-biased coin $M$ times. Letting $\tilde{p}$ denote the estimate, which has mean $p$ and variance $p(1-p)/M$, Chebyshev's inequality implies that $|p - \tilde{p}| \leq \epsilon$ with probability at least $8/\pi^2$ as long as $M \geq Cp(1-p)/\epsilon^2$ where $C = 1/(1 - 8/\pi^2)$. Thus, when $p \approx \mathcal{O}(\epsilon)$ or $p \approx 1 - \mathcal{O}(\epsilon)$, the classical approach achieves $M \sim 1/\epsilon$, and the quantum speedup is never more than quadratic.

[51]In order to achieve bias $\leq \epsilon\eta$, the algorithm of [2] pays a multiplicative cost overhead $\sim \frac{1}{\eta}$ which, up to logarithmic factors, could also be achieved by merely improving the precision to $\epsilon\eta$. The additive $\sim \log(\frac{1}{\epsilon\eta})$ cost overhead of [3, 4] is much more satisfactory.





**Example use cases**

- Approximate counting of solutions marked by an oracle (e.g., topological data analysis, combinatorial optimization).

- Amplitude estimation provides a quadratic speedup for Monte Carlo estimation [6, 7] with uses in pricing financial assets. The general idea is to prepare a state $|\psi\rangle = \sum_x \sqrt{p(x)f(x)}|x\rangle|0\rangle + |\phi 0^\perp\rangle$ where $\mathbb{E}[f(x)] = \sum_x p(x)f(x)$ represents the expectation value we wish to evaluate using Monte Carlo sampling and corresponds to the probability that we measure the second register in state $|0\rangle$. Hence, amplitude estimation provides a quadratic speedup for estimating this quantity.

- A special case of amplitude estimation is overlap estimation [8], where given two states $|\psi\rangle, |\psi_0\rangle$ and a unitary such that $|\psi\rangle = U|\psi_0\rangle$, the goal is to measure $\langle\psi_0|U|\psi_0\rangle = \langle\psi_0|\psi\rangle$. This can be viewed as an application of amplitude amplification, where $|\psi_g\rangle = |\psi_0\rangle$. As a result, we only require the ability to implement $R_{\psi_0} = I - 2|\psi_0\rangle\langle\psi_0|$, $U, U^\dagger$ (or equivalently $R_{\psi_0}$ and $R_\psi$). Note that in overlap estimation, one additionally wants to determine the phase of $a$, which can be obtained by applying amplitude estimation on a controlled variant of $U$, as outlined in [8]. Overlap estimation can be used for estimating observables, for example, in quantum chemistry.

- A generalization of amplitude estimation, via the quantum gradient algorithm, forms a core subroutine in some approaches for quantum state tomography [3]. Pure state tomography can be thought of as a generalization of amplitude estimation, in which we seek to learn all amplitudes individually, rather than only a single aggregate quantity. Closely related work on multivariate amplitude estimation [9] has broad applicability, including in convex optimization [10] and finance [11].

**Further reading**

- Variants of amplitude estimation using fewer ancilla qubits (including ancilla-free approaches), or with depth-repetition tradeoffs have been proposed [12], including work to make these methods nonadaptive [13]. For a summary of these approaches and their unification within the QSVT framework, see [2].

- There has been some work on computing, optimizing, and comparing the constant prefactor of the $M = \mathcal{O}(1/\epsilon)$ relation using different approaches to amplitude estimation, relevant for concrete resource estimates. For example, building on the analysis of [12], the method from [2] was estimated to scale roughly as $M \approx 4.7/\epsilon$ based on numerical experiments on a range of choices for $\epsilon$ and with fixed $a = 0.5$. This was observed to be about an order of magnitude better than the textbook method from [1] described above.[52] The method from [14] furthermore showed that a comparable total query complexity could be obtained while parallelizing across multiple processors, with maximum query depth roughly $0.4/\epsilon$.

---

[52] Asymptotically speaking, the complexity of the methods from [2, 12] scales suboptimally, as $\mathcal{O}(\log\log(1/\epsilon)/\epsilon)$, but the extra $\log\log(1/\epsilon)$ factor grows sufficiently slowly that for practical values of $\epsilon$ it can be bounded by a small constant.

# 15 Gibbs sampling

*The authors are grateful to Rolando Somma for reviewing this section of the survey.*

**Rough overview (in words)**

Gibbs sampling is the task of preparing a quantum state in thermal equilibrium. This task is interesting in its own right as a means of testing the thermodynamic properties of quantum systems in a controlled way, but it is also a subroutine that is surprisingly useful within other quantum algorithms. Formally, given a Hamiltonian and a temperature, the task is to prepare the *Gibbs state* (also known as the *thermal state*) of that Hamiltonian at the associated temperature, or equivalently, to sample eigenstates of the Hamiltonian with probability proportional to their Boltzmann weights (motivating the name Gibbs *sampling*).

Physically, Gibbs sampling is routinely achieved in experiments via cooling as a manifestation of open-system thermodynamics, although theoretical understanding of such processes has been largely heuristic. Computationally, quantum Gibbs sampling is the quantum analog of the same classical task in the computational basis, often achieved by Markov chain Monte Carlo (MCMC) methods. As a representative example, the Metropolis–Hastings algorithm [1] performs local accept-reject steps to construct a Markov chain whose stationary state is the classical Gibbs distribution; the Gibbs distribution can be efficiently sampled if the Markov chain mixes rapidly. Nowadays, Monte Carlo methods have already outgrown their original intent and found ubiquitous applications in optimization and machine learning due to their simplicity and versatility. It is natural to wonder if the same features will be present for quantum Gibbs sampling.

The most direct quantum algorithms for Gibbs sampling (for noncommuting Hamiltonians) suffer from an explicit cost exponential in the size of the system. Another approach is to quantize classical Monte Carlo algorithms [2], but this approach has faced serious technical challenges rooted in quantum mechanics: the energy-time uncertainty principle (for imposing the Boltzmann weights) and the no-cloning theorem (for "rejecting" a quantum state). Recently, a new wave [3, 4, 5, 6, 7] of proposals revisits the issue from the angle of open-system thermodynamics and gives nature-inspired algorithms for Gibbs sampling. These more directly resemble the dynamical process of thermalization and have the potential to achieve better runtimes for specific systems where thermalization is expected to be fast.

**Rough overview (in math)**

Given a Hamiltonian $H = \sum_i E_i |\psi_i\rangle\langle\psi_i|$ over $n$ qubits, a desired inverse temperature $\beta$, and an error parameter $\epsilon$, the Gibbs sampling task is to prepare an $n$-qubit quantum state $\rho$ such that

$$\|\rho - \sigma_\beta\|_{\mathrm{tr}} \leq \epsilon \,,$$

where

$$\sigma_\beta := \frac{\mathrm{e}^{-\beta H}}{\mathcal{Z}} \propto \sum_i \mathrm{e}^{-\beta E_i} |\psi_i\rangle\langle\psi_i| \quad \text{and} \quad \mathcal{Z} := \mathrm{tr}[\mathrm{e}^{-\beta H}].$$





The quantity $\mathcal{Z}$ is known as the partition function. The above uses the convenient error metric given by the trace norm $\|\cdot\|_{\text{tr}}$, which controls the error for arbitrary bounded (possibly nonlocal) observables. In some applications, it could be sufficient to give a state $\rho$ that approximates all *local* observables to high precision, even if the global distance between $\rho$ and $\sigma_\beta$ is large. Note that $\sigma_\beta$ corresponds to an ensemble of eigenstates of $H$, where an eigenstate with energy $E_i$ occurs with probability proportional to the Boltzmann weight $\mathrm{e}^{-\beta E_i}$ ($\beta$ has units of inverse energy so that $\beta E_i$ is dimensionless).

To solve this problem, the quantum algorithm requires access to $H$, for example, through a block-encoding of $H$. Block-encodings can often be efficiently constructed, for instance, when $H$ is a sparse matrix or when $H$ is given as a sum of poly($n$) local interaction terms. Henceforth, assume that $H$ is offset such that it is guaranteed to be a non-negative operator (no negative energies).

An early approach [8] for Gibbs sampling relied on quantum phase estimation (QPE) and amplitude amplification. In particular, one starts with a $2n$-qubit maximally entangled state (for which the reduced density matrix on the first $n$ qubits is the maximally mixed state) and applies QPE to the first $n$ qubits, reading an estimate of the energy into an ancilla register. Under the simplification that QPE has perfect resolution, one now has the state

$$\frac{1}{\sqrt{2^n}}\sum_i |\psi_i\rangle|\phi_i\rangle|E_i\rangle\,,$$

where $|\psi_i\rangle$ is the $i$-th eigenstate of $H$, $E_i$ is the associated energy, and the states $|\phi_i\rangle$ form an arbitrary (unimportant) orthonormal basis. Next, one coherently rotates an ancilla qubit to put the correct Boltzmann weight into the amplitude:

$$\frac{1}{\sqrt{2^n}}\sum_i |\psi_i\rangle|\phi_i\rangle|E_i\rangle\Big(\mathrm{e}^{-\beta E_i/2}|0\rangle + \sqrt{1-\mathrm{e}^{-\beta E_i}}|1\rangle\Big)\,.$$

Note that the probability of measuring the final qubit in $|0\rangle$ is precisely $\mathcal{Z}/2^n$. Rather than measure and postselect, one now performs amplitude amplification on the ancilla being $|0\rangle$ to produce

$$\frac{1}{\sqrt{\mathcal{Z}}}\sum_i \mathrm{e}^{-\beta E_i/2}|\psi_i\rangle|\phi_i\rangle|E_i\rangle$$

up to small error, which is a purification of the Gibbs state $\sigma_\beta = \mathcal{Z}^{-1}\sum_i \mathrm{e}^{-\beta E_i}|\psi_i\rangle\langle\psi_i|$. While QPE does not exactly produce the operation described above, a more complete analysis in [8, 9] shows the idea still works. This approach is akin to classical rejection sampling (see also [10]), where a state is chosen at random and accepted with probability $\mathrm{e}^{-\beta E_i}$, such that repeating until acceptance yields a sample from the correct distribution. Due to amplitude amplification, the quantum algorithm enjoys a quadratic speedup.

More advanced methods that have exponentially better $\epsilon$ dependence have since been developed. Reference [11] used a linear combination of unitaries approach to perform the imaginary-time evolution operator $\mathrm{e}^{-\beta H}$, again followed by amplitude amplification. Technically, that work assumed access to an operator similar to $\sqrt{H}$, but this requirement was removed in Gibbs samplers appearing in [12, 13], which employ a method for implementing smooth Hamiltonian functions. Alternatively, one can use the quantum singular value transformation along with a polynomial approximation to the function $\mathrm{e}^{-\beta(1-x)/2}$ on the interval $x \in [-1, 1]$ [14, Section 5.3] and combine this with (fixed-point) amplitude amplification [15].





Another family of quantum algorithms is closer in spirit to classical Monte Carlo methods. They quantize the Metropolis–Hastings algorithm (quantum Metropolis sampling [2]) or simulate the dynamics arising from a system-bath interaction [3, 4, 5, 6, 7]. These algorithms make fundamental usage of QPE (or the quantum operator Fourier transform [6, 7]) for probing the energy, but most importantly (and most nontrivially), they construct a detailed-balance "quantum Markov chain" via either discretely or continuously "rejecting" the quantum state. Intuitively, the algorithm implements the following rule:

"Apply a jump. If the energy increases, reject with some probability."

Care must be taken to perform the rejection step coherently and to handle the fact that the energies cannot be learned to infinite precision. Abstractly, Monte Carlo–style quantum algorithms emulate a continuous-time quantum Markov chain (i.e., a Lindbladian $\mathcal{L}$)[53] that converges to the Gibbs state after evolution time $t_{\mathrm{mix}}$ (often called the *mixing time*)

$$\mathcal{L}[\sigma_\beta] \approx 0 \quad \text{and} \quad \|\mathrm{e}^{\mathcal{L}t_{\mathrm{mix}}}[\rho_0] - \sigma_\beta\|_{\mathrm{tr}} \leq \epsilon$$

for some initial state $\rho_0$. Like the classical Metropolis–Hastings algorithm, for some systems, $t_{\mathrm{mix}}$ can be exponentially large (or worse) in the system size $n$, while for other systems, it can be much smaller (polynomial or logarithmic). It is a generally difficult problem to determine $t_{\mathrm{mix}}$. As a consequence, classical Monte Carlo algorithms rarely have convergence guarantees in practice. Rather, they are employed in conjunction with a variety of heuristic convergence tests.

Note that such a process can be further quantized to gain quadratic speedup [16, 6].

**Dominant resource cost (gates/qubits)**

Assuming one has access to a block-encoding of the Hamiltonian $H$, that is, a unitary whose upper left block is the operator $H/\alpha$, where $\alpha$ is a normalization constant at least as large as the spectral norm of $H$, one can accomplish the Gibbs sampling task using [12, Lemma 44] (see also [13, Corollary 16])

$$\alpha\beta\sqrt{\frac{2^n}{\mathcal{Z}}} \cdot \mathrm{poly}(\log(1/\epsilon), n) \tag{54}$$

calls to the block-encoding and a similar number of other gates. Note that since we have assumed $H$ is non-negative, we have $\mathcal{Z} \leq 2^n$. In the case that $H$ is $d$-sparse, we can take $\alpha = d\|H\|_{\mathrm{max}}$, where $\|H\|_{\mathrm{max}}$ is the maximum absolute value of any entry of the matrix $H$. In the case one has access to $\sqrt{H}$, the $\beta$ dependence can be reduced from $\beta$ to $\sqrt{\beta}$ [11]. This complexity statement might be regarded as a *quadratic speedup* compared to the classical method of rejection sampling, which requires $2^n/\mathcal{Z}$ samples on average; however, note that this classical method only directly applies to diagonal (classical) Hamiltonians $H$. Otherwise, a classical approach may need to resort to exact diagonalization of $H$, which has $\mathcal{O}(2^n)$ space complexity and even worse time complexity.

Monte Carlo–style quantum Gibbs sampling algorithms have complexity determined by

$$(\text{mixing time}) \cdot (\text{cost per unit evolution time } \mathrm{e}^{\mathcal{L}}). \tag{55}$$

The mixing time is expected to vary significantly for different systems of interest (based on classical Monte Carlo intuition), but for systems appearing in nature, one may be optimistic

---

[53] Most constructions are continuous-time quantum Markov chains as they are mathematically easier to work with than discrete-time quantum channels [2].





based on the observed fast thermalization of physically relevant observables. In fact, in many cases, rapidly thermalizing metastable states will have physically relevant characteristics; in analogy with classical ferromagnetic systems below the critical temperature. The cost per unit evolution time is dominated by the QPE subroutine, which then scales with a certain energy resolution. Typically, we may work in a convention where $H$ is given by poly$(n)$ Hamiltonian terms each of strength $\mathcal{O}(1)$, enabling a block-encoding of $H$ with normalization $\alpha = $ poly$(n)$ to be implemented in gate complexity poly$(n)$, or alternatively, the ability to perform black-box Hamiltonian simulation for time $t$ at gate cost $t \cdot$ poly$(n)$.[54] In this case, an overall gate complexity for a rapidly mixing system (i.e., $t_{\mathrm{mix}} \leq$ poly$(n)$) can be poly$(n, \beta, 1/\epsilon)$; see [6, Table I] for a catalog of existing constructions. A recent construction [7] improved the asymptotic cost per unit Lindbladian evolution time to $\tilde{\mathcal{O}}(\beta)$ Hamiltonian simulation time assuming black-box access to Hamiltonian simulation; for lattice Hamiltonians, this further simplifies as one merely needs to simulate lattice patches of diameter $\tilde{\mathcal{O}}(\beta)$. To put together a practically relevant end-to-end resource estimate, one needs to design better algorithms (see, e.g., [17] for a single-ancilla variant for ground states) to reduce the per unit time cost, as well as to estimate the mixing time (e.g., by exact diagonalization of the map for small system sizes). Of course, if Gibbs sampling is employed as a heuristic (as in many classical applications of Monte Carlo methods), the cost will be empirical.

**Caveats**

On the one hand, the superpolynomial $\mathcal{O}(\sqrt{2^n})$ complexity for Gibbs sampling that appears explicitly in Eq. (54) is necessary in general (for sufficiently large $\beta$ it allows one to solve NP-hard or even QMA-hard problems in the general case). On the other hand, most physical Hamiltonians (if they appear to thermalize in nature) should be simulable without exponential hidden prefactors. The Monte Carlo–style approach to Gibbs sampling attempts to mimic nature more closely than the other algorithms with guaranteed complexities mentioned above; hence, it looks more promising for obtaining polynomial runtimes, but this must be verified through system-specific analysis or hardware demonstrations.

Finally, if the Hamiltonian comes from classical problems (such as solving semidefinite programs), loading the instance may have exponential cost ($e^{\Omega(n)}$), which in the above presentation is hidden in the assumption of a block-encoding of classical data. Additionally, it is unclear whether Hamiltonians arising from classical data—which would likely lack the local interaction structure that one sees in chemical and physical systems—should be expected to "thermalize" quickly (i.e., whether Monte Carlo–style algorithms converge in a small number of iterations).

**Example use cases**

- Multiplicative weights update (MWU) method and conic programming: Gibbs sampling is the main source of quantum speedup in the MWU method, which is used to solve semidefinite programs and other conic programs [18, 19, 12, 13, 20]. Existing analyses in this direction have employed Gibbs samplers with a guaranteed quadratic (but no larger) speedup, rather than the more heuristic and recent Monte Carlo–style algorithms.

---

[54]In the model of black-box Hamiltonian simulation, one can apply the unitary $U_i = e^{iHt_i}$ for user-specified choices of $t_i$, and the goal is to minimize the cumulative total $\sum_i t_i$ over the course of the algorithm. In this model, it is sensible for $t$ and $\beta$ to have the same units (interpreted as time), as $\beta H$ and $iHt$ are both unitless.





- **Quantum chemistry**: An important step of estimating the ground state energy of electronic structure Hamiltonians is generating an ansatz state that has a large overlap with the ground state. This might be done via Gibbs sampling at sufficiently low temperatures; the overlap with the ground state is $\mathrm{e}^{-\beta E_0}/\mathcal{Z}$, which can be large when $1/\beta$ is sufficiently small compared to the spectral gap between the ground and excited space of the Hamiltonian.

- **Condensed matter physics**: Similar to quantum chemistry, Gibbs sampling provides a method for producing ansatz states for ground state energy calculations, which often capture the relevant physics. However, condensed matter physicists are also interested in material properties at finite temperatures so that the Gibbs state itself might equally be of interest.

- Computing partition functions: One of the early references to develop quantum Gibbs samplers [8] applied it to the problem of estimating the partition function $\mathcal{Z}$ up to small relative error. The partition function contains all the relevant thermodynamic information of the system.

- **Combinatorial optimization**: Many combinatorial optimization problems can be viewed as finding the ground state of classical Hamiltonians (i.e., Hamiltonians that are diagonal in the computational basis), or finding a low-energy state that achieves a high approximation ratio with the ground state. Classical Monte Carlo algorithms are a key technique in this area, and quantization of these methods can sample the same classical thermal distribution with a quadratic speedup in the mixing time [21]. The full power of Gibbs sampling for general nondiagonal Hamiltonians could be useful in situations where one adds a noncommuting transverse field to the classical Hamiltonian and wishes to prepare a low-energy state, such as in quantum annealing or for training **quantum Boltzmann machines** [22].

### Further reading

Gibbs sampling has been studied in several specific cases. For example, [23] studied Gibbs sampling of local Hamiltonians in 1D. Moreover, [24] studied *commuting* spatially local Hamiltonians and showed conditions under which they thermalize in polynomial time, suggesting efficient Gibbs sampling via Monte Carlo–style methods. These conditions hold for any 1D system at any temperature, and in any higher-spatial dimension above a certain threshold temperature.

# 16    Quantum adiabatic algorithm

*The authors are grateful to Dong An for reviewing this section of the survey.*

**Rough overview (in words)**

The *quantum adiabatic algorithm* (QAA) [1], sometimes referred to as *adiabatic state preparation*, is a continuous-time procedure for (approximately) preparing an eigenstate (typically the ground state) of a particular Hamiltonian of interest on a quantum device. The QAA also forms the basis for a model of quantum computation called *adiabatic quantum computation* which acts as an alternative to the standard quantum circuit model.

The main idea of the QAA is to begin in an eigenstate of a simpler Hamiltonian that is easy to prepare, and then slowly change the Hamiltonian to be equal to the more complex Hamiltonian of interest. The adiabatic theorem (see [2] and references therein), a celebrated concept from physics, dictates that if the evolution is sufficiently slow, the system will evolve to (approximately) remain in the instantaneous eigenstate of the continuously varying Hamiltonian and thus finish in the desired state. The length of time required for the evolution to succeed depends on the spectral properties of the Hamiltonian path and in particular on the minimum spectral gap. The adiabatic algorithm can be simulated on a gate-based quantum computer with time-dependent Hamiltonian simulation.

**Rough overview (in math)**

Let $H(s)$, where $s$ varies as $0 \leq s \leq 1$, denote a single-parameter path through the space of Hamiltonians, and let $|\phi_j(s)\rangle$ and $E_j(s)$ denote the eigenstates and eigenvalues of $H(s)$, indexed by $j$ in increasing order. The goal of the QAA is to prepare a certain eigenstate $|\phi_j(1)\rangle$ of $H(1)$. Let $|\psi(t)\rangle$ denote the state of our system at time $t$ and let $T$ be the total evolution time. The procedure calls for beginning in the state $|\psi(0)\rangle = |\phi_j(0)\rangle$ and allowing $|\psi(t)\rangle$ to evolve by the Schrödinger equation according to the Hamiltonian $H(t/T)$, that is, $i\frac{d}{dt}|\psi(t)\rangle = H(t/T)|\psi(t)\rangle$ from $t = 0$ to $t = T$. Thus, as $T$ is made larger, the path from $H(0)$ to $H(1)$ is traversed increasingly slowly.

**Dominant resource cost (gates/qubits)**

The main resource for the continuous-time QAA is the total evolution time $T$. The adiabatic theorem suggests that if $T$ is chosen sufficiently large, and as long as eigenvalue $E_j$ is nondegenerate along the entire path, then $|\psi(T)\rangle \approx |\phi_j(1)\rangle$ will hold. The often-quoted heuristic condition [2] for success is that

$$T \gg \max_{0 \leq s \leq 1} \frac{\left\| \frac{dH}{ds} \right\|}{\Delta(s)^2},  \tag{56}$$

where $\Delta(s)$ is the spectral gap, that is, $\min_{i \neq j} |E_i(s) - E_j(s)|$, and $\|\cdot\|$ denotes the spectral norm. Thus, the runtime needed for the QAA to have small error is primarily governed by the minimum size of the spectral gap along the adiabatic path. This aspect of the QAA is a common sticking point as it is often difficult to produce lower bounds on $\Delta(s)$ that would suffice for proving upper bounds on $T$. In practice, the value of $T$ can be chosen heuristically,





or by trial-and-error, but a more detailed understanding of $\Delta(s)$ would inform smarter choices of Hamiltonian path $H(s)$.

While Eq. (56) is nonrigorous and potentially loose in specific scenarios, the polynomial dependence of $T$ on the inverse spectral gap is an essential feature of the QAA. For example, it was shown in [3] that in general, any rigorous bound on the QAA runtime must scale at least linearly in the inverse spectral gap.

The QAA is typically formulated as a continuous-time procedure, but a gate-based quantum computer can simulate the QAA by discretizing the path and approximately implementing the evolution from time $t$ to $t + \delta t$ with product formulas or with more advanced techniques for time-dependent Hamiltonian simulation. This incurs error in addition to the adiabatic error of the continuous-time QAA. The number of gates needed to do this can be made proportional to $T$ (up to logarithmic corrections), polynomial in the number of qubits needed to hold the state $|\psi(t)\rangle$, and logarithmic in the approximation error incurred (e.g., [4]).

**Caveats**

A technical caveat of the QAA is that rigorous formulations of sufficient conditions for success (e.g., [5, 6]) are more complex than Eq. (56) and likely looser than what is necessary in practice. Also, in most cases, the dependence of the runtime $T$ on the final approximation error $\epsilon = \||\psi(T)\rangle - |\phi_j(1)\rangle\|$ goes as $T = \text{poly}(1/\epsilon)$, rather than $T = \text{polylog}(1/\epsilon)$. To circumvent this and achieve $\text{polylog}(1/\epsilon)$ dependence, one can choose more sophisticated Hamiltonian paths $H(s)$ for which all time derivatives vanish at $s = 0$ and $s = 1$ [7, 2].

A practical caveat of the QAA is that the spectral gap—the main determiner of the resource cost—is difficult to study theoretically. Numerically, it can often be computed only for small system sizes, and it is unclear whether extrapolations to larger system sizes would be accurate.

Furthermore, in many end-to-end applications the spectral gap is not only unknown, but also expected to be extremely small, implying that the QAA requires a large runtime (see §Example use cases, below). For the QAA to be efficient, the spectral gap must decay only like an inverse polynomial of the system size, but such scaling requires the problem instance to have a special structure. Such structure should not be assumed to exist without justification.

**NISQ implementations**

The QAA is closely related to the concept of *quantum annealing* [8], a term used especially in the context of near-term implementations on existing quantum hardware. In quantum annealing, the system is exposed to a time-dependent Hamiltonian, typically a transverse-field Ising model. The strength of the transverse field is slowly reduced, eventually to zero, where the Hamiltonian is equal to a classical Ising model encoding a hard combinatorial optimization problem. If implemented perfectly and sufficiently slowly, this would be a manifestation of the QAA, and one would obtain the solution to the problem. However, the typical setting of quantum annealing is to consider faster implementations, and to possibly allow for some amount of control noise and finite-temperature effects (rather than evolving under a closed system at zero temperature), which induce transitions from the ground state to excited states. The goal is relaxed from ending in the exact ground state of the final Hamiltonian to ending in a low-energy state that can be considered an *approximately* optimal solution to the problem. The success metric is often the quality of the solution produced rather than the runtime required to find the best solution. As such, it is a heuristic algorithm and must be compared with classical heuristic algorithms, where





evidence of a scalable advantage is mixed. See, for example, [9] for a perspective on quantum annealing and the most promising related directions.

Separately, the QAA can be related to variational quantum algorithms, which are NISQ friendly. In particular, by applying product formulas to the QAA, one obtains alternating time evolutions by $H(0)$ and by $H(1)$; in the case that $H(0)$ is a transverse field and $H(1)$ is a classical cost function, this is precisely an instance of the quantum approximate optimization algorithm (QAOA) [10], a leading NISQ algorithm. In the limit of large depth, the QAOA can fully simulate the QAA to arbitrarily small precision. However, in a NISQ setting, the depth of the QAOA would need to be restricted, and the QAOA would not exactly follow the QAA.

**Example use cases**

- Combinatorial optimization: The QAA was first invented [1] as a way to solve hard classical combinatorial optimization problems on a quantum computer. An example is constraint satisfaction problems, where one is given a Hamiltonian $H(1)$ that is diagonal in the computational basis (i.e., "classical") and equal to the sum of various constraints on $n$ bits. The ground state of $H(1)$ is the bit string that violates the fewest constraints. One typically chooses the initial Hamiltonian to be $H(0) = -\sum_{i=1}^{n} X_i$, where $X_i$ denotes the Pauli-$X$ operator on qubit $i$, whose ground state is an easy-to-prepare product state. The QAA is guaranteed to find the ground state of $H(1)$ if it is run with sufficiently large evolution time. However, in general, it is expected that the spectral gaps along the adiabatic path become exponentially small in $n$ [11, 12, 13, 14, 15], indicating that the QAA requires exponentially long runtime.

- Quantum chemistry and condensed matter physics: A central problem of quantum chemistry and computational condensed matter physics is the problem of finding the ground state energy of a molecule, material, or lattice model. This can be solved efficiently with quantum phase estimation (QPE) so long as one can prepare a state that has substantial overlap with the ground state of the Hamiltonian. Adiabatic state preparation has been proposed as a method for producing such a state (see, e.g., [16, 17, 18, 19, 20, 21, 22]). This initial state preparation is often the bottleneck in the end-to-end quantum solution, as it can require exponential time for systems of interest (see, e.g., [23]).

- Quantum linear system solvers: The state-of-the-art quantum linear system solvers [24] leverage the QAA to produce a quantum state $|x\rangle$ corresponding to the solution of a linear system $Ax = b$ (see also [25, 26, 27, 28]). In particular, this method allows the runtime to scale linearly in the condition number of the matrix $A$.

**Further reading**

- See [2] for a comprehensive 2018 review of the QAA and adiabatic quantum computation more generally.

- See [29] for a digital version of the QAA for a gate-based quantum computer, but distinct from a direct simulation of the QAA. The idea is to choose a sequence of $s$ values $0 = s_0 < s_1 < s_2 < \cdots < s_T = 1$ and perform measurements of $H(s_t)$ for $t = 0, \ldots, T$ in sequence using QPE. As long as the difference between consecutive values of $s$ is sufficiently small, the quantum Zeno effect guarantees that each measurement will project onto the correct





eigenstate $|\phi_j(s_t)\rangle$ with high probability (see also [30, 31]). One can also take larger jumps, and amplify their success probability with fixed-point amplitude amplification. The resource cost has a similar dependence on the spectral gap as the continuous-time QAA: if the "path length" traced by the eigenstate $|\phi_j(s)\rangle$ is $L$, the minimum gap is $\Delta$, and the target error is $\epsilon$, then the gate cost of the algorithm is $\mathcal{O}(L\log(L/\epsilon)/\Delta)$. The path length $L$ can be upper bounded by $\|dH/ds\|/\Delta$, which roughly recovers Eq. (56).

- Along the lines of the previous bullet, [32] gives an alternative way to effect adiabatic state preparation on a gate-based computer with polylog$(1/\epsilon)$ overall error dependence, via quasi-adiabatic continuation.

# 17 Loading classical data

The end-to-end quantum applications covered in this survey have classical inputs and classical outputs, in the sense that the problem is specified by some set of classical data, and the solution to the problem should be a different set of classical data. In some cases, the input data is relatively small, and loading it into the algorithm does not contribute significantly to the cost of the algorithm. In other cases—for example, "big data" problems within the areas of machine learning and finance—the dominant costs, both for classical and quantum algorithms, can be related to how the algorithms load and manipulate this large quantity of input data. Consequently, the availability of quantum speedups for these problems is often dependent on the ability to quickly and coherently access this data. The true cost of this access is the source of significant subtlety in many end-to-end quantum algorithms.

*The authors are grateful to Thomas Häner, Damian Steiger, and Xiao Yuan for reviewing this section of the survey.*

**This primitive area contains:**







## 17.1  Quantum random access memory

**Rough overview (in words)**

Quantum random access memory (QRAM) is a construction that enables coherent access to a database, such that multiple different elements can be read in superposition. The ability to rapidly access large, unstructured datasets in this way is crucial to the speedups of certain quantum algorithms, for example in quantum machine learning based on quantum linear algebra. QRAM is commonly invoked to circumvent data-input bottlenecks [1] in situations where loading input data could dominate the end-to-end runtime of an algorithm. It remains an open question, however, whether a large-scale dedicated QRAM device will ever be practical, casting doubt on quantum speedups that rely on QRAM. Note that, while here we focus on the more common use case of loading *classical* data with QRAM, certain QRAM architectures can be adapted to also load *quantum* data [2].

**Rough overview (in math)**

Consider a length-$N$, unstructured classical data vector $x$, and denote the $i$-th entry as $x_i$. Let the number of bits of $x_i$ be denoted by $d$. Given an input quantum state $|\psi\rangle = \sum_{i=0}^{N-1} \sum_{b \in \{0,1\}^d} \alpha_{ib} |i\rangle |b\rangle$, QRAM is defined [3] as a unitary operation $Q$ with the action

$$Q|\psi\rangle = Q \sum_{i=0}^{N-1} \sum_{b \in \{0,1\}^d} \alpha_{ib} |i\rangle |b\rangle = \sum_{i=0}^{N-1} \sum_{b \in \{0,1\}^d} \alpha_{ib} |i\rangle |b \oplus x_i\rangle. \qquad (57)$$

Here, the first $\log_2(N)$-qubit register stores the "address" (assuming for simplicity that $N$ is a power of 2), while the second $d$-qubit register stores the corresponding "data." Note that the unitary $Q$ can also be understood as an oracle (or black box) providing access to $x$, as $Q(\sum_i \alpha_i |i\rangle |0\rangle) = \sum_i \alpha_i |i\rangle |x_i\rangle$.

   Let $T_Q$ denote the "cost" of implementing the operation $Q$, where $T_Q$ can be measured in wall time, circuit depth, total number of gates, total circuit spacetime occupancy, total number of $T$ gates, etc., depending on the context. Algorithms that rely on QRAM to claim exponential speedups over their classical counterparts frequently assume that $T_Q = \mathrm{polylog}(N)$. However, as discussed in §Caveats, below, it is crucial to emphasize that this assumption can only hold when $T_Q$ is interpreted as the circuit depth or wall time (or something similar) to implement $Q$; whereas, if $T_Q$ is taken to be the total gate cost or the spacetime occupied by the computation, simple gate counting arguments imply a lower bound of $T_Q \geq \Omega(dN)$. In a discrete gate set, each unit of spacetime can be occupied with only a finite number of unique gates; since there are $2^{dN}$ different possible data vectors $x$, the circuit must have at least $\Omega(dN)$ spacetime to be able to implement all possibilities (see also [4, Section V] for a more detailed discussion).

**Dominant resource cost (gates/qubits)**

Let us consider for simplicity the $d = 1$ case, that is, when each data entry is a single bit. The QRAM operation $Q$ can be implemented as a quantum circuit that uses $\mathcal{O}(N)$ gates. Assuming gates acting on disjoint qubits can be parallelized, a circuit depth of $T_Q = \mathcal{O}(\log(N))$ can be achieved at the expense of using $\mathcal{O}(N)$ ancillary qubits; explicit circuits can be found in, for example, [5, 6]. The number of ancillary qubits can be traded off for increased circuit depth; circuits implementing $Q$ can be constructed using $\mathcal{O}(N/M)$ ancillary qubits and depth





$\mathcal{O}(M \log(N))$, where $M \in [1, N]$; see examples in [7, 8, 5, 6] (the setting of $M = N/\log(N)$ is sometimes referred to as "QROM"—see terminology caveats below—and its fault-tolerant cost of implementation is well established [9]).

If the data vector $x$ is sparse—that is, only $s$ of its $N$ entries are nonzero—then there exist circuit implementations with depth as shallow as $T_Q = \mathcal{O}(\log(s \log(N)))$, using $\mathcal{O}(s \log(s) \log(N))$ ancillary qubits, both of which for $s = \mathcal{O}(1)$ are exponentially better than the general case of a dense vector $x$ [10].

Each of the above constructions can be generalized to the $d > 1$ case with various space-time tradeoffs. For example, the $d$ bits of a data entry can be queried in series, requiring $\mathcal{O}(N)$ ancillary qubits with depth $T_Q = \mathcal{O}(d \log(N))$ (improvement to $T_Q = \mathcal{O}(d + \log(N))$ is possible for certain QRAM architectures [11]). Alternatively, the $d$ bits can be accessed in parallel, with depth $T_Q = \mathcal{O}(\log(N))$, but at the price of $\mathcal{O}(Nd)$ ancillary qubits.

**Caveats**

The main concern for QRAM's practicality is the large hardware overhead that is necessary to realize fast queries with depth $T_Q = \mathcal{O}(\log(N))$. This cost is likely to be prohibitive for big-data applications where $N$ can be millions or billions. The cost will also be magnified by additional overhead associated with error correction and fault tolerance [5], especially considering that circuits implementing $Q$ are composed of $\mathcal{O}(N)$ non-Clifford gates. Indeed, this observation together with the assumption that magic state distillation is expensive to run in a massively parallel fashion, has led some to argue that $T_Q = \mathcal{O}(\log(N))$ is not realistic in a fault-tolerant setting (see, e.g., [4]). However, it is possible that alternative approaches to fault tolerance tailored to QRAM could help alleviate this large hardware overhead.

The fault-tolerance overhead may be reduced for the so-called bucket-brigade QRAM (BBQRAM) [3, 12, 6], which is a family of circuits implementing $Q$ that are intrinsically resilient to noise. More precisely, [6] shows that if $\epsilon$ is the per-gate error rate, BBQRAM circuits can implement $Q$ with leading-order fidelity $F \sim 1 - \epsilon \operatorname{polylog}(N)$, while generic circuits implementing $Q$ have leading-order fidelity $F \sim 1 - \epsilon \mathcal{O}(N)$. Nevertheless, at the scale necessary for useful end-to-end applications, some amount of error correction will almost certainly be required even for BBQRAM circuits.

Even if depth $T_Q = \operatorname{polylog}(N)$ is practically achievable, some have argued that any fair comparison with state-of-the-art classical methods should then allow for classical parallel computation. After all, the parallel classical hardware necessary to operate the circuit $Q$ (including the quantum error correction) could in principle be repurposed directly toward solving the end-to-end computational problem. For example, many linear algebra tasks such as matrix-matrix and matrix-vector multiplication of size-$N$ objects are amenable to parallelization, and can also have cost scaling as $\operatorname{polylog}(N)$ in some parallel classical models of computation [13, 4]. Under such a comparison, it becomes difficult to identify conditions where QRAM-based quantum algorithms can give rise to a significant scaling advantage [4].

Some terminology caveats:

- The unitary $Q$ in Eq. (57) is referred to by some as quantum read-only memory (QROM) [9], reflecting the fact that $Q$ corresponds only to reading data. Some algorithms also require the ability to write to the vector $x$ during computation, but the writing of classical (i.e., not in superposition) data need not be implemented via a quantum circuit.





- The term QRAM is used by different authors to refer to the unitary $Q$, families of circuits that implement $Q$, or quantum hardware that runs said circuits.

- The terms QRAM and QROM are sometimes used for distinguishing the cases of $T_Q = \text{polylog}(N)$ and $T_Q = \text{poly}(N)$, respectively, even though $T_Q$ is unrelated to the distinction between reading and writing. The term QROAM has also been used to describe intermediate circuits that trade off depth and width [8].

- Some use the term QRAM to refer exclusively to the case $N \gg 1$ and $T_Q = \text{polylog}(N)$ depth, where the implementation challenges for QRAM are most pronounced.

Elsewhere in this document, we follow the convention described in the final bullet point above: usage of the term QRAM, unless specified otherwise, refers to the ability to implement $Q$ at cost $\text{polylog}(N)$.

**Example use cases**

- Quantum linear algebra: QRAM can be used as an oracle for implementing linear algebra algorithms operating on unstructured data (e.g., by acting as a subroutine in a block-encoding), with applications in machine learning, finance, etc. For example, the quantum recommendation systems algorithm [14] (now dequantized [15]) uses QRAM as a subroutine to efficiently encode rows of an input data matrix in the amplitudes of quantum states (see Appendix A of [14] for details).

- Hamiltonian simulation, quantum chemistry, condensed matter physics: In the linear combination of unitaries input model, QRAM can be used as a subroutine for "PREPARE" oracles that encode coefficients of the simulated Hamiltonian into the amplitudes of quantum states [9]. These use cases typically consider the hybrid QROM/QRAM constructions with $\mathcal{O}(K \log(N))$ ancillary qubits and depth $\mathcal{O}(N/K)$ (with the parameter $K$ to be optimized), because the amount of data (and thus the size of $N$) scales only polynomially with the system size.

- Grover search: QRAM can be used as an implementation for Grover's oracle in the context of an unstructured database search; see Chapter 4 of [16]. This appears for example in quantum algorithms that utilize dynamic programming to give polynomial speedups for combinatorial optimization problems like the traveling salesperson problem [17]. However, it has been argued that a quantum computer running Grover's algorithm with a QRAM-based oracle would not provide a speedup over a classical computer with comparable hardware resources [13].

- Topological data analysis (TDA): A small QRAM (i.e., not exponentially larger than the main quantum data register) is used in some quantum algorithms for TDA [18, 19] in order to load the positions of the data points for computing whether simplices are present in the complex at a given length scale.

**Further reading**

- Reference [4] focuses on various fundamental and practical concerns for large-scale QRAM, while also providing a comprehensive survey.





- Reference [20] provides an overview of practical concerns facing QRAM in the context of big-data applications (though the discussions of noise resilience there and in [12] are somewhat outdated, cf. [6]).

## 17.2   Preparing quantum states from classical data

**Rough overview (in words)**

An important subroutine in many quantum algorithms is preparing a quantum state given a list of its amplitudes stored, for example, in a classical database.[55]  The upshot is that $N$ amplitudes, which require $\mathcal{O}(N)$ classical bits to write down, can be encoded in a quantum state with only $\log_2(N)$ qubits, an exponential compression in memory.  However, there are caveats; for example, simple information-theoretic bounds [1] dictate that the quantum circuit that prepares the $\log_2(N)$-qubit state must still have at least $\mathcal{O}(N)$ gates, so no exponential advantage in gate complexity is possible.  Additionally, reading out the $N$ amplitudes from the encoded quantum state generally requires full quantum state tomography, requiring $\Omega(N)$ preparations of the state.  Depending on which resource is being optimized, the best protocol for state preparation will look different, and optimal state preparation methods are known for several natural choices of metric.

**Rough overview (in math)**

Let $x = (x_0, \ldots, x_{N-1}) \in \mathbb{C}^N$ be a vector of $N$ complex numbers, where $N$ is a power of 2, and let

$$|\psi\rangle = \frac{1}{\|x\|} \sum_{i=0}^{N-1} x_i |i\rangle \tag{58}$$

be the associated normalized quantum state, where $\|x\|$ denotes the standard Euclidean vector norm. Let $n = \log_2(N)$ denote the number of qubits of $|\psi\rangle$. The goal is to prepare the state $|\psi\rangle$ by applying a quantum circuit to the initial state $|0\rangle^{\otimes n}$. This problem has been extensively studied in the literature; a common approach, originating in [2], is to iterate through each of the $n$ qubits and perform a single-qubit rotation, with the angle of rotation depending on the setting of the previous qubits. The rotation on the first qubit creates the 1-qubit state

$$\left( \sqrt{\sum_{i=0}^{N/2-1} |x_i|^2} \right)|0\rangle + \left( \sqrt{\sum_{i=N/2}^{N-1} |x_i|^2} \right)|1\rangle$$

by performing a single-qubit rotation (about the $Y$ axis) on the state $|0\rangle$ by an appropriate angle. Next, a similar kind of single-qubit rotation is performed on the second qubit, where the angle of rotation depends on whether the first qubit is $|0\rangle$ or $|1\rangle$. The $(m+1)$st rotation is by one of $2^m$ angles, depending on the setting of the first $m$ qubits. Thus, in total there are $1 + 2 + \cdots + 2^{n-1} = N - 1$ total angles that might be used for single-qubit rotations. This sequence of operations prepares the state $\|x\|^{-1} \sum_{i=0}^{N-1} |x_i||i\rangle$. To apply the phases, a single- (or zero-) qubit phase gate with the appropriate phase "angle" is performed—the angle depends on the setting of all $n$ qubits, corresponding to the $N = 2^n$ different phases that might be needed. Thus, the total number of angles that define the protocol is $2N - 1$, exactly corresponding to the number of real parameters needed to describe the general state in Eq. (58).

It remains to describe how the controlled single-qubit rotations are performed when there are many control bits and different angles for each setting of the control. Here, one has many choices and the exact method will depend on how one has access to the data in $x$ and what

---

[55]When the amplitudes are given by some well-behaved function, rather than being arbitrarily chosen, different (related) protocols are used; see §Further reading, below.





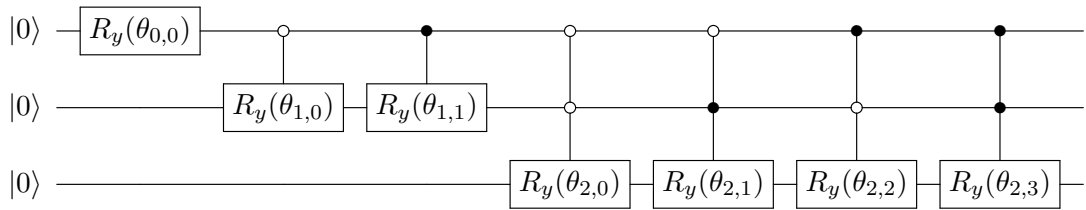

Figure 9: A simple quantum circuit to prepare an arbitrary state $|\psi\rangle$ with non-negative real amplitudes on $n = 3$ qubits. The gate $R_y(\theta)$ denotes a single-qubit rotation by angle $\theta$ about the $Y$ axis. The $2^n - 1$ angles $\theta_{s,p}$ indexed by $s \in \{0, 1, \ldots, n-1\}$ and $p \in \{0, 1, \ldots, 2^s - 1\}$ can be calculated from the amplitudes $x_i$. To account for negative or complex amplitudes, $2^n$ additional controlled phase gates would be needed. More sophisticated proposals can reduce the depth for ancilla-free constructions from $\mathcal{O}(2^n)$ to $\mathcal{O}(2^n/n)$ [3].

resource is being optimized. The most straightforward way is to iterate through each possible setting of the control bits and perform a multiply controlled rotation by a fixed angle for each in sequence. This approach requires $\mathcal{O}(N)$ $n$-qubit gates applied sequentially, as depicted in Fig. 9. Assuming one can perform arbitrary single-qubit gates to exact precision, it is possible to prepare the state $|\psi\rangle$ exactly. However, it is often useful to work with finite precision angles stored in binary, moreover one often needs to design circuits from a discrete gate set, such as the Clifford + $T$ gate set, when compiling into a gate sequence that can be implemented fault tolerantly. When this is the case, single-qubit rotations must be performed approximately: to approximate a single-qubit rotation to error $\epsilon$, a Clifford + $T$ sequence of length $\mathcal{O}(\log(1/\epsilon))$ can be applied [4].

When ancilla qubits are available, one can design protocols that have shallower depth (but about the same total number of gates). For example, one can store the $2N - 1$ angles in a quantum random access memory (QRAM) data structure. In this case, for $m = 1, \ldots, n$, to perform the required controlled rotations on the $m$th qubit in superposition, one can (i) read in the (approximate) binary value of the rotation angle (depending on the setting of the first $m - 1$ qubits) into an ancilla register using a single query to the QRAM, and then (ii) perform fixed-angle single-qubit rotations on the $m$th qubit, controlled by the qubits of the ancilla register storing the binary representation of the rotation angle, and finally (iii) reset (uncompute) the ancilla register with another query to the QRAM. This way, one applies the correct angle in parallel, rather than iterating through all possible $2^{m-1}$ angles.

**Dominant resource cost (gates/qubits)**

In Table 7, we collect several state preparation results in the model where any single-qubit gate can be performed exactly and the only multiqubit gates allowed are CNOTs. The circuit size (i.e., the total number of single-qubit and CNOT gates) and depth (i.e., the number of parallel-acting layers of gates), as well as the number of ancilla qubits (i.e., the number of qubits beyond the $n$ qubits needed to hold the state $|\psi\rangle$) are listed.

Note that the result of [5], which shows depth $\mathcal{O}(2^n/(m + n))$ using $m$ ancilla qubits for $m \leq \mathcal{O}(2^n/n)$, encompasses all other results in the table (and is superior to the third row as it





| Ref. | Circuit size | Circuit depth | Ancilla qubits |
|---|---|---|---|
| [3, 5] | $\mathcal{O}(2^n)$ | $\mathcal{O}(2^n/n)$ | none |
| [3, 5] | $\mathcal{O}(2^n)$ | $\mathcal{O}(2^n/(m+n))$ | $m \in [0, \mathcal{O}(2^n/n)]$ |
| [3, 6, 7] | $\mathcal{O}(2^n)$ | $\mathcal{O}(n)$ | $\mathcal{O}(2^n)$ |

Table 7: Asymptotic resource cost (and tradeoffs therein) of exact state preparation of abitrary states in a gate set with CNOT gates and arbitrary single-qubit gates.

uses $\mathcal{O}(2^n/n)$ ancilla qubits instead of $\mathcal{O}(2^n)$). We include the other results for completeness, as they are distinct constructions and can have other potential upsides.

A lower bound of $\Omega(2^n)$ is known for circuit size [1], so all of the results above are size optimal up to constant factors. Moreover, for any $m$ ancillas, a lower bound of $\Omega(\max(n, 2^n/(n+m)))$ is known for the circuit depth [3], so all of the results above are also optimal in circuit depth, up to constant factors.

For approximate state preparation using a discrete gate set such as Clifford + $T$, the state $|\psi\rangle$ is prepared up to $\epsilon$ error in $\ell_2$-norm, so that the circuit size and depth depends on $\epsilon$. Results in this model are collected in Table 8.

| Ref. | Circuit size | Circuit depth | Ancilla qubits |
|---|---|---|---|
| [3] | $\mathcal{O}(2^n \log(2^n/\epsilon))$ | $\mathcal{O}\left(\frac{2^n}{n} \log(2^n/\epsilon)\right)$ | none |
| [3] | $\mathcal{O}(2^n \log(2^n/\epsilon))$ | $\mathcal{O}\left(\frac{2^n}{m+n} \log(2^n/\epsilon)\right)$ | $m \in [0, \mathcal{O}\left(\frac{2^n}{n\log(n)}\right)]$ |
| [8] | $\mathcal{O}(2^n \log(1/\epsilon))$ | $\mathcal{O}\left(\frac{2^n}{m} \log(m) \log(\log(m)/\epsilon)\right)$ | $m \in [0, \mathcal{O}(2^n)]$ |
| [7] | $\mathcal{O}(2^n \log(n/\epsilon))$ | $\mathcal{O}(n + \log(1/\epsilon))$ | $\mathcal{O}(2^n)$ |

Table 8: Asymptotic resource cost (and tradeoffs therein) of approximate state preparation using the Clifford + $T$ gate set.

If the state $|\psi\rangle$ is sparse, meaning that only $s$ of the $N$ amplitudes are nonzero, then more efficient state preparation methods are known. In particular, [6, 3, 9] have studied shallow-depth circuits for sparse state preparation, achieving circuit depth $\mathcal{O}(\log(ns))$ using only $\mathcal{O}(ns/\log(s))$ ancilla qubits [9], a great improvement over the general case when $s \ll N$.

In some fault-tolerant implementation schemes, such as lattice surgery using surface codes, Clifford gates can be performed cheaply, while $T$ gates require the expensive process of magic state distillation. While $\Omega(2^n \log(1/\epsilon)/\log(n))$ total gates are necessary [10, Eq. 4.85] to approximately create $|\psi\rangle$, [11] noted that it is possible to reduce the number of $T$ gates to $\sqrt{2^n} \log(2^n/\epsilon)$ using $\sqrt{2^n} \log(1/\epsilon)$ ancillas (in fact, there is a smooth tradeoff between the $T$ count and the number of ancillas). Furthermore, these ancillas can be *dirty*, meaning that they can be initialized into any quantum state and they are returned to their (potentially unknown) initial state at the end of the procedure.

All of the above constructions are "garbage-free" state preparation protocols, because they prepare the state $|\psi\rangle$ exactly and all ancilla qubits are returned to their initial state. However, in some applications, it is allowed to leave an ancilla register entangled with the data as long as





the amplitudes are correct. That is, one might prepare the state

$$\frac{1}{\|x\|} \sum_{i=0}^{N-1} x_i |i\rangle \otimes |\text{garbage}_i\rangle .$$

In this setting, en route to giving better algorithms for the electronic structure problem, [12, Section IIID] gave a construction that approximately prepares the above state using only $\mathcal{O}(N + \log(1/\epsilon))$ $T$ gates, albeit still requiring $\mathcal{O}(N \log(1/\epsilon))$ Clifford gates and $\mathcal{O}(\log(N/\epsilon))$ ancillas. In [12], the construction is presented with $\mathcal{O}(N)$ depth, but it could be improved to $\mathcal{O}(\log(N))$ depth at the expense of additional ancillas, using log-depth constructions for QRAM, and it could also be combined with the space-time tradeoffs mentioned above, as discussed in [11, 13].

**Caveats**

- Classical preprocessing: Computing the circuits for preparing $|\psi\rangle$ given the list of $N$ coefficients $x$ can be a non-negligible classical cost. For example, computing each of the $\mathcal{O}(N)$ single-qubit rotation angles requires computing sums and evaluating trigonometric functions, which can be done to precision $\epsilon$ in polylog$(1/\epsilon)$ classical time. Moreover, computing Clifford + $T$ gate sequences that approximate given rotation angles to error $\epsilon$ likewise requires polylog$(1/\epsilon)$ classical time [4]. The total classical work scales as $O(N \text{polylog}(1/\epsilon))$, although this cost can be parallelized.

- Coherent arithmetic: To avoid some of the classical preprocessing, one might try to perform the arithmetic coherently. This might be unavoidable if the entries of $x$ arrive in an online fashion and rotation angles and other quantities need to be computed after superpositions have been created. Formally, the scaling of coherent arithmetic is mild, generally requiring just polylog$(N, 1/\epsilon)$ number of gates and ancilla qubits, but in practice this is likely to be expensive. For example, known methods for coherently computing $\arcsin(\cdot)$ to nine bits of precision use order-$10^4$ Toffoli gates and more than 100 ancilla qubits [14]. See [15] for a general black-box approach that avoids coherent arithmetic.

- Too many ancilla qubits: Achieving depths that scale logarithmically with $N$ requires $\mathcal{O}(N)$ ancilla qubits, which limits the size of $N$ that might be practical. This could be mitigated if it is possible to develop a large-scale hardware element specialized for performing the sort of circuits that arise in these protocols, similar to a QRAM.

- Long-range gates: Achieving polylog$(N)$ depth for state preparation requires $\mathcal{O}(N)$ ancilla qubits and $\mathcal{O}(N)$ gates, many of which act in parallel and on far-separated qubits. If spatial locality were imposed, it would likely be difficult to avoid $\mathcal{O}(N)$ overhead in depth.

- Dequantization: Consider the task of drawing samples from the same probability distribution induced by measuring $|\psi\rangle$ in the computational basis in time polylog$(N)$ time. Preparing $|\psi\rangle$ as described is a quantum method of doing so, but the same can be done classically by first constructing a certain classical data structure and assuming access to classical RAM [16]. In some machine learning applications, this idea leads to classical algorithms that effectively dequantize quantum algorithms that utilize the state preparation primitive [17, 18].





**Example use cases**

- Hamiltonian simulation via linear combination of unitaries (LCU) requires a PREPARE step where a state is prepared with certain classically computed coefficients. Relatedly, the same PREPARE gadget is used to construct block-encodings of such Hamiltonians. However, in this application, state preparation with garbage is generally allowable.

- In certain quantum machine learning protocols, classical data (e.g., image pixel values) are encoded into a quantum state via the so-called "amplitude encoding," where $N$ classical features are stored in a quantum state of $\log_2(N)$ qubits [19]. Following the preparation of the amplitude-encoded data, the state is processed with the goal of, for example, classifying the image.

- Creating a block-encoding of a matrix of classical data is performed using state preparation as a subroutine (more precisely, block-encoding classical data requires controlled state preparation). The block-encoding is then useful in a variety of contexts, for example in quantum interior point methods, and certain quantum machine learning algorithms.

**Further reading**

- When the amplitudes $x_i$ correspond to an efficiently computable function $f(i)$, the complexity of state preparation can be reduced. In this case, the oracle access to $x_i$ can be replaced by a reversible computation of $f(i)$, up to $t$ bits of precision, using coherent arithmetic $|i\rangle|0^t\rangle \to |i\rangle|f(i)\rangle$ [14, 20, 21]. The value of $f(i)$ can be *transduced* into the amplitude using the methods of [22, 15, 23, 24], and the success probability boosted to unity using quantum amplitude amplification. There is an alternative method [25], based on quantum singular value transformation (QSVT) that circumvents the need for the coherent evaluation of $f(i)$ by implementing a low-cost block-encoding of $\sin(i)$, and then using QSVT to apply $f(\arcsin(\cdot))$ to this block-encoding. The complexity of both of these approaches depends on an "$\ell_2$-norm filling-fraction" $\mathcal{F}_f^{[N]} := \|f(i)\|_2/(\sqrt{N}|f(i)|_{\max})$ as $\mathcal{O}(1/\mathcal{F}_f^{[N]})$ (see [25] for more detail). There is also an approach [26] based on the adiabatic algorithm which has a worse dependence on $\mathcal{F}_f^{[N]}$. For efficiently integrable probability distributions, one can use the approach of [2], which has complexity independent of $\mathcal{F}_f^{[N]}$. However, this approach requires coherent arithmetic to reversibly evaluate the integral of the desired function (when applied to functions for which an analytic expression for the integral is not available, this can nullify the quadratic speedup in quantum-accelerated Monte Carlo estimation [27]). There also exist methods specialized for certain target states, such as Gaussians [28, 29].

- A related problem asks to synthesize an arbitrary $2^n \times 2^n$ unitary. Without ancillas, this requires depth and size $\mathcal{O}(4^n)$, for which there are upper [30] and lower [31] bounds that match up to constant factors. With ancillas, it is an open question whether or not the depth can be reduced to poly($n$); this is related to the "unitary synthesis problem" from the list of open problems in [32], and it has been studied in several works, for example, [3, 33, 5]. A depth lower bound of $\Omega(n + 4^n/(m + n))$ is known for $m$ ancilla qubits [3], but the shallowest upper bound is depth $\mathcal{O}(n2^{n/2})$, using $m = \mathcal{O}(4^n/n)$ ancilla qubits [5].

## 17.3 Block-encoding dense matrices of classical data

**Rough overview (in words)**

Many potential applications of quantum algorithms, especially in the area of machine learning, require access to large amounts of classical data, and in order to process this data on quantum devices, one needs coherent query access to the data. Block-encoding is a technique for importing classical data into quantum computers that provides exactly this type of coherent query access. Block-encodings work by encoding the matrices of classical data as blocks within larger matrices, which are defined such that the full encoding is a unitary operator. One way of thinking of this process is by "brute-force" compiling a unitary with the right structure, and then postselecting measurement outcomes to ensure the desired block of the unitary was applied. In general, block-encoding a dense matrix is not an efficient process, as both the normalization factor of the block-encoding and the circuit complexity to implement it can scale with the size of the matrix (e.g., poly($N$) for an $N \times N$ matrix). Nonetheless, end-to-end applications often assign *polylogarithmic* cost to the block-encoding, which is achievable if the relevant cost metric is the circuit depth (rather than the circuit size)—this is similar to the assumption that one has access to large-scale log-depth quantum random access memory (QRAM). For a general treatment not restricted to dense classical data, see Section 10.1 on block-encoding.

**Rough overview (in math)**

Given an $N \times N$ matrix $A$, a block-encoding is a way of encoding the matrix $A$ as a block in a larger unitary matrix:

$$U_A = \begin{pmatrix} A/\alpha & \cdot \\ \cdot & \cdot \end{pmatrix}.$$

If $A$ is not square, one can pad it with zeros such that it becomes square. Let $n = \lceil \log_2(N) \rceil$. We say that the $(n + a)$-qubit unitary $U_A$ is an $(\alpha, a, \epsilon)$-block-encoding of the matrix $A \in \mathbb{C}^{N \times N}$ if

$$\left\| A - \alpha(\langle 0|^{\otimes a} \otimes I) U_A (|0\rangle^{\otimes a} \otimes I) \right\| \leq \epsilon,$$

where $a \in \mathbb{N}$ represents the number of ancilla qubits needed, $\alpha \in \mathbb{R}_+$ is a normalization constant, and $\epsilon \in \mathbb{R}_+$ is an error parameter. The fact that $U_A$ is unitary implies that the normalization constant $\alpha$ must satisfy $\alpha \geq \|A\|$, where $\|\cdot\|$ denotes the spectral norm.

In this section, we consider the case where the $N^2$ entries of $A$ are arbitrary values provided to us in a classical database. In general, all of these entries can be nonzero; that is, $A$ is a dense matrix. The goal is to provide a quantum circuit implementing a unitary $U_A$ as above while minimizing quantities such as the circuit depth and circuit size, as well as the number of ancilla qubits $a$ and the normalization constant $\alpha$. The block-encoding construction we focus on performs $U_A$ using a pair of state preparation unitaries [1, 2, 3], following the more general method of block-encoding Gram matrices [1, Lemma 47]. In particular, the product

$$U_A = U_R^\dagger U_L$$

is an exact $(\alpha, a, 0)$-block-encoding of $A$, where $U_L$ and $U_R$ are unitaries that perform (controlled) state preparation. Specifically, the $(n + a)$-qubit unitaries $U_L$ and $U_R$ prepare a different $2n$-qubit state for each of $2^n$ possible settings of the final $n$-qubit register, with the assistance of





$a - n$ additional ancilla qubits, as follows:

$$U_L|0\rangle^{\otimes(a-n)}|0\rangle^{\otimes n}|i\rangle = |0\rangle^{\otimes(a-n)}|\psi_i\rangle$$
$$U_R|0\rangle^{\otimes(a-n)}|0\rangle^{\otimes n}|j\rangle = |0\rangle^{\otimes(a-n)}|\phi_j\rangle\,, \tag{59}$$

where the $2n$-qubit states $|\psi_i\rangle$ and $|\phi_j\rangle$ are chosen such that $\langle\psi_i|\phi_j\rangle = A_{ij}/\alpha$, where $A_{ij}$ is the matrix entry of $A$ in row $i$ and column $j$—the states $|\psi_i\rangle$ and $|\phi_j\rangle$ encode the (normalized) rows of $A$ and norms of those rows, respectively. For this construction, the normalization constant $\alpha$ satisfies

$$\alpha = \|A\|_F\,,$$

where $\|\cdot\|_F$ is the Frobenius norm. Note that for an $N \times N$ matrix the Frobenius norm satisfies $\|A\| \leq \|A\|_F \leq \sqrt{N}\|A\|$—thus, the value of $\alpha$ achieved by this method can be larger than its minimal possible value ($\|A\|$) by a factor as large as $\sqrt{2^n}$.[56]

There are several methods of implementing the (controlled) state preparation unitaries $U_L$ and $U_R$, offering tradeoffs between various metrics, as discussed in Section 17.2 on state preparation. Of particular relevance is the $T$-count and $T$-depth of the circuit when it is decomposed into a Clifford + $T$ gate set, as the $T$ gate is the most difficult to implement in many fault-tolerant schemes. A general strategy for implementing $U_L$ and $U_R$ involves constructing binary trees representing the amplitudes in the states $|\psi_i\rangle$ and $|\phi_j\rangle$ in Eq. (59), and building the state preparation unitaries out of controlled $Y$ rotations by angles defined in those binary trees (the controlled $Y$ rotations are performed *approximately* in a discrete gate set like Clifford + $T$, leading to a nonexact block-encoding). Tradeoffs involving the $T$-depth, $T$-count, and number of ancilla qubits are established based on the manner in which these controlled $Y$ gates are performed. For example, the shallowest implementation [4] requires a large number of ancilla qubits, which are used to perform all the controlled $Y$ rotations in one parallel layer, before "injecting" a subset of these ancillas into the main data qubits using a controlled SWAP network and then uncomputing the ancillas.

**Dominant resource cost (gates/qubits)**

The shallowest implementations are able to achieve polylog($N$) depth for $U_L$ and $U_R$ (and hence for $U_A$), at the expense of $a = \mathcal{O}(N^2)$ ancilla qubits. On the other hand, the fewest number of ancillas needed by this family of methods would be $a = n$, in which case the circuit depth would scale as $\mathcal{O}(N^2)$. While the total circuit size must be at least $\Omega(N^2)$, the number of gates in the circuit that are $T$ gates can be as small as $\mathcal{O}(N)$ using the techniques in [5].

Detailed resource counts (including the constant prefactors for key metrics) and implementations of block-encodings were studied in [4]. We reproduce their resource counts optimized for minimum $T$-depth and for $T$-count in Table 9.

**Caveats**

An important caveat is that the total gate complexity of $U_A$ must be at least $\Omega(N^2)$, reflecting the $N^2$ degrees of freedom in the arbitrary $N \times N$ matrix $A$. Thus, while $A$ operates on a quantum system of only $\log_2(N)$ qubits, achieving depth polylog($N$) requires parallel-acting gates across

---

[56]See [1, Lemma 50] for a variant of this method yielding normalization factor $\alpha = \sqrt{n_q(A)n_{2-q}(A^\dagger)}$ for $q \in [0, 2]$, where $n_q(A) = \max_i \|A_{i,\cdot}\|_q^q$, with $\|\cdot\|_q$ the vector $q$-norm and $A_{i,\cdot}$ the $i$-th row of $A$.





| | Optimized for min depth | Optimized for min count |
|---|---|---|
| # Qubits | $4N^2$ | $N \log(1/\epsilon)$ |
| $T$-Depth | $10 \log(N) + 24 \log(1/\epsilon)$ | $8N + 12 \log(N)(\log(1/\epsilon))^2$ |
| $T$-Count | $12N^2 \log(1/\epsilon)$ | $16N \log(1/\epsilon) + 12 \log(N)(\log(1/\epsilon))^2$ |

Table 9: Explicit resource counts for block-encoding circuits of arbitrary matrices of classical data. These expressions omit subleading terms; the full expressions can be found in [4].

at least $\Omega(N^2)$ ancilla qubits. In many algorithms, it is assumed that the cost of implementing $U_A$ is polylog$(N)$, in order to preserve an end-to-end runtime that is polylog$(N)$ and claims of exponential speedup. This is only defensible if the key metric is the circuit depth and if many ancilla qubits are available. Furthermore, the normalization constant $\alpha$ can introduce poly$(N)$ factors into an end-to-end analysis, owing to the fact that $\|A\|_F$ can be larger than $\|A\|$ by a $\sqrt{N}$ factor.

Another caveat to note is that if the matrix being block-encoded is sparse and if the values and locations of its nonzero entries can be computed efficiently, or if the matrix enjoys some structure in the data in addition to sparsity, then more efficient block-encoding methods can be employed—see Section 10.1 on block-encoding for details. In those cases, the results stated here may not be applicable.

**Example use cases**

In financial portfolio optimization, classical data representing average historical returns and covariance matrices for a universe of assets is needed in a quantum algorithm for optimizing a portfolio. See, for example, [6]. Similarly, in quantum machine learning based on quantum linear algebra, the algorithm often requires fast coherent access to large matrices of classically stored data.

**Further reading**

- An excellent overview of block-encodings and quantum linear algebra: [7].

- A detailed resource count of block-encoding with explicit circuits: [4].

- Select-SWAP QRAM and a tradeoff between qubit count and $T$ gates: [5].

- For sparse matrices of classical data, or matrices expressed as a linear combination of Pauli matrices, more efficient methods for block-encoding exist. Asymptotic resource expressions for varying number of ancilla qubits are reported in [8].

# 18  Quantum linear system solvers

*The authors are grateful to Dong An for reviewing this section of the survey.*

**Rough overview (in words)**

The goal is to solve linear systems of equations with quantum subroutines. More precisely, a *quantum linear system solver* (QLSS) takes as input an $N \times N$ complex matrix $A$ together with a complex vector $b$ of size $N$, and outputs a pure quantum state $|\tilde{x}\rangle$ that is an $\varepsilon$-approximation of the normalized solution vector of the linear system of equations $Ax = b$. In basic versions, QLSSs do so by loading the normalized entries of the matrix $A$ and the normalized entries of the vector $b$ into a unitary quantum circuit, either from a quantum random access memory (QRAM) data structure, or—if the structure of $A$ and $b$ allows for this—by efficiently computing the corresponding entries on the fly.

Crucially, the number of algorithmic qubits of the linear system solver itself is only roughly $\log_2(N)$, which is exponentially smaller than the matrix size. While for general systems the number of QRAM qubits still scales with the matrix/vector size, QRAM encodings can be made more space efficient for sparse systems or can even be avoided when the corresponding entries are efficiently computable. The complexity of QLSSs depends on the condition number $\kappa(A) = \|A^{-1}\| \cdot \|A\|$ of the matrix $A$, and one then aims to give circuits with minimal quantum resource costs—such as ancilla qubits, total gate count, circuit depth, etc.—in terms of $\kappa(A)$ and the desired accuracy $\varepsilon \in (0,1)$.

**Rough overview (in math)**

There are different standard input models on how the classical data from $(A, b)$ is loaded into the quantum processing unit, which are equivalent up to small polylogarithmic overhead for general matrices. We state the complexities in terms of query access of a unitary $U_b$ preparing the $n = \lceil \log_2(N) \rceil$-qubit pure quantum state $|b\rangle = \|b\|^{-1} \cdot \sum_{i=1}^{N} b_i |i\rangle$ for $b = (b_1, \ldots, b_N)$, where $\|\cdot\|$ for vector arguments denotes the standard Euclidean norm, together with an $(\alpha, a, 0)$-block-encoding $U_A$ of the matrix $A$. The QLSS problem is then stated as follows: for a triple $(U_A, U_b, \varepsilon)$ as above, the goal is to create an $n$-qubit pure quantum state $|\tilde{x}\rangle$ such that

$$\left\| |\tilde{x}\rangle - |x\rangle \right\| \leq \varepsilon$$

with

$$|x\rangle = \frac{\sum_{i=1}^{N} x_i |i\rangle}{\left\| \sum_{i=1}^{N} x_i |i\rangle \right\|} \text{ defined by } Ax = b \text{ with } x = (x_1, \ldots, x_N), \tag{60}$$

by employing as few times as possible the unitary operators $U_A, U_b, U_A^\dagger, U_b^\dagger$, controlled versions of $U_A, U_b, U_A^\dagger, U_b^\dagger$, and additional quantum gates on potentially additional ancilla qubits. An alternative (and closely related) error metric studied in some works is based on the trace norm, requiring $\frac{1}{2}\||x\rangle\langle x| - |\tilde{x}\rangle\langle\tilde{x}|\|_1 \leq \varepsilon$.

One way to think of the QLSS problem is that we seek the matrix inverse $A^{-1}$, and that this can be implemented by, for example, quantum singular value transformation (QSVT) acting





on $A$ (via its block-encoding) with a polynomial approximation of the inverse function on the interval $[\|A\|/\kappa(A), \|A\|]$. The complexity of the corresponding scheme thereby depends on the degree of the polynomial needed for a good approximation of the inverse function on the relevant interval, and as such on the condition number $\kappa(A)$, the normalization factor $\alpha$, and the approximation error $\varepsilon$ of the resulting QLSS. In fact, it turns out that the complexity of most quantum algorithms depends on the following combined quantity

$$\kappa'(A) := \kappa(A) \cdot \frac{\alpha}{\|A\|} = \alpha \cdot \|A^{-1}\|,$$

which is no smaller than $\kappa(A)$, because $\alpha \geq \|A\|$ due to the unitarity of the block-encoding. Note that in QRAM-based implementations for dense matrices $A$, one naturally gets $\alpha = \|A\|_F$, which then leads to linear complexity dependence on the Frobenius norm $\|A\|_F$.

As noted in [1, 2], in general, we need not assume that $A$ is invertible nor that it is a square matrix, but can instead use the Moore–Penrose pseudoinverse $A^+$ of the matrix to solve the regression problem Eq. (60) in a least-squares sense, in which case one needs to appropriately change the definition of $\kappa(A)$ to $\|A^+\| \cdot \|A\|$. In fact, the above QSVT-based approach directly solves this more general version of the problem [3].

**Dominant resource cost (gates/qubits)**

The performance of different QLSSs is typically compared based on how their query complexity (to $U_A$ and $U_b$) grows with the condition number, where a lower bound of $\Omega(\kappa(A))$ is known; see [4, 5]. Methods achieving $\mathcal{O}(\kappa'(A))$ dependence are termed "optimal" and methods achieving $\kappa'(A)\,\mathrm{polylog}(\kappa'(A))$ are termed "near-optimal."[57]

The first optimal method was given in [6] (for invertible matrices), which does not directly employ the QSVT for the inverse function. Instead, it is based on discrete adiabatic methods together with quantum eigenstate filtering based on the QSVT for a minimax polynomial [8]. In particular, the adiabatic portion prepares an "ansatz" state $|x_\mathrm{ans}\rangle$ for which $|\langle x_\mathrm{ans}|x\rangle|^2 \geq 1/2$, using at most $\mathcal{O}(\kappa'(A))$ (controlled) queries to $U_A$ and $U_b$. Then, the eigenstate filtering step refines this state by approximately projecting it onto $|x\rangle$: one obtains the state $|\tilde{x}\rangle$ that is $\varepsilon$-far from $|x\rangle$ at additional query cost $\mathcal{O}(\kappa'(A)\log(1/\varepsilon))$. The projection succeeds with probability $p \geq 1/2$, so the whole procedure must be repeated no more than twice on average. Overall, the expected number of queries made by the algorithm is $Q$ controlled queries to each of $U_A$ and $U_A^\dagger$ and $2Q$ queries to each of $U_b$ and $U_b^\dagger$, where

$$Q = \kappa'(A)\Big(C + D\ln(2\varepsilon^{-1})\Big) + o(\kappa'(A)) = \mathcal{O}\big(\kappa'(A)\log(\varepsilon^{-1})\big). \tag{61}$$

Here, $o(\kappa'(A))$ denotes terms growing sublinearly in $\kappa'(A)$, and $C, D$ are constants. The algorithm operates on $n + \mathcal{O}(1)$ qubits ($n + 5$ in the case of [6]), plus the additional qubits used for the block-encoding, discussed in more detail below. There is an additional constant quantum gate complexity for each query to $U_A$ and $U_b$. For the discrete adiabatic method in [6], the constant $C$ can be rigorously bounded as $C \leq 117{,}235$[58] and the constant $D$ is at most 2. Note

---

[57]Regarding the optimal $\varepsilon$ dependence, it has additionally been claimed [6] that a complexity of $\mathcal{O}(\kappa\log(1/\varepsilon))$ is jointly optimal in both $\kappa$ and $\varepsilon$, based on forthcoming work by Harrow and Kothari; see also [7, Appendix A].

[58]This number is derived from applying [6, Theorem 9] with $\sqrt{2-\sqrt{2}} \times 44{,}864 \times \kappa$ steps, each of which incurs one call to the block-encoding, such that the output is guaranteed to have overlap at least $1/\sqrt{2}$ with the ideal state. Eigenstate filtering then succeeds with probability at least $1/2$; accounting for the need to repeat twice on average, one arrives at a constant 117,235, matching [9, Eq. (L2)].





that when $C$ is this large, the corresponding term will actually dominate the $D\kappa'(A)\log(\varepsilon^{-1})$ term for practical scenarios.

Subsequent work has given alternative methods that achieve optimal asymptotic complexity [10, 11]. Reference [10] achieves this by a small modification to and improved analysis of the adiabatic path-following method of [12]. Meanwhile, [11] replaces the step of ansatz state preparation via adiabatic methods with a simpler norm-estimation step, where one seeks a constant-factor approximation of the Euclidean norm $\|x\|$, following that with an eigenstate filtering–like step. An optimized version of this approach was reported to have complexity following Eq. (61) with $C = 56$ and $D = 1.05$. Additionally, this method does not require $A$ to be invertible, but does require $b$ to be in the column space of $A$ [11].

Other known QLSSs with suboptimal asymptotic complexities are based on other versions of adiabatic ansatz state preparation [12, 13, 8], QSVT [3, 14], linear combination of unitaries (LCU) [15], or variable-time amplitude amplification (VTAA) [16, 17, 2]. While the known bounds on the asymptotic complexities of these methods are slightly worse, it remains open if finite-size performance could be competitive (see, e.g., [9, 11]). Moreover, to date, the VTAA-based algorithms are the only variants that are proven to solve the generic least-squares (pseudoinverse) problem while achieving a near-optimal asymptotic scaling [2].

Note that if the matrix $A$ is given in a classical data structure in the computational basis, then standard ways to create the block-encoding $U_A$ make use of a QRAM structure. For general (dense) matrices $A$, the requirement is then size $\mathcal{O}(N^2)$ (number of qubits) with circuit depth $\mathcal{O}(n)$ for each query—or alternatively, as few as $\mathcal{O}(n)$ ancilla qubits could suffice, but at the expense of using $\mathcal{O}(N^2)$ circuit depth [18, 19]. Initializing the depth-efficient QRAM data structure will in general also take $\mathcal{O}(N^2)$ time. However, if $A$ is sparse, either in the computational basis [20], Pauli basis [21], or any orthonormal basis with efficiently implementable basis transformation, there are more efficient direct constructions for block-encoding $A$. Moreover, for Pauli basis access, there exist randomized QLSSs with complexity scaling as the $\ell_1$-norm of the Pauli coefficients [22], completely avoiding the use of block-encodings (and as such QRAM and ancilla qubits).

**Caveats**

QLSSs are an important subroutine for a variety of application areas of quantum algorithms. However, it is crucial to keep track of all the quantum and classical resources required and to compare these to state-of-the-art classical methods. In particular, the following factors should be taken into account:

- The classical precomputation complexities for the eigenstate filtering routine are neglected, but can be kept efficient in practice [23].

- The rigorous upper bound on the size of the complexity constant $C$ has been reduced by several orders of magnitude [11] since the first optimal QLSS was given in [6], but nevertheless remains larger than ideal for usage in applications where QLSS plays a heavy role. However, numerical investigations of two adiabatic methods on small random matrices gave evidence that the empirical performance of those methods is significantly better than the rigorous upper bounds [7].





- When needed, the QRAM cost can be prohibitive, if it requires the full overhead of quantum error correction and fault tolerance [18], especially for QRAMs of maximum size $\mathcal{O}(N^2)$ qubits, required for general (dense) matrices.

- In the formulation of the QLSS problem, the pure quantum state $|x\rangle$ corresponds to the normalized solution vector of the linear system $Ax = b$. While the normalization factor $\|x\|$ can be obtained as well, this comes at the price of added query complexity scaling as $\mathcal{O}(\kappa'(A)\varepsilon^{-1}\log(\varepsilon^{-1}))$ [11] (see also [2, Corollary 32]). This nearly achieves the lower bound of $\Omega(\kappa(A)\varepsilon^{-1})$ [11] (note that norm estimation necessarily has worse $\varepsilon$ dependence than the QLSS itself).

- QLSSs do not produce a classical description of the solution vector $x$ or an approximation thereof, but rather the pure quantum state $|\tilde{x}\rangle$. In order to obtain a classical approximation of the vector $x$, one needs to combine QLSSs with pure state quantum tomography, which can be performed using $\mathcal{O}(N\varepsilon^{-2})$ samples. If poly($n$) query-cost QRAM is also available, then the complexity can be quadratically improved in terms of the precision using optimized pure state tomography [24], or alternatively the overall complexity may be further improved using *iterative refinement* to $\mathcal{O}(Ns^2 + Ns\kappa^2(A)/\|A\|) \cdot \text{polylog}(N/\varepsilon)$, as described in [25], where $s$ is the maximum number of nonzero elements of $A$ in any row or column. In the special case of Laplacian, or more generally symmetric, weakly diagonally dominant (SDD) matrices, [26] gives a quantum algorithm with complexity $\widetilde{\mathcal{O}}(\sqrt{Ns}/\varepsilon)$ that outputs an $\varepsilon$-approximate solution $\tilde{x}$ with respect to the $A$-induced norm. (Measuring error in this norm enables their algorithm not to have a condition number dependence.) The algorithm uses QRAM and provides a subquadratic speedup compared to the classical complexity $\mathcal{O}(N\log(1/\varepsilon))$, but uses rather different techniques compared to standard QLSS algorithms [4].

- The overall complexities $\widetilde{\mathcal{O}}(N\kappa'(A)\varepsilon^{-1})$ and $\mathcal{O}(Ns^2 + Ns\kappa^2(A)/\|A\|) \cdot \text{polylog}(N/\varepsilon)$ (where we generously allow poly($n$) query-cost QRAM) to obtain a classical description of the solution can be compared to classical textbook Gaussian elimination–based computation, which leads to complexity $\mathcal{O}(N^3)$ or more precisely $\mathcal{O}(N^\omega)$ with $\omega \in [2, 2.372]$ denoting the matrix multiplication exponent. Further, QLSSs should also be compared with state-of-the-art randomized solvers. For example, the randomized Kaczmarz method [27] with standard classical access to the matrix elements returns an $\varepsilon$-approximation of the vector $x$, while scaling as $\mathcal{O}(s\kappa_F^2(A)\log(\varepsilon^{-1}))$ for $s$ row-sparse matrices and $\kappa_F(A) = \|A^{-1}\| \cdot \|A\|_F$. Moreover, if $A$ is $s$-sparse and positive semidefinite (PSD), then using the conjugate gradient method one can obtain a solution in time $\widetilde{\mathcal{O}}(Ns\sqrt{\kappa(A)}\log(\varepsilon^{-1}))$ [28, Chapter 10.2], which can be generalized to the least-squares problem (and thus non-Hermitian matrices) at the cost of a quadratically worse condition number dependence $\mathcal{O}(Ns\kappa\log(\kappa(A)/\varepsilon))$ by considering the modified equation $A^\dagger A x = A^\dagger b$. As such, it seems that the QLSS may not provide a superquadratic speedup when a full classical solution is to be extracted, and even subquadratic speedups seem to be limited to a narrow parameter regime.

- Quantum-inspired methods [29, 30] that start from a classical data structure intended to mimic QRAM—allowing one to sample from probability distributions with probabilities proportional to the squared magnitudes of elements in a given row of $A$—give samples from an $\varepsilon$-approximation to the solution vector in ($N$-independent) complexity





$\mathcal{O}(\kappa_F^4(A)\kappa^2(A)\varepsilon^{-2})$ [31, 30], and can be used to compute an approximate solution by repeated sampling. Note that while the required data structure is classical, it might still be prohibitively expensive to build when the matrix $A$ is huge.

- When it comes to classical methods, solvers that depend on the condition number are useful in practice whenever combined with preconditioners [32]. However, the performance of preconditioners in the quantum setting (see, e.g., [33, 34, 35, 36]) is often only heuristic, or applies only to restricted situations. This topic would benefit from further exploration.

### Example use cases

- Quantum interior point methods in convex optimization and corresponding applications [37, 38].

- Quantum machine learning applications [1, 39].

- Solving differential equations and corresponding applications, for example, for the finite element method that does not require a tomography step [40].

### Further reading

- Original QLSS (termed HHL) [4].

- For an overview discussion of QLSS, see [13].

- Optimal-in-$\kappa$ QLSSs are given in [6, 11, 10].

- There are also known (polynomial) speedups in case one needs a full classical description of the output vector in linear equation solving and in some regression variants [25, 41].

# 19 Quantum gradient estimation

*The authors are grateful to Nikitas Stamatopoulos for reviewing this section of the survey.*

**Rough overview (in words)**

Estimating the gradient of a high-dimensional function is a widely useful subroutine of classical and quantum algorithms. The function's gradient at a certain point can be classically estimated by querying the value of the function at many nearby points. However, the number of evaluations will scale with the number of dimensions in the function, which can be very large. By contrast, the quantum gradient estimation algorithm evaluates the function a *constant* number of times (in superposition over many nearby points) and uses interference effects to produce the estimate of the gradient. While there are caveats related to the precise access model and the classical complexity of gradient estimation in specific applications, this procedure can potentially lead to significant quantum speedups.

**Rough overview (in math)**

Let $f: \mathbb{R}^d \to \mathbb{R}$ be a real function on $d$-dimensional inputs, and assume that it is differentiable at a specific input of interest, taken to be the origin $\mathbf{0} = (0, 0, \ldots, 0)$ for simplicity (the algorithm works equally well elsewhere). Let $g = (g_1, \ldots, g_d)$ denote the gradient of $f$ at $\mathbf{0}$, that is, $g = \nabla f(\mathbf{0})$. We wish to produce a classical estimate $\tilde{g}$ of $g$ that satisfies $|g_j - \tilde{g}_j| < \varepsilon$ for all $j = 1, \ldots, d$.

Ignoring higher-order terms, the function may be approximated near the origin as $f(x) \approx f(\mathbf{0}) + \langle g, x \rangle$, where $\langle \cdot, \cdot \rangle$ denotes the normal inner product. The original gradient estimation algorithm by Jordan [1] then considers a $d$-dimensional grid of points near the origin denoted by $G$. For simplicity, suppose on each of the $d$ dimensions, the grid has $N$ evenly spaced points on the interval $[-\ell/2, \ell/2]$, for a certain parameter $\ell$ related to the precision requirements of the algorithm, where $N$ is assumed to be a power of 2. Let $m$ be an upper bound on the magnitude of the components of $g$. Define $g' = Ng/2m$ to have components between $-N/2$ and $N/2$. Similarly, let $\tilde{g}' = N\tilde{g}/2m$ be the desired normalized-and-shifted output.

The quantum algorithm prepares a superposition of the grid points $x \in G$ and computes function $f(x)$ (times a constant $\pi N/m\ell$) into the phase, producing the state

$$\frac{1}{\sqrt{N^d}} \sum_{x \in G} e^{\mathrm{i}\pi N f(x)/m\ell} |x\rangle \approx \frac{e^{\mathrm{i}\pi N f(\mathbf{0})/m\ell}}{\sqrt{N^d}} \sum_{x \in G} e^{\mathrm{i}\pi N \langle g, x \rangle/m\ell} |x\rangle,$$

where $|x\rangle$ denotes the product state $|l_1\rangle|l_2\rangle \cdots |l_d\rangle$, where $l_j$ is a binary string of length $\log_2(N)$ containing a representation of the $j$-th component $x_j$ of the vector $x$, with the identification $x_j = -\ell/2 + \ell l_j/N$. With this in mind, the latter state is rewritten as the product state, up to a global phase and normalization constant

$$\left( e^{-\pi \mathrm{i} g_1'} \sum_{l_1=0}^{N-1} e^{2\pi \mathrm{i} l_1 g_1'/N} |l_1\rangle \right) \left( e^{-\pi \mathrm{i} g_2'} \sum_{l_2=0}^{N-1} e^{2\pi \mathrm{i} l_2 g_2'/N} |l_2\rangle \right) \cdots \left( e^{-\pi \mathrm{i} N g_d'} \sum_{l_d=0}^{N-1} e^{2\pi \mathrm{i} l_d g_d'/N} |l_d\rangle \right).$$





Due to the approximated linearity of $f$, each of the product state constituents is observed to be close to a basis state in the Fourier basis (see Eq. (50)). By performing an inverse quantum Fourier transform (QFT) in parallel for each of the $d$ dimensions and measuring in the computational basis, a computational basis state

$$|\tilde{g}'\rangle = |\tilde{g}'_1\rangle|\tilde{g}'_2\rangle \cdots |\tilde{g}'_d\rangle$$

is retrieved (up to an unimportant global phase), where with high probability $\tilde{g}'_j$ approximates $g'_j$ to $\log_2(N)$ bits of precision. The coordinate $\tilde{g}_j$ is then recovered as $\tilde{g}_j = 2m\tilde{g}'_j/N$. Assuming $m = \mathcal{O}(1)$, taking $N = \mathcal{O}(1/\varepsilon)$ suffices to solve the problem. In a full analysis, one must make sure not to choose $\ell$ too large (else the linearity approximation breaks down).

In [1], the unitary $U_f$ sending $|x\rangle \mapsto e^{i\pi N f(x)/m\ell}|x\rangle$ was performed using a constant number of calls to the evaluation oracle that computes an approximation to $f(x)/m$ to precision $\mathcal{O}(\varepsilon^2/\sqrt{d})$ into an ancilla register. In [2], the precision required was improved to $\mathcal{O}(\varepsilon/\sqrt{d})$ using finite difference formulas to put the gradient into the phase. Additionally, it was shown how $U_f$ can be implemented using $\mathcal{O}(\sqrt{d}/\varepsilon)$ calls to a "probability oracle" that (assuming $0 \leq f(x) \leq 1$) performs the map $|x\rangle|0\rangle \mapsto \sqrt{f(x)}|x\rangle|1\rangle + \sqrt{1 - f(x)}|x\rangle|0\rangle$.

The gradient estimation algorithm can be viewed as a generalization of the Bernstein–Vazirani algorithm [3], which considers binary functions $f : \{0,1\}^n \to \{0,1\}$, and promised that $f(x) = \langle g, x\rangle \bmod 2$ for some unknown vector $g$, determines $g$ with one query to $f$.

**Dominant resource cost (gates/qubits)**

The superposition over grid points can be easily accomplished with Hadamard gates. Likewise, the inverse QFT operation is relatively cheap. The number of qubits is $\mathcal{O}(d\log(N))$, and the number of elementary operations for each of the $d$ parallel QFTs is polylog($N$)—thus, the gate depth is independent of $d$, while the total gate complexity is linear in $d$. Additionally, an important component of the complexity comes from performing the unitary $U_f$, which requires implementing either an evaluation oracle or a probability oracle for the function $f$. If one has access to an evaluation oracle, the function must be evaluated to precision $\mathcal{O}(\varepsilon/\sqrt{d})$. Thus, if function evaluations can be made to precision $\delta$ in circuit depth polylog($d, 1/\delta$), the overall circuit depth of the quantum gradient estimation algorithm will be polylog($d, 1/\varepsilon$), a potentially exponential speedup over the at least $\Omega(d)$ classical query complexity to learn the gradient. In the case that one has access to a probability oracle, a number of oracle calls scaling as $\mathcal{O}(\sqrt{d}/\varepsilon)$ must be made.

For some functions, it is possible to classically compute $f(x)$ to precision $\delta$ with gate complexity poly($d, \log(1/\delta)$). This can be turned into a quantum circuit $U_f$ with a comparable gate complexity. For other functions, computing $f(x)$ may be much harder. For example, if $f(x)$ is defined as the output probability of a quantum circuit described by $d$ parameters, then computing $f(x)$ to precision $\delta$ might be difficult for a classical computer, and even on a quantum computer, it generally requires $\mathcal{O}(1/\delta)$ complexity. However, in this case, implementing a probability oracle is simple, leading to the motivation for the work of [2].

**Caveats**

Jordan's formulation of the algorithm [1] appears to offer a large quantum speedup by accomplishing in a single quantum query what requires $\Omega(d)$ classical queries. However, this requires a fairly strong access model where one has access to an oracle for computing the value of the





function $f$ to high precision. For an exponential speedup to be possible, precision $\varepsilon$ must be achievable at cost polylog($d, 1/\varepsilon$). Unfortunately, for actual functions $f$ that show up in applications where this is possible, it is often the case that one can classically compute the gradient much more efficiently than simply querying the value of $f$ at many nearby points. Indeed, the "cheap gradient principle" [4, 5] asserts that (in many practical situations) computing the gradient has roughly the same cost as computing the function itself. This principle limits the scope of application of the large speedup of Jordan's algorithm.

By contrast, [2] shows how the gradient can alternatively be computed using a probability oracle rather than an evaluation oracle, which makes the algorithm compatible with computing gradients in the setting of variational quantum algorithms. However, $\mathcal{O}(\sqrt{d}/\varepsilon)$ calls to the oracle are required, which represents a (much less dramatic) *quadratic* speedup compared to the strategy of using the probability oracle to estimate $f(x)$ at many nearby points and subsequently estimating the gradient classically.

**Example use cases**

- Convex optimization: In convex optimization, local optima are also global optima, and thus a global optimum can be found by greedy methods such as gradient descent. When one can efficiently compute the function $f$ much more cheaply than computing its gradient, the quantum gradient estimation algorithm can give rise to a speedup over classical optimization procedures [6, 7].

- Pure state tomography: Given access to a unitary $U$ that prepares the pure state $|\psi\rangle$, [8] utilizes the gradient estimation algorithm to estimate the amplitudes of $|\psi\rangle$ in the computational basis using an optimal number of queries to $U$.

- Estimating multiple expectation values: Amplitude estimation can be used to estimate an expectation value to precision $\epsilon$ at cost $\mathcal{O}(1/\epsilon)$. In [9, 8], it is shown how the gradient estimation algorithm further allows $M$ expectation values to be simultaneously estimated at cost $\widetilde{\mathcal{O}}(\sqrt{M}/\epsilon)$ calls to a state preparation unitary, considered the most expensive part of the circuit.

- Computing molecular forces: While ground state energies are the object most often studied in algorithms for quantum chemistry, other interesting quantities such as molecular forces can be related to gradients of molecular energies. Reference [10] studies how the gradient estimation algorithm can be leveraged into a quantum algorithm for computing such quantities.

- Escaping saddle points: Although not the essential ingredient, the gradient estimation algorithm was used in the algorithm of [11] for escaping saddle points.

- Variational quantum algorithms: Variational quantum algorithms involve optimizing the parameters of a quantum circuit under some cost function. The ability to estimate the gradient of the cost function with respect to the parameters might allow acceleration of this loop.

- Financial market risk analysis: In [12], the quantum gradient estimation subroutine was utilized to compute the Greeks, parameters associated with financial market sensitivity.





**Further reading**

See [2] for a full discussion of the state of the art with respect to the quantum gradient estimation algorithm.

# 20 Variational quantum algorithms

*The authors are grateful to Marco Cerezo and Xiao Yuan for reviewing this section of the survey.*

**Rough overview (in words)**

The so-called noisy intermediate-scale quantum (NISQ) era is a term used to describe the regime in which the best quantum processors have fifty to a few hundred noisy qubits [1]. In this regime, one does not have enough qubits or low enough error rates to carry out fault-tolerant quantum computation, and so one is constrained to run low-depth quantum circuits. Under these constraints, structured quantum algorithms with prescribed circuits and provable guarantees are unknown. In light of this, variational quantum algorithms (VQAs) have been proposed. We remark that, despite this original setting, it would also be possible to run VQAs on fault-tolerant devices. While many VQAs have been proposed for a wide range of applications, they all share the same core primitive, which we describe below.

The main idea is to encode the target problem into an optimization task of minimizing the expectation value of some parameterized quantum circuit, or a function thereof. In each optimization step, a quantum computer is used to evaluate expectation values at chosen parameter values, which are read by a classical optimizer that updates the parameters for the next step. The motivation for this framework is to offload some of the computational complexity onto the classical optimization algorithm, with an aim for the quantum subroutines to perform classically intractable calculations.

**Rough overview (in math)**

Given some parameterized unitary $U(\boldsymbol{\theta})$ with adjustable parameters $\boldsymbol{\theta}$, input state $\rho$, measurement operator $O$, and function $f(\cdot)$, one evaluates $C(\boldsymbol{\theta}) = f\big(\mathrm{Tr}\big[OU(\boldsymbol{\theta})\rho U^\dagger(\boldsymbol{\theta})\big]\big)$ on a quantum computer, which is known as a cost function. A classical optimizer is then tasked to solve the problem $\boldsymbol{\theta}_* = \mathrm{argmin}_{\boldsymbol{\theta}} f(\mathrm{Tr}[OU(\boldsymbol{\theta})\rho U^\dagger(\boldsymbol{\theta})])$. By careful choice of $f(\cdot)$, $\rho$, and $O$, one can encode a problem of interest such that $U(\boldsymbol{\theta}_*)$ enables an (approximate) solution to the problem. For instance, the solution could correspond to the computational basis state with which the output state $U(\boldsymbol{\theta}_*)\rho U(\boldsymbol{\theta}_*)^\dagger$ has the largest overlap, or to the value of $f(\mathrm{Tr}[OU(\boldsymbol{\theta}_*)\rho U(\boldsymbol{\theta}_*)^\dagger])$ itself. In general, one can also construct a more elaborate cost function comprising a sum of observable-dependent functions with different input states and measurement operators.

The parameterized circuit $U(\boldsymbol{\theta})$ is commonly referred to as the "ansatz circuit." The choice of cost function and ansatz are key components in designing a VQA. Namely, they should ideally satisfy the following properties:

(i) Smaller values of the cost function should correspond to better quality of solution.

(ii) The ansatz should be sufficiently expressive to contain a unitary $U(\boldsymbol{\theta}_*)$, which yields an acceptable solution.

(iii) The ansatz should lead to a trainable cost landscape in parameter space, such that a sufficiently good solution can be found efficiently by the classical optimizer.

(iv) The cost function should be classically hard to simulate, given the choice of ansatz.





It should be noted that while one would expect any VQA to satisfy the first point by design, in general, it can be hard to satisfy all of the above requirements simultaneously via theoretical guarantees or even heuristically in practice. These caveats are discussed in more detail below.

**Dominant resource cost (gates/qubits)**

The gate complexity is wholly dependent on the choice of ansatz. Satisfying properties (ii) and (iv) may place lower bounds on the required circuit depth. In addition, the connectivity of the device may also significantly affect the depth of the circuit. For instance, compilation of a generic two-qubit gate acting on an $n$-qubit state on hardware with 1D nearest-neighbor connectivity incurs $\mathcal{O}(n)$ circuit depth.

Throughout the optimization, the cost function is evaluated at different parameter settings $\boldsymbol{\theta}$, chosen adaptively based on the outcome of prior evaluations. (In the case of gradient-based optimization, one can use the parameter shift rule [2, 3, 4, 5] or finite difference methods.) Each evaluation of the cost function corresponds to approximating an expectation value to some additive error $\varepsilon$ using finite measurement shots, where $\varepsilon$ should be chosen to be sufficiently small for accurate optimization over the landscape. Specifically, it should be expected that $\varepsilon$ is at most $\mathcal{O}(\sqrt{\mathbb{V}_{\boldsymbol{\theta}}[C(\boldsymbol{\theta})]})$ in order to accurately distinguish arbitrarily chosen points on the parameter landscape, where $\mathbb{V}_{\boldsymbol{\theta}}$ denotes the variance over uniformly distributed parameter settings.

**Caveats**

The optimization of certain parameterized quantum circuits is known to be subject to the detrimental phenomena of "barren plateaus," in which deviations between different cost values with high probability (or deterministically, depending on the setting) vanish exponentially with increasing number of qubits [6, 7, 8, 9, 10, 11, 12, 13, 14]. This is often characterized by observing that $\mathbb{V}_{\boldsymbol{\theta}}[C(\boldsymbol{\theta})] = \mathcal{O}(2^{-\beta n})$ for some $\beta > 0$ [15]. This mandates an exponential shot complexity for each evaluation of a cost value in order to reliably navigate the cost landscape (when probabilistic, this should be considered an average-case phenomenon). Note that this affects both gradient-based and gradient-free optimization strategies. Moreover, it has been found that many standard techniques to avoid vanishing gradients render the VQA classically simulable [16].

If VQAs are run on noisy devices, the effects of noise are known to severely restrict the scope for computation [17, 18, 19, 20, 21]. This effect is amplified on devices with limited hardware connectivity, where one has to use additional circuit depth to compile generic gates [20, 19].

Finally, in general, there is a lack of end-to-end theoretical guarantees for variational quantum algorithms. In order to show advantage over classical algorithms, at minimum one has to satisfy all of the properties laid out above. In particular, the classical parameter optimization is generally left as a heuristic subroutine. This optimization task is in general NP-hard, and can be burdened by many local minima of poor quality [22, 23]. This leads to a slow optimization process and many cost values may need to be evaluated.

**Example use cases**

- Quantum chemistry and condensed matter physics (ground state energy): The ground state and ground state energy of a given Hamiltonian $H$ can be found by minimizing the cost $\langle \psi(\boldsymbol{\theta})|H|\psi(\boldsymbol{\theta})\rangle$, where $|\psi(\boldsymbol{\theta})\rangle = U(\boldsymbol{\theta})|\psi_0\rangle$ for some input state $|\psi_0\rangle$ [24]. This is





known as the variational quantum eigensolver (VQE) algorithm. A widely used ansatz for fermionic Hamiltonians is the unitary coupled cluster (UCC) ansatz [25, 24, 26, 27, 28, 29, 30, 31]. Beyond direct variational optimization on a quantum computer, alternative hybrid classical-quantum algorithms have been proposed to use a variational circuit as trial states for quantum Monte Carlo [32, 33] or combining the use of VQE with density matrix embedding theory [34, 35].

- **Combinatorial optimization**: In the quantum approximate optimization algorithm (QAOA), combinatorial problems on bit-strings can be encoded in the Pauli-$Z$ basis with Hamiltonian $H_P$ [36]. By finding the state that minimizes $\langle\phi(\boldsymbol{\theta})|H_P|\phi(\boldsymbol{\theta})\rangle$, where $|\phi(\boldsymbol{\theta})\rangle = U(\boldsymbol{\theta})|0\rangle$, the optimal bit-string can be extracted by sampling the optimized state in the computational basis. A widely studied ansatz for this problem is the quantum alternating operator ansatz (which bears the same acronym as the algorithm), inspired by Trotterized adiabatic evolution [37]. The ansatz takes the form $U(\boldsymbol{\gamma}, \boldsymbol{\beta}) = \prod_{l=1}^{p} e^{-i\beta_l H_M} e^{-i\gamma_l H_P}$ where $H_M$ is a specific "mixing" Hamiltonian. This ansatz is known to be computationally universal (when $p \to \infty$) for certain classes of Hamiltonians [38, 39]. Moreover, under reasonable complexity-theoretic assumptions, it is known that sampling from the output of the QAOA at $p = 1$ is classically hard [40]. On the other hand, there is evidence that shallow (small $p$) QAOA does not perform well [41, 42, 43, 44], leading to intuition that $p$ may need to grow with problem size to produce better approximate solutions than what can be easily found classically. Alternatively, there is some evidence that an exponential number of samples from shallow QAOA circuits may yield polynomial speedups over classical methods for finding exactly optimal solutions [45, 46]; see Section 4.2 on beyond-quadratic speedups for combinatorial optimization.

- **Linear system solvers**: Given matrix $A$ and vector $b$ encoded in a quantum state $|b\rangle$, the goal is to variationally prepare a quantum state $|x\rangle$ with amplitudes proportional to elements of the vector $x = A^{-1}b$ [47, 48, 49]. The strategy employed is to minimize the cost $\langle\tilde{x}(\boldsymbol{\theta})|H_L|\tilde{x}(\boldsymbol{\theta})\rangle$, where $|\tilde{x}(\boldsymbol{\theta})\rangle = U(\boldsymbol{\theta})|0\rangle$ and $H_L = A^{\dagger}(I - |b\rangle\langle b|)A$. These approaches require the assumption that $A$ has a decomposition into a sum of a small number of efficiently implementable unitaries. Here the absolute value of the cost function bounds the approximation error. A numerical study up to 30 qubits showed favorable scaling in the time to solution with respect to the matrix size, condition number, and precision [47].

- Compiling: An interesting near-term application could be to approximate a given unitary $V$ with native gate sequence $U(\boldsymbol{\theta})$. This can lead to a compressed approximate implementation of the unitary. One option is to construct a cost function via the Hilbert–Schmidt test circuit to evaluate $1 - |\langle\Phi|V^* \otimes U(\boldsymbol{\theta})|\Phi\rangle|^2 = 1 - \left|2^{-n}\operatorname{tr}[V^{\dagger}U(\boldsymbol{\theta})]\right|^2$, where $|\Phi\rangle$ is the maximally entangled state [50].

- Quantum dynamics: VQAs can also be constructed to simulate real-time and imaginary-time evolution. We discuss two examples. First, by using the compiling primitives as outlined above, compressed gate sequences for short-time evolution can be found [51, 52, 53, 54, 55]. This in turn can also be used to fast forward and simulate evolution at long times [56, 57, 58]. Second, a line of work has considered simulating open and closed dynamics, as well as imaginary-time evolution, via McLachlan's variational principle [59], which gives a linear equation for parameter dynamics $\sum_j M_{i,j}\dot{\theta}_j = V_i$ describing the evolution of a parameterized state $|\psi(\boldsymbol{\theta})\rangle$ [60, 61, 62]. Here, $M_{i,j}$ and $V_i$ can both be found





on a quantum computer with basic circuit primitives, leading to a hybrid classical-quantum algorithm. While a full review of variational quantum algorithms for dynamics is beyond the scope of this section, we refer the reader to more complete reviews in [61, 63].

- **Factoring**: Variational methods for factoring have been proposed that exploit a mapping between the factoring problem and that of finding the ground state of an Ising Hamiltonian [64]. The authors use the QAOA ansatz and heuristically find that $p = \mathcal{O}(n)$ rounds of the ansatz can lead to a good solution overlap for small system sizes.

- **Machine learning**: Here one employs a parameterized quantum circuit to construct a hypothesis family. Variational methods have been proposed for both classical and quantum data for classification [65, 2, 66, 67, 68], generative models [69, 70, 71], autoencoders [72, 73, 74], and beyond [75, 76]. Specific ansatzes have been proposed in these contexts, sometimes referred to as *quantum neural networks*, in analogy with their classical counterparts. "Classically inspired" quantum neural networks have been proposed, such as perceptron-based QNNs [77, 73, 78, 79] and a quantum analog to the convolutional neural network [68], as well as approaches based on tensor networks [80, 81].

**Further reading**

- See [63, 82] for extensive reviews of VQAs, including a summary of different widely studied ansatzes, applications, and challenges.

# 21  Quantum tomography

*The authors are grateful to Richard Kueng for reviewing this section of the survey.*

### Rough overview (in words)

In quantum tomography, often also termed quantum state estimation, we are given repeated copies of an unknown quantum state (or quantum channel) and the goal is to obtain a full classical description of the quantum state (or quantum channel) by extracting information by means of performing measurements. Here, we focus on quantum state tomography, with multiple independent and identical copies of an unknown quantum state $\rho$ provided—that is fixed and of known dimension—and the task is to find an estimate of the density matrix of the quantum state up to an approximation error in some distance measure (and up to some failure probability). We are then typically interested in the optimal sample complexity in terms of the number of copies $n$, the quantum state dimension $d$, the approximation error $\varepsilon$, and the overall failure probability $\delta$. Additionally, algorithmic complexity aspects of the used schemes might be of importance as well.

### Rough overview (in math)

Given (many copies of) an unknown quantum state $\rho$ of known dimension $d$, the goal is to give a description of $\tilde{\rho}$ with the statistical estimate $\tilde{\rho} \approx_\varepsilon \rho$, up to some distance measure with corresponding approximation parameter $\varepsilon \geq 0$. This is achieved by extracting classical information by applying measurements $\mathcal{M}^n(\cdot)$ via $\rho^{\otimes n}$. To start with, one has to distinguish tomography schemes based on different types of measurements used. This includes in particular:

(i) Independent and identical (IID) measurements, where the choice of measurement $\mathcal{M}^n = \mathcal{M}^{\otimes n}$ is fixed and the same for each copy.

(ii) Adaptive measurements, where the choice of measurement $\mathcal{M}_2$ on the second copy can depend on the outcomes of measurement $\mathcal{M}_1$ on the first copy, and so on.

(iii) Entangled measurements, where one measurement $\mathcal{M}^k$ with $1 < k \leq n$ is performed on $k$ copies at once.

Further, if one has some information about the type of quantum state provided, then tomography schemes can become more efficient. This includes, for example, pure state tomography, low-rank-$k$ state tomography, matrix product state tomography, or ground/thermal state tomography of Hamiltonians (some references on tight schemes are given later on). For some schemes, one a priori has certain information about the state in question and under this assumption the scheme is then promised to work (e.g., low-rank tomography [1]). Other schemes work generally, but are only *a posteriori* guaranteed to be more efficient if the unknown state happens to be approximately of the type sought after (e.g., matrix product state tomography [2]). Finally, for maximum likelihood estimates or Bayesian statistical estimates and alike, priors could be added as well.





Note that the best understood case of pure state tomography can also be used for general quantum states, if one has access to the relevant purification. Specifically for pure state tomography, one then also needs to specify in what form access is given to the quantum state. Possible access models for pure state tomography include:

- Via samples of computational basis measurements $p(x) = \langle x|\rho|x\rangle$ for estimating the probabilities in the computational basis (not yet the pure state amplitudes).

- Via copies of the state that can be processed before measurement.

- Via the state preparation unitary $U|0^n\rangle\langle 0^n|U^\dagger = \rho$ (with $\rho$ pure).

- Via the controlled version of aforementioned state preparation unitary $U$.

- Via aforementioned state preparation unitary $U$ and its inverse $U^\dagger$.

Typically studied distance metrics to measure closeness of the statistical estimate to the true quantum state are the trace distance $T(\rho, \sigma) = \frac{1}{2}\text{tr}[|\rho - \sigma|]$, the quantum fidelity $F(\rho, \sigma) = \left(\text{tr}\left[\left|\sqrt{\rho}\sqrt{\sigma}\right|\right]\right)^2$, and for pure quantum states also the $\ell_2$-norm of the difference $\||\psi\rangle - |\phi\rangle\|$ of the pure states $|\psi\rangle, |\phi\rangle$ corresponding (up to global phase) to $\rho$ and $\sigma$, respectively.

**Dominant resource cost (gates/qubits)**

Besides some potential ancilla qubits (few for typical tomographic schemes), the number of qubits is fixed by the dimension of the quantum state (of course, whenever entangled measurements are used, the corresponding number of copies is needed). As such, the sample complexity is typically the relevant figure of merit. In the following, the notation $\Theta(\cdot)$ stands for simultaneous upper $\mathcal{O}(\cdot)$ and lower $\Omega(\cdot)$ bounds on the asymptotic sample complexity, and the variant $\widetilde{\Theta}(\cdot)$ denotes the same up to factors that scale polylogarithmically in the relevant parameters. Tight sample and query complexity characterizations, in terms of an approximation error $\varepsilon \in [0, 1]$, then include the following noteworthy results:

- $\widetilde{\Theta}(d\varepsilon^{-2})$ sample complexity for pure state tomography in $\ell_2$-norm up to global phase [3] ([4] gave an algorithm with similar complexity, but requiring a state preparation unitary). The main idea is to use computational basis measurements to recover the absolute values of the amplitudes and then create some interference pattern for learning the phases.

- $\widetilde{\Theta}(d\varepsilon^{-1})$ query complexity for pure state tomography in $\ell_2$-norm with access to controlled state preparation unitary and its inverse [3], featuring a quadratic speedup in $1/\varepsilon$ reminiscent of amplitude estimation. The achievability results are based on the subroutine of quantum gradient estimation via an unbiased version of quantum phase estimation. Note that [5] used an alternative simpler algorithm based on iterative refinement and amplitude amplification that achieves the same query complexity but comes with improved gate complexity.

- $\Theta(dk^2\varepsilon^{-2})$ sample complexity for rank-$k$ state tomography in trace distance with IID measurements [6, 7, 8]. The achievability results are based on low-rank matrix recovery techniques, where semidefinite programs have to be solved for reconstructing the quantum state from the collected measurement statistics. Note that the special case $k = 1$ corresponds to pure state tomography as in the setting of the first bullet point.





- $\widetilde{\Theta}(dk\varepsilon^{-2})$ sample complexity for rank-$k$ state tomography in trace distance with entangled measurements [9, 7, 10]. The achievability results are based on representation-theoretic techniques around the Schur transform.

- $\widetilde{\Theta}(dk\varepsilon^{-1})$ query complexity for rank-$k$ state tomography in trace distance with access to controlled state preparation unitary of a purification and its inverse [3], featuring a quadratic speedup in $1/\varepsilon$ reminiscent of amplitude estimation.

For variations of the above, additional results in terms of lower and upper bounds are known. The derivations of the sample complexity lower bounds are often based on information-theoretic methods, exploiting the monotonicity of quantum-entropy-based measures. For sample complexity upper bounds, it is in practice additionally important that the algorithmic complexities of the underlying schemes become efficient (in particular for entangled measurements performed on all $n$ copies at once). Relevant metrics for the algorithmic complexity include quantum gate depth and number of measurement outcomes needed, as well as runtime and memory requirements of the classical postprocessing stage. We refer to [11] for a recent discussion on these computational aspects.

### Caveats

As shown by the presented information-theoretic lower bounds, the sample complexity for general quantum state tomography grows exponentially in the number of qubits. As such, whenever quantum tomography is invoked as a subroutine in quantum algorithms, one has to carefully analyze if this step does not eliminate any claimed speedups of the quantum algorithm compared to state-of-the-art classical methods. One also has the inverse polynomial scaling in terms of the approximation parameter from the finite statistics, which is often prohibitively expensive for certain applications.

Additionally, on top of sample complexity for tomography schemes, the accompanying gate complexity should be considered as well. We refer to [3] for a discussion.

An alternative is to resort to only revealing partial classical information about quantum states, which might still be informative for the (algorithmic) task at hand. One such example with favorable scaling is shadow tomography, where the task is to not estimate the density matrix itself, but (very) many observables thereof. Shadow tomography—also known as quantum data analysis—can achieve exponential sample complexity improvements in terms of Hilbert space dimension compared to full state tomography and is guaranteed to yield exponential improvements in the number of target observables (compared to directly measuring all of them sequentially). The strongest result of this kind [12, 13, 14] requires entangled measurements across many state copies, as well as prohibitively large gate counts. More hardware-friendly protocols have been derived, known as classical shadow tomography [15, 16, 17]. In more detail, there exist algorithmically efficient and universal schemes that can simultaneously $\varepsilon$-approximate $M$ linear functions $\mathrm{tr}[O_i\rho]$ of an unknown quantum state $\rho$ by only using $\mathcal{O}(\log(M) \cdot \max_i \|O_i\|_s^2 \varepsilon^{-2})$ IID measurements. Note the scaling with $\log(M)$ instead of the standard $M$ scaling. The shadow norm term $\|O_i\|_s^2$ scales in general as $d$, leading to the worst-case query complexity $\mathcal{O}(d\log(M)\varepsilon^{-2})$. However, for observables with bounded Hilbert–Schmidt norm or for local observables, the overall dimension-free query complexity $\mathcal{O}(\log(M)\varepsilon^{-2})$ is achievable.





**Example use cases**

Quantum tomographic or related data collection schemes are omnipresent in quantum algorithms. Some applications include:

- Quantum linear system solvers that output the full classical solution vector, where such solvers are, for example, employed for quantum interior point methods or for solving differential equations.

- Classical data about quantum states for variational quantum algorithms.

- Characterizing the performance of physical devices.

- Probing entanglement dynamics throughout a quantum simulation.

- Characterizing quantum processes.

**Further reading**

- Short perspective article, entitled "Focus on quantum tomography" [18].

- Recent overview on query complexity aspects [3].

- Recent overview on computational complexity aspects [11].

- Shadow tomography of quantum states [12].

- Review article on classical shadows and randomized measurements [17].

# 22 Quantum interior point methods

*The authors are grateful to Sander Gribling for reviewing this section of the survey.*

**Rough overview (in words)**

Interior point methods (IPMs) are a type of efficient classical algorithm for solving convex optimization problems such as linear programs (LPs), second-order cone programs (SOCPs), and semidefinite programs (SDPs). IPMs are the basis for effective optimization software tools (e.g., [1, 2]), which are widely used for solving convex optimization problems that arise in industry. They are called *interior* point methods because, in contrast to the simplex method, they iteratively generate a sequence of points that lie in the interior of the convex region; this sequence of points is guaranteed to rapidly approach the optimal point (which, when it exists and the objective function is convex, is guaranteed to lie at the boundary of the convex region). At each iteration, the next point is produced by solving a system of linear equations. See, for example, [3, 4, 5, 6] for context on how IPMs fit into the history of methods for optimization.

*Quantum interior point methods* (QIPMs) are quantum algorithms that leverage a similar approach as classical IPMs, but perform certain aspects of the algorithm in a quantum manner. For example, QIPMs were first introduced in [7], where the quantum algorithm is identical to classical IPMs, except that it determines the next point using a quantum linear system solver (QLSS) combined with quantum state tomography, rather than a classical linear system solver. Subsequent work has explored other forms of quantizing classical IPMs that do not rely on the QLSS [8, 9].

Classical IPMs are generally efficient in the sense that they can solve convex optimization problems in time scaling as a polynomial in the number of variables. The exact degree of the polynomial depends on which kind of convex optimization problem is being solved, as well as certain choices about the IPM. Since QIPMs often rely on state tomography, they are generally expected to require time that scales at least linearly in the number of variables, and lower bounds along these lines are known [10]; thus, *the best one can hope for is a polynomial speedup* over classical IPMs. The exact runtime of the quantum algorithm depends on instance-specific parameters, such as the condition number of matrices that appear during the course of the algorithm, which makes it difficult to determine whether a speedup exists in practice.

**Rough overview (in math)**

For simplicity, we focus on LPs, the simplest kind of optimization problem where QIPMs can be applied. An LP is specified by an $m \times n$ matrix $A$, an $n$-dimensional vector $c$, and an $m$-dimensional vector $b$, and it is given by

$$
\begin{aligned}
&\min_{x \in \mathbb{R}^n} \langle c, x \rangle \\
&\text{subject to } Ax = b \\
&\qquad x_i \geq 0 \text{ for } i = 1, \ldots, n
\end{aligned}
\tag{62}
$$

where $\langle u, v \rangle$ denotes the standard dot product between vectors $u$ and $v$.

The function $\langle c, x \rangle$ is called the objective function, and a point $x$ is called feasible if it satisfies $Ax = b$ and $x_i \geq 0$ for all $i$. Inequality constraints of the form $Ax \leq b$ can be handled





by introducing slack variables. We denote the feasible point that optimizes the objective function by $x^*$.

An important concept in mathematical optimization is duality, where given one optimization problem, an equivalent "dual" optimization problem can be generated through the method of Lagrange multipliers (see [11, Section 5]). The dual of the LP in Eq. (62) is given by

$$\max_{y \in \mathbb{R}^m} \langle b, y \rangle$$
$$\text{subject to } A^\mathsf{T} y + s = c$$
$$s_i \geq 0 \text{ for } i = 1, \ldots, n$$
$$(63)$$

Alternatively, one can drop the $s$ variable and constraints that $s_i$ are positive, and simply write $A^\mathsf{T} y \leq c$. Denote the optimal feasible points for the dual by $(y^*, s^*)$.

It can be shown that the optimal point lies at the boundary of the feasible region and satisfies the relationship $x_i s_i = 0$ for all $i$. A key concept in IPMs is the *central path*, a set of points parameterized by $\mu > 0$. The central point with parameter $\mu$ is the feasible point for which $x_i s_i = \mu$ for all $i$. In general, this point will be in the interior of the feasible region, but as $\mu \to 0$, the central path approaches the optimal point on the boundary.

The most effective classical IPMs are "primal-dual path-following methods," which generate a length-$T$ sequence of primal-dual point pairs $(x^{(t)}, y^{(t)}, s^{(t)}) \in \mathbb{R}^n \times \mathbb{R}^m \times \mathbb{R}^n$ for $t = 0, \ldots, T-1$ that approximately follows the central path toward the optimum. Given $(x^{(t)}, y^{(t)}, s^{(t)})$, the point $(x^{(t+1)}, y^{(t+1)}, s^{(t+1)}) = (x^{(t)} + \Delta x, y^{(t)} + \Delta y, s^{(t)} + \Delta s)$ is formed by solving the following linear system of equations, which is called the *Newton system*, as it corresponds to one iteration of Newton's method.

$$\begin{pmatrix} A & 0 & 0 \\ 0 & A^\mathsf{T} & I \\ S & 0 & X \end{pmatrix} \begin{pmatrix} \Delta x \\ \Delta y \\ \Delta s \end{pmatrix} = \begin{pmatrix} b - Ax^{(t)} \\ c - A^\mathsf{T} y^{(t)} - s^{(t)} \\ \sigma \frac{x^{(t)\mathsf{T}} s^{(t)}}{n} \mathbf{1} - X s^{(t)} \end{pmatrix}, \qquad (64)$$

where $\sigma < 1$, $\mathbf{1}$ denotes the all 1s vector, and $S = \text{diag}(s^{(t)}), X = \text{diag}(x^{(t)})$ are diagonal $n \times n$ matrices formed from the entries of $s^{(t)}$ and $x^{(t)}$. Note that there are alternative ways to formulate the Newton system (see, e.g., [12, 13]). To understand Eq. (64), note that if the point $(x^{(t)}, y^{(t)}, s^{(t)})$ is feasible, then the first two entries on the right-hand side are zero. Furthermore, if it is on the central path, then $X s^{(t)} = \frac{x^{(t)\mathsf{T}} s^{(t)}}{n} \mathbf{1}$, so if we were to choose $\sigma = 1$, then the entire right-hand side would be zero, and the solution to the system would be $\Delta x = \Delta y = \Delta s = 0$. If instead we set $\sigma = 1 - \delta$ for sufficiently small $\delta$, the solution will correspond to taking a small step along the central path in the direction of decreasing $\mu$. Technically, we do not exactly follow the central path, but it can be guaranteed that the sequence of points stays within a small neighborhood of it. As $\mu \to 0$, the central path approaches the optimal point $(x^*, y^*, s^*)$, so by following the path toward $\mu = 0$, a classical or quantum IPM can guarantee success.

The classical IPM can solve the Newton system exactly using Gaussian elimination in $\mathcal{O}(n^3)$ operations, or it can solve the system approximately using a variety of iterative solvers such as the conjugate gradient method. In contrast, the standard approach for a QIPM is to solve the Newton system by using a QLSS to repeatedly prepare the $\mathcal{O}(\log(n))$-qubit state $|\Delta x, \Delta y, \Delta s\rangle$ whose amplitudes encode the solution to the Newton system. By preparing many copies, the algorithm can perform (pure state) quantum state tomography to yield an estimate $(\overline{\Delta x}, \overline{\Delta y}, \overline{\Delta s})$ for the amplitudes $(\Delta x, \Delta y, \Delta s)$ to some desired precision $\xi$ (in 2-norm), that is,

$$\|(\overline{\Delta x}, \overline{\Delta y}, \overline{\Delta s}) - (\Delta x, \Delta y, \Delta s)\| \leq \xi \|(\Delta x, \Delta y, \Delta s)\|.$$





Due to the tomography step, the QIPM is only able to generate solutions to the Newton system that are *inexact*. There has been some question in the literature whether the (classical or quantum) IPMs with the fastest guaranteed convergence rate (i.e., the number of iterations needed to reduce $\mu$ to $\epsilon$) are applicable even when inexact solutions are used, as this causes intermediate points to be (slightly) infeasible [12]. However, if $\xi$ is sufficiently small, the method appears to work empirically even using the inexact solutions that would be output by a quantum solver [14]. Alternatively, there exist workarounds [12] that ensure feasibility is maintained even when linear systems are solved inexactly, at the expense of some additional classical cost.

The IPMs and QIPMs for SOCPs [15, 13] are quite similar to those for LPs described above: the main difference is that the matrices $X$ and $S$ are no longer strictly diagonal matrices. QIPMs have also been proposed for SDPs [7, 12, 16], which are more complex but have more expressive power; here, additional considerations must be taken to guarantee that the intermediate solutions continue to be symmetric even after experiencing errors due to tomography.

The above exposition represents the original approach to quantizing the classical IPM, which has so far garnered the most study. An alternative to this approach was proposed in [8], which focuses on the case that the LP constraint matrix $A$ is "tall," that is, $m \gg n$. As above, they follow the central path to the optimal point; however, they adopt a primal-only approach, where the Newton linear system takes on the form $(B^\mathsf{T} B)g = h$, with $g$ and $h$ length-$n$ vectors and $B$ an $m \times n$ matrix. Rather than using the QLSS and quantum state tomography, their quantum algorithm performs a Grover search–like step to identify the "important" rows of $B$ and thus produce an $\mathcal{O}(n) \times n$ matrix $\tilde{B}$ for which $\tilde{B}^\mathsf{T} \tilde{B} \approx B^\mathsf{T} B$. This enables a quadratic speedup in the large parameter $m$. To obtain the right-hand side vector $h$, which is the gradient of the objective function, they require the multivariate mean-estimation algorithm of [17], which is related to the quantum gradient estimation primitive developed in [18, 19]—this is key for avoiding a dependence on the condition number of $B^\mathsf{T} B$. Matrix inversion is then performed classically at cost polynomial in $n$, independent of $m$ and not depending on the condition number of any matrix.

Meanwhile, another quantum algorithm inspired by IPMs was proposed in [9]. Where the standard QIPM encodes the variable $x$ into the amplitudes of the quantum state, requiring readout with quantum state tomography, the method of [12] encodes the components of $x$ into separate binary registers, truncated to some finite number of bits of precision. It constructs a Hamiltonian, parameterized by $\mu$, whose ground state is a wavefunction localized near the associated point on the central path. By slowly decreasing $\mu$ and invoking the adiabatic theorem, the wavepacket follows the central path to $\mu = 0$, where the optimal point can be recovered by a measurement. Thus, the main primitive required is time-dependent Hamiltonian simulation.

**Dominant resource cost (gates/qubits)**

The outer loop of QLSS-based QIPMs is purely classical; at each iteration a small step is taken to form the next point in the sequence. For LPs, SOCPs, and SDPs, the number of iterations $T$ required to yield a point for which the objective function is within $\epsilon$ of optimal is $\mathcal{O}(\sqrt{n}\log(1/\epsilon))$. The main cost of each iteration is solving the Newton system. In the complexity statements that follow, we assume the number of constraints $m$ is on the order of the number of degrees of freedom (i.e., $m = \mathcal{O}(n)$ in the case of LPs and SOCPs, and $m = \mathcal{O}(n^2)$ in the case of SDPs).

The QIPM solves the Newton system by preparing many copies of the state corresponding to the solution to the linear system. This state can be prepared in time polylog$(n) \cdot \zeta\kappa$, where $\kappa$ is the condition number of the matrix in Eq. (64) and $\zeta$ is the ratio $\|\cdot\|_F/\|\cdot\|$ of the Frobenius and





spectral norms of the matrix, assuming that one can perform a block-encoding of the Newton matrix in polylog($n$) time, a task that requires access to large-scale quantum random access memory (QRAM).[59] For LPs and SOCPs, the number of copies that must be prepared scales as $\mathcal{O}(n/\xi^2)$ when using the basic version (see [7, Section 4] and [14, Section IVD]) of pure state tomography that simply measures each copy in the computational basis. A more recent and complex version of tomography [20] can achieve this task using $\mathcal{O}(n/\xi)$ copies along with additional gates. For SDPs, since the variables are matrices rather than vectors, the number of copies is $\mathcal{O}(n^2/\xi^2)$ or $\mathcal{O}(n^2/\xi)$. Overall, using the more efficient version of tomography and ignoring the additional gates, the runtime of the QIPM is expected to scale as

$$
\begin{aligned}
\text{LP, SOCP:} \quad & \widetilde{\mathcal{O}}\left(\frac{n^{1.5}\zeta\kappa}{\xi}\log(1/\epsilon)\right) \\
\text{SDP:} \quad & \widetilde{\mathcal{O}}\left(\frac{n^{2.5}\zeta\kappa}{\xi}\log(1/\epsilon)\right),
\end{aligned}
\tag{65}
$$

where $\kappa$ denotes the maximum condition number, $\zeta$ the maximum ratio of Frobenius to spectral norm, and $\xi$ the minimum tomographic precision required across all iterations. There may be an additional purely classical cost of $\mathcal{O}(n^{2.5})$ for LPs/SOCPs and $\mathcal{O}(n^{4.5})$ for SDPs, deriving from classical matrix-vector multiplications necessary for setting up the Newton system at each iteration.

In the worst case, it may be necessary to take $\xi$ as small as $\mathcal{O}(1/\kappa)$, and $\zeta$ can be as large as $\sqrt{n}$ (SOCP/LP) or $n$ (SDP)—complexity statements in the literature, such as [12], often assume these worst-case values for those parameters, but we refrain from doing so as these worst-case values may be overly pessimistic in practice. The hidden constant prefactors are dependent primarily on the implementation of the QLSS and tomography. It is clear that the viability of the QIPM is highly dependent on the value and scaling of the parameters $\kappa$ and $\xi$. Unfortunately, it is believed that for some LP/SOCP/SDP instances, the value of $\kappa$ will diverge as the target precision $\epsilon$ is made smaller, perhaps as $\mathcal{O}(1/\epsilon)$ [15, 12], although this may not be the case in every instance (see, e.g., the numerical results of [14]).

The QIPM only requires a register of $\mathcal{O}(\log(n))$ qubits to hold the solution of the linear system; however, achieving the runtimes quoted requires queries to QRAM. In this case, the explicit QRAM circuits that achieve shallow depths of $\mathcal{O}(\log(n))$ necessarily require $\mathcal{O}(n^2)$ total gates across $\mathcal{O}(n^2)$ total qubits.

The alternative approach of [8] is best suited for the case where $m \gg n$, and requires $\sqrt{m} \cdot$ poly($n, \log(1/\epsilon)$) queries to the entries of the matrix $A$, where the $\sqrt{m}$-dependence fundamentally comes from Grover-like primitives with quadratic speedup. Like the standard QIPM formulation, this approach requires a QRAM to implement the queries in polylog($m$) time. However, since it does not use QLSS or tomography, it avoids polynomial dependence on the instance-specific parameters $\kappa$ and $1/\xi$.

## Caveats

There are several important caveats that must be considered when evaluating a speedup claimed by QIPM.

---

[59]It is worth emphasizing that the origin of the dependence on the Frobenius norm of the Newton matrix here is the normalization factor that arises when block-encoding a dense classical matrix. If the matrix were sparse or had some compact representation, this normalization factor could potentially be improved—but for Newton matrices in QIPMs we do not expect this to be the case.





- Even in a best case scenario, the quantum speedup is at most polynomial (and even subquadratic). Since quantum computation requires significant constant-factor overheads due to slower clock speeds and error correction, the value of $n$ for which a QIPM would be faster than a classical IPM on actual hardware is likely to be large (see [14] for further discussion).

- Since $n$ must be large for a quantum speedup to be obtained, a very large QRAM, corresponding to millions or billions of (logical) qubits, would be needed for any speedup to be realized.

- QIPMs are most effective when the matrices that need to be inverted over the course of the algorithm are well conditioned, due to their reliance on the QLSS. However, when the condition number $\kappa$ is small, iterative classical methods may also be effective, limiting the advantage of the quantum algorithm. In particular, a linear system with $\mathcal{O}(n)$ dense constraints on $n$ variables can be solved to error $\xi$ in time $\mathcal{O}(n\zeta^2\kappa^2\log(1/\xi))$ using the randomized Kaczmarz method [21]. In comparison, the QIPM utilizes QLSS and tomography to solve the same task (once per iteration) in time $\mathcal{O}(n\zeta\kappa/\xi)$. Even if $\xi = \Omega(1)$, this limits the magnitude of the quantum speedup to a factor of $\mathcal{O}(\zeta\kappa)$. Thus, for the quantum speedup to be maximized, $\kappa$ can be neither too small nor too large. While we are not aware of any IPM implementations based on the Kaczmarz method, its complexity allows for clean comparison with quantum algorithms involving the QLSS for dense matrices, since both depend directly on the quantity $\zeta\kappa$. Here it is also worth mentioning that there exist other approximate classical linear system solvers for which the complexity depends on $\kappa$, but not on $\zeta$. One example is the conjugate gradient method [22]. Another straightforward example is to solve the system $Gu = v$ by finding a degree-$\mathcal{O}(\kappa\log(1/\xi))$ polynomial approximation $p(x) \approx 1/x$, and then computing $p(G)v \approx G^{-1}v = u$ via a sequence of $\mathcal{O}(\kappa\log(1/\xi))$ matrix-vector products—this is a classical analog of the quantum approach based on the quantum singular value transformation [23]. Classically, each matrix-vector product costs $\mathcal{O}(n^2)$ when $G$ is dense and $\mathcal{O}(ns)$ for when $G$ is $s$-sparse.

- If the matrices that define the convex problem have a certain structure (e.g., sparsity), this could be exploited to potentially reduce the overhead from block-encoding—in particular, the value of $\zeta$ and the size of the QRAM required. However, this can help the quantum algorithm only to a limited extent, as the vectors $(\Delta x, \Delta y, \Delta s)$ will still be dense and reading out estimates for all $\mathcal{O}(n)$ amplitudes with quantum tomography will be necessary.

**Example use cases**

- Portfolio optimization, the canonical optimization problem that appears in finance, can be formulated as an SOCP and solved with a QIPM; a study of the condition number of the matrices that appear in this application was consistent with a small quantum speedup [24]; however, a follow-up study did not replicate this finding [14] and also pointed out that in any case large constant-factor overheads would make achieving practical advantage challenging.

- Support vector machines, a common task in machine learning, can be reduced to SOCPs and solved with a QIPM; a study of the condition number of the matrices that appear in this application was consistent with a small quantum speedup [15].





- Sample-efficient protocols for mixed-state tomography reduce the problem of reconstructing an estimate of the quantum state to solving an SDP. This SDP could be solved with a QIPM (note that the tomography needed within the QIPM is always on *pure states* and does not require solving an SDP, thus avoiding an issue of circular logic).

- Nonconvex optimization is often solved approximately by relaxing the problem into a convex problem like an SDP. For example, the MAX-CUT problem is a combinatorial optimization problem over the nonconvex space $\{+1, -1\}^n$, but by solving the associated SDP relaxation and rounding, an approximate solution can be obtained.

**Further reading**

- See Boyd and Vandenberghe [11] for an accessible book on convex optimization including (classical) IPMs.

- QIPMs are an active area of research. A QIPM for LPs and SDPs was originally proposed by Kerenidis and Prakash in [7]. This was followed up by a QIPM for SOCPs in [15], along with numerical simulations for specific applications [15, 24]. Later, [12] pointed out a potential error in the convergence analysis of previous works, and they presented two possible workarounds called the "inexact-infeasible" and "inexact-feasible" IPMs. Note also the work in [16] for another way to avoid this issue, giving a QIPM for SDP.

- See [8, 9] for quantum methods related to IPMs that do not rely on the QLSS.

# 23 Multiplicative weights update method

*The authors are grateful to Sander Gribling for reviewing this section of the survey.*

**Rough overview (in words)**

The multiplicative weights update (MWU) method is an algorithmic strategy, sometimes referred to as a "meta-algorithm," with varying applications in classical and quantum algorithms. Reference [1] gives an overview of the MWU strategy. The introductory example problem where the MWU method is used is the problem of making predictions for a binary outcome given advice from a panel of $n$ "experts." The MWU approach assigns a weight to each of the $n$ experts, and the weight is reduced by a multiplicative factor whenever the expert makes an incorrect prediction. The outcome of the process can be shown to give an approximately optimal strategy.

This general approach can be applied to convex programs including linear programs (LPs) and semidefinite programs (SDPs). The SDP version generalizes the MWU method to allow for matrix-valued weights and matrix-valued costs. These weight matrices are positive semidefinite operators with trace equal to one, that is, density matrices. In fact, the states that arise in the SDP-solving algorithm are Gibbs states. Thus, they can be naturally represented as quantum states on a logarithmic number of qubits and generated through the process of Gibbs sampling. The existence of fast Gibbs samplers can lead to a quantum speedup in certain circumstances.

**Rough overview (in math)**

We present an example problem. Let $\mathbf{1}$ denote the all-ones vector. Consider the following set of linear constraints on the vector $x = (x_1, \ldots, x_n) \in \mathbb{R}^n$

$$\langle a^{(j)}, x \rangle \geq 0 \qquad j = 1, \ldots, m$$
$$\langle \mathbf{1}, x \rangle = 1$$
$$x_i \geq 0 \qquad i = 1, \ldots, n$$

for $m$ fixed vectors $a^{(j)} \in \mathbb{R}^n$ with entries in $[-1, 1]$, for $j = 1, \ldots, m$, where $\langle \cdot, \cdot \rangle$ denotes the standard dot product between vectors. Suppose we are given a value of $\epsilon$ and promised either that there is no choice of $x$ that satisfies all the constraints or that there exists an $x^*$ such that $\langle a^{(j)}, x^* \rangle \geq \epsilon$ for all $j$, with $\langle \mathbf{1}, x^* \rangle = 1$ and $x_i^* \geq 0$ for all $i$. We wish to determine which is the case and find a vector $x^*$ in the second case. This problem is equivalent to the machine learning problem of finding a linear classifier for a set of $m$ labeled $n$-dimensional training points, similar to a support vector machine [1, 2]. The problem is also similar to the form of an LP and to the problem of solving for the optimal point of a zero-sum game [1, 3], and the MWU meta-algorithm can also be straightforwardly applied to solve these problems.

A classical solution to this problem is given by the multiplicative weights method [1]. The algorithm iteratively updates the vector $x$, with initialization $x = \mathbf{1}/n$. At each iteration, the algorithm finds a constraint $j$ for which $\langle a^{(j)}, x \rangle < 0$ (or if no such $j$ exists, it terminates and outputs $x$). Let $\eta = \mathcal{O}(\epsilon)$ be a fixed constant. Once $j$ is found, the entries of the vector $x$ are updated according to

$$x_i \leftarrow \frac{x_i \mathrm{e}^{\eta a_{ij}}}{\sum_\ell x_\ell \mathrm{e}^{\eta a_{\ell j}}}, \tag{66}$$





where $a_{ij}$ denotes entry $i$ of vector $a^{(j)}$, and the denominator works to enforce $\langle \mathbf{1}, x \rangle = 1$. By upweighting $x$ in the direction of the violated constraint $a^{(j)}$, this update rule brings the $x$ closer to satisfying the constraint. The magic of the multiplicative weights method is that the promise problem described above can be solved after only $\mathcal{O}(\log(n)/\epsilon^2)$ iterations [1]. By searching for a violated constraint using a Grover search, the runtime of each iteration can be sped up quantumly, giving rise to polynomial speedups for solving zero-sum games and LPs more generally [3].

In the analogy to a panel of experts, we may view the above problem as follows. Each expert $i \in [n]$ produces a prediction for each of $j \in [m]$ data points, denoted by $a_{ij}$. We wish to produce a weighting $x$ over the $n$ experts such that the weighted majority of the $n$ experts yields a positive value for all $m$ data points. The multiplicative weights method solves this iteratively by beginning with a uniform weighting over the $n$ experts, and repeatedly observing an index $j$ where the weighted majority misclassifies (i.e., predicts a negative value for) data point $j$, assessing a multiplicative penalty of $e^{\eta a_{ij}}$ to the weight of expert $i$. In a machine learning context, the weighting of experts produced by the algorithm can then be used as a classifier that allows us to predict a label for a new data point, by following the opinion of the weighted majority of the experts.

The *matrix* MWU method generalizes the $n$-dimensional vector $x$ to an $n \times n$ symmetric matrix $X$. An example problem generalizing the above is

$$\langle A^{(j)}, X \rangle \geq 0 \qquad j = 1, \ldots, m$$
$$\langle \mathbf{I}, X \rangle = 1$$
$$X \succeq 0,$$

where $A^{(j)}$ are fixed symmetric constraint matrices and the notation $\langle U, V \rangle := \mathrm{Tr}(UV)$ generalizes the dot product from vectors to matrices. Here $\mathbf{I}$ denotes the identity matrix, and $X \succeq 0$ denotes that $X$ is positive semidefinite. The problem above is related to the general form of an SDP, and the matrix MWU approach can be applied to solve SDPs. Note that we recover the vector example if we specify that the matrices $A^{(j)}$ and $X$ are diagonal. The final two constraints indicate that $X$ is a density matrix and is associated with a quantum state on $\log_2(n)$ qubits. When $X$ is updated by a generalization of the rule in Eq. (66), then at every iteration of the MWU method, $X$ will be a Gibbs state for a certain Hamiltonian that is a weighted sum of the symmetric constraint matrices $A^{(j)}$. Thus, the quantum state $X$ can be prepared on a quantum computer using algorithms for Gibbs sampling. Taking this approach, quantum algorithms can achieve guaranteed polynomial speedups for performing an iteration of the MWU method compared to classical approaches, and it is conceivable that larger speedups could be available if the associated quantum systems admit faster-than-worst-case Gibbs sampling.

**Dominant resource cost (gates/qubits)**

The MWU method, both in the classical and quantum setting, consists of some number $T$ of iterations, where each iteration updates a classical data structure. In typical applications, $T = \mathrm{poly}(\log(n)/\epsilon)$, where $n$ is the problem size and $\epsilon$ is a precision parameter related to how close to optimal the solution has to be. This contrasts with other approaches to solving optimization problems, such as interior point methods, for which the number of iterations can scale as $\mathcal{O}(\mathrm{poly}(n) \log(1/\epsilon))$.





Each iteration typically takes $\text{poly}(n, m, 1/\epsilon)$ time and is carried out with subroutines that can often be sped up with quantum algorithms. These subroutines can include Grover search / amplitude amplification and, in the case of the matrix MWU method, Gibbs sampling, which end up dominating the quantum cost of the algorithm.

Here it is important to point out that, especially in the quantum case, the MWU method can benefit from keeping an implicit representation of the $n$-dimensional vector $x$ (or in the case of matrix MWU, the $n \times n$ matrix $X$). For instance, in the example problem above, we need not explicitly write down the vector $x$; rather, we can keep track of the indices $j_1, j_2, \ldots, j_t$ corresponding to the penalties assessed at iterations $1, 2, \ldots, t$. These indices can be organized into a $t$-sparse vector $y \in \mathbb{R}^m$ from which $x$ is defined implicitly by $x_i \propto e^{\eta \sum_{j=1}^{m} a_{ij} y_j}$. In the context of optimization problems like LPs and SDPs, the vector $y$ can often be related to the *dual* version of the optimization problem (see, e.g., [4]), where the goal is to find an optimal $y$, and each value $y_j$ may be interpreted as the weight assigned to decision $j \in [m]$. Given $y$, the Gibbs sampling primitive can then produce a quantum state on $\mathcal{O}(\log(n))$ qubits encoding the vector $x$. At iteration $t + 1$, this quantum state is used to find an index $j_{t+1}$ corresponding to a violated constraint without ever explicitly writing down the vector $x$. This implicit representation is essential if the quantum algorithm is to achieve complexity sublinear in $n$. The same situation arises in algorithms for SDP based on matrix MWU, where Gibbs sampling is used to produce a $\mathcal{O}(\log(n))$-qubit mixed quantum state $\rho = e^{-H}/\text{tr}(e^{-H})$, where on iteration $t + 1$, the Hamiltonian $H = \sum_j y_j A^{(j)}$ is given by a weighted sum of at most $t$ distinct input matrices. By keeping track only of the sparse vector $y \in [m]$, one avoids needing to write down the $n^2$ entries of $\rho$. In fact, ignoring dependence on $\epsilon$, the Gibbs state $\rho$ can typically be prepared using only $\widetilde{\mathcal{O}}(s\sqrt{n})$ queries to the input data, where $s \leq n$ is the sparsity of the $n \times n$ input matrices $A^{(j)}$ (see, e.g., [5, 4]). This represents a polynomial speedup in the per-iteration cost compared to classical methods.

Additionally, there is also an appealing possibility that, for specific cases, the Gibbs sampling step for the $\log_2(n)$-qubit system could be accomplished in $\text{polylog}(n)$ time if the system thermalizes rapidly, allowing quantum algorithms based on the matrix MWU method to have faster runtime, perhaps as fast as $\text{poly}(\log(n), 1/\epsilon)$, representing an exponential speedup over their $\text{poly}(n, 1/\epsilon)$-time classical counterparts.

### Caveats

One caveat is that the best outlook for quantum advantage occurs when the constraint matrices $A^{(j)}$ that appear in applications are sparse matrices (and especially if they correspond to physical local Hamiltonians). However, this sparsity constraint may not be satisfied often in practice. There can in principle still be a speedup for dense matrices, but in this case, access to a large quantum random access memory might be required, which has its own caveats.

Another caveat to achieving a practically useful algorithm with either the classical or the quantum version of the MWU method is that the theoretical dependence of the runtime on the error parameter $\epsilon$ may lead to poor practical runtimes. The original quantum SDP solver based on MWU had $\mathcal{O}(\epsilon^{-18})$ dependence [6], and this was later improved to $\mathcal{O}(\epsilon^{-5})$ [4]. While this is technically $\text{poly}(1/\epsilon)$ scaling, the large power would likely lead the algorithm to be worse than alternatives, such as classical or quantum interior point methods which have $\text{polylog}(1/\epsilon)$ scaling, unless it is tolerable for $\epsilon$ to be essentially constant. In the case of zero-sum games, the quantum algorithm based on the MWU method has a slightly more tolerable $\mathcal{O}(\epsilon^{-3})$ dependence.





**Example use cases**

- The MWU method can be used to gain an asymptotic quantum speedup in solving zero-sum games, and relatedly, solving LPs [3, 7]. This speedup is generated by Grover-like methods and does not require Gibbs sampling of quantum states. Many interesting optimization problems can be reduced to an LP.

- The MWU method was used in [8] to give an algorithm for online portfolio optimization, relevant to applications in finance.

- The matrix MWU method can be used to gain an asymptotic speedup for solving SDPs in the regime where the precision parameter $\epsilon$ to which the program should be optimized is large. Many interesting optimization problems can be reduced to an SDP. One notable example is that approximate solutions to (discrete) binary optimization problems can be found by solving the (continuous) SDP relaxation of the problem and performing a rounding procedure on the solution (see, e.g., [9, 10]).

**Further reading**

- See Arora, Hazan, and Kale [1] for an overview of the MWU method from a classical perspective, including its matrix generalization.

- The quantum algorithm for SDP based on the MWU method was introduced by Brandão and Svore [6]. This was improved in subsequent works [11, 5, 4]. The method was applied to the specific application of solving SDP relaxations of binary optimization problems in [9, 10], and to the specific application of computing optimal strategies of zero-sum games in [3].

- In [7], a "dynamic Gibbs sampling" method is proposed to improve the complexity of the MWU algorithm for zero-sum games. It would be interesting if this method can be extended to other applications of the MWU method.

- For an overview of multiplicative weights methods within quantum algorithms, see [12].

# 24  Approximate tensor network contraction


*The authors are grateful to Glen Evenbly, Johnnie Gray, Daniel Malz, and Ashley Milsted for reviewing this section of the survey.*


**Rough overview (in words)**

Tensor network algorithms are a versatile tool that is playing an increasingly important role in problems both within and outside of physics and quantum computation [1], whenever the size of the underlying linear space is exponentially large in some appropriately defined dimension (i.e., tensor decomposition of the space). Their application to exponentially large linear systems is ultimately limited by the ability to contract (i.e., sum over repeated indices) large networks of tensors, in particular when the network forms a graph with many loops. Quantum approximate contraction of tensor networks [2] is a quantum algorithm for contracting arbitrary tensor networks up to a constant additive error. Estimating partition functions up to an additive error is a special case of the general problem, where all elements of the tensor network are positive.

This quantum approach to approximate tensor network contraction is of particular interest since many commercially relevant problems do not care about asymptotic speedups, but rather time-to-solution on smaller or medium problem sizes, and oftentimes approximate solutions found with heuristics are good enough. Tensor network (sometimes called quantum-inspired) algorithms for industrially relevant problems can be used heuristically, and the quantum approximate contraction backend might be used in cases where the classical algorithms do not provide sufficient accuracy, speed, or scale. Quantum-inspired classical algorithms based on tensor networks might allow for the identification of promising heuristic applications of quantum computing.

At this time, however, the only known problems where the quantum backend provides substantial speedup is for problems originating from quantum computing itself, such as quantum computational supremacy experiments based on random quantum circuits [3].

**Rough overview (in math)**

We define a tensor network as an abstract object $T(G, M)$ defined on a graph $G = (V, E)$, where to each vertex $v \in V$ we associate a tensor $M^{(v)}$ with one index for each adjacent edge. The tensor network $T(G, M)$ is closed, in that all edges are contracted. This means that for any specific set of tensors $M$ on $G$, $T(G, M)$ maps to a scalar. Given the graph $G$, we define a contraction pathway as a sequence of edges in $G$, where at each step the pair of vertices adjacent to the edge are merged into one vertex, until, at the end of the sequence, only one vertex remains in the graph. When merging two vertices, it can occur that edges that were previously distinct now coincide. It is important to keep track of the multiplicity of such edges. The optimal contraction pathway, in terms of cost scaling, is the merge sequence for which the maximum number of edges (counting multiplicity) incident to any vertex at any of the steps—a quantity called the contraction width—is minimized. The optimal contraction width is related to the tree width of the graph [4]. Classical exact contraction algorithms typically scale exponentially in the contraction width [5, 6, 7, 8]. For generic, loopy networks, the contraction width is expected





to be polynomially related to $|V|$; thus, the exact contraction algorithm will quickly become intractable with growing $|V|$. However, many approximate contraction methods exist [9, 10].

Suppose we fix a contraction pathway for which the sequence of edges forms a path in $G$, known as a "bubbling" because one can draw a bubble around contracted vertices, which sequentially expands as the path is traversed. This may not be the pathway with optimal contraction width (for which the sequence of edges need not always form a path). Then, for any $\epsilon > 0$, there exists a quantum algorithm that runs in $\mathcal{O}(|V|\epsilon^{-2}\text{poly}(q^d))$ quantum time and outputs a complex number $r$ such that [2]

$$\Pr(|T(G, M) - r| \geq \epsilon\Delta) \leq \frac{1}{4},$$

where $d$ is the maximum degree of the graph and $q$ is the dimension of the edge Hilbert space (or bond dimension). The parameter $\Delta$ is the sequential norm of the operations in the contraction path: $\Delta = \prod_{v \in V} \|O_v\|$, where $O_v$ are called swallowing operators (see Definitions 3.1 and 3.2 in [2]), which control the sequential contraction of the tensor network.

Intuitively, one can think of contracting the network one edge at a time along a connected path, such as a snake covering a 2D lattice. At each step of the way, the contracted vertices—which form a potentially large tensor—are encoded as a quantum state, and each new vertex is contracted by a local operator $O_v$ (the process is called bubbling in [2]). The dimension of the "state" can increase or decrease with every operation. Each operator $O_v$ in the contraction pathway is approximately mapped onto a unitary operator on the linear space ($q^d$ dimensional) connecting vertex $v$ in the network plus one ancilla qubit. The approximation comes from the Solovay–Kitaev theorem. This way, the exact contraction of the tensor network is approximately mapped onto a quantum circuit of volume roughly equal to the graph "volume." The output state of the quantum circuit encodes the result of the tensor network contraction into one of its amplitudes. In [2], they show how to estimate this amplitude using the Hadamard test, contributing the factor of $\epsilon^{-2}$ in the runtime. Alternatively, using the amplitude estimation subroutine, the $\epsilon$-dependence could be reduced to $\mathcal{O}(\epsilon^{-1})$.

The algorithm can be thought of as the reverse process of mapping a quantum circuit to a tensor network.

**Dominant resource cost (gates/qubits)**

The dominant cost of the algorithm is, on the one hand, the $\text{poly}(q^d)$ scaling, which can be substantial for highly connected graphs. More importantly though, for problems of interest is the value of $\Delta$, which can grow exponentially with $|V|$ and require extremely high precision $\epsilon$ to give a meaningful answer. In other words, $\Delta$ sets the scale of the approximation.

The complexity of the quantum algorithm depends sensitively on the structure of the graph $G(V, E)$, on the tensors $\{M_v\}_{v \in V}$ and on the choice of the contraction pathway. A number of limiting cases are known [2]:

- There are tensor networks for which it is NP-hard to obtain a classical additive approximation of the full contraction, suggesting the classical hardness of the problem.

- There exist families of tensor networks for which the additive approximation in Eq. (67) is BQP-hard, suggesting that there exists a complexity separation between the classical and quantum problem.





- There are specific examples of tensor networks representing partition functions, for which the quantum approximation scale $\Delta$ is exponential in $|V|$, but with a smaller exponent than the best known classical additive approximation scheme. There exist other examples where the converse is true [2].

Furthermore, approximate contraction of a tensor network representing a quantum partition function of a positive semidefinite Hamiltonian has been shown to be complete for the one clean qubit (DQC1) model of quantum computation [11], which suggests that approximate contraction is likely classically hard, at least for certain specific instances. Classical algorithms for this problem that are not based on tensor networks have also been examined [12].

### Caveats

The main caveat at present is that we do not have a good understanding of the structure of the network that allows for significant speedup on a quantum computer, due in part to the appearance of the complicated parameter $\Delta$ in the complexity statement. It is possible that the only situations where this is possible is when the tensor network can be mapped directly to a quantum circuit, without significant overhead. For example, in [13], a specific kind of tensor network called DMERA was shown to admit an exponential quantum speedup for approximate contraction because it arises from a specific kind of quantum circuit. The speedup is in the depth of the disentangling layer of the DMERA. A more critical caveat is that we do not understand when classical contraction algorithms are inefficient in practice. For instance, certain specific classes of tensor networks can be contracted efficiently [14, 15].

Even quantum computational supremacy experiments [16], which were designed specifically to maximize the separation between quantum and classical simulation, allow for tractable tensor network simulations up to large system sizes ($\sim 50$) and circuit depths ($\sim 30$) [3], though these simulations become much more challenging if we allow for nonlocal gates. Another subtle point is that it is often, or even usually, not clear what magnitude to expect for the value of a contracted tensor network from just looking at tensor norms. For example, terms may have partially canceling phases. The additive error bound for the quantum algorithm might not be very helpful in such cases.

Finally, it is likely difficult to make a proper comparison between classical approximate methods (e.g., the corner transfer matrix) and the above quantum approximation schemes, as the classical and quantum approximation errors have very different origins, and the quantum algorithm cannot be simulated at scale. The quantum algorithm might thus be regarded as a new heuristic to be be tested on a case to case basis once sufficiently powerful quantum hardware is available.

### Example use cases

There is an obvious case where the quantum algorithm provides an advantage, and that is if you prepare a quantum circuit, and map it onto a tensor network. Less trivial examples involve estimating partition functions of classical statistical mechanics models—although for this problem, good classical methods exist for the additive approximation [11]. Other applications involving large-scale tensor network contractions include condensed matter physics and molecular simulations. Inference problems [17] or differential equations simulation [18] might benefit from a quantum backend in some regimes, but a careful analysis has not yet been performed.





**Further reading**

- Pedagogical introductions to tensor networks [19, 1, 20].

- Quantum-inspired tensor network algorithms [21, 22, 23, 24].

- Complexity analysis of the quantum partition function problem [25].

# Fault-tolerant quantum computation

Throughout this survey, we predominantly restrict our attention to the circuit model of quantum computation. Within this paradigm, any quantum algorithm can be expressed as a sequence of basic operations, such as product state preparation, unitary single- and two-qubit gates, and single-qubit Pauli measurements. In order to accurately determine complete end-to-end resource estimates for quantum algorithms it is essential to understand the costs of (i) decomposing quantum algorithms into basic operations and (ii) realizing these basic operations reliably with the physical hardware. In other parts of this survey we assume noiseless logical qubits and operations (unless otherwise noted) and focus on item (i). In this part, we take into account that physical qubits and operations are noisy and discuss item (ii). We first review the fundamental ideas behind the theory of fault tolerance. We then illustrate them with concrete realizations in the paradigm of the surface code and lattice surgery.

**This part contains:**







# 25 Basics of fault tolerance


*The authors are grateful to Earl Campbell, Andrew Cross, and John Preskill for reviewing this section of the survey.*


**Rough overview (in words)**

The error rates of all known realizations of physical qubits and basic operations are too high to enable implementation of the majority of quantum algorithms considered in this survey. Even if the probability $p$ for each basic operation to malfunction was minute, we would nevertheless expect an error to occur in any quantum circuit comprising more than $\mathcal{O}(1/p)$ operations. One may optimistically assume that in the foreseeable future $p = 10^{-6}$ might be achieved by certain quantum architectures, such as trapped ions [1, 2]. This, in turn, limits the size of any quantum circuit that one may hope to reliably execute to roughly one million basic operations. Such a bound places a severe restriction on the algorithms that could be run and is orders of magnitude smaller than the resources needed to implement the quantum algorithms described in the other parts of this survey.

The theory of quantum fault tolerance [3] and quantum error correction [4, 5, 6] provides a collection of techniques to deal with imperfect operations and unavoidable noise afflicting the physical hardware, at the expense of moderately increased resource overheads. In the basic model for fault tolerance, one assumes that each elementary component of a quantum circuit (including the identity gate) may fail with some small but nonzero probability, independently of the other components, and classical information processing is noiseless. For concreteness and simplicity, one may choose to model any noisy component as an ideal component followed by (or, in the case of measurements, preceded by) some Pauli channel acting on the same subset of qubits. Let $\mathcal{C}$ be a quantum circuit (possibly with classical input and output) describing a desired quantum algorithm. Since each component of $\mathcal{C}$ may fail, one should not implement $\mathcal{C}$ directly; rather, one needs to implement a different quantum circuit $\mathcal{F}(\mathcal{C})$, which is a fault-tolerant (FT) version of $\mathcal{C}$. This, in turn, can be achieved by replacing each qubit in $\mathcal{C}$ with a logical qubit encoded in some quantum error correcting (QEC) code and each elementary component of $\mathcal{C}$ with a corresponding FT gadget; see Fig. 10. The desired quantum computation will then be realized on the logical level of $\mathcal{F}(\mathcal{C})$ without leaving the protective encoding guaranteed by the QEC code.

To realize universal FT quantum computation, it suffices to have state preparation gadgets (for at least one type of state), measurement gadgets (for at least one type of measurement), gate gadgets (for a universal set of gates), and QEC gadgets. One requires that all of these gadgets satisfy certain FT conditions; see, for instance, [7, 8]. Although the asymptotic scaling of resource overheads associated with FT gadgets is manageable (e.g., polylogarithmic in the inverse of the target logical error rate), the constant prefactors tend to be large, resulting in the qubit and time overheads that currently constitute one of the main bottlenecks to practical FT quantum computation. We will discuss this point in more detail for the implementation of logical gates and quantum error correction with the planar architecture based on the surface code [9, 10].





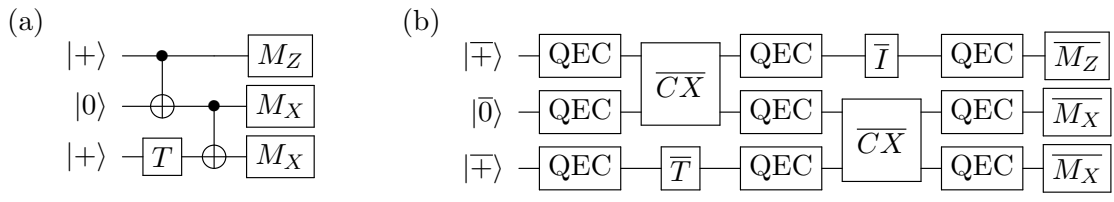

Figure 10: (a) A quantum circuit $\mathcal{C}$ consists of state preparation, unitary gates, and measurements. (b) An FT realization of $\mathcal{C}$ is a quantum circuit $\mathcal{F}(\mathcal{C})$ obtained by replacing each qubit in $\mathcal{C}$ with a logical qubit encoded in some QEC code and using appropriate FT gadgets interspersed with QEC gadgets in place of each basic component of $\mathcal{C}$. Note that some gadgets may require considerable resources (not shown in the picture); see Section 26 on quantum error correction and Section 27 on logical gates with the surface code for more details.

**Rough overview (in math)**

Designing FT gadgets is a challenging task for several reasons. First, FT gadgets are usually developed and optimized for a specific QEC code. Second, even though they comprise imperfect basic components, they are required to work reliably as long as the number of malfunctioning components is limited. Third, FT gadgets may spread errors, however, they must not do so in an uncontrollable way.

Given a set of FT gadgets, one can reliably perform an arbitrarily long quantum computation as long as the physical error rate of each basic component is below some constant value, often referred to as the FT threshold. This result is established by the celebrated threshold theorem [11, 12, 13, 7]. To be more precise, consider the basic model for FT. The threshold theorem asserts that there exists a constant $p_{\mathrm{FT}} > 0$, such that for any $\epsilon > 0$ and any quantum circuit $\mathcal{C}$ there exists a quantum circuit $\widetilde{\mathcal{C}}$ that produces an output with statistical distance at most $\epsilon$ from the output of $\mathcal{C}$, provided the physical error rate $p$ is below $p_{\mathrm{FT}}$. Moreover, $\widetilde{\mathcal{C}}$ uses a number of qubits and time steps that are at most polylog($|\mathcal{C}|/\epsilon$) times bigger than the number of qubits and time steps in $\mathcal{C}$, where $|\mathcal{C}|$ denotes the number of basic components in $\mathcal{C}$.

The basic idea behind the proof of the threshold theorem proceeds as follows. Consider a quantum circuit $\mathcal{F}(\mathcal{C})$, which is an FT implementation of $\mathcal{C}$. Assuming the basic model for fault tolerance described above, for sufficiently small physical error rate $p$, the logical error rate for $\mathcal{F}(\mathcal{C})$ should be smaller than $p$, since $\mathcal{F}(\mathcal{C})$ is an FT implementation of $\mathcal{C}$. One can then consider a quantum circuit $\mathcal{F} \circ \mathcal{F}(\mathcal{C})$, which is an FT implementation of $\mathcal{F}(\mathcal{C})$, reducing the logical error rate even further. By repeating this process, one eventually obtains a quantum circuit $\widetilde{\mathcal{C}} = \mathcal{F} \circ \cdots \circ \mathcal{F}(\mathcal{C})$ with the logical error rate below $\epsilon$. The resulting FT protocol is based on concatenated QEC codes.

One may improve the scaling of the resource overheads from the threshold theorem with concatenated QEC codes. In particular, in the asymptotic limit of large quantum circuits, the ratio of qubits in $\mathcal{C}$ and $\widetilde{\mathcal{C}}$ can be a constant [14]. In this construction, the FT protocol requires a family of QEC codes that satisfies certain properties, including the desired scaling of code parameters, computationally efficient decoding algorithms, and constant-weight parity checks. Such a family of QEC codes was first provided in [15].





**Dominant resource cost (gates/qubits)**

At the heart of FT quantum computation, there is usually some QEC code. The choice of a QEC code affects the relationship between a quantum circuit $\mathcal{C}$ and its FT realization $\widetilde{\mathcal{C}}$, and the subsequent scaling of the resource overheads. Therefore, we would like to choose a QEC code for which the encoding rate (defined as the ratio $k/n$, where $k$ and $n$ are the number of logical and physical qubits, respectively) as well as the relative code distance (defined as the ratio $d/n$, where $d$ is the minimum weight of any nontrivial logical operator) are as high as possible. Although for concatenated QEC codes (that feature in the threshold theorem), both $k/n$ and $d/n$ go to zero as $n$ goes to infinity, we know that there exist QEC codes with good parameters, that is, for which $k/n$ and $d/n$ are asymptotically constant [16]. Moreover, recent groundbreaking results [17, 18, 19, 20] provided constructions of QEC codes that not only have good parameters but also constant-weight parity checks (thus their name—quantum low-density parity check codes). The latter property is particularly important from the perspective of fault tolerance. In fact, all quantum low-density parity check codes with code distance $d = \mathcal{O}(\log n)$ have a nonzero FT threshold [21]. However, experimental realization of these constructions (in contrast to the surface code) seems to become extremely challenging as the number of physical qubits increases, at least within the realm of solid-state qubits constrained by geometric locality of their physical entangling gates.

Another aspect of FT quantum computation that affects the resource overheads are the FT gadgets being used. One of the easiest ways to implement FT gadgets for gates is via transversal gates. By definition, transversal gates are implemented via a tensor product of single-qubit unitaries (or, more generally, via a depth-one quantum circuit) and therefore do not spread errors in an uncontrollable way. Unfortunately, transversal gates are limited by the Eastin–Knill theorem [22, 23, 24, 25], which rules out the existence of a (finite-dimensional) QEC code with a universal set of transversal logical gates. One strategy to circumvent this limitation is to prepare certain magic states and use them to realize FT gates [26]; see Section 27 on implementing logical gates for more details and a discussion of other strategies.

To realize FT gadgets for state preparation, QEC, and measurement, one typically chooses among three FT schemes: Shor's [3], Steane's [27], or Knill's [28]. Roughly speaking, Shor's scheme uses simple states (verified cat states) of the ancilla qubits at the expense of implementing many gates on the data qubits, whereas Steane's and Knill's schemes trade highly complex states of the ancilla qubits (logical states encoded in the underlying QEC code) for minimizing the number of gates on the data qubits. To determine the best choice, one needs to consider the underlying QEC code (e.g., Steane's scheme is applicable only to the subset of codes known as CSS codes [16, 5]) and the quantum hardware restrictions (e.g., lack of extra ancilla qubits). For an illuminating and detailed discussion of FT schemes, see [8]. There are FT schemes that do not squarely fit in the aforementioned classification, for example, [29, 30, 31]. Also, for QEC codes with additional structure, such as quantum low-density parity check codes, one may pursue different approaches toward FT quantum computation; see Section 26 on QEC with the surface code.

**Caveats**

Rigorous proofs provide lower bounds on the FT threshold $p_{\mathrm{FT}}$. For instance, for an FT scheme based on the 7-qubit code, one finds $p_{\mathrm{FT}} > 2.73 \times 10^{-5}$ [7]. For an FT scheme by Knill [28] that relies on complex ancilla preparation techniques, one finds $p_{\mathrm{FT}} > 1.04 \times 10^{-3}$ [32]. However, these





values can differ by orders of magnitude from the values estimated in numerical simulations. For instance, the FT scheme by Knill is estimated to have an FT threshold $p_{FT}$ as high as $5 \times 10^{-2}$, constituting one of the highest-known FT thresholds. We remark that these values depend sensitively on the details of the FT schemes and the assumptions about noise. In particular, to obtain the aforementioned values we assume the ability to implement gates between any qubits. On the other hand, if we arrange qubits on a geometric lattice and restrict gates to be local, then while FT thresholds still exist, their values are significantly reduced.

One can expand the threshold theorem in many ways. Even using the basic model for fault tolerance, one may choose the failure probabilities for each elementary component of a quantum circuit differently, for example, the failure probability of a measurement to be an order of magnitude higher than that of a gate. One can consider more general noise (which includes systematic errors, such as overrotations) arising due to a weak interaction between the system and a non-Markovian environment [7, 33]. In general, although experimental realizations of quantum computation may not satisfy exactly the assumptions of the threshold theorem, we expect the main conclusions to hold as long as the assumptions are not violated too much.

To simplify the analysis of FT schemes, we often assume unlimited classical computational power that one needs to, for example, process the error syndrome and infer an appropriate recovery operator in a QEC gadget; a number of such decoding algorithms have been developed for QEC with the surface code. It is important not to abuse this assumption by, for instance, solving the initial problem with an inefficient classical algorithm. At some point, however, one needs to take into account the finite speed of classical information processing. If the classical unit that processes the error syndrome is unable to keep pace with the rate at which this syndrome is being produced, then the error syndrome will start to accumulate and one will suffer from the so-called backlog problem [34]. Subsequently, the speed of quantum computing will be exponentially reduced and the computational advantage of quantum computing will be annulled. This issue will be especially prominent for quantum algorithms with only polynomial speedups.

**Further reading**

- An accessible introduction to quantum error correction and the theory of fault tolerance can be found in [35].

- A detailed introduction to quantum error correction and fault-tolerant quantum computation can be found in [8].

- A perspective on roads toward fault-tolerant universal quantum computation can be found in [36].

- The error correction zoo provides a catalog of error correcting codes.

# 26  Quantum error correction with the surface code

*The authors are grateful to Earl Campbell, Andrew Cross, and John Preskill for reviewing this section of the survey.*

**Rough overview (in words)**

To protect quantum information from the detrimental effects of noise, we can encode it into a code space of some quantum error correcting (QEC) code [1, 2]. Oftentimes, we choose to work with stabilizer codes [3]. By definition, a code space of a stabilizer code is the simultaneous $(+1)$-eigenspace of a set of commuting Pauli operators, commonly referred to as parity checks.

The surface code [4, 5, 6] is one of the most studied stabilizer codes. It can be implemented with a planar layout of qubits and entangling gates only between neighboring qubits. For that reason, the surface code is particularly appealing for quantum hardware architectures with restricted qubit layout and connectivity, such as superconducting circuits [7, 8]. The most common realization of the surface code uses $n = L^2$ data qubits to encode $k = 1$ logical qubit and has code distance $d = L$, where $L$ is the linear size of the $L \times L$ square lattice with open boundary conditions. Additionally, $n_A = L^2 - 1$ ancilla qubits are typically used to measure parity checks; see Fig. 11(a).

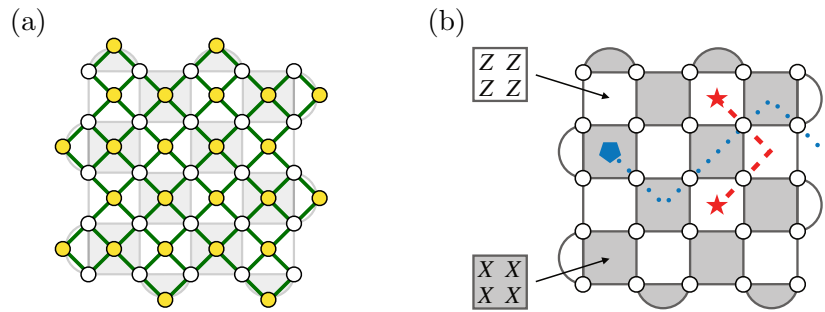

Figure 11: (a) A planar layout of data and ancilla qubits (empty and filled dots, respectively) with entangling gates (thick edges) only between neighboring qubits. This layout gives rise to the $L \times L$ square lattice with open boundary conditions, where $L = 5$ here. (b) The surface code can be realized by measuring Pauli $Z$- and $X$-type parity checks (light and dark faces, respectively). The error syndrome (stars and pentagons) can be interpreted as the endpoints of string-like Pauli $X$ and $Z$ errors (dashed and dotted lines, respectively).

In order to perform QEC, we have to be able to detect errors without revealing the encoded information. For stabilizer codes, we can achieve that by measuring their parity checks to obtain the error syndrome (which comprises the measurement outcomes returning $-1$). Then, the error syndrome is processed by specialized classical algorithms, also known as "decoders," to find an appropriate recovery operator that attempts to remove errors afflicting the encoded information. For generic stabilizer codes, the problem of optimal decoding is computationally hard, even for simple noise models [9]. However, for QEC codes with some underlying structure, such as the surface code, there exist a variety of computationally efficient (albeit not optimal) decoding





algorithms. In particular, the three most popular classes of decoders for the surface code are as follows:

- Matching decoders, including the minimum-weight perfect matching algorithm [6] and its follow-up improvements, such as the belief-matching algorithm [10, 11]. These decoders phrase the problem of surface code decoding as a graph-theoretic problem of perfect matching, which can be efficiently solved [12].

- Clustering decoders, such as the renormalization-group decoder [13, 14] and the union-find decoder [15]. These decoders primarily exploit the structure of the error syndrome in the surface code; see Fig. 11(b).

- Tensor network decoders [16, 17, 18]. These decoders phrase the problem of surface code decoding as a numerical problem of contracting tensor networks.

In order to assess the usefulness of decoders, one usually considers two criteria: runtime and accuracy. The first criterion, runtime, is defined as the time needed for the decoder to process the error syndrome. It is crucial that any practical decoder is able to operate at a rate compatible with the rate of parity check measurements; otherwise, the error syndrome will start to accumulate, leading to the backlog problem [19]. The second criterion, accuracy, is typically defined for a given noise model in terms of the logical error rate, that is, the failure rate of the decoder to successfully undo the effects of noise on the encoded information. From the perspective of reducing runtime and improving accuracy, matching and clustering decoders stand out. Namely, they can achieve almost-linear runtime [20, 15] and good accuracy. Combined with the techniques of parallelizable real-time decoding [21, 22], these decoders will likely play a key role in scalable QEC with the surface code. To achieve optimal accuracy, one can use tensor network decoders, however, they are often not computationally efficient, with runtime that scales unfavorably.

**Rough overview (in math)**

In addition to being compatible with planar layouts of qubits and admitting computationally efficient decoders with good accuracy, the surface code also exhibits one of the highest QEC thresholds. Recall that a QEC threshold is specified for the following triple: a QEC code family of growing distance $d$, a decoder, and a noise model. It is defined as the highest value $p_{\text{th}}$ such that for any error rate $p < p_{\text{th}}$ the probability that the decoder fails to undo the effects of noise goes to zero as $d$ goes to infinity. For example, the QEC threshold for the surface code, using the minimum-weight perfect matching algorithm for decoding, with a circuit noise model based on depolarizing noise, is around 1% [23, 11].

Typically, if the error rate $p$ describing noise is sufficiently low and below the threshold $p_{\text{th}}$, then the logical error rate $p_{\text{fail}}$ roughly scales as

$$p_{\text{fail}} \sim \left(\frac{p}{p_{\text{th}}}\right)^{\left\lceil \frac{d}{2} \right\rceil}. \tag{67}$$

This implies that in order to achieve the target error rate $\epsilon$, it suffices to implement the surface code with code distance $d = \mathcal{O}(\log(1/\epsilon)/\log(p_{\text{th}}/p))$ using $n + n_A = \mathcal{O}(d^2) = \mathcal{O}(\log^2(1/\epsilon)/\log^2(p_{\text{th}}/p))$ data and ancilla qubits. Subsequently, qubit overhead associated with QEC based on the surface code scales polylogarithmically in the inverse target error rate $1/\epsilon$.





**Dominant resource cost (gates/qubits)**

Performing reliable QEC in the presence of measurement errors becomes challenging since the error syndrome can be corrupted. A straightforward solution to the problem of unreliable error syndrome is to repeatedly measure the parity checks in order to gain enough confidence in their measurement outcomes [24, 6]. If this approach is applied to the surface code with code distance $d$, then one needs to perform $\mathcal{O}(d)$ rounds of parity check measurements, incurring relatively large time overhead.

To reduce time overhead, one can pursue single-shot QEC [25], which does not require repeated measurement rounds. It is possible to realize single-shot QEC with the surface code [26, 27, 28], however, in addition to the parity checks in Fig. 11(b), one would need to measure nonlocal high-weight parity checks, which is a serious limitation. A more streamlined approach is to consider a different realization of the surface code, the 3D subsystem toric code [29, 30], which can be implemented with qubits arranged on the cubic lattice and local low-weight parity checks. Although this approach is natively defined in three spatial dimensions, it can be emulated with planar layouts of qubits and either a limited number of nonlocal gates or the ability to reshuffle qubits (which is available with, e.g., Rydberg atoms [31, 32]). In order to realize code distance $d$, one incurs qubit overhead of $\mathcal{O}(d^3)$ (compared to qubit overhead of $\mathcal{O}(d^2)$ for the surface code). From that perspective, single-shot QEC with the subsystem toric code can be viewed as trading time overhead for qubit overhead.

**Caveats**

There have been efforts to improve surface code decoders by incorporating various machine learning methods, including neural networks [33, 34, 35, 36, 37] and reinforcement learning [38]. At the current stage, decoders solely based on machine learning methods seem to be of limited applicability, mostly due to high training costs and scalability issues. Nevertheless, these approaches are likely to be immensely beneficial for QEC in the settings where (possibly correlated) noise is unknown and may have to be learned first.

Typically, in QEC analysis one considers simple Pauli noise, such as depolarizing noise acting independently and identically on each qubit. If noise exhibits bias between the $X$, $Y$, and $Z$ components of Pauli noise, then this structure can be exploited, leading to dramatically increased QEC thresholds, as exemplified by variants of the surface code [39, 40, 41]. Similarly, noise that is biased toward erasure errors can be beneficial from the perspective of QEC [42, 43, 44, 45]. On the other hand, realistic noise may be coherent or correlated and thus not only difficult to correct, but also to numerically simulate. For instance, the logical error rates for coherent noise may be orders of magnitude higher than the estimates of the logical error rates for simple Pauli noise (assuming both types of noise have the same error rate) [46]. In certain cases of coherent noise, however, QEC with the surface code may be efficiently simulable [47].

In addition to the 3D subsystem toric code, one can also consider other higher-dimensional versions of the surface code. With these codes, roughly speaking, one improves the QEC capabilities at the expense of increased qubit overhead. Moreover, for the higher-dimensional surface code, it may suffice to use arguably the least complex decoders that are based on cellular automata (which, by definition, are parallelizable and only use local information about the error syndrome) [6, 48, 49, 50].





**Example use cases**

- Decoders for the surface code can be used for other QEC code families, such as the color code [51, 52, 53]. In fact, due to a close connection between the color codes and the surface codes [54, 55], any surface code decoder can be used as a subroutine in the restriction decoder for any color code (in two or more spatial dimensions) [56, 57].

**Further reading**

- The seminal paper by Dennis et al. [6] is a thorough introduction to QEC with the surface code.

- A recent perspective [58] discusses how to use matching decoders to decode stabilizer codes.

- Open-source software packages have been developed for implementing QEC with the surface code, such as Stim [59] and PyMatching [60].

# 27 Logical gates with the surface code

*The authors are grateful to Earl Campbell, Andrew Cross, and John Preskill for reviewing this section of the survey.*

**Rough overview (in words)**

The ability to implement an arbitrary unitary operation, either exactly or approximately, is a prerequisite for performing quantum computation. It can be achieved with unitary gates that form a universal gate set [1, 2]. A commonly considered gate set contains two Clifford gates, the Hadamard gate $H$ and the controlled $X$ gate $CX$ (also known as the controlled NOT gate), and one non-Clifford gate, the $T = Z^{1/4}$ gate. One can consider other non-Clifford gates, such as the Toffoli gate $CCX$. Note that non-Clifford gates are essential for quantum computation, as any quantum circuit comprising only Clifford gates, state preparation, and measurement in the computational basis can be simulated in polynomial time on a probabilistic classical computer [3, 4].

Since we are interested in fault-tolerant quantum computation, we would like to implement a universal set of logical gates $\overline{H}$, $\overline{CX}$, and $\overline{T}$ on information encoded in some quantum error correcting (QEC) code, such as the surface code. We can implement these gates with a planar layout of qubits and nearest-neighbor entangling gates. To be more precise, we consider a simple architecture [5] that comprises $N$ surface code patches, each encoding a logical qubit into the surface code with code distance $d$, and the routing space in between; see Fig. 12(a). In such an architecture, the total number of data and ancilla qubits is $\mathcal{O}(Nd^2)$.

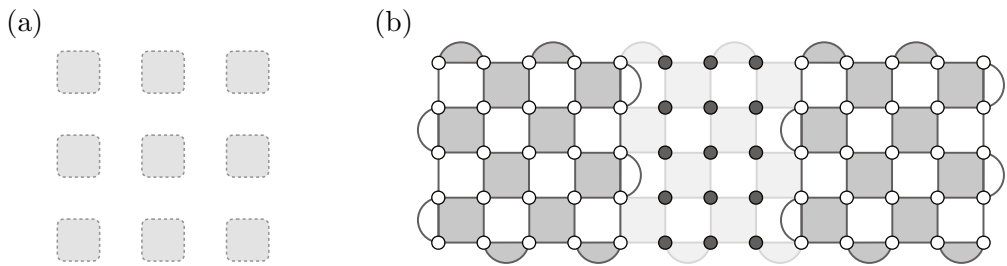

Figure 12: (a) A planar layout of qubits comprises surface code patches (shaded), each using the layout depicted in Fig. 11(a) and encoding a logical qubit, with routing space in between the patches. (b) The logical Pauli measurement $\overline{M_{XX}}$ is implemented by preparing the routing space qubits (filled dots) in the state $|0\rangle$ and repeatedly measuring parity checks (lightly shaded) in the routing space spanning between the two surface code patches. Other logical Pauli measurements, for example, $\overline{M_{ZZ}}$ and $\overline{M_{YZ}}$, require connecting different boundaries of the two patches.

**Rough overview (in math)**

The logical $\overline{H}$ does not pose any challenges. From a practical standpoint, it is transversal, since it can be realized by applying the Hadamard gate $H$ to every data qubit in the surface code patch, followed by swapping of the roles of Pauli $Z$- and $X$-type parity checks in the subsequent





QEC rounds. As such, the logical $\overline{H}$ takes constant time and the surface code patch is effectively rotated (which may alter how subsequent operations are implemented).

The logical $\overline{CX}$ is more challenging than the logical $\overline{H}$, since it is impossible to implement it transversally with the planar layout of qubits and nearest-neighbor entangling gates shown in Fig. 12(a). Instead, one can use the following quantum circuit, where the first qubit (top wire) is the control and the third qubit (bottom wire) is the target of the logical $\overline{CX}$ gate:

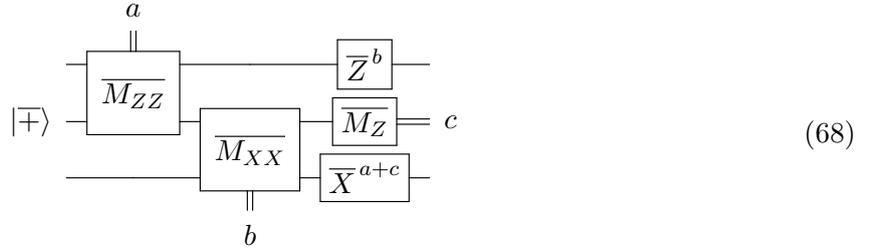

$$(68)$$

It is straightforward to fault-tolerantly realize preparation of the logical state $|\overline{+}\rangle$, logical Pauli measurement $\overline{M_Z}$, and logical Pauli operators $\overline{Z}$ and $\overline{X}$. In addition, the required logical Pauli measurements $\overline{M_{ZZ}}$ and $\overline{M_{XX}}$ can be implemented fault-tolerantly via "lattice surgery" techniques [5, 6, 7]; see Fig. 12(b) for an illustration of how to realize $\overline{M_{XX}}$. Unlike the logical $\overline{H}$, logical Pauli measurements $\overline{M_{ZZ}}$ and $\overline{M_{XX}}$ and, subsequently, the logical $\overline{CX}$ cannot be realized in constant time; rather, due to the need to account for measurement errors, they typically incur time overhead of $\mathcal{O}(d)$.

The logical $\overline{T}$ can be implemented using gate teleportation [8] via the following quantum circuit:

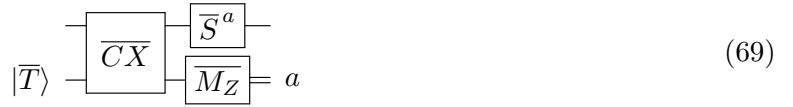

$$(69)$$

Here, the logical resource state $|\overline{T}\rangle$ is defined as $\left(|\overline{0}\rangle + e^{i\pi/4}|\overline{1}\rangle\right)/\sqrt{2}$, the logical gate $\overline{S}$ is defined as $\overline{Z}^{1/2}$, and the first qubit (top wire) is the control and the second qubit (bottom wire) is the target of the logical $\overline{CX}$ gate. One can fault-tolerantly implement the logical $\overline{S}$ with a planar layout of qubits [9, 10] (or even in a transversal way given access to nonlocal entangling gates [11, 12]). However, the need to apply the logical $\overline{S}$ conditioned on the measurement outcome of $\overline{M_Z}$ may slow down quantum computation, and, for this reason, it may be beneficial to use the following quantum circuit from [7, Fig. 17(b)], which is an alternative to the one in Eq. (69) that uses one additional logical qubit but requires only logical Pauli corrections, rather than logical Clifford corrections.

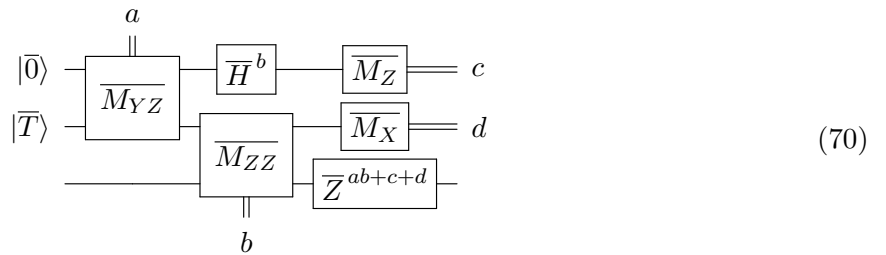

$$(70)$$

In either case, given the logical resource state $|\overline{T}\rangle$, the logical $\overline{T}$ typically incurs time overhead of $\mathcal{O}(d)$. We conclude that implementing the logical $\overline{T}$ reduces to the problem of preparing the logical state $|\overline{T}\rangle$, which, in turn, can be realized via state distillation [13, 14]; see [15] for a brief overview of state distillation.





**Dominant resource cost (gates/qubits)**

State distillation provides a fault-tolerant method to prepare high-fidelity logical resource states, such as the logical state $|\overline{T}\rangle$. The basic idea is to convert some number of noisy resource states into fewer but, crucially, less noisy resource states. Importantly, this task can be accomplished with quantum circuits comprising only Clifford gates (together with state preparation and measurement in the computational basis) and postselection. Typically, state distillation circuits are based on some QEC code, for example, the 15-qubit Reed–Muller code.

State distillation is often described as a resource-intensive method that contributes the most to the resource overhead of fault-tolerant quantum computation with the surface code [16] (assuming many state distillation circuits working in parallel). For that reason, numerous efforts have been devoted to finding possible alternatives [17, 18, 19, 20, 15]. However, recent results indicate that state distillation may not be as costly as one may think [7, 21], especially when one allows only a few state distillation circuits to run in parallel and optimizes them for specific quantum hardware and noise that exhibits some bias [22]. In the task of estimating the ground state energy density of the Fermi–Hubbard model, state distillation of logical Toffoli resource states injected one at a time uses less than 10% of the total resources and is never a bottleneck on runtime of the quantum algorithm [23].

Often, a quantum algorithm is expressed as a quantum circuit $\mathcal{C}$ comprising Clifford and $T$ gates. Thus, by using the aforementioned logical gates $\overline{H}$, $\overline{CX}$, and $\overline{T}$, we can fault-tolerantly implement the logical quantum circuit $\overline{\mathcal{C}}$ with the surface code of code distance $d$ and a planar layout of qubits in Fig. 12(a). However, from the perspective of reducing the resource overheads, it may be beneficial to consider a quantum circuit $\mathcal{C}'$ equivalent to the circuit $\mathcal{C}$, which is obtained from $\mathcal{C}$ by commuting all Clifford gates to the end of $\mathcal{C}$ [7]. As a result, the circuit $\mathcal{C}'$ only comprises multiqubit Pauli $\pi/8$ rotations (which are a generalization of the $T$ gate and can be realized via, e.g., quantum circuits analogous to the one in Eq. (70)). Consequently, fault-tolerant implementation of the logical circuit $\overline{\mathcal{C}'}$ incurs qubit overhead of $\mathcal{O}(Nd^2)$ and time overhead of $\mathcal{O}(Md)$, where $N$ and $M$ are the number of qubits and $T$ gates in $\mathcal{C}$, respectively. We remark that the time overhead can be reduced at the expense of increased qubit overhead—first, by distilling more resource states and being able to use them faster, then, by implementing them in parallel [7].

**Caveats**

Lattice surgery is not necessary to realize fault-tolerant quantum computation with a planar layout of qubits and nearest-neighbor gates. An alternative approach (which actually preceded the development of lattice surgery) relies on the surface code with defects and braiding [24, 25, 16, 9]. However, resource overhead estimates strongly suggest that this approach is not competitive with lattice surgery [6].

The simple architecture depicted in Fig. 12(a) can be improved in a couple of ways to reduce the qubit overhead. First, it is possible to pack surface code patches more densely, resulting in more logical qubits for the given total number of qubits and target code distance [26, 7]. Second, one can designate certain regions, commonly referred to as magic state factories, to solely produce resource states, such as the logical state $|\overline{T}\rangle$, and optimize their design [27, 7, 21].

To simplify implementation of logical gates, one can consider other QEC codes, for example, the 3D color code [28, 29]. The gauge color code has redundant degrees of freedom, commonly





referred to as gauge qubits. For different states of its gauge qubits, the gauge color code admits transversal implementation of different logical gates, which, *combined*, form a universal gate set (thus circumventing the Eastin–Knill theorem [30, 31]). Importantly, changing the state of gauge qubits can be done fault tolerantly in constant time. However, to realize this construction one needs, for instance, a 3D layout of qubits with nearest-neighbor gates or a planar layout of qubits with a limited number of nonlocal gates, which are more challenging to engineer compared to the simple architecture in Fig. 12(a). To achieve logical distance $d$ with the gauge color code, one incurs qubit overhead of $\mathcal{O}(d^3)$ (compared to qubit overhead of $\mathcal{O}(d^2)$ for the surface code), so, similarly to single-shot QEC described in Section 26 on QEC with the surface code, this approach trades time overhead for qubit overhead.

**Example use cases**

- Lattice surgery techniques developed for the surface code can be straightforwardly adapted to, for example, the color code [32] or the surface code with a twist [33], leading to fault-tolerant quantum computation with potentially reduced qubit overhead. Recent work [34] proposed a generalization to the setting of quantum low-density parity check codes. In addition, lattice surgery techniques can also be used for the fault-tolerant transfer of encoded information between arbitrary topological quantum codes [35].

- Now, we are ready to present a rough, order-of-magnitude estimate of the resource overheads needed to realize fault-tolerant quantum computation in the architecture based on the surface code and lattice surgery. For concreteness, we consider the circuit noise of strength $p = 0.001$, where each basic operation, including state preparation, CNOT gate, and measurement, can fail with probability $p$. Assume that we want to implement a quantum circuit $\mathcal{C}$ comprising $N = 10^3$ qubits and a certain number $M = 10^{10}$ of $T$ gates. These resource counts are in the ballpark of estimates for various quantum algorithms in the application areas of quantum chemistry, condensed matter physics, and cryptanalysis. First, following the procedure from [7], we compile $\mathcal{C}$ into a new circuit $\mathcal{C}'$ of depth $M$ that comprises $N$ qubits and $M$ multiqubit Pauli $\pi/8$-rotations implemented one at a time. Since there are $NM$ possible fault locations in the circuit $\mathcal{C}'$, the error rate for each qubit of $\mathcal{C}'$ should not exceed

$$\epsilon \approx 1/(NM).$$

Since each qubit of $\mathcal{C}'$ is realized as a logical qubit of the surface code with distance $d$, then its logical error rate $p_{\text{fail}}$ can be approximated by

$$p_{\text{fail}} \approx \alpha(p/p_{\text{th}})^{d/2},$$

where we can crudely set $\alpha = 0.05$ and $p_{\text{th}} = 0.01$; see Section 26 on QEC with the surface code for more details. Note that these values are empirical and depend heavily on the choice of the decoder, in our case, the belief-matching algorithm [36]. Thus, in order for the logical error rate $p_{\text{fail}}$ to reach the target error rate $\epsilon$ we need the surface code distance at least

$$d \approx \lceil 2\log(\alpha NM)/\log(p_{\text{th}}/p) \rceil.$$

Assuming that half of all required qubits are devoted to realizing $N$ surface code patches (each comprising $2d^2 - 1$ data and ancilla qubits), with the other half used for resource





state distillation and routing [7], we obtain that the fault-tolerant implementation of $\mathcal{C}'$ incurs qubit overhead of

$$n_{\mathcal{C}'} \approx 4Nd^2$$

and time overhead of

$$t_{\mathcal{C}'} \approx Md\tau,$$

where we crudely set $\tau = 1\,\mu s$ to be the time needed to implement one syndrome measurement round with the superconducting circuits architecture. Finally, our order-of-magnitude resource estimate gives $2.3 \times 10^6$ physical qubits and 67 hours of runtime. This general approach to resource estimation has been applied to a number of specific quantum algorithms in a variety of application areas; see, for example, [37, 38, 39, 40, 41]. These references often go beyond a back-of-the-envelope calculation and provide a more meticulous analysis that accounts for exact qubit layouts and the physical footprint of resource state distillation factories. They also pursue optimizations to how the circuit is implemented (e.g., exploiting space-time tradeoffs) in light of these considerations.

**Further reading**

- An accessible overview of fault-tolerant quantum computation based on the surface code and lattice surgery can be found in [7].

- A convenient way to describe and optimize lattice surgery operations is via the ZX calculus, which is a diagrammatic language for quantum computing [42, 43].

- A direct comparison of the resource overhead associated with preparation of the logical resource state $|\overline{T}\rangle$ using either state distillation or transversal gates (with the 3D color code) can be found in [15].

- To read about a framework for estimating resources required to realize large-scale fault-tolerant quantum computation, see [40].

- The recently introduced paradigm of algorithmic fault tolerance [44] may significantly reduce the space-time overhead of FT quantum computation with the surface code.

# Appendix: Background, conventions, and notation

This survey aims to be as modular as possible, where each numbered section and subsection can be read on its own—as such, specific notational definitions, technical terms, and acronyms are redefined the first time they appear in each numbered (sub)section.

Nevertheless, the mathematical presentation throughout this survey does assume familiarity with certain concepts, techniques, and conventions that are ubiquitous within the field of theoretical quantum information science. This survey is targeting an interdisciplinary community of researchers; thus, the assumed conventions and understanding of "common knowledge" will vary from reader to reader, depending on one's background and experience. In order to make the material as widely accessible as possible, here we collect some of the concepts and notational choices that are commonly used throughout this survey.

For readers interested in a more complete introduction to the field and its standard conventions, we recommend the following resources:

- The definitive reference in the field of quantum computation is the book by Nielsen and Chuang [1]. Other classic textbooks include those by Kitaev, Shen, and Vyalyi [2] and Kaye, Laflamme, and Mosca [3]. A similarly general set of topics is covered in the lecture notes of Preskill [4].

- Several sets of more recent lecture notes have a specific focus on quantum algorithms, for example, by de Wolf [5], Childs [6], and Lin [7]. See also the review article on quantum algorithms by Montanaro [8].

- Some online resources include a website containing lecture notes on quantum algorithms for data analysis and quantum machine learning by Luongo [9], the Pennylane codebook [10], and the quantum algorithm zoo [11].

## This appendix contains:

# A  Quantum systems and bra-ket notation

Basic concepts from linear algebra are an essential prerequisite for understanding quantum computation and thus for much of the technical discussion in this survey. We adopt bra-ket notation to denote quantum states and the linear algebraic objects they correspond to. The state of a quantum system, such as a collection of qubits, is labeled by a *ket*, such as $|\psi\rangle$—this object corresponds to a vector in a finite-dimensional vector space. For instance, the state of a single qubit is an element of the 2D complex vector space $\mathbb{C}^2$, and can be represented by a length-2 column vector or by a *superposition* over orthonormal basis states, denoted $|0\rangle$ and $|1\rangle$:

$$|\psi\rangle = \begin{pmatrix} \alpha_0 \\ \alpha_1 \end{pmatrix} = \alpha_0|0\rangle + \alpha_1|1\rangle, \qquad \alpha_0, \alpha_1 \in \mathbb{C}\,.$$

The state of a system of $n$ qubits is an element of the $2^n$-dimensional complex vector space $\mathbb{C}^{2^n}$—the tensor product of the $n$ individual 2D vector spaces—and can be represented in any of the following equivalent ways:

$$|\psi\rangle = \begin{pmatrix} \alpha_0 \\ \alpha_1 \\ \vdots \\ \alpha_{2^n-1} \end{pmatrix} = \sum_{x=0}^{2^n-1} \alpha_x|x\rangle = \sum_{x=0}^{2^n-1} \alpha_x|x_0\rangle \otimes |x_1\rangle \otimes \cdots \otimes |x_n\rangle, \qquad \alpha_x \in \mathbb{C}\ \forall x\,,$$

where $x_i \in \{0, 1\}$ denotes the $i$-th bit of the integer $x$ when $x$ is written in binary (leading zeros are added such that $x$ has $n$ digits, and $x_0$ corresponds to the most significant bit). The orthonormal basis $|0\rangle, |1\rangle, \ldots, |2^n - 1\rangle$ is called the *computational basis*. An $n$-qubit quantum state is said to be a *product state* if it can be written as a tensor product of 2D states on each of the $n$ systems

$$|\psi\rangle = |\phi_1\rangle \otimes |\phi_2\rangle \otimes \cdots \otimes |\phi_n\rangle, \qquad |\phi_i\rangle \in \mathbb{C}^2\ \forall i\,,$$

and it is said to be *entangled* if it cannot be written as a product state.

For each quantum state $|\psi\rangle$ corresponding to a column vector as above, we denote its Hermitian adjoint (i.e., the complex conjugate of its transpose) by the *bra* $\langle\psi|$, which corresponds to the row vector

$$\langle\psi| = \begin{pmatrix} \alpha_0^* & \alpha_1^* & \cdots & \alpha_{2^n-1}^* \end{pmatrix},$$

where $\alpha_x^*$ denotes the complex conjugate of $\alpha_x$. A bra $\langle\phi| = \sum_x \beta_x^*\langle x|$ and ket $|\psi\rangle = \sum_x \alpha_x|x\rangle$ together form a braket

$$\langle\phi|\psi\rangle = \sum_{x=0}^{2^n-1} \beta_x^* \alpha_x \in \mathbb{C}\,,$$

which is simply the standard Hermitian inner product between vectors $|\phi\rangle$ and $|\psi\rangle$. The *norm* of the state $|\psi\rangle = \sum_x \alpha_x|x\rangle$ refers to the standard Euclidean vector norm, or 2-norm of the vector, given by

$$\|\,|\psi\rangle\,\| = \sqrt{\langle\psi|\psi\rangle} = \sqrt{\sum_{x=0}^{2^n-1} |\alpha_x|^2}\,.$$





A state for which $\| |\psi\rangle \| = 1$ is said to be *normalized*; in this survey, kets are usually (but not always) normalized.

The above corresponds to the case for *pure* quantum states; in some instances, we consider the more general case that the state of the quantum system is *mixed*—that is, it is drawn from a probabilistic ensemble of multiple pure quantum states. In this case, an $n$-qubit quantum state is represented by a $2^n \times 2^n$ matrix called a *density matrix*, typically denoted by a lowercase Greek letter such as $\rho$. A matrix $\rho$ is a valid quantum state if it is Hermitian and positive semidefinite. Furthermore, it is a normalized quantum state if it satisfies $\mathrm{tr}(\rho) = 1$. In this language, a pure state $|\psi\rangle$ corresponds to the rank-1 Hermitian matrix $|\psi\rangle\langle\psi|$ given by the outer product of the vector with itself.

Linear transformations of an $n$-qubit quantum system correspond to $2^n \times 2^n$ matrices, called *operators*. Given an operator $M$, there is always a singular value decomposition (SVD)

$$M = \sum_{i=0}^{2^n-1} \sigma_i |w_i\rangle\langle v_i|\,,$$

where the *singular values* $\sigma_i$ are non-negative real numbers, and each of the sets $\{|w_i\rangle\}$ and $\{|v_i\rangle\}$ are orthonormal bases for the vector space. If $M$ is Hermitian, then we may call $M$ an *observable*, and in this case, $M$ is guaranteed to have an eigenvalue decomposition

$$M = \sum_{i=0}^{2^n-1} \lambda_i |\psi_i\rangle\langle\psi_i|, \qquad M|\psi_i\rangle = \lambda_i |\psi_i\rangle\ \forall i$$

for which the eigenvalues $\lambda_0$, $\lambda_1$, ..., $\lambda_{2^n-1}$ are real, and the eigenvectors (also known as *eigenstates*) $|\psi_0\rangle$, $|\psi_1\rangle$, ..., $|\psi_{2^n-1}\rangle$ form an orthonormal set.

Many end-to-end problems solved by quantum algorithms boil down to estimating the expectation value of an observable, which correspond to a physical property of the system. Given an observable $M$ and a mixed state $\rho$, the expectation value of $M$ is given by $\mathrm{tr}(M\rho)$. For a pure state $\rho = |\psi\rangle\langle\psi|$, this reduces to $\langle\psi|M|\psi\rangle$.

An important observable is the *Hamiltonian*, which corresponds to the energy of the physical system. The Hamiltonian generates time evolution of the state; that is, denoting the state at time $t$ by $|\psi(t)\rangle$ and the time-dependent Hamiltonian by $H(t)$, the state obeys the time-dependent Schrödinger equation

$$\mathrm{i}\frac{\mathrm{d}|\psi(t)\rangle}{\mathrm{d}t} = H(t)|\psi(t)\rangle\,.$$

Here we have set the physical constant $\hbar$ to 1, which is the typical convention in the literature that we cite. The specification of an initial state $|\psi(0)\rangle$ uniquely determines $|\psi(t)\rangle$ for all other $t$. If $H(t) = H$ is independent of $t$, this is given exactly by the matrix exponential

$$|\psi(t)\rangle = \mathrm{e}^{-\mathrm{i}Ht}|\psi(0)\rangle\,, \tag{71}$$

and if $H(t)$ is time dependent, it is given by a time-ordered matrix exponential.

The eigenvalues of the Hamiltonian are often called the *energies*. The eigenstate corresponding to the minimal eigenvalue is called the *ground state*, and its eigenvalue is called the *ground state energy*; the eigenstates corresponding to larger energies are called *excited states*. In actual quantum systems like atomic nuclei, molecules, and materials, the system's lower energies—and especially its ground state and ground state energy—often determine its key properties; the





higher excited states are rarely populated due to energy exchange with the environment favoring lower energy levels. As such, many of the relevant end-to-end problems for which quantum computing may be helpful relate to computing ground state energies and other properties of low-energy states.

One complication is that actual quantum systems in nature typically do not directly correspond to a collection of two-level qubit systems. Instead, they are modeled as discrete or continuous systems with a larger (possibly infinite) number of levels. However, the states of these systems are still described by vectors in a well-defined vector space. For example, the position of an electron in 3D space is given by an element of the vector space of square integrable functions on $\mathbb{R}^3$—in this context, the state vector $|\psi\rangle$ is often called the *wavefunction*, a term that is sometimes also used in discrete situations as well. The position of $\eta$ particles in 3D space has $3\eta$ continuous degrees of freedom, and states correspond to square integrable functions on $\mathbb{R}^{3\eta}$. However, particles found in nature, such as electrons, are indistinguishable, and quantum mechanics dictates that the corresponding wavefunctions must either be antisymmetric (if the particles are fermions) or symmetric (if they are bosons) under particle exchange, which restricts the accessible vector space. For generic multiqubit systems, no such symmetry is naturally imposed. Fermionic and bosonic systems are the subjects of quantum algorithms for chemistry, condensed matter physics, and nuclear and particle physics—to simulate these and other non-qubit systems on a quantum computer, algorithmic choices must be made on how to embed the relevant vector space into a tensor product of qubit systems. It may also be required to truncate the (possibly infinite-dimensional) vector space, incurring errors in the calculation.





# B   The quantum circuit model

We follow standard convention and work in the quantum circuit model of quantum computation. In this paradigm, quantum computations process the information in quantum states by applying a sequence of unitary operators to the state, known as *gates*, which generalize classical Boolean logic gates—unitarity ensures that the norm of the state is preserved by the operator being applied. A single-qubit gate is given by a $2 \times 2$ unitary matrix. There are a few essential examples that occur throughout the survey.

- The Pauli matrices (which are both unitary and Hermitian):

$$\sigma_x = X = \begin{pmatrix} 0 & 1 \\ 1 & 0 \end{pmatrix}, \qquad \sigma_y = Y = \begin{pmatrix} 0 & -\mathrm{i} \\ \mathrm{i} & 0 \end{pmatrix}, \qquad \sigma_z = Z = \begin{pmatrix} 1 & 0 \\ 0 & -1 \end{pmatrix},$$

- The Hadamard gate $H$ and the phase gate $S$:

$$H = \frac{1}{\sqrt{2}} \begin{pmatrix} 1 & 1 \\ 1 & -1 \end{pmatrix}, \qquad S = \sqrt{Z} = \begin{pmatrix} 1 & 0 \\ 0 & i \end{pmatrix},$$

- The $T$ gate:

$$T = \sqrt{S} = \begin{pmatrix} 1 & 0 \\ 0 & \mathrm{e}^{\mathrm{i}\pi/4} \end{pmatrix}.$$

A $k$-qubit gate is given by a $2^k \times 2^k$ unitary matrix, and some of the essential multiqubit gates include the 2-qubit controlled NOT (CNOT) gate and the 3-qubit Toffoli gate

$$\mathrm{CNOT} = \begin{pmatrix} 1 & 0 & 0 & 0 \\ 0 & 1 & 0 & 0 \\ 0 & 0 & 0 & 1 \\ 0 & 0 & 1 & 0 \end{pmatrix} \qquad \mathrm{TOFFOLI} = \begin{pmatrix} 1 & 0 & 0 & 0 & 0 & 0 & 0 & 0 \\ 0 & 1 & 0 & 0 & 0 & 0 & 0 & 0 \\ 0 & 0 & 1 & 0 & 0 & 0 & 0 & 0 \\ 0 & 0 & 0 & 1 & 0 & 0 & 0 & 0 \\ 0 & 0 & 0 & 0 & 1 & 0 & 0 & 0 \\ 0 & 0 & 0 & 0 & 0 & 1 & 0 & 0 \\ 0 & 0 & 0 & 0 & 0 & 0 & 0 & 1 \\ 0 & 0 & 0 & 0 & 0 & 0 & 1 & 0 \end{pmatrix}.$$

Given as input a 2-qubit state $|x_0 x_1\rangle$ with $x_0, x_1 \in \{0, 1\}$, the CNOT gate flips the value of $x_1$ conditioned on ("controlled on") $x_0$ being 1; the first qubit is the control and the second is the target. The Toffoli gate is a doubly controlled NOT gate in the sense that it flips the third qubit controlled on both the first and second qubits being set to 1.

When a $k$-qubit gate described by the $2^k \times 2^k$ matrix $V$ acts on a subset of qubits in an $n$-qubit system (with $n > k$), the gate enacted on the $n$-qubit system is given by a tensor product of $V$ with the identity matrix. For example, if a single-qubit $X$ gate acts on the second qubit of an $n$-qubit system, then the full $2^n \times 2^n$ unitary operator $U$ for the gate may be decomposed as

$$U = I \otimes X \otimes I \otimes I \otimes \cdots \otimes I,$$

where $I$ is the $2 \times 2$ identity matrix.





Given a fixed discrete (i.e., finite) set of gates, we may consider the set of all gates generated by the discrete gate set—that is, gates that can be formed by multiplying a sequence of gates drawn from the discrete gate set. A gate set is *universal* if it generates a dense subset of the set of all $n$-qubit unitary operators on the system, or equivalently, if any $n$-qubit unitary may be approximated to arbitrary precision by a product of gates drawn from the generating set.

For an $n$-qubit system, the discrete gate set formed from single-qubit gates $\{X, Y, Z, H, S\}$ on each of the $n$ qubits combined with 2-qubit CNOT gates between any pair of qubits generates the *Clifford* group, which is a finite group that is not universal. Importantly, quantum computations on $n$ qubits involving only Clifford gates can be efficiently simulated on a classical computer; thus, significant quantum computational speedups cannot be achieved using only Clifford gates. By adding either the $T$ gate or the Toffoli gate to the generating set, the gate set becomes universal. The Clifford $+$ $T$ gate set is the most common discrete gate set considered in compilations of quantum algorithms. The Toffoli gate can be exactly decomposed into Clifford gates and $T$ gates.

In this language, a quantum computation consists of the initialization of a quantum state (typically an $n$-qubit product state such as $|0\rangle^{\otimes n}$), the application of a prespecified sequence of gates, and finally, a measurement of the $n$ qubits in the computational basis. If the normalized $n$-qubit initial state is $|\psi_0\rangle$ and the sequence of gates is $U_1, \ldots, U_\ell$, then the quantum state prior to the measurement is given by $|\phi\rangle = U_\ell U_{\ell-1} \cdots U_1 |\psi_0\rangle$. The measurement then produces an outcome $x$ with probability equal to $|\langle x|\phi\rangle|^2$. This procedure can be depicted in a *quantum circuit* diagram, such as Fig. 13.

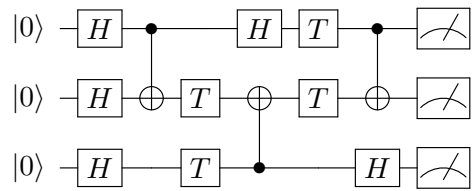

Figure 13: Example of a quantum circuit with gates drawn from the Clifford $+$ $T$ gate set. Time flows from left to right. The two qubit gates are CNOT gates with the control indicated by the symbol ● and the target indicated by the symbol ⊕.

When performing a *resource estimate* of a quantum algorithm in the Clifford $+$ $T$ gate set, the key quantities are the total number of qubits and gates that appear in the associated quantum circuit diagram. Occasionally, one is also interested in the circuit *depth*. For example, the circuit in Fig. 13 acts on 3 qubits and has a total gate count equal to 12. The circuit depth is 6 since the gates can be *parallelized* into 6 sequential layers. When working in the Clifford $+$ $T$ gate set, it is common to ignore the Clifford gates and only count the non-Clifford gates, that is, the $T$ gates (or the Toffoli gates). The main reason for this is that non-Clifford gates are more difficult to implement than Clifford gates in many (but not all) schemes for fault-tolerant quantum computing. The circuit in Fig. 13 has a $T$-count of 4 and a $T$-depth of 2, since only 2 of the 6 layers contain $T$ gates.

A quantum algorithm for a certain computational problem is a procedure that takes as input an instance of the computational problem and determines a quantum circuit (or multiple quantum circuits), as well as a procedure to convert the measurement result(s) into the answer





to the computational problem. Since measurement outcomes are random, the answer need only be correct with high probability.





# C   Noise in quantum gates and the NISQ era

The quantum circuit model is an idealized and abstract depiction of a quantum computer. Actual quantum computers attempt to realize this model by using physical 2-level quantum systems as qubits—several options can be considered including the electronic states of ions or neutral atoms, the spin states of electrons, the polarization states of photons, and the number of excitations in superconducting electrical circuits. Gates are applied by turning on and off external fields. While experimental control of these quantum systems has improved dramatically over time, one cannot expect gates to be performed perfectly. Much of the work in theoretical quantum information science deals with how to characterize, detect, and correct the errors that occur in noisy quantum computations.

Specifically, methods for fault-tolerant quantum computation have been developed, whereby a quantum computation can be accomplished correctly using faulty components, provided that the noise in the system meets certain conditions. The ideal quantum circuit is referred to as the *logical* circuit, composed of logical gates and logical qubits. Each logical qubit is realized using a larger number of noisy *physical qubits*, and each logical gate requires the action of multiple faulty *physical gates*. A resource estimate for the ideal logical circuit can be converted into a resource estimate for the actual physical quantum computer, a calculation that depends on the specifics of the hardware and the fault-tolerance scheme being utilized.

Due to the resource overhead required by fault-tolerant quantum computation, researchers have also investigated the question of whether noisy quantum computers can solve interesting problems without correcting the errors. The era of quantum computing where quantum devices of tens or hundreds of qubits exist, but large-scale fault-tolerant quantum computation is not yet possible, has been referred to as the noisy intermediate-scale quantum (NISQ) era. Generally speaking, algorithms for NISQ-era quantum computers should possess a certain resilience to the inevitable occurrence of errors in the computation, often by restricting to quantum circuits with a limited gate count or gate depth. While the focus of this survey is on quantum algorithms for fault-tolerant quantum computers and logical resource estimates, we comment in passing on NISQ algorithms for many of the tasks.





# D  Big-$\mathcal{O}$ notation

Analyses of (classical or quantum) algorithms often focus on how the computational cost, also referred to as the *complexity*, scales with the size of the input. Inputs to a computational problem are assigned an integer size $n$—for example, the number of digits in a number one wishes to factor—and resource metrics such as the qubit count, gate count, and circuit depth are expressed as functions of $n$. Often, of primary interest is the asymptotic scaling of these complexities with $n$. To facilitate this, we adopt big-$\mathcal{O}$ notation. Given two positive-valued real functions $f(n)$ and $g(n)$, we use the following definitions:

- $f(n) = \mathcal{O}(g(n))$ if there exist $n_0$, $c$, such that $f(n) \leq cg(n)$ whenever $n \geq n_0$.

- $f(n) = \Omega(g(n))$ if there exist $n_0$, $c$, such that $f(n) \geq cg(n)$ whenever $n \geq n_0$.

- $f(n) = \Theta(g(n))$ if $f(n) = \Omega(g(n))$ and $f(n) = \mathcal{O}(g(n))$.

- $f(n) = \widetilde{\mathcal{O}}(g(n))$ if there exists $c$ such that $f(n) = \mathcal{O}(g(n) \cdot \log^c(g(n)))$.

- $f(n) = \text{poly}(n)$ if there exists $c$ such that $f(n) = \mathcal{O}(n^c)$.

- $f(n) = \text{polylog}(n)$ if there exists $c$ such that $f(n) = \mathcal{O}(\log(n)^c)$.

Above, $n_0$ and $c$ are always constants, independent of $n$. Intuitively, $\mathcal{O}$, $\Omega$, and $\Theta$ are used to indicate that the asymptotic growth rate of $f(n)$ is upper bounded, lower bounded, and exactly equal to that of $g(n)$, respectively. Tildes are added to suppress logarithmic factors and simplify the expressions.

We also occasionally utilize little-$o$ notation, which has the following definitions:

- $f(n) = o(g(n))$ if *for any* constant $c$ there exists $n_0$ for which $f(n) \leq cg(n)$ whenever $n \geq n_0$.

- $f(n) = \omega(g(n))$ if *for any* constant $c$ there exists $n_0$ for which $f(n) \geq cg(n)$ whenever $n \geq n_0$.

Thus, little-$o$ and little-$\omega$ communicate instances where the growth rate of $f(n)$ is *strictly* smaller than $g(n)$ and larger than $g(n)$, respectively.

While big-$\mathcal{O}$ and little-$o$ notation carries the formal mathematical definitions above, in some contexts this notation is utilized in a less mathematically precise fashion. For example, big-$\mathcal{O}$ is occasionally used simply to indicate that constant prefactors have been omitted or that a certain quantity is roughly of the same order as another. The expression $\mathcal{O}(1)$ is often used as a placeholder for an unspecified constant, even when there is not a well-defined growing parameter $n$. Meanwhile, the expression $o(1)$ is used for functions $f(n)$ that approach 0 as $n \to \infty$. The usage of $\Omega$ is often chosen to add emphasis to the fact that a certain quantity is a *lower bound* for another, even when $\Theta$ would also have been mathematically appropriate.

We also employ big-$\mathcal{O}$ notation for functions of multiple independent parameters. For example, if the input is an $m \times n$ matrix, we might be interested in the complexity dependence on both $m$ and $n$. Another common scaling parameter is the target precision $\epsilon$ to which a certain quantity should be estimated by the quantum algorithm. Smaller $\epsilon$ typically incurs greater resources, and thus we wish to compute how the complexity scales with growing $1/\epsilon$. When two multivariate functions $f$ and $g$ are monotonically nondecreasing in all of the scaling parameters, there is little ambiguity about how to extend the definitions. For example:





- $f(n, m) = \mathcal{O}(g(n, m))$ if there exist $n_0$, $c$, such that $f(n, m) \leq cg(n, m)$ whenever $n, m \geq n_0$.

- $f(n, m) = \text{poly}(n, m)$ if there exists $c$ such that $f(n, m) = \mathcal{O}((nm)^c)$.

Ambiguity may arise if it is possible for the function to decrease when certain scaling parameters are increased; in this case, the limiting behavior of the function can depend on the rates at which the parameters grow relative to one another. Additional context about the range or relationship of the scaling parameters may be required to understand what is being communicated by the big-$\mathcal{O}$ notation.

Big-$\mathcal{O}$ notation enables a determination of the magnitude of a quantum speedup. Suppose the complexity of the quantum algorithm is $Q(n)$ and the complexity of the classical algorithm is $C(n)$.

- We say that the quantum algorithm has an *exponential speedup* if $Q(n) = \mathcal{O}(\log(C(n)))$.

- We say that the quantum algorithm has a *polynomial speedup* of degree $d$ if $Q(n) = \mathcal{O}(C(n)^{1/d})$. If $Q(n) = \widetilde{\mathcal{O}}(C(n)^{1/d})$, we say that the speedup is *essentially* (or *nearly*) degree-$d$, and often we drop these qualifiers for ease of discussion.

- If $Q(n)$ and $C(n)$ meet the criterion for a polynomial speedup for all $d \geq 1$ but not the criterion for an exponential speedup, then we say the speedup is *superpolynomial*.

Here are some examples:

- If $Q(n) = 3n$ and $C(n) = 2n^3$, there is a degree-3, or *cubic*, polynomial speedup.

- If $Q(n) = n2^{n/4}$ and $C(n) = 2^n$, there is a nearly degree-4, or *quartic*, polynomial speedup.

- If $Q(n) = n^2$ and $C(n) = e^{n^{1/3}}$, then there is a superpolynomial speedup.

- If $Q(n) = 10n$ and $C(n) = 2^{n/1000}$, then there is an exponential speedup.

End-to-end analyses should ideally also assess the constant prefactors that are omitted when using big-$\mathcal{O}$ notation (and polylogarithmic prefactors when using big-$\widetilde{\mathcal{O}}$), as these can still contribute significantly to the outlook of a certain application if they are especially large.





# E  Complexity theory background

Occasionally, we make reference to concepts and results from complexity theory. Complexity theory aims to classify different computational problems based on the quantity of computational resources required to solve them. These computational complexities are often categorized solely based on whether they scale polynomially or superpolynomially with the size of the input. If the complexity of an algorithm is polynomial in the relevant parameter, it is called *efficient*.

Let $x$ be an instance of a computational problem, which is associated with an integer length $n$. The desired output of the problem on input $x$ is denoted by $M(x)$. If $M(x) \in \{0,1\}$ is a single bit, the problem is called a *decision problem*. This enables a definition of the following important complexity classes:

- The set P contains decision problems where $M(x)$ can be computed by a deterministic classical algorithm with time complexity $\text{poly}(n)$.

- The set BPP contains decision problems where $M(x)$ can be computed with high probability by a randomized classical algorithm (i.e., an algorithm that can make coin flips) with time complexity $\text{poly}(n)$.

- The set BQP contains decision problems where $M(x)$ can be computed with high probability by a quantum algorithm with gate complexity $\text{poly}(n)$.

To arrive at a precise form for the time or gate complexity, one must specify a particular computational model; in the case of the quantum circuit model, one also needs to specify a gate set, such as Clifford $+ T$. However, for the purpose of these complexity classes, these details are generally unimportant, as they do not change which problems are in P and BQP.

Complexity theory also defines classes of problems for which the solutions are efficient to verify, even if they are not efficient to compute. Specifically, we may fix a verification algorithm with time complexity $\text{poly}(n)$ that computes a function $M'(x,y)$ of two inputs, where $x$ has size $n$ and $y$ has size $\text{poly}(n)$. We say that $y$ is a witness for $x$ if $M'(x,y) = 1$. The verification algorithm can also be a quantum algorithm, in which case $y$ can be a quantum state $|y\rangle$, which acts as the initial state for a quantum circuit in the quantum circuit model. A state $|y\rangle$ is a witness for $x$ if the quantum verification algorithm produces output $M'(x, |y\rangle) = 1$ with high probability.

- The set NP contains decision problems for which there exists a deterministic classical verification algorithm where, on input $x$, there exists a witness $y$ if and only if $M(x) = 1$.

- The set QMA contains decision problems for which there exists a quantum verification algorithm where, on input $x$, there exists a witness $|y\rangle$ if and only if $M(x) = 1$.

The most famous outstanding open question in complexity theory is whether P = NP, that is, whether or not there exist problems that cannot be solved efficiently, but for which solutions can be verified efficiently given a witness. It is widely believed that P $\neq$ NP. The prototypical example of a problem in NP that is believed not to be in P is the Boolean satisfiability problem. Here, the input is a Boolean formula $\varphi$ (referred to as $x$ above, and specified by a description of length $n$). The formula $\varphi$ maps an input string $z$ consisting of $m$ bits to an output bit $\varphi(z) \in \{0,1\}$, where $n = \text{poly}(m)$. Given $z$, the output bit $\varphi(z)$ can be evaluated in time complexity $\text{poly}(n)$, and the question is whether there exists a $z$ for which $\varphi(z) = 1$, in which





case we call $\varphi$ satisfiable. This problem is efficient to verify since whenever $\varphi(z) = 1$, the string $z$ acts as a witness to the fact that $\varphi$ is satisfiable. However, there are $2^m$ possible inputs $z$, so without a witness, it naively requires trying all possible inputs to determine if $\varphi$ is satisfiable, a procedure which has superpolynomial-in-$n$ time complexity.

In fact, it has been shown that the Boolean satisfiability problem is NP-hard, a term that means it is as hard as any other problem in NP. Specifically, we can make the following (slightly informal) definitions:

- A problem is NP-hard if the existence of an efficient deterministic classical algorithm for the problem would imply that P = NP.

- A problem is QMA-hard if the existence of an efficient quantum algorithm for the problem would imply that BQP = QMA.

- A problem is BQP-hard if the existence of an efficient randomized classical algorithm for the problem would imply that BPP = BQP.[60]

A problem that is both NP-hard and in NP is called NP-complete. The Boolean satisfiability problem is one example of an NP-complete problem. Similar definitions follow for the terms QMA-complete and BQP-complete.

The conjecture that P $\neq$ NP entails that all NP-hard problems do not admit efficient classical algorithms. Thus, one way to give evidence that a problem cannot be solved in polynomial time is to prove that it is NP-hard. Since it is also widely believed that BQP $\neq$ QMA and that NP $\not\subset$ BQP, showing that a problem is NP-hard or QMA-hard is strong evidence that it does not admit an efficient quantum algorithm. In the search for good quantum algorithms, these hardness results establish limits on what we expect to be possible.

On the other hand, if one can show that a problem is BQP-complete, then this is evidence that the problem exhibits a superpolynomial quantum speedup over the best possible classical algorithm. If it did not, then this would imply that BPP = BQP, which is widely believed to be false.

These complexity-theoretic results are useful guides for navigating quantum algorithms, but it is worth emphasizing that they typically deal with worst-case hardness and may not always be relevant for real-world instances of a problem. For example, preparing the ground state of a Hamiltonian consisting of local terms—which is in a sense a quantum analog of the Boolean satisfiability problem—is well known to be QMA-complete; thus, it is not expected to admit an efficient quantum algorithm in all instances. Nevertheless, for many specific Hamiltonians that arise in nature, we do expect efficient ground state preparation to be possible, and this forms the basis for many proposed applications of quantum computing.

We conclude with a discussion of the concept of *oracles*. A (classical or quantum) algorithm is said to have access to an oracle $g$ if it can *query* the oracle in a black-box fashion, by fixing an $n$-bit input string $z$ and receiving the corresponding $m$-bit output string $g(z)$. The *query complexity* of an algorithm is the number of times it requests an output from the oracle. In quantum algorithms, one typically allows the oracle to be queried in superposition, in the sense of performing the unitary map $U_g$, defined by

$$U_g \left( \sum_{z=0}^{2^n-1} \sum_{w=0}^{2^m-1} \alpha_{z,w} |z\rangle |w\rangle \right) = \sum_{z=0}^{2^n-1} \sum_{w=0}^{2^m-1} \alpha_{z,w} |z\rangle |w \oplus g(z)\rangle \,,$$

---

[60]Technically, due to their probabilistic nature, BPP and BQP should be defined as classes of *promise* problems to correctly formalize the notion of BQP-hard; see [1].

where $\oplus$ denotes bitwise addition modulo 2, and the coefficients $\alpha_{z,w}$ are arbitrary complex numbers. The ability to query in superposition may give a quantum algorithm an advantage over classical algorithms that cannot do so.

Oracles play multiple conceptual roles. For one, they enable modular accounting of the costs of an algorithm. For example, in algorithms where oracle calls represent the dominant computational burden, one may count the number of oracle queries made by the algorithm and multiply this by the computational cost (e.g., time complexity or gate complexity) required to implement a single oracle query. In end-to-end analyses, it is important to instantiate all oracles using elementary operations and account for the costs of implementing them in final resource expressions.

In complexity theory, oracles also provide a mechanism for establishing more definitive separations between different models of computation. Oracles are black-box objects, and algorithms may only interact with oracles by querying them on different inputs—this makes it easier to prove a lower bound on the query complexity of an algorithm than to prove a lower bound on the time complexity or gate complexity. For example, there exists an oracle $g$ for which it can be shown that P $\neq$ NP *relative to $g$*, even though P = NP remains possible in the non-oracle setting (in fact, there also exist oracles relative to which P = NP).

Similarly, there are specific computational problems involving oracles, such as Simon's problem, where one can show an unconditional exponential separation between the quantum and classical query complexity required to solve the problem. Relative to this oracle, it holds that BPP $\neq$ BQP. Such separations do not alone constitute a definitive quantum advantage in end-to-end complexity, but they may capture the core mechanism by which an end-to-end analysis aims to achieve an advantage.

# Consolidated bibliography

# Index